\documentclass[fleqn,10pt]{article}
\usepackage{latexsym, graphicx, epsfig, amsmath, amssymb,amsfonts}
\usepackage{natbib,amsthm,version}
\usepackage{amsbsy,bm,multirow,enumerate}
\usepackage[titletoc,page]{appendix}
\usepackage[mathscr]{eucal}
\usepackage{mathtools}
\usepackage{color}
\usepackage{subfigure}
\usepackage[utf8]{inputenc}
\usepackage[english]{babel}
\usepackage{float}
\usepackage{caption}

\newcommand{\bs}[1]{\boldsymbol{#1}}

\graphicspath{{./Figures/capillarywave/}{./Figures/oiljet/}{./Figures/oiljet/rho1/}{./Figures/oiljet/rho2/}{./Figures/standardtest/}    }

\title{
Multiphase Flows of $N$ Immiscible Incompressible Fluids:
An Outflow/Open Boundary Condition and Algorithm
} 
\author{Zhiguo Yang,\,
  Suchuan Dong\thanks{Author of correspondence, Email: sdong@purdue.edu} \\
  Center for Computational and Applied Mathematics \\
  Department of Mathematics \\
  Purdue University 
 } 

\date{(\today)}
\begin{document}
\maketitle



\begin{abstract}

  We present a set of effective outflow/open boundary
  conditions and an associated  algorithm
  for simulating the dynamics of multiphase flows
  consisting of $N$ ($N\geqslant 2$) immiscible incompressible
  fluids in domains involving outflows or open
  boundaries. These boundary conditions are devised based on
  the properties of energy stability and reduction consistency. 
  The energy stability property ensures that
  the contributions of these boundary conditions to
  the energy balance will not cause the total energy
  of the N-phase system to increase over time.
  Therefore, these open/outflow boundary conditions are very effective
  in overcoming the backflow instability in multiphase systems.
  The reduction consistency property
  ensures that if some fluid components are absent from
  the N-phase system then these N-phase boundary conditions
  will reduce to those corresponding boundary conditions
  for the equivalent smaller system. 
  Our numerical algorithm for the proposed boundary conditions
  together with the N-phase governing equations involves only the
  solution of a set of de-coupled individual Helmholtz-type equations
  within each time step, and the resultant linear algebraic systems
  after discretization involve only constant and
  time-independent coefficient matrices which can be pre-computed.
  Therefore, the algorithm is computationally very efficient and attractive.
  We present extensive numerical experiments for flow problems
  involving multiple fluid components and inflow/outflow
  boundaries to test the proposed method.
  In particular, we compare in detail the simulation results 
  of a three-phase capillary wave problem with Prosperetti's exact
  physical solution 
  and demonstrate that the method developed herein produces
  physically accurate results.

\end{abstract}


\vspace{0.05cm}

\section{Introduction}
\label{sec:intro}


In this work we focus on the dynamics and interactions of a system of 
$N$ ($N\geqslant 2$) immiscible incompressible fluids
in an unbounded flow domain. 
In order to numerically simulate such problems it is necessary
to truncate the domain to a finite size. Consequently,
part of the boundary in the computational domain will be open, in the sense that
the fluids can freely leave (or even enter) the domain through
such boundaries, and appropriate
boundary conditions will be required on the open (or outflow)
portions of the domain boundary.
We are particularly concerned with situations in which the multitude
of fluid interfaces formed in the system will pass through the open
 domain boundaries.
Following the notation of our previous works~\cite{Dong2014,Dong2015,Dong2017},
we refer to such problems as N-phase outflows. Here
$N$ denotes the number of different fluid components in the system, not
necessarily the number of material phases.


N-phase outflows and open boundaries pose a number of
issues to numerical simulations.
First, the problem involves multiple fluid interfaces at
the open/outflow boundary, which are associated with
multiple surface tensions and the contrasts in densities and
viscosities of these fluids. How to deal with
the surface tensions, and the density and viscosity contrasts
in the N-phase open/outflow boundary conditions (OBC) poses
the foremost issue.
Second, backflow instability is another crucial issue confronting
N-phase outflow simulations. Backflow instability refers to
the numerical instability associated with strong vortices or
backflows at the open/outflow boundary,
which causes computations to blow up instantly when strong vortices
or backflows occur at the outflow boundary.
The backflow instability issue is not unique to multiphase flows.
This issue is well-known in single-phase
outflow problems~\cite{DongKC2014,DongS2015,Dong2015clesobc}, but it becomes much
worse for two-phase~\cite{Dong2014obc,DongW2016} and multiphase
outflows because of the density contrasts and viscosity
contrasts at the outflow boundary.
Third, N-phase problems with $N\geqslant 3$ pose the so-called
reduction consistency issue on the design of outflow/open boundary
conditions~\cite{Dong2017}. Reduction consistency refers to
the property that, if only $M$ ($2\leqslant M\leqslant N-1$)
fluid components are present in the N-phase system (while the
other fluid components are absent),
the governing equations and the boundary conditions for
the N-phase system should reduce to those for the corresponding
smaller M-phase system~\cite{Dong2017}.
The reduction consistency of N-phase outflow/open
boundary conditions is an issue unique to multiphase outflow
and open-boundary problems.


The development of effective outflow/open boundary conditions
is an important problem  in
computational fluid dynamics. For single-phase
problems, this has been under intensive investigations
for decades and a large volume of literature exists;
see e.g.~\cite{Gresho1991,SaniG1994}
for a comprehensive review of related literature
and~\cite{DongKC2014,Dong2015clesobc} and the
references therein for a sample of more recent works.
On the other hand, 
for two-phase ($N=2$)
outflows and open boundaries
the existing work in the literature is very limited,
and for  multiphase outflow and open-boundary problems involving
three or more ($N\geqslant 3$) fluid components, 
there is no existing work available in the literature
to the best of our knowledge. 
The zero-flux (Neumann) and extrapolation boundary conditions
from single-phase flows have been used
for the two-phase Lattice-Boltzmann equation
in \cite{LouGS2013}. The zero-flux condition
has also been employed for the outflow boundary
with a level-set type method
in \cite{AlbadawiDRMD2013,Son2001}.
The outflow condition for two immiscible fluids is considered
for a porous medium in \cite{LenzingerS2010},
and for one-dimensional two-phase compressible
flows in \cite{Munkejord2006,DesmaraisK2014}.
In \cite{Dong2014obc,DongW2016} we have developed
a set of two-phase open boundary conditions having
the attractive property that these conditions ensure
the energy stability of the two-phase system, which is
therefore effective for dealing with two-phase open boundaries. 

%

In the current paper we consider the multiphase outflow
and open-boundary problem with $N$ ($N\geqslant 3$)
immiscible incompressible fluid components in the system,
and present a set of effective outflow/open boundary
conditions and an associated numerical algorithm for such
problems within the phase field framework. The proposed open boundary
conditions are designed based on considerations of two properties:
energy stability and reduction consistency.
By looking into the energy balance of the N-phase system,
we design the open boundary conditions in such a way to ensure 
that their contributions shall not cause the total energy of
the N-phase system to increase over time,
regardless of the flow state at the outflow/open boundary.
This energy-stable property holds
even in situations where strong vortices or backflows occur
at the open boundary. 
As a result, these  boundary conditions are very effective
in overcoming the backflow instability.
We then look into the reduction consistency of these
boundary conditions, and study how these
conditions transform if some fluid components 
are absent from the N-phase system.
The reduction consistency property limits the choice
and the form of those boundary conditions  that
ensure the energy stability.
The N-phase outflow/open boundary conditions and also the inflow
boundary conditions proposed herein satisfy both
the energy stability and the reduction consistency.


The outflow/open boundary conditions proposed herein
are developed in the context of an N-phase physical formulation
we developed recently in \cite{Dong2017}.
This formulation is based on a phase field model for the N-fluid
mixture that is more general than a previous model \cite{Dong2014}.
The thermodynamic consistency and the reduction
consistency of this formulation
have been extensively studied in \cite{Dong2017}.
The formulation rigorously satisfies the mass conservation,
momentum conservation, the second law of thermodynamics,
and the Galilean invariance principle.
This formulation is fully reduction consistent, provided that
an appropriate potential free energy density function
satisfying certain properties is employed
for the N-phase system~\cite{Dong2017}.
The reduction consistency of a set of Cahn-Hilliard type
equations for a three-component and multi-component
system (without hydrodynamic interactions) has previously been
considered in \cite{BoyerL2006,BoyerM2014}.
The thermodynamic consistency of two-phase and multiphase
systems has also been considered
in~\cite{LowengrubT1998,KimL2005,AbelsGG2012,HeidaMR2012,Dong2014,LiW2014,Dong2015,WuX2017}.
We refer the reader
to e.g.~\cite{AndersonMW1998,LiuS2003,YueFLS2004,BoyerLMPQ2010,Kim2012,ZhangW2016,BanasN2017,YangZWS2017,ZhaoLWY2017}
for other contributions to two-phase and multiphase flow problems.


We further present an efficient numerical algorithm
for the proposed outflow and inflow boundary conditions 
together with the N-phase governing equations.
This is a semi-implicit splitting type scheme.
Special care is taken in the numerical treatments of
the open/ouflow boundary conditions such that
the computations for different flow variables
and the computations for the ($N-1$) phase field
functions have all been de-coupled.
The algorithm involves only the solution of
a set of individual de-coupled Helmholtz-type equations (including Poisson) within
each time step. The resultant linear algebraic
systems after discretization involves only constant
and time-independent coefficient matrices,
which can be pre-computed during pre-processing,
even when large density contrasts and large viscosity
contrasts are involved in the N-phase system.


The novelties of this paper lie in two aspects:
(i) the set of N-phase energy-stable and reduction-consistent
outflow/open boundary conditions and inflow boundary conditions,
and (ii) the numerical algorithm for treating the proposed set
of outflow and inflow boundary conditions.


The rest of this paper is structured as follows.
In the rest of this section we provide a summary of
the general phase field model developed in \cite{Dong2017}
for the N-fluid mixture.
This model provides the basis for the N-phase energy balance
relation and the development of energy-stable boundary conditions.
In Section \ref{sec:method} we propose a set of outflow and
inflow boundary conditions based on considerations of energy
stability and reduction consistency of the N-phase system, and present an efficient
algorithm for numerically treating these boundary conditions
together with the N-phase governing equations.
In Section \ref{sec:tests} we present several representative
numerical examples involving multiple fluid components
and inflow/outflow boundaries to demonstrate the effectiveness of
the proposed outflow/open boundary conditions and
the performance of the numerical algorithm  herein.
Section \ref{sec:summary} then concludes the discussion
with some closing remarks.



\subsection{A Thermodynamically Consistent N-Fluid Mixture Model}


We summarize below the phase field model proposed in \cite{Dong2017}
 for an isothermal mixture
of $N$ ($N\geqslant 2$) immiscible incompressible fluids.
This model modifies and generalizes the N-phase model
developed in \cite{Dong2014}, and it
satisfies the conservations of mass and momentum,
the second law of thermodynamics, and the Galilean invariance
principle. This model forms the basis for the development
of outflow/open boundary conditions in subsequent sections.

Consider a mixture of $N$ ($N\geqslant 2$) immiscible incompressible
fluids contained in some flow domain $\Omega$ with boundary $\partial\Omega$. 
Let $\tilde{\rho}_i$ and $\tilde{\mu}_i$
($1\leqslant i\leqslant N$) denote 
the constant densities and constant dynamic 
viscosities of these $N$ pure fluids (before mixing).
Define auxiliary parameters
\begin{equation}
\tilde{\gamma}_i = \frac{1}{\tilde{\rho}_i}, \ 1\leqslant i\leqslant N;
\quad
\Gamma = \sum_{i=1}^N \tilde{\gamma}_i;
\quad
\Gamma_{\mu} = \sum_{i=1}^N \frac{\tilde{\mu}_i}{\tilde{\rho}_i}.
\end{equation}
Let $\phi_i$ ($1\leqslant i\leqslant N-1$) denote
the ($N-1$) independent order parameters,
or interchangeably the phase field variables,
that characterize the system,
and $\vec{\phi}=(\phi_1,\dots,\phi_{N-1})$.
Let $\rho_i(\vec{\phi})$ and $c_i(\vec{\phi})$
($1\leqslant i\leqslant N$)
denote the density and volume fraction of
fluid $i$ {\em within the mixture}, and
let $\rho(\vec{\phi})$ denote the density of
the N-phase mixture. Then we have the relations \cite{Dong2014}
\begin{equation}
c_i = \frac{\rho_i}{\tilde{\rho}_i}, \ 1\leqslant i\leqslant N; \quad
\sum_{i=1}^N c_i = 1; \quad
\rho = \sum_{i=1}^N \rho_i.
\label{equ:volfrac_expr}
\end{equation}

Let $W(\vec{\phi},\nabla\vec{\phi})$ denote
the free energy density function of the system
which satisfies the condition,
$ 
\sum_{i=1}^{N-1}\nabla\phi_i \otimes
\frac{\partial W}{\partial(\nabla\phi_i)}
= \sum_{i=1}^{N-1}\frac{\partial W}{\partial(\nabla\phi_i)}
\otimes \nabla\phi_i,
$ 
where $\otimes$ denote the tensor product.
Then the motion of this N-phase system
is described by the following equations~\cite{Dong2017}:
\begin{subequations}
\begin{equation}
\rho(\vec{\phi})\left(
  \frac{\partial\mathbf{u}}{\partial t}
  + \mathbf{u}\cdot\nabla\mathbf{u}
\right)
+ \tilde{\mathbf{J}}\cdot\nabla\mathbf{u}
= 
-\nabla p
+ \nabla\cdot\left[
  \mu(\vec{\phi}) \mathbf{D}(\mathbf{u})
\right]
- \sum_{i=1}^{N-1} \nabla\cdot\left(
  \nabla\phi_i \otimes \frac{\partial W}{\partial(\nabla\phi_i)}
\right),
\label{equ:nse_original}
\end{equation}
\begin{equation}
\nabla\cdot\mathbf{u} = 0,
\label{equ:continuity_original}
\end{equation}
\begin{equation}
\sum_{j=1}^{N-1}\frac{\partial\varphi_i}{\partial\phi_j}\left(
  \frac{\partial\phi_j}{\partial t} + \mathbf{u}\cdot\nabla\phi_j
\right)
=
\sum_{j=1}^{N-1}\nabla\cdot\left[
  \tilde{m}_{ij}(\vec{\phi}) \nabla \mathcal{C}_j
\right],
\qquad 1 \leqslant i \leqslant N-1,
\label{equ:CH_original}
\end{equation}
\end{subequations}
where $\mathbf{u}(\mathbf{x},t)$ is velocity,
$p(\mathbf{x},t)$ is pressure, 
$
\mathbf{D}(\mathbf{u}) = \nabla\mathbf{u} + \nabla\mathbf{u}^T
$ 
(superscript $T$ denoting transpose),
$\mathbf{x}$ and $t$
are respectively the spatial and temporal coordinates.
$\tilde{m}_{ij}$ ($1\leqslant i,j\leqslant N-1$)
are coefficients and the matrix formed by these coefficients
\begin{equation}
\tilde{\mathbf{m}} = \begin{bmatrix} \tilde{m}_{ij} \end{bmatrix}_{(N-1)\times (N-1)}
\end{equation}
is required to be symmetric positive definite (SPD)~\cite{Dong2017}.
$\varphi_i(\vec{\phi})$ are defined by
\begin{equation}
\varphi_i \equiv  \rho_i(\vec{\phi}) - \rho_N(\vec{\phi}) = \varphi_i(\vec{\phi}),
\quad 1\leqslant i\leqslant N-1.
\label{equ:varphi_expr}
\end{equation}
The chemical potentials
$\mathcal{C}_i(\vec{\phi},\nabla\vec{\phi})$ ($1\leqslant i\leqslant N-1$)
are given by the following linear
algebraic system
\begin{equation}
\sum_{j=1}^{N-1} \frac{\partial\varphi_j}{\partial\phi_i} \mathcal{C}_j
=
\frac{\partial W}{\partial \phi_i}
- \nabla\cdot \frac{\partial W}{\partial(\nabla\phi_i)},
\quad 1\leqslant i\leqslant N-1,
\label{equ:chem_potential}
\end{equation}
which can be solved given $W(\vec{\phi},\nabla\vec{\phi})$
and $\varphi_i(\vec{\phi})$.
$\tilde{\mathbf{J}}(\vec{\phi},\nabla\vec{\phi})$ takes the form
\begin{equation}
\tilde{\mathbf{J}} = -\sum_{i,j=1}^{N-1}\left(  
1 - \frac{N}{\Gamma}\tilde{\gamma}_i
\right)
\tilde{m}_{ij}(\vec{\phi})\nabla \mathcal{C}_j.
\label{equ:J_expr}
\end{equation}
The density of fluid $i$ within the mixture $\rho_i$, 
the volume fraction $c_i$,
and the mixture density $\rho$ and dynamic viscosity $\mu$, 
are given by 
\begin{equation}
\left\{
\begin{split}
&
\rho_i(\vec{\phi}) = 
\frac{1}{\Gamma} + \sum_{j=1}^{N-1}\left(
  \delta_{ij} - \frac{\tilde{\gamma}_j}{\Gamma}
\right)\varphi_j(\vec{\phi}), 
\quad 1\leqslant i\leqslant N, \\
&
c_i(\vec{\phi}) = \tilde{\gamma}_i\rho_i(\vec{\phi}) = 
\frac{\tilde{\gamma}_i}{\Gamma} + \sum_{j=1}^{N-1}\left(
  \tilde{\gamma}_i\delta_{ij} - \frac{\tilde{\gamma}_i\tilde{\gamma}_j}{\Gamma}
\right)\varphi_j(\vec{\phi}), 
\quad 1\leqslant i\leqslant N, \\
&
\rho(\vec{\phi})=\sum_{i=1}^N \rho_i = 
\frac{N}{\Gamma} + \sum_{i=1}^{N-1}\left(1 - \frac{N}{\Gamma}\tilde{\gamma}_i \right)\varphi_i(\vec{\phi}), \\
&
\mu(\vec{\phi}) = \sum_{i=1}^N \tilde{\mu}_i c_i(\vec{\phi})
= \frac{\Gamma_{\mu}}{\Gamma} + \sum_{i=1}^{N-1}\left(
  \tilde{\mu}_i - \frac{\Gamma_{\mu}}{\Gamma}
\right) \tilde{\gamma}_i \varphi_i(\vec{\phi})
\end{split}
\right.
\label{equ:density_expr}
\end{equation}
where $\delta_{ij}$ is the Kronecker delta.

In this model the functions $\varphi_i(\vec{\phi})$,
the free energy density function $W(\vec{\phi},\nabla\vec{\phi})$,
and the coefficients $\tilde{m}_{ij}$ ($1\leqslant i,k\leqslant N-1$)
remain to be specified. Once they are known, all the other
quantities can be computed.
Note that the equation \eqref{equ:varphi_expr} is
to define the set of order parameters $\vec{\phi}$.
Once $\varphi_i(\vec{\phi})$ is given, the set of order parameters
$\phi_i$ ($1\leqslant i\leqslant N-1$) will be fixed.

\section{N-Phase Energy-Stable Open Boundary Conditions}
\label{sec:method}

In this section
we propose a set of N-phase outflow/open (and also inflow) boundary conditions
based on considerations of energy stability and reduction consistency,
and develop an algorithm for numerically treating the proposed
boundary conditions together with the N-phase governing equations.

\subsection{N-phase Energy Balance and Energy-Stable Boundary Conditions}
\label{sec:energy_balance}

We first derive the energy balance relation for
the N-phase model represented by \eqref{equ:nse_original}--\eqref{equ:CH_original},
and then based on this relation
look into possible forms for the boundary conditions
to ensure the energy stability of the N-phase
system.

It is straightforward to verify that
the $\rho(\vec{\phi})$ given by \eqref{equ:density_expr}
and $\tilde{\mathbf{J}}(\vec{\phi},\nabla\vec{\phi})$
given by \eqref{equ:J_expr}
satisfy the following relation
\begin{equation}
  \frac{\partial\rho}{\partial t} + \mathbf{u}\cdot\nabla\rho
  = -\nabla\cdot\tilde{\mathbf{J}}
  \label{equ:mass_balance}
\end{equation}
where we have used equations \eqref{equ:CH_original}
and \eqref{equ:varphi_expr}.
Let $\mathbf{T} = -p\mathbf{I} + \mu\mathbf{D}(\mathbf{u})$ denote
the stress tensor,
where $\mathbf{I}$ is the identity tensor. Then
equation \eqref{equ:nse_original} can be written as
\begin{equation}
\rho\frac{D\mathbf{u}}{Dt}
+ \tilde{\mathbf{J}}\cdot\nabla\mathbf{u}
= \nabla\cdot\mathbf{T}
- \sum_{i=1}^{N-1} \nabla\cdot\left(
  \nabla\phi_i \otimes \frac{\partial W}{\partial(\nabla\phi_i)}
\right),
\label{equ:nse_trans_1}
\end{equation}
where $\frac{D}{Dt}=\frac{\partial}{\partial t}+\mathbf{u}\cdot\nabla$ 
denote the material derivative.
Taking the $L^2$ inner product between equation \eqref{equ:nse_trans_1}
and $\mathbf{u}$ leads to
\begin{equation}
\begin{split}
\frac{\partial}{\partial t}\int_{\Omega} \frac{1}{2}\rho|\mathbf{u}|^2 =&
-\int_{\Omega}\frac{\mu}{2}\|\mathbf{D}(\mathbf{u}) \|^2
-\int_{\Omega}\sum_{i=1}^{N-1}\left[\nabla\cdot\left(
  \nabla\phi_i\otimes\frac{\partial W}{\partial\nabla\phi_i}
\right)\right]\cdot\mathbf{u} \\
& +\int_{\partial\Omega}\left[
  \mathbf{n}\cdot\mathbf{T}\cdot\mathbf{u}
  -\frac{1}{2}(\mathbf{n}\cdot\tilde{\mathbf{J}})|\mathbf{u}|^2
  -\frac{1}{2}\rho|\mathbf{u}|^2\mathbf{n}\cdot\mathbf{u}
\right]
\end{split}
\label{equ:kinetic_energy_balance}
\end{equation}
where $\mathbf{n}$ is the outward-pointing unit vector
normal to $\partial\Omega$, and we have used the divergence theorem, the equations
\eqref{equ:continuity_original} and \eqref{equ:mass_balance},
and the following relations
\begin{equation}
\left\{
\begin{split}
&
(\nabla\cdot\mathbf{T})\cdot\mathbf{u}=\nabla\cdot(\mathbf{T}\cdot\mathbf{u})
-\mathbf{T}:(\nabla\mathbf{u})^T
=\nabla\cdot(\mathbf{T}\cdot\mathbf{u})+p\nabla\cdot\mathbf{u}
-\frac{\mu}{2}\|\mathbf{D}(\mathbf{u}) \|^2, \\
&
\rho\frac{D\mathbf{u}}{Dt}\cdot\mathbf{u} = \frac{D}{Dt}\left(\frac{1}{2}\rho|\mathbf{u}|^2  \right)
-\frac{D\rho}{Dt}\left(\frac{1}{2}|\mathbf{u}|^2  \right), \\
&
\left(\tilde{\mathbf{J}}\cdot\nabla\mathbf{u}  \right)\cdot\mathbf{u}  
=\nabla\cdot\left(\frac{1}{2}|\mathbf{u}|^2\tilde{\mathbf{J}}  \right)
-\nabla\cdot\tilde{\mathbf{J}}\left(\frac{1}{2}|\mathbf{u}|^2  \right).
\end{split} 
\right.
\end{equation}

Take the $L^2$ inner product between equation \eqref{equ:CH_original}
and $\mathcal{C}_i$, sum over $i$ from $1$ to $(N-1)$, and we arrive at
\begin{equation}
\int_{\Omega}\sum_{j=1}^{N-1}\left(
  \frac{\partial W}{\partial \phi_j} 
  - \nabla\cdot\frac{\partial W}{\partial\nabla\phi_j}
\right) \frac{D\phi_j}{Dt}
= -\int_{\Omega}\sum_{i,j=1}^{N-1}\tilde{m}_{ij}\nabla\mathcal{C}_i\cdot\nabla\mathcal{C}_j
+ \int_{\partial\Omega}\sum_{i,j=1}^{N-1}\tilde{m}_{ij}(\mathbf{n}\cdot\nabla\mathcal{C}_j)\mathcal{C}_i
\label{equ:free_eng_1}
\end{equation}
where we have used the integration by part, the divergence theorem,
and the equation \eqref{equ:chem_potential}.
By noting the relations
\begin{equation}
\left\{
\begin{split}
&
\frac{\partial W}{\partial t} = \sum_{i=1}^{N-1}\frac{\partial W}{\partial\phi_i}\frac{\partial\phi_i}{\partial t}
+ \sum_{i=1}^{N-1}\frac{\partial W}{\partial\nabla\phi_i}\cdot\nabla\frac{\partial\phi_i}{\partial t} \\
&
\nabla\cdot\frac{\partial W}{\partial\nabla\phi_i}\frac{\partial\phi_i}{\partial t}
=\nabla\cdot\left(\frac{\partial W}{\partial\nabla\phi_i} \frac{\partial\phi_i}{\partial t} \right)
-\frac{\partial W}{\partial\nabla\phi_i}\cdot\nabla \frac{\partial\phi_i}{\partial t},
\end{split}
\right.
\end{equation}
equation \eqref{equ:free_eng_1} can be transformed into
\begin{multline}
  \int_{\Omega}\frac{\partial W}{\partial t}
  + \int_{\Omega}\sum_{i=1}^{N-1}\left(
  \frac{\partial W}{\partial\phi_i} - \nabla\cdot\frac{\partial W}{\partial\nabla\phi_i}
  \right)\mathbf{u}\cdot\nabla\phi_i
  -\int_{\partial\Omega}\sum_{i=1}^{N-1}\mathbf{n}\cdot\frac{\partial W}{\partial\nabla\phi_i}
  \frac{\partial\phi_i}{\partial t} \\
  = -\int_{\Omega}\sum_{i,j=1}^{N-1}\tilde{m}_{ij}\nabla\mathcal{C}_i\cdot\nabla\mathcal{C}_j
  + \int_{\partial\Omega}\sum_{i,j=1}^{N-1}\tilde{m}_{ij}(\mathbf{n}\cdot\nabla\mathcal{C}_j)\mathcal{C}_i
  \label{equ:free_eng_2}
\end{multline}
where we have used the divergence theorem.
With the help of the relations
\begin{equation}
  \left\{
  \begin{split}
    &
    \nabla\cdot\left(\nabla\phi_i\otimes\frac{\partial W}{\partial\nabla\phi_i} \right)\cdot\mathbf{u}
    = \nabla\cdot\left(\frac{\partial W}{\partial\nabla\phi_i}\otimes\nabla\phi_i \right)\cdot\mathbf{u}
    = \nabla\cdot\frac{\partial W}{\partial\nabla\phi_i}(\mathbf{u}\cdot\nabla\phi_i)
    +\frac{\partial W}{\partial\nabla\phi_i}\cdot\nabla\nabla\phi_i\cdot\mathbf{u} \\
    &
    \mathbf{u}\cdot\nabla W = \sum_{i=1}^{N-1}\frac{\partial W}{\partial\phi_i}\mathbf{u}\cdot\nabla\phi_i
    + \sum_{i=1}^{N-1}\frac{\partial W}{\partial\nabla\phi_i}\cdot\nabla\nabla\phi_i\cdot\mathbf{u} \\
    &
    \mathbf{u}\cdot\nabla W = \nabla\cdot(\mathbf{u} W)
  \end{split}
  \right.
\end{equation}
we can further transform \eqref{equ:free_eng_2} into
\begin{equation}
  \begin{split}
    \frac{\partial}{\partial t}\int_{\Omega} W =&
    -\int_{\Omega}\sum_{i,j=1}^{N-1}\tilde{m}_{ij}\nabla\mathcal{C}_i\cdot\nabla\mathcal{C}_j
    + \int_{\Omega} \sum_{i=1}^{N-1} \nabla\cdot\left(\nabla\phi_i\otimes\frac{\partial W}{\partial\nabla\phi_i} \right)\cdot\mathbf{u} \\
    &
    + \int_{\partial\Omega}\sum_{i,j=1}^{N-1}\tilde{m}_{ij}(\mathbf{n}\cdot\nabla\mathcal{C}_j)\mathcal{C}_i
    + \int_{\partial\Omega}\sum_{i=1}^{N-1}\mathbf{n}\cdot\frac{\partial W}{\partial\nabla\phi_i}\frac{\partial\phi_i}{\partial t}
    -\int_{\partial\Omega}W (\mathbf{n}\cdot\mathbf{u})
  \end{split}
  \label{equ:free_eng_3}
\end{equation}
where we have used the divergence theorem and equation \eqref{equ:continuity_original}.

Summing up equations \eqref{equ:kinetic_energy_balance}
and \eqref{equ:free_eng_3}, we obtain the energy balance
equation for the N-phase system described
by \eqref{equ:nse_original}--\eqref{equ:CH_original}:
\begin{equation}
  \begin{split}
    \frac{\partial}{\partial t}\int_{\Omega}\left[\frac{1}{2}\rho|\mathbf{u}|^2 + W  \right]
    =& -\int_{\Omega}\frac{\mu}{2}\|\mathbf{D}(\mathbf{u}) \|^2
    - \int_{\Omega}\sum_{i,j=1}^{N-1}\tilde{m}_{ij}\nabla\mathcal{C}_i\cdot\nabla\mathcal{C}_j \\
    & + \int_{\partial\Omega} \underbrace{ \left[ 
      \mathbf{n}\cdot\mathbf{T}\cdot\mathbf{u}
      -\frac{1}{2}(\mathbf{n}\cdot\tilde{\mathbf{J}})|\mathbf{u}|^2
      -\frac{1}{2}\rho|\mathbf{u}|^2\mathbf{n}\cdot\mathbf{u}
      -W \mathbf{n}\cdot\mathbf{u} 
      \right] }_{\text{boundary term (I)}} \\
    & + \int_{\partial\Omega} \underbrace{ 
      \sum_{i,j=1}^{N-1}\tilde{m}_{ij}(\mathbf{n}\cdot\nabla\mathcal{C}_j)\mathcal{C}_i
      }_{\text{boundary term (II)}}
    + \int_{\partial\Omega}\underbrace{
    \sum_{i=1}^{N-1}\mathbf{n}\cdot\frac{\partial W}{\partial\nabla\phi_i}\frac{\partial\phi_i}{\partial t} }_{\text{boundary term (III)}}.
  \end{split}
  \label{equ:energy_balance}
\end{equation}
Since the free energy form $W(\vec{\phi},\nabla\vec{\phi})$ and
the order parameters $\phi_i$ ($1\leqslant i\leqslant N-1$) are
unspecified,
the above energy balance holds
for any specific form of $W(\vec{\phi},\nabla\vec{\phi})$
and any specific choice of the order parameters.

In the above energy balance equation, the left hand side (LHS) is
the time derivative of the total energy of the N-phase system.
On the right hand side (RHS), the volume-integral terms
are always dissipative by noting the symmetric positive definiteness
of the matrix formed by $\tilde{m}_{ij}$ ($1\leqslant i,j\leqslant N-1$).
The boundary-integral terms, on the other hand, can be
positive or negative, depending on the boundary conditions.


We are interested in boundary conditions for the flow 
and phase field variables which ensure that
the boundary-integral terms (I), (II) and (III) in the energy balance equation
\eqref{equ:energy_balance} are non-positive.
In other words, the contributions of the boundary terms will
be dissipative under these conditions. 
As such, the total energy of the system will not
increase over time, and this ensures the energy stability
of the N-phase system.
We refer to such boundary conditions
as energy-stable  boundary conditions.

We look into the following 
choices that ensure the dissipativeness
of the boundary term (I) in equation \eqref{equ:energy_balance}:
\begin{subequations}
  \begin{equation}
    \mathbf{u}=0, \quad \text{on} \ \partial\Omega;
    \label{equ:bc_vel_1}
  \end{equation}
  \begin{equation}
    \mathbf{n}\cdot\mathbf{T} - W\mathbf{n}
    -\frac{1}{2}(\mathbf{n}\cdot\tilde{\mathbf{J}})\mathbf{u}
    -\frac{1}{2}\rho|\mathbf{u}|^2\mathbf{n} = 0, \quad \text{on} \ \partial\Omega;
    \label{equ:bc_vel_2}
  \end{equation}
  \begin{multline}
    \mathbf{n}\cdot\mathbf{T} - W\mathbf{n}
    -\frac{1}{2}(\mathbf{n}\cdot\tilde{\mathbf{J}})\mathbf{u} \\
    -\rho\left[\theta \frac{1}{2}(\mathbf{u}\cdot\mathbf{u})\mathbf{n}
      + (1-\theta)\frac{1}{2}(\mathbf{n}\cdot\mathbf{u})\mathbf{u}
      -C_1(\mathbf{n},\mathbf{u})\mathbf{u} + C_2(\mathbf{n},\mathbf{u})\mathbf{n}
      \right]\Theta_0(\mathbf{n},\mathbf{u}) = 0, \quad \text{on} \ \partial\Omega;
    \label{equ:bc_vel_3}
  \end{multline}
\end{subequations}
where $\theta$ is a constant parameter
satisfying $0\leqslant\theta\leqslant 1$,
and $C_1(\mathbf{n},\mathbf{u})\geqslant 0$
and $C_2(\mathbf{n},\mathbf{u})\geqslant 0$ are
two non-negative constants or functions.
$\Theta_0(\mathbf{n},\mathbf{u})$ is a smoothed
step function given in \cite{DongS2015}, expressed as follows,
\begin{equation}
  \Theta_0(\mathbf{n},\mathbf{u}) = \frac{1}{2}\left(
  1-\tanh\frac{\mathbf{n}\cdot\mathbf{u}}{U_0\delta}
  \right),
  \quad \lim_{\delta\rightarrow 0}\Theta_0(\mathbf{n},\mathbf{u})
  = \Theta_{s0}(\mathbf{n},\mathbf{u})
  =\left\{
  \begin{array}{ll}
    1, & \text{if} \ \mathbf{n}\cdot\mathbf{u}<0 \\
    0, & \text{otherwise}
  \end{array}
  \right.
\end{equation}
where $U_0$ is a velocity scale, and
$\delta>0$ is a small positive parameter that controls the
sharpness of the smoothed step function.
As $\delta\rightarrow 0$, $\Theta_0$ approaches
the step function $\Theta_{s0}$, taking unit value when
$\mathbf{n}\cdot\mathbf{u}<0$ and zero otherwise.
The boundary condition \eqref{equ:bc_vel_2}
ensures the energy stability. But it prohibits
the kinetic energy from being convected out of the domain
in the presence of inflow/outflows, resulting
in poor physical results.
The form of the $\Theta_0$ term in
 condition \eqref{equ:bc_vel_3}
is inspired by the boundary condition developed in \cite{DongS2015}
for the single-phase incompressible Navier-Stokes equations;
see also \cite{Dong2014obc,DongW2016} for two-phase flows.
The condition \eqref{equ:bc_vel_3}
ensures the energy dissipation
of the boundary term (I) as $\delta\rightarrow 0$,
i.e.~when $\delta$ is sufficiently small, because
with this condition
\begin{equation}
  \begin{split}
    \mathbf{n}\cdot\mathbf{T}\cdot\mathbf{u} 
      -\frac{1}{2}(\mathbf{n}\cdot\tilde{\mathbf{J}})|\mathbf{u}|^2 &
      -\frac{1}{2}\rho|\mathbf{u}|^2\mathbf{n}\cdot\mathbf{u}
      -W \mathbf{n}\cdot\mathbf{u} \\
      &=\left\{
      \begin{array}{ll}
        -C_1\rho|\mathbf{u}|^2 + C_2\rho\mathbf{n}\cdot\mathbf{u}\leqslant 0, &
        \text{where} \ \mathbf{n}\cdot\mathbf{u}<0, \\
        -\frac{1}{2}\rho|\mathbf{u}|^2\mathbf{n}\cdot\mathbf{u}\leqslant 0, &
        \text{where} \ \mathbf{n}\cdot\mathbf{u}\geqslant 0,
      \end{array}
      \right.
      \quad \text{on} \ \partial\Omega, \ \text{as} \ \delta\rightarrow 0.
      \end{split}
\end{equation}

We look into the following
choices that ensure the energy dissipation of the
boundary term (II) in equation \eqref{equ:energy_balance}:
\begin{subequations}
  \begin{equation}
    \sum_{i=1}^{N-1}\tilde{m}_{ij}\mathcal{C}_i = 0, \quad 1\leqslant j\leqslant N-1, \quad \text{on} \ \partial\Omega;
    \label{equ:bc_phi_A_1}
  \end{equation}
  \begin{equation}
    \sum_{j=1}^{N-1}\tilde{m}_{ij}\mathbf{n}\cdot\nabla\mathcal{C}_j=0, \quad 1\leqslant i\leqslant N-1, \quad \text{on} \ \partial\Omega;
    \label{equ:bc_phi_A_2}
  \end{equation}
\begin{equation}
\sum_{j=1}^{N-1}\tilde{m}_{ij}\mathbf{n}\cdot\nabla\mathcal{C}_j = -\sum_{j=1}^{N-1}d_{ij}\mathcal{C}_j,
\quad 1\leqslant i\leqslant N-1, \quad \text{on\
} \ \partial\Omega;
\label{equ:bc_phi_A_3}
\end{equation}
\end{subequations}
In \eqref{equ:bc_phi_A_3} $d_{ij}$ ($1\leqslant i,j\leqslant N-1$) are
chosen coefficients, and the $(N-1)\times(N-1)$ matrix formed by $d_{ij}$
is required to be symmetric semi-positive definite.
Because the matrix $\tilde{\mathbf{m}}$ formed by 
$\tilde{m}_{ij}$ is SPD, the boundary conditions \eqref{equ:bc_phi_A_1}
and \eqref{equ:bc_phi_A_2} are equilvalent
to $\mathcal{C}_i=0$ and $\mathbf{n}\cdot\nabla\mathcal{C}_i=0$ 
($1\leqslant i\leqslant N-1$) on $\partial\Omega$,
respectively.

We look into the following
choices to ensure the dissipativeness of the
boundary term (III) in equation \eqref{equ:energy_balance}:
\begin{subequations}
\begin{equation}
\frac{\partial\phi_i}{\partial t} = 0, \quad 1\leqslant i\leqslant N-1,
\quad \text{on} \ \partial\Omega;
\label{equ:bc_phi_B_1}
\end{equation}
\begin{equation}
  \mathbf{n}\cdot\frac{\partial W}{\partial\nabla\phi_i} = 0, \quad 1\leqslant i\leqslant N-1,
  \quad \text{on}\ \partial\Omega;
\label{equ:bc_phi_B_2}
\end{equation}
\begin{equation}
\mathbf{n}\cdot\frac{\partial W}{\partial\nabla\phi_i} 
= -\sum_{j=1}^{N-1}q_{ij}\frac{\partial\phi_j}{\partial t},
\quad 1\leqslant i\leqslant N-1, \quad \text{on}\ \partial\Omega;
\label{equ:bc_phi_B_3}
\end{equation}
\end{subequations}
In \eqref{equ:bc_phi_B_3} $q_{ij}$ ($1\leqslant i,j\leqslant N-1$)
are chosen coefficients, and the matrix formed by $q_{ij}$ is
required to be symmetric semi-positive definite.

The boundary conditions \eqref{equ:bc_vel_1}--\eqref{equ:bc_vel_3},
\eqref{equ:bc_phi_A_1}--\eqref{equ:bc_phi_A_3},
and \eqref{equ:bc_phi_B_1}--\eqref{equ:bc_phi_B_3} are
favorable from the energy stability standpoint.
Additionally, the boundary conditions should satisfy
the reduction consistency property for the N-phase systems,
as pointed out by \cite{Dong2017}.
The reduction consistency
consideration can place restrictions on the form of these boundary
conditions. In the subsequent section
we look into 
the implications of the reduction consistency
property on these boundary conditions, and
in particular we suggest conditions for the inflow and
outflow boundaries taking account of both reduction consistency and
energy stability.

\subsection{Reduction Consistency and Inflow/Outflow Boundary Conditions}
\label{sec:reduction_consistency}

%
%
%

The reduction consistency of N-phase formulations has been investigated
extensively in \cite{Dong2017}.
Let us first define reduction consistency according to \cite{Dong2017},
and then apply this requirement to the energy-stable
boundary conditions from the previous subsection.
A physical entity (e.g.~variable, equation, or condition)
for the N-phase system
is said to be reduction consistent if it has the following property:
  If only a set of $M$ ($2\leqslant M\leqslant N-1$) fluid components are
  present in the N-phase system, then the physical entity
  for the N-phase system reduces to that for the corresponding
  equivalent M-phase system. 

We insist that 
the formulation for the N-phase system should honor
the reduction consistency property, namely, the N-phase formulation
should be reduction consistent.
Issues of reduction consistency have been considered recently in
\cite{Dong2017} for the N-phase governing equations (coupled
system of momentum and phase-field equations);
see also \cite{BoyerL2006,BoyerM2014} for an investigation of
the consistency issues of a system of Cahn-Hilliard type equations
(without hydrodynamic interaction).
The consistency properties
explored in \cite{Dong2017} can be summarized as
the following three:
\begin{enumerate}[($\mathscr{C}$1):]

\item
  The N-phase free energy
  density function should be reduction consistent;

\item
  The N-phase governing equations should be reduction consistent;

\item
  The boundary conditions for the N-phase system should
  be reduction consistent.

\end{enumerate}
The goal of this subsection is to investigate
the implications of the consistency property ($\mathscr{C}$3)
on the energy-stable boundary conditions
from the previous subsection.


To make the presentation more concrete, hereafter
we will specifically employ the volume fractions of the first
($N-1$) fluids as the set of order parameters,
namely,
\begin{equation}
  \phi_i \equiv c_i, \ \phi_i\in[0,1], \quad 1\leqslant i\leqslant N-1; \quad
  \vec{\phi} = \vec{c} = (c_1,c_2,\dots,c_{N-1})^T.
\end{equation}
Then with this choice, 
equation \eqref{equ:varphi_expr} is given by (see \cite{Dong2015}
for details)
\begin{equation}
  \varphi_i(\vec{c}) = \sum_{j=1}^{N-1}a_{ij}c_j - \tilde{\rho}_N, \ \
  1\leqslant i\leqslant N-1; \quad
  a_{ij} = \tilde{\rho}_{i}\delta_{ij}+\tilde{\rho}_N, \ \
  1\leqslant i,j\leqslant N-1
  \label{equ:varphi_volfrac_expr}
\end{equation}
where $\delta_{ij}$ is the Kronecker delta.
Let $\mathbf{A}_1=[a_{ij}]_{(N-1)\times(N-1)}$.
It is straightforward to verify that $\mathbf{A}_1$ is
symmetric positive definite and thus non-singular. 
It should be noted that
the boundary conditions and numerical algorithms
presented below can be formulated similarly
in terms of the class of general order parameters
introduced in \cite{Dong2015}.

Following \cite{Dong2017}, we employ the following general
form for the free energy density function
\begin{equation}
  W(\vec{c},\nabla\vec{c}) =
  \sum_{i,j=1}^{N-1}\frac{\lambda_{ij}}{2}\nabla c_i\cdot\nabla c_j
  + H(\vec{c})
  \label{equ:free_energy}
\end{equation}
where the constants $\lambda_{ij}$ ($1\leqslant i,j\leqslant N-1$) are
referred to as the mixing energy density coefficients,
and the matrix
$\mathbf{A}=[\lambda_{ij}]_{(N-1)\times(N-1)}$ is required
to be symmetric positive definite.
$H(\vec{c})$ is referred to as the potential energy density
function, and is to be specified later.
In this work we assume that the coefficients
$\tilde{m}_{ij}$ ($1\leqslant i,j\leqslant N-1$) in
\eqref{equ:CH_original} are constants.


The following are the conditions obtained in \cite{Dong2017}
about $\lambda_{ij}$, $H(\vec{c})$,
and $\tilde{m}_{ij}$ based on
the reduction consistency properties
($\mathscr{C}$1) and ($\mathscr{C}$2):
\begin{enumerate}[(DC-1):]

\item
  $\lambda_{ij}$ are given by
  \begin{equation}
    \lambda_{ij} = \frac{3}{\sqrt{2}}\eta(\sigma_{iN} + \sigma_{jN} - \sigma_{ij}),
    \quad 1\leqslant i,j\leqslant N-1
    \label{equ:lambda_ij_expr}
  \end{equation}
  where $\eta$ is the characteristic interfacial thickness,
  $\sigma_{ij}$ ($1\leqslant i\neq j\leqslant N$) is the surface
  tension between fluids $i$ and $j$, and $\sigma_{ii}=0$ ($1\leqslant i\leqslant N$).

\item
  $\tilde{m}_{ij}$ are given by
  \begin{equation}
    [\tilde{m}_{ij}]_{(N-1)\times(N-1)} = \tilde{\mathbf{m}}
    = m_0\mathbf{A}_1\mathbf{A}^{-1}\mathbf{A}_1^T
    \label{equ:mij_expr}
  \end{equation}
  where the constant $m_0>0$ is the mobility coefficient,
  $\mathbf{A}_1$ is the matrix formed by $a_{ij}$ as given
  in \eqref{equ:varphi_volfrac_expr}, and $\mathbf{A}$ is
  the matrix formed by $\lambda_{ij}$.

\item
  $H(\vec{c})$ is reduction consistent.

\item
  If any one fluid $k$ ($1\leqslant k\leqslant N$) is absent
  from the N-phase system, i.e.~$c_k\equiv 0$, then $H(\vec{c})$ is chosen such that
  \begin{equation}
    \left\{
    \begin{split}
      &
      L_k^{(N)} = 0 \\
      &
      L_i^{(N-1)} = L_i^{(N)}, \quad 1\leqslant i\leqslant k-1, \\
      &
      L_i^{(N-1)} = L_{i+1}^{(N)}, \quad k\leqslant i\leqslant N-1,
    \end{split}
    \right.
    \label{equ:L_cond}
  \end{equation}
  where 
  $L_i^{(N)}$ ($1\leqslant i\leqslant N$) is defined by 
  \begin{equation}
    \left\{
    \begin{split}
      &
      L_i^{(N)} = \sum_{j=1}^{N-1}\zeta_{ij}^{(N)}\frac{\partial H^{(N)}}{\partial c_j^{(N)}},
      \quad 1\leqslant i\leqslant N-1,
      \quad \text{where} \ \left[\zeta_{ij}^{(N)}\right]_{(N-1)\times(N-1)} = \mathbf{A}^{-1};
      \\
      &
      L_N^{(N)} = -\sum_{i=1}^{N-1}L_i^{(N)}.
    \end{split}
    \right.
    \label{equ:L_def}
  \end{equation}
  In the above equations the superscript $N$ in $(\cdot)^{(N)}$ accentuates
  the point that the variable is with respect to the N-phase system. 
  
\end{enumerate}
It is shown in \cite{Dong2017} that, with $\lambda_{ij}$ and  $\tilde{m}_{ij}$
given by \eqref{equ:lambda_ij_expr} and
\eqref{equ:mij_expr} respectively, and $H(\vec{c})$ satisfying 
(DC-3) and (DC-4), the N-phase governing equations represented by
\eqref{equ:nse_original}--\eqref{equ:CH_original} and
the free energy density function given by \eqref{equ:free_energy}
satisfy the reduction consistency properties ($\mathscr{C}$1)
and ($\mathscr{C}$2).

In subsequent discussions, whenever necessary,
we will use the superscript notation
$(\cdot)^{(N)}$ to signify that the variable is with respect to
the N-phase system, 
but will drop the superscript
where no confusion arises.

\subsubsection{Reduction Consistency of Boundary Conditions}

We employ the $\lambda_{ij}$ and $\tilde{m}_{ij}$ values
given by \eqref{equ:lambda_ij_expr} and \eqref{equ:mij_expr}, and
assume that the potential energy density function
$H(\vec{c})$ satisfies the conditions (DC-3) and (DC-4).
Let us now look into
the energy-stable boundary conditions
from Section \ref{sec:energy_balance}
in light of the reduction consistency requirement ($\mathscr{C}$3).

We insist that the boundary conditions \eqref{equ:bc_vel_1}--\eqref{equ:bc_vel_3},
\eqref{equ:bc_phi_A_1}--\eqref{equ:bc_phi_A_3}
and \eqref{equ:bc_phi_B_1}--\eqref{equ:bc_phi_B_3}
should satisfy the consistency property ($\mathscr{C}$3).
To ensure the reduction consistency between
the N-phase and M-phase ($2\leqslant M\leqslant N-1$) systems,
it suffices to consider only the reduction between
N-phase and ($N-1$)-phase systems, i.e.~if only one fluid
component is absent from the system.

Consider first the conditions \eqref{equ:bc_vel_1}--\eqref{equ:bc_vel_3}.
The condition \eqref{equ:bc_vel_1}
is evidently reduction consistent because no phase field variable is
involved.
The conditions \eqref{equ:bc_vel_2} and \eqref{equ:bc_vel_3}
are reduction consistent because, as shown in
\cite{Dong2017},
the variables
$\rho(\vec{c})$ and $\mu(\vec{c})$ given by \eqref{equ:density_expr}
are reduction consistent, and the $\tilde{\mathbf{J}}$ given by \eqref{equ:J_expr}
is also reduction consistent under
the condition (DC-4).
Note also that the free energy density function
$W(\vec{c},\nabla\vec{c})$ given by \eqref{equ:free_energy}
satisfies the consistency property ($\mathscr{C}$1) under
the condition (DC-3), as mentioned earlier.

We next consider the boundary
conditions \eqref{equ:bc_phi_A_1}--\eqref{equ:bc_phi_A_3}.
Define
\begin{equation}
  \vec{\mathcal{C}}=\left[\mathcal{C}_i  \right]_{(N-1)\times 1}, \ \
  \frac{\partial H}{\partial \vec{c}} = \left[\frac{\partial H}{\partial c_i}  \right]_{(N-1)\times 1}, \ \
  \mathbf{D} = \left[d_{ij}  \right]_{(N-1)\times(N-1)}.
\end{equation}
In light of the equations \eqref{equ:varphi_volfrac_expr},
\eqref{equ:free_energy} and \eqref{equ:lambda_ij_expr},
the chemical potentials $\mathcal{C}_i$ can be obtained from
equation \eqref{equ:chem_potential}
in a matrix form,
\begin{equation}
  \vec{\mathcal{C}} = \mathbf{A}_1^{-T}\left( \frac{\partial H}{\partial\vec{c}}
  - \mathbf{A}\nabla^2\vec{c} \right).
  \label{equ:chempot_expr}
\end{equation}
So boundary condition \eqref{equ:bc_phi_A_1} is transformed into
\begin{equation}
\left\{
\begin{split}
&
  -\nabla^2\vec{c} + \mathbf{A}^{-1}\frac{\partial H}{\partial \vec{c}}=0, 
  \ \ \text{on} \ \partial\Omega, \ \text{or equivalently}
  \\
&
  -\nabla^2 c_i + \sum_{j=1}^{N-1}\zeta_{ij}\frac{\partial H}{\partial c_j} = 0,
  \ \ 1\leqslant i\leqslant N-1, \ \ \text{on} \ \partial\Omega
\end{split}
\right.
  \label{equ:bc_phi_A_1_trans}
\end{equation}
where we have used \eqref{equ:mij_expr}.
Boundary conditions \eqref{equ:bc_phi_A_2} can be transformed into
\begin{equation}
\left\{
\begin{split}
&
  \mathbf{n}\cdot\nabla\left(-\nabla^2\vec{c} + \mathbf{A}^{-1}\frac{\partial H}{\partial \vec{c}}\right)=0, 
  \ \ \text{on} \ \partial\Omega, \ \text{or equivalently}
\\
&
  \mathbf{n}\cdot\nabla\left(-\nabla^2 c_i + \sum_{j=1}^{N-1}\zeta_{ij}\frac{\partial H}{\partial c_j}\right) = 0,
  \ \ 1\leqslant i\leqslant N-1, \ \ \text{on} \ \partial\Omega.
\end{split}
\right.
  \label{equ:bc_phi_A_2_trans}
\end{equation}

Equations \eqref{equ:bc_phi_A_1_trans} and \eqref{equ:bc_phi_A_2_trans}
are reduction consistent under
the condition (DC-4). It suffices to consider only \eqref{equ:bc_phi_A_1_trans}.
Suppose the fluid $k$ (for any $1\leqslant k\leqslant N$) is absent
from the N-phase system, i.e.~$c_k^{(N)}\equiv 0$.
Let
$\chi_i$ denote a variable from the set of variables
$\{ c_i, \rho_i, \tilde{\rho}_i, \tilde{\mu}_i, \tilde{\gamma}_i \}$, and
the following correspondence relations  hold between the N-phase system and
the ($N-1$)-phase system without fluid $k$:
\begin{equation}
  \chi_i^{(N-1)} = \left\{
  \begin{array}{ll}
    \chi_i^{(N)}, & 1\leqslant i< k \\
    \chi_{i+1}^{(N)}, & k\leqslant i\leqslant N-1.
  \end{array}
  \right.
  \label{equ:correspond_relation}
\end{equation}
Therefore, for $1\leqslant i=k\leqslant N-1$,
the equation \eqref{equ:bc_phi_A_1_trans} becomes
an identity,
\begin{equation}
  -\nabla^2 c_k^{(N)} + \sum_{j=1}^{N-1}\zeta_{kj}^{(N)}\frac{\partial H^{(N)}}{\partial c_j^{(N)}} = -\nabla^2 c_k^{(N)} + L_k^{(N)} = 0
\end{equation}
in light of the equation \eqref{equ:L_cond} under
the condition (DC-4).
For $1\leqslant i\leqslant  k-1$, the equation \eqref{equ:bc_phi_A_1_trans} becomes
\begin{equation}
  \begin{split}
  0 &= -\nabla^2c_i^{(N)}
  + \sum_{j=1}^{N-1}\zeta_{ij}^{(N)}\frac{\partial H^{(N)}}{\partial c_j^{(N)}}
  = -\nabla^2c_i^{(N)} + L_i^{(N)} 
  = -\nabla^2c_i^{(N-1)} + L_i^{(N-1)} \\ 
  &= -\nabla^2c_i^{(N-1)} + \sum_{j=1}^{N-2}\zeta_{ij}^{(N-1)}\frac{\partial H^{(N-1)}}{\partial c_j^{(N-1)}}, \quad
  1\leqslant i\leqslant k-1
  \end{split}
\end{equation}
where we have used the correspondence relation \eqref{equ:correspond_relation}
and the equation \eqref{equ:L_cond} under the condition (DC-4).
Therefore,
\begin{equation}
-\nabla^2c_i^{(N)}
+ \sum_{j=1}^{N-1}\zeta_{ij}^{(N)}\frac{\partial H^{(N)}}{\partial c_j^{(N)}}
=0 \ \ \Longrightarrow \ \
-\nabla^2c_i^{(N-1)} + \sum_{j=1}^{N-2}\zeta_{ij}^{(N-1)}\frac{\partial H^{(N-1)}}{\partial c_j^{(N-1)}} = 0, \quad
1\leqslant i\leqslant k-1.
\end{equation}
For $k+1\leqslant i+1\leqslant N$, equation \eqref{equ:bc_phi_A_1_trans} becomes
\begin{equation}
  \begin{split}
0 &= -\nabla^2c_{i+1}^{(N)}
+ \sum_{j=1}^{N-1}\zeta_{i+1,j}^{(N)}\frac{\partial H^{(N)}}{\partial c_j^{(N)}}
= -\nabla^2c_{i+1}^{(N)} + L_{i+1}^{(N)} 
= -\nabla^2c_i^{(N-1)} + L_i^{(N-1)} \\
&= -\nabla^2c_i^{(N-1)} + \sum_{j=1}^{N-2}\zeta_{ij}^{(N-1)}\frac{\partial H^{(N-1)}}{\partial c_j^{(N-1)}}, \quad
k\leqslant i\leqslant N-1
\end{split}
\end{equation}
where we have used \eqref{equ:correspond_relation}
and \eqref{equ:L_cond}. Therefore,
\begin{equation}
-\nabla^2c_{i+1}^{(N)}
+ \sum_{j=1}^{N-1}\zeta_{i+1,j}^{(N)}\frac{\partial H^{(N)}}{\partial c_j^{(N)}}
=0 \ \ \Longrightarrow \ \
-\nabla^2c_i^{(N-1)} + \sum_{j=1}^{N-2}\zeta_{ij}^{(N-1)}\frac{\partial H^{(N-1)}}{\partial c_j^{(N-1)}} = 0, \quad
k\leqslant i\leqslant N-1.
\end{equation}
Combining the above results, we conclude that
if any fluid is absent then the boundary condition
\eqref{equ:bc_phi_A_1_trans} for the N-phase system
will reduce to that for the corresponding $(N-1)$-phase
system. So it is reduction consistent. It follows that
the boundary condition \eqref{equ:bc_phi_A_2_trans}
is also reduction consistent
under the condition (DC-4).

The boundary condition \eqref{equ:bc_phi_A_3}
can be written in matrix form as
\begin{equation}
  \tilde{\mathbf{m}}(\mathbf{n}\cdot\nabla\vec{\mathcal{C}})
  =-\mathbf{D}\vec{\mathcal{C}} \ \Longrightarrow \
  m_0\mathbf{n}\cdot \nabla \left(-\nabla^2\vec{c}+\mathbf{A}^{-1}\frac{\partial H}{\partial\vec{c}}  \right)
  =-\mathbf{A}_1^{-1}\mathbf{DA}_1^{-T}\mathbf{A}\left(
 -\nabla^2\vec{c}+\mathbf{A}^{-1}\frac{\partial H}{\partial\vec{c}} \right) 
\end{equation}
where we have used \eqref{equ:chempot_expr} and \eqref{equ:mij_expr}.
Let
$\mathbf{A}_1^{-1}\mathbf{DA}_1^{-T}\mathbf{A} = \left[b_{ij}  \right]_{(N-1)\times(N-1)}$.
Then the above equation can be written in terms of the
component terms as
\begin{equation}
  m_0\mathbf{n}\cdot\nabla\left(-\nabla^2c_i + \sum_{j=1}^{N-1}\zeta_{ij}\frac{\partial H}{\partial c_j}  \right)
  = - \sum_{j=1}^{N-1}b_{ij}\left(-\nabla^2c_j
  + \sum_{k=1}^{N-1}\zeta_{jk}\frac{\partial H}{\partial c_k} \right),
  \quad 1\leqslant i\leqslant N-1.
  \label{equ:bc_phi_A_3_trans}
\end{equation}
Note that the terms
$
\left( -\nabla^2c_i + \sum_{j=1}^{N-1}\zeta_{ij}\frac{\partial H}{\partial c_j} \right)
$
for $1\leqslant i\leqslant N-1$
are reduction consistent, as shown in the above discussions.
Therefore, a sufficient condition for
the equation \eqref{equ:bc_phi_A_3_trans} to
be reduction consistent is that
the matrix formed by $b_{ij}$ be diagonal,
\begin{equation}
  \mathbf{A}_1^{-1}\mathbf{DA}_1^{-T}\mathbf{A}
  = \text{diag}(\hat{e}_1, \dots,\hat{e}_{N-1}) = \mathbf{G}
\end{equation}
for some $\hat{e}_i$ ($1\leqslant i\leqslant N-1$).
It then follows that
\begin{equation}
\mathbf{A}_1^{-1}\mathbf{DA}_1^{-T} = \mathbf{G} \mathbf{A}^{-1}
\end{equation}
The left hand side of this equation is a symmetric semi-positive
definite matrix,
because $\mathbf{A}_1$ is non-singular and
$\mathbf{D}$ is required to be symmetric semi-positive definite.
Note that on the right hand side $\mathbf{G}$ is diagonal
and $\mathbf{A}$ is a general SPD
matrix. We therefore conclude that
\begin{equation}
\mathbf{G} = e_0\mathbf{I} 
\end{equation}
where $\mathbf{I}$ is the identity matrix and $e_0\geqslant 0$ is
a non-negative constant. Consequently
\begin{equation}
  \mathbf{D} = \mathbf{A}_1\mathbf{GA}^{-1}\mathbf{A}_1^T
  = e_0\mathbf{A}_1\mathbf{A}^{-1}\mathbf{A}_1^T
  = \frac{e_0}{m_0}\tilde{\mathbf{m}}.
\end{equation}
So the boundary condition \eqref{equ:bc_phi_A_3}
is transformed into
\begin{equation}
  \mathbf{n}\cdot\nabla\left(-\nabla^2c_i + \sum_{j=1}^{N-1}\zeta_{ij}\frac{\partial H}{\partial c_j}  \right)
  = -\frac{e_0}{m_0}\left(-\nabla^2c_i
  + \sum_{k=1}^{N-1}\zeta_{ik}\frac{\partial H}{\partial c_k} \right),
  \quad 1\leqslant i\leqslant N-1,
  \label{equ:bc_phi_A_3_trans_1}
\end{equation}
and these conditions are reduction consistent.

Let us now consider the boundary conditions
\eqref{equ:bc_phi_B_1}--\eqref{equ:bc_phi_B_3}.
The condition \eqref{equ:bc_phi_B_1} implies
that
\begin{equation}
  c_i(\mathbf{x},t) = c_{bi}(\mathbf{x}), \ \ 1\leqslant i\leqslant N-1; \ \
  c_N(\mathbf{x},t) = 1-\sum_{i=1}^{N-1} c_{bi}(\mathbf{x}) = c_{bN}(\mathbf{x}),
  \ \ \text{on} \ \partial\Omega.
  \label{equ:bc_phi_B_1_trans}
\end{equation}
If a fluid $k$ is absent from the N-phase system throughout time,
then the reduction consistency requires that $c_{bk}(\mathbf{x})\equiv 0$.
Indeed, if $c_{bi}(\mathbf{x})$ is non-zero on the boundary
for any fluid $i$, that fluid cannot be absent from
the  system.

In light of \eqref{equ:free_energy},
the boundary condition \eqref{equ:bc_phi_B_2} 
is transformed into
\begin{equation}
  \sum_{j=1}^{N-1}\lambda_{ij}\mathbf{n}\cdot\nabla c_j = 0, \ \
  1\leqslant i\leqslant N-1 \ \
  \Longrightarrow \ \
  \mathbf{n}\cdot\nabla c_i = 0, \ \
  1\leqslant i\leqslant N-1, \ \ \text{on} \ \partial\Omega
  \label{equ:bc_phi_B_2_trans}
\end{equation}
by noting that 
the matrix $\mathbf{A}$ formed by $\lambda_{ij}$
($1\leqslant i,j\leqslant N-1$) is non-singular.
The boundary condition \eqref{equ:bc_phi_B_2_trans} is reduction consistent.
Note that this boundary condition implies
$
\mathbf{n}\cdot\nabla c_N = -\sum_{i=1}^{N-1}\mathbf{n}\cdot\nabla c_i=0.
$
Let us suppose a fluid $k$ ($1\leqslant k\leqslant N$)
is absent from the N-phase system, i.e.~$c_k^{(N)}\equiv 0$.
Then $\mathbf{n}\cdot\nabla c_k^{(N)} = 0$ becomes an
identity. Based on the correspondence relation \eqref{equ:correspond_relation}, for $1\leqslant i\leqslant k-1$,
\begin{equation}
  \mathbf{n}\cdot\nabla c_i^{(N)} =0 \ \
  \Longrightarrow \ \
  \mathbf{n}\cdot\nabla c_i^{(N-1)} =0;
\end{equation}
for $k\leqslant i\leqslant N-1$,
\begin{equation}
  \mathbf{n}\cdot\nabla c_{i+1}^{(N)} = 0 \ \ \Longrightarrow \ \
  \mathbf{n}\cdot\nabla c_{i}^{(N-1)} = 0.
\end{equation}
Therefore, if any one fluid is absent,
the boundary condition \eqref{equ:bc_phi_B_2_trans} (together with
$\mathbf{n}\cdot\nabla c_N=0$)
is reduced to
$
\mathbf{n}\cdot\nabla c_i^{(N-1)} = 0
$
for $1\leqslant i\leqslant N-1$.

The boundary condition \eqref{equ:bc_phi_B_3} can be
transformed into
\begin{equation}
  \sum_{j=1}^{N-1}\lambda_{ij}\mathbf{n}\cdot\nabla c_j
  = -\sum_{j=1}^{N-1} q_{ij}\frac{\partial c_j}{\partial t},
  \quad \text{or} \quad
  \mathbf{A}(\mathbf{n}\cdot\nabla\vec{c})
  =-\mathbf{Q}\frac{\partial \vec{c}}{\partial t}
  \label{equ:bc_phi_B_3_trans}
\end{equation}
where the matrix $\mathbf{Q} = [ q_{ij} ]_{(N-1)\times(N-1)}$ 
is required to be symmetric semi-positive definite.
Let $\mathbf{A}^{-1}\mathbf{Q} = [r_{ij} ]_{(N-1)\times(N-1)}$.
The above condition can be further transformed into
\begin{equation}
  \mathbf{n}\cdot\nabla c_i
  = - \sum_{j=1}^{N-1} r_{ij}\frac{\partial c_j}{\partial t},
  \quad 1\leqslant i\leqslant N-1.
  \label{equ:bc_phi_B_3_trans_1}
\end{equation}
Noting that both $\mathbf{n}\cdot\nabla c_i=0$ ($1\leqslant i\leqslant N$)
and $\frac{\partial c_i}{\partial t}=0$ ($1\leqslant i\leqslant N$)
are reduction consistent,
we impose the  condition
that the matrix $\mathbf{A}^{-1}\mathbf{Q}$ be diagonal 
in order to facilitate 
the reduction consistency of
equation \eqref{equ:bc_phi_B_3_trans_1}, i.e.
\begin{equation}
  \mathbf{A}^{-1}\mathbf{Q} = \text{diag}(\hat{r}_1,\dots,\hat{r}_{N-1})
  = \mathbf{E},
  \quad \text{or} \quad
  \mathbf{Q} = \mathbf{AE}
\end{equation}
for some $\hat{r}_i$ ($1\leqslant i\leqslant N-1$).
Note that $\mathbf{Q}$ is required to be symmetric semi-positive definite,
$\mathbf{A}$ is a general SPD
matrix, and $\mathbf{E}$ is diagonal.
We then conclude that
\begin{equation}
\mathbf{E} = d_0\mathbf{I}
\end{equation}
where $d_0\geqslant 0$ is a non-negative constant.
Therefore, the boundary condition \eqref{equ:bc_phi_B_3}
is reduced to
\begin{equation}
  \mathbf{n}\cdot\nabla c_i = -d_0 \frac{\partial c_i}{\partial t},
  \quad 1\leqslant i\leqslant N-1, \quad
  \text{on} \ \partial\Omega.
  \label{equ:bc_phi_B_3_trans_2}
\end{equation}
This implies that
\begin{equation}
  \mathbf{n}\cdot\nabla c_N = -\sum_{i=1}^{N-1}\mathbf{n}\cdot\nabla c_i
  = d_0\sum_{i=1}^{N-1}\frac{\partial c_i}{\partial t}
  = -d_0\frac{\partial c_N}{\partial t}, \quad \text{on} \ \partial\Omega.
  \label{equ:bc_phi_B_3_trans_2A}
\end{equation}

The condition \eqref{equ:bc_phi_B_3_trans_2}, together
with \eqref{equ:bc_phi_B_3_trans_2A}, is reduction consistent.
To demonstrate this point, let us assume that
fluid $k$ ($1\leqslant k\leqslant N$) is absent from
the system, i.e.~$c_k^{(N)}\equiv 0$.
Then for $1\leqslant i\leqslant k-1$,
\begin{equation}
\mathbf{n}\cdot\nabla c_i^{(N)} + d_0\frac{\partial c_i^{(N)}}{\partial t} = 0 \ \ \Longrightarrow \ \
\mathbf{n}\cdot\nabla c_i^{(N-1)} + d_0\frac{\partial c_i^{(N-1)}}{\partial t} = 0
\end{equation}
where we have used the correspondence relation \eqref{equ:correspond_relation}.
For $k\leqslant i\leqslant N-1$,
\begin{equation}
\mathbf{n}\cdot\nabla c_{i+1}^{(N)} + d_0\frac{\partial c_{i+1}^{(N)}}{\partial t} = 0 \ \ \Longrightarrow \ \
\mathbf{n}\cdot\nabla c_i^{(N-1)} + d_0\frac{\partial c_i^{(N-1)}}{\partial t} = 0
\end{equation}
where the correspondence relation \eqref{equ:correspond_relation}
is again used.
One also notes that the condition
$\mathbf{n}\cdot\nabla c_k^{(N)} + d_0\frac{\partial c_k^{(N)}}{\partial t}=0$ becomes
an identity.

\subsubsection{Outflow and Inflow Boundary Conditions}


The above discussions involve general
considerations of the energy stability and reduction consistency properties of
the N-phase system and the implications of these properties
on the boundary conditions. 
The resultant boundary conditions are applicable to any type of boundary.
We next focus on the outflow and inflow boundaries specifically, and use 
these results to suggest specific outflow and inflow boundary conditions.


With $\lambda_{ij}$ given by \eqref{equ:lambda_ij_expr},
$\tilde{m}_{ij}$ given by \eqref{equ:mij_expr}
and the free energy density given by \eqref{equ:free_energy},
the governing equations \eqref{equ:nse_original} and
\eqref{equ:CH_original} are reduced into,
in terms of volume fractions $c_i$ ($1\leqslant i\leqslant N-1$)
as the order parameters,
\begin{equation}
\rho\left(
  \frac{\partial\mathbf{u}}{\partial t}
  + \mathbf{u}\cdot\nabla\mathbf{u}
\right)
+ \tilde{\mathbf{J}}\cdot\nabla\mathbf{u}
=
-\nabla p
+ \nabla\cdot\left[
  \mu \mathbf{D}(\mathbf{u})
\right]
- \sum_{i,j=1}^{N-1} \nabla\cdot\left(\lambda_{ij}
  \nabla c_i \otimes \nabla c_j
\right)
+ \mathbf{f}(\mathbf{x},t),
\label{equ:nse}
\end{equation}
\begin{equation}
\frac{\partial c_i}{\partial t}
+ \mathbf{u}\cdot\nabla c_i = m_0\nabla^2\left(
-\nabla^2 c_i + \sum_{j=1}^{N-1}\zeta_{ij}\frac{\partial H}{\partial c_j}
\right) + g_i(\mathbf{x},t),
\quad 1\leqslant i\leqslant N-1
\label{equ:CH}
\end{equation}
where we have added an external body
force $\mathbf{f}$ to the momentum equation,
and a source term $g_i$ to each of the $N-1$
phase field equations.
$g_i$ ($1\leqslant i\leqslant N-1$) are 
for the purpose of numerical testing only,
and will be set to $g_i=0$ in actual simulations.
$\tilde{\mathbf{J}}$ is given by (simplifed from equation \eqref{equ:J_expr})
\begin{equation}
\tilde{\mathbf{J}} = -m_0\sum_{i=1}^{N-1}\left(
\tilde{\rho}_i - \tilde{\rho}_N
\right)\nabla\left(  
-\nabla^2 c_i + \sum_{j=1}^{N-1}\zeta_{ij}\frac{\partial H}{\partial c_j}
\right).
\label{equ:J_expr_1}
\end{equation}

We assume that the domain boundary consists of three
types which are non-overlapping with one another:
$\partial\Omega = \partial\Omega_i \cup \partial\Omega_w \cup \partial\Omega_o$, where
\begin{itemize}

\item
$\partial\Omega_i$ is the inflow boundary, on which the velocity distribution
and the fluid-material distributions are known.

\item
$\partial\Omega_w$ is the wall boundary with certain
wetting properties, on which the velocity distribution (e.g.~zero velocity)
and the contact angles are known.

\item
$\partial\Omega_o$ is the outflow (or open) boundary, on which
none of the flow variables (velocity, pressure, phase field variables) is
known.

\end{itemize}
Since the phase field equations \eqref{equ:CH} are of fourth spatial order,
two independent boundary conditions will be needed on
each type of boundary for the phase field
variables $c_i$.


On the outflow/open boundary $\partial\Omega_o$ 
we propose the boundary conditions \eqref{equ:bc_phi_A_2_trans}
and \eqref{equ:bc_phi_B_3_trans_2} for the phase field equations, i.e.
\begin{subequations}
\begin{equation}
  \mathbf{n}\cdot\nabla\left(-\nabla^2 c_i + \sum_{j=1}^{N-1}\zeta_{ij}\frac{\partial H}{\partial c_j}\right) = 0,
  \ \ 1\leqslant i\leqslant N-1, \ \ \text{on} \ \partial\Omega_o,
\label{equ:obc_phi_1}
\end{equation}
\begin{equation}
\mathbf{n}\cdot\nabla c_i = -d_0 \frac{\partial c_i}{\partial t},
  \quad 1\leqslant i\leqslant N-1,
\quad \text{on} \ \partial\Omega_o.
\label{equ:obc_phi_2}
\end{equation}
\end{subequations}
For the momentum equation we propose 
the boundary condition \eqref{equ:bc_vel_3} on $\partial\Omega_o$.
Note that the combination of equations \eqref{equ:J_expr_1} and \eqref{equ:obc_phi_1}
leads to $\mathbf{n}\cdot\tilde{\mathbf{J}} = 0$ on $\partial\Omega_o$.
We will consider the following choice for $C_1(\mathbf{n},\mathbf{u})$ and
$C_2(\mathbf{n},\mathbf{u})$ in \eqref{equ:bc_vel_3} in
the present work,
analogous to the outflow condition for single-phase Navier-Stokes
equations in \cite{DongS2015},
\begin{equation*}
C_1(\mathbf{n},\mathbf{u}) = -\frac{\alpha_1}{2}\mathbf{n}\cdot\mathbf{u}, \quad
C_2(\mathbf{n},\mathbf{u}) = \frac{\alpha_2}{2} \mathbf{u}\cdot\mathbf{u}
\end{equation*}
where $\alpha_1\geqslant 0$ and $\alpha_2\geqslant 0$ are constants.
Therefore, the boundary condition \eqref{equ:bc_vel_3} is reduced to
\begin{multline}
-p\mathbf{n} + \mu\mathbf{n}\cdot\mathbf{D}(\mathbf{u})
- \left[\sum_{i,j=1}^{N-1}\frac{\lambda_{ij}}{2}\nabla c_i\cdot\nabla c_j + H(\vec{c})  \right]\mathbf{n} \\
- \rho\left[ \frac{1}{2}
  (\theta + \alpha_2)(\mathbf{u}\cdot\mathbf{u})\mathbf{n}
  + \frac{1}{2}(1-\theta+\alpha_1)(\mathbf{n}\cdot\mathbf{u})\mathbf{u}
\right]\Theta_0(\mathbf{n},\mathbf{u}) = 0, \quad
\text{on} \ \partial\Omega_o
\label{equ:obc_vel}
\end{multline}
where $0\leqslant \theta\leqslant 1$, $\alpha_1\geqslant 0$ and $\alpha_2\geqslant 0$
are constant parameters.
The open boundary conditions \eqref{equ:obc_phi_1}--\eqref{equ:obc_vel}
are reduction consistent, and they
ensure the energy dissipativity on the open/outflow boundary
$\partial\Omega_o$ even when strong vortices or backflows occur on $\partial\Omega_o$.


Equation \eqref{equ:obc_vel} represents a family of boundary conditions
for $\partial\Omega_o$ with $(\theta,\alpha_1,\alpha_2)$ as
the parameters.
The term involving $\Theta_0$ in \eqref{equ:obc_vel}
is critical to the energy stability when strong vortices or backflows occur
at the open boundary. This term is similar in form to that of
the open boundary conditions developed in \cite{DongS2015} for single-phase
flows. It is observed from single-phase flow simulations of \cite{DongS2015} that,
among the family represented by $(\theta,\alpha_1,\alpha_2)$,
the condition corresponding to $(\theta,\alpha_1,\alpha_2)=(1,1,0)$
produces overall the best results in terms of the smoothness
of the velocity field at the outflow boundary and the distortion of
flow structures when they exit the domain.
We specifically list below this particular
open boundary condition corresponding to $(\theta,\alpha_1,\alpha_2)=(1,1,0)$
among those given by \eqref{equ:obc_vel},
\begin{multline}
-p\mathbf{n} + \mu\mathbf{n}\cdot\mathbf{D}(\mathbf{u})
- \left[\sum_{i,j=1}^{N-1}\frac{\lambda_{ij}}{2}\nabla c_i\cdot\nabla c_j + H(\vec{c})  \right]\mathbf{n} \\
- \frac{1}{2}\rho\left[ 
  (\mathbf{u}\cdot\mathbf{u})\mathbf{n}
  + (\mathbf{n}\cdot\mathbf{u})\mathbf{u}
\right]\Theta_0(\mathbf{n},\mathbf{u}) = 0, \quad
\text{on} \ \partial\Omega_o.
\label{equ:obc_vel_best}
\end{multline}
The majority of numerical simulations presented in
Section \ref{sec:tests} will be performed with
this boundary condition for $\partial\Omega_o$.


Let us make a comment on the boundary condition \eqref{equ:obc_phi_2}.
This condition is analogous to
a convective type condition on the outflow boundary
if $d_0>0$,
\begin{equation}
\frac{\partial c_i}{\partial t} + U_c \mathbf{n}\cdot\nabla c_i = 0,
\quad 1\leqslant i\leqslant N-1, \quad \text{on} \ \partial\Omega_o,
\quad \text{where} \ U_c = \frac{1}{d_0}.
\label{equ:obc_convective}
\end{equation}
Therefore, $\frac{1}{d_0}$ plays the role of a convection velocity
at the open/outflow boundary. In practical simulations, one could first
estimate a convection velocity scale $U_c>0$ at the outflow
boundary based on physical considerations (e.g.~mass conservation)
or by preliminary simulations using e.g.~$d_0=0$.
Then one can determine $d_0$ based on $d_0 = \frac{1}{U_c}$.


On the inflow boundary $\partial\Omega_i$ 
the material distribution is known, implying a Dirichlet type
condition
\begin{equation}
  c_i = c_{bi}(\mathbf{x},t), \quad 1\leqslant i\leqslant N-1,
  \quad \text{on} \ \partial\Omega_i
  \label{equ:ibc_phi_1}
\end{equation}
where $c_{bi}$ is boundary volume-fraction distribution.
For the other boundary condition on $\partial\Omega_i$ for
the phase field equations,
we propose the condition \eqref{equ:bc_phi_A_1_trans}, i.e.
\begin{equation}
-\nabla^2 c_i + \sum_{j=1}^{N-1}\zeta_{ij}\frac{\partial H}{\partial c_j} = 0,
\ \ 1\leqslant i\leqslant N-1, \ \ \text{on} \ \partial\Omega_i.
\label{equ:ibc_phi_2}
\end{equation}


When a solid-wall boundary $\partial\Omega_w$ is present,
in the current paper we will assume that
the wall is of neutral wettability to all fluids,
that is, the contact angles for all fluid interfaces
are $90^0$. This corresponds to the condition \eqref{equ:bc_phi_B_2_trans},
namely,
\begin{equation}
  \mathbf{n}\cdot\nabla c_i = 0,
  \quad 1\leqslant i\leqslant N-1, \quad \text{on} \ \partial\Omega_w.
  \label{equ:wbc_phi_1}
\end{equation}
For N-phase flows bounded by solid walls
with more general wetting properties we refer the
reader to \cite{Dong2017} for a method
to deal with general contact angles.
We employ the condition \eqref{equ:bc_phi_A_2_trans} for
the other boundary condition for the phase field function
on $\partial\Omega_w$, i.e.
\begin{equation}
\mathbf{n}\cdot\nabla\left(-\nabla^2 c_i + \sum_{j=1}^{N-1}\zeta_{ij}\frac{\partial H}{\partial c_j}\right) = 0,
\ \ 1\leqslant i\leqslant N-1, \ \ \text{on} \ \partial\Omega_w.
\label{equ:wbc_phi_2}
\end{equation}

In addition, the velocity distribution on the inflow and wall boundaries
are assumed to be known, leading to a Dirichlet type condition
\begin{equation}
  \mathbf{u} = \mathbf{w}(\mathbf{x},t), \quad
  \text{on} \ \partial\Omega_i\cup\partial\Omega_w,
  \label{equ:dbc_vel}
\end{equation}
where $\mathbf{w}$ is the boundary velocity.

Finally, the initial distributions for the velocity ($\mathbf{u}^{in}$) and
the phase field functions ($c_i^{in}$) are assumed to be known,
\begin{subequations}
  \begin{align}
    &
    \mathbf{u}(\mathbf{x},0) = \mathbf{u}^{in}(\mathbf{x}), \label{equ:ic_vel} \\
    &
    c_i(\mathbf{x},0) = c_i^{in}(\mathbf{x}), \quad 1\leqslant i\leqslant N-1.
    \label{equ:ic_phi}
  \end{align}
\end{subequations}


\subsection{Algorithm Formulation}
\label{sec:algorithm}

%

The equations \eqref{equ:nse}--\eqref{equ:CH} and \eqref{equ:continuity_original},
supplemented by the boundary conditions
\eqref{equ:dbc_vel}, \eqref{equ:ibc_phi_1}--\eqref{equ:ibc_phi_2},
\eqref{equ:wbc_phi_1}, \eqref{equ:wbc_phi_2}, \eqref{equ:obc_vel},
\eqref{equ:obc_phi_1}--\eqref{equ:obc_phi_2},
together with the initial conditions
\eqref{equ:ic_vel}--\eqref{equ:ic_phi},
constitute the system to be solved  in numerical
simulations.
In the current paper, we employ the same potential
energy density function $H(\vec{c})$ as in \cite{Dong2017}
(originally suggested by \cite{BoyerM2014}),
given by
\begin{equation}
H(\vec{c}) = \frac{3}{\sqrt{2}\eta} \sum_{i,j=1}^N 
\frac{\sigma_{ij}}{2}\left[
  f(c_i) + f(c_j) - f(c_i+c_j)
  \right], \quad
\text{with} \ f(c) = c^2(1-c)^2
\label{equ:potential_energy}
\end{equation}
where $\eta$ is the characteristic interfacial thickness
of the diffuse interfaces.
As pointed out in \cite{Dong2017}, this function is reduction
consistent, but satisfies only a subset of the (DC-4) property.
It ensures the reduction consistency between N-phase
systems and M-phase systems for $M=2$.

To numerically test with manufactured analytic solutions,
we will modify several boundary conditions by adding
certain prescribed source terms. Define $h_i={\partial H}/{\partial c_i},$ $1\leqslant i \leqslant N-1.$
We modify \eqref{equ:ibc_phi_2} as
\begin{equation}
-\nabla^2 c_i + \sum_{j=1}^{N-1}\zeta_{ij}h_j = g_{ai}(\mathbf{x},t),
\ \ 1\leqslant i\leqslant N-1, \ \ \text{on} \ \partial\Omega_i,
\label{equ:ibc_phi_2_mod}
\end{equation}
where $g_{ai}$ ($1\leqslant i\leqslant N-1$) are prescribed functions.
We combine \eqref{equ:obc_phi_1} and \eqref{equ:wbc_phi_2}
and re-write them as
\begin{equation}
\mathbf{n}\cdot\nabla\Big(-\nabla^2 c_i + \sum_{j=1}^{N-1}\zeta_{ij}h_j\Big) = g_{bi}(\mathbf{x},t),
\ \ 1\leqslant i\leqslant N-1, \ \ \text{on} \ \partial\Omega_w\cup\partial\Omega_o
\label{equ:bc_chempot}
\end{equation}
where $g_{bi}$ ($1\leqslant i\leqslant N-1$) are prescribed functions.
We modify \eqref{equ:wbc_phi_1} as
\begin{equation}
\mathbf{n}\cdot\nabla c_i = g_{ci}(\mathbf{x},t),
  \quad 1\leqslant i\leqslant N-1, \quad \text{on} \ \partial\Omega_w
  \label{equ:wbc_phi_1_mod}
\end{equation}
where $g_{ci}$ ($1\leqslant i\leqslant N-1$) are prescribed functions.
The boundary condition \eqref{equ:obc_phi_2} is modified
as
\begin{equation}
\mathbf{n}\cdot\nabla c_i = -d_0 \frac{\partial c_i}{\partial t} + g_{ei},
  \quad 1\leqslant i\leqslant N-1,
\quad \text{on} \ \partial\Omega_o
\label{equ:obc_phi_2_mod}
\end{equation}
where $g_{ei}$ ($1\leqslant i\leqslant N-1$) are prescribed functions.
The prescribed source terms $g_{ai}$, $g_{bi}$, $g_{ci}$ and
$g_{ei}$ in the above equations \eqref{equ:ibc_phi_2_mod}--\eqref{equ:obc_phi_2_mod}
are for numerical testing only and will be set to zero
in actual simulations.

We re-write the momentum equation \eqref{equ:nse} as
\begin{equation}
  \frac{\partial\mathbf{u}}{\partial t}
  + \mathbf{u}\cdot\nabla\mathbf{u}
+ \frac{1}{\rho}\tilde{\mathbf{J}}\cdot\nabla\mathbf{u}
=
-\frac{1}{\rho}\nabla P
+ \frac{\mu}{\rho}\nabla^2\mathbf{u}
+ \frac{1}{\rho}\nabla\mu\cdot
   \mathbf{D}(\mathbf{u})
- \frac{1}{\rho}\sum_{i,j=1}^{N-1} \lambda_{ij}(\nabla^2 c_j)
  \nabla c_i 
+ \frac{\mathbf{f}}{\rho}
\label{equ:nse_trans}
\end{equation}
where $P = p+\frac{1}{2}\sum_{i,j=1}^{N-1}\lambda_{ij}\nabla c_i\cdot\nabla c_j$
and will also be loosely referred to as the pressure.
The boundary condition \eqref{equ:obc_vel} is re-written as
\begin{equation}
-P\mathbf{n} + \mu\mathbf{n}\cdot\mathbf{D}(\mathbf{u})
-  H(\vec{c}) \mathbf{n} - \mathbf{E}(\mathbf{n},\mathbf{u},\rho)
= \mathbf{f}_{b}(\mathbf{x},t), \quad
\text{on} \ \partial\Omega_o
\label{equ:obc_vel_mod}
\end{equation}
where
$
\mathbf{E}(\mathbf{n},\mathbf{u},\rho)
= \frac{1}{2}\rho\left[ 
  (\theta + \alpha_2)(\mathbf{u}\cdot\mathbf{u})\mathbf{n}
  + (1-\theta+\alpha_1)(\mathbf{n}\cdot\mathbf{u})\mathbf{u}
\right]\Theta_0(\mathbf{n},\mathbf{u}),
$
and $\mathbf{f}_b$ is a prescribed function for numerical testing only
and will be set to $\mathbf{f}_b=0$ in actual simulations.


We next present an algorithm for solving
the equations consisting of \eqref{equ:nse_trans},
\eqref{equ:continuity_original} and \eqref{equ:CH},
the boundary conditions consisting of \eqref{equ:dbc_vel},
\eqref{equ:obc_vel_mod}, \eqref{equ:ibc_phi_1},
\eqref{equ:ibc_phi_2_mod}, \eqref{equ:wbc_phi_1_mod},
\eqref{equ:bc_chempot} and \eqref{equ:obc_phi_2_mod},
together with the initial conditions
\eqref{equ:ic_vel} and \eqref{equ:ic_phi}.
The treatment for the governing equations follows a similar scheme as
in \cite{Dong2017}. Our emphasis below is on the numerical treatment and implementation of
various outflow and inflow boundary conditions.

Let $J$ ($J=1$ or $2$) denote the temporal order of accuracy,
$\Delta t$ denote the time step size,
and $n$ ($n\geqslant 0$) denote the time step index.
Let $\chi$ denote a generic variable. Then
$\chi^n$ represents the variable at time step $n$ in
the following, and we define
\begin{equation}
  \chi^{*,n+1} = \left\{
  \begin{array}{ll}
    \chi^n, &  J=1, \\
    2\chi^n - \chi^{n-1}, &  J=2;
  \end{array}
  \right.
  \ \
  \hat{\chi} = \left\{
  \begin{array}{ll}
    \chi^n, & J=1, \\
    2\chi^n-\frac{1}{2}\chi^{n-1}, & J=2;
  \end{array}
  \right.
  \ \
  \gamma_0 = \left\{
  \begin{array}{ll}
    1, &  J=1, \\
    3/2, &  J=2.
  \end{array}
  \right.
  \label{equ:param_def}
\end{equation}

Given $(\mathbf{u}^n, P^n, c_i^n)$, we compute
$c_i^{n+1}$, $P^{n+1}$ and $\mathbf{u}^{n+1}$
successively in a de-coupled fashion as follows. \\
\underline{For $c_i^{n+1}$:}
\begin{subequations}
  \begin{multline}
    \frac{\gamma_0 c_i^{n+1}-\hat{c}_i}{\Delta t}
    + \mathbf{u}^{*,n+1}\cdot\nabla c_i^{*,n+1} \\
    = m_0\nabla^2\left[
      -\nabla^2 c_i^{n+1}
      + \frac{S}{\eta^2}\left(c_i^{n+1} - c_i^{*,n+1}  \right)
      + \sum_{j=1}^{N-1} \zeta_{ij} h_j(\vec{c}^{*,n+1})
      \right] + g_i^{n+1},
    \quad 1\leqslant i\leqslant N-1
    \label{equ:phi_1}
  \end{multline}
  \begin{equation}
    -\nabla^2 c_i^{n+1} + \sum_{j=1}^{N-1}\zeta_{ij}h_j(\vec{c}^{n+1})=g_{ai}^{n+1},
    \quad 1\leqslant i\leqslant N-1, \ \text{on} \ \partial\Omega_i
    \label{equ:phi_2}
  \end{equation}
  \begin{equation}
    c_i^{n+1} = c_{bi}^{n+1}, \quad  1\leqslant i\leqslant N-1,
    \ \text{on} \ \partial\Omega_i
    \label{equ:phi_3}
  \end{equation}
  \begin{multline}
    \mathbf{n}\cdot\nabla\left[
      -\nabla^2c_i^{n+1} + \frac{S}{\eta^2}(c_i^{n+1}-c_i^{*,n+1})
      + \sum_{j=1}^{N-1} \zeta_{ij} h_j(\vec{c}^{*,n+1})
      \right] = g_{bi}^{n+1}, \\
    \ 1\leqslant i\leqslant N-1,
    \ \text{on} \ \partial\Omega_w\cup\partial\Omega_o
    \label{equ:phi_4}
  \end{multline}
  \begin{equation}
    \mathbf{n}\cdot\nabla c_i^{n+1} = g_{ci}^{n+1},
    \quad 1\leqslant i\leqslant N-1,
    \ \text{on} \ \partial\Omega_w
    \label{equ:phi_5}
  \end{equation}
  \begin{equation}
    \mathbf{n}\cdot\nabla c_i^{n+1} = -d_0\left.\frac{\partial c_i}{\partial t}\right|^{n+1}_{exp} + g_{ei}^{n+1},
    \quad 1\leqslant i\leqslant N-1, \ \text{on} \ \partial\Omega_o
    \label{equ:phi_6}
  \end{equation}
  \begin{equation}
    \mathbf{n}\cdot\nabla c_i^{n+1} = -d_0\frac{\gamma_0 c_i^{n+1}-\hat{c}_i}{\Delta t}
    + g_{ei}^{n+1}, \quad
    1\leqslant i\leqslant N-1, \ \text{on} \ \partial\Omega_o
    \label{equ:phi_7}
  \end{equation}
\end{subequations}
\underline{For $P^{n+1}$:}
\begin{subequations}
  \begin{equation}
    \begin{split}
      \frac{\gamma_0\tilde{\mathbf{u}}^{n+1}-\hat{\mathbf{u}}}{\Delta t}
      &+ \mathbf{u}^{*,n+1}\cdot\nabla\mathbf{u}^{*,n+1}
      + \frac{1}{\rho^{n+1}}\tilde{\mathbf{J}}^{n+1}\cdot\nabla\mathbf{u}^{*,n+1}
      + \frac{1}{\rho_0}\nabla P^{n+1} = \\
      &
      \left(\frac{1}{\rho_0} - \frac{1}{\rho^{n+1}}  \right)\nabla P^{*,n+1}       
      - \frac{\mu^{n+1}}{\rho^{n+1}}\nabla\times\nabla\times\mathbf{u}^{*,n+1}
      + \frac{1}{\rho^{n+1}}\nabla\mu^{n+1}\cdot\mathbf{D}(\mathbf{u}^{*,n+1}) \\
      &
      - \frac{1}{\rho^{n+1}}\sum_{i,j=1}^{N-1}\lambda_{ij}\nabla^2 c_i^{n+1}\nabla c_i^{n+1} 
      + \frac{1}{\rho^{n+1}}\mathbf{f}^{n+1}
    \end{split}
    \label{equ:pressure_1}
  \end{equation}
  \begin{equation}
    \nabla\cdot\tilde{\mathbf{u}}^{n+1} = 0
    \label{equ:pressure_2}
  \end{equation}
  \begin{equation}
    \mathbf{n}\cdot\tilde{\mathbf{u}}^{n+1} = \mathbf{n}\cdot\mathbf{w}^{n+1},
    \quad \text{on} \ \partial\Omega_i\cup\partial\Omega_w
    \label{equ:pressure_3}
  \end{equation}
  \begin{equation}
    P^{n+1} = \mu^{n+1}\mathbf{n}\cdot\mathbf{D}(\mathbf{u}^{*,n+1})\cdot\mathbf{n}
    -H(\vec{c}^{n+1}) - \mathbf{n}\cdot\mathbf{E}(\mathbf{n},\mathbf{u}^{*,n+1},\rho^{n+1})
    -\mathbf{f}_b^{n+1}\cdot\mathbf{n},
    \quad \text{on} \ \partial\Omega_o
    \label{equ:pressure_4}
  \end{equation}
\end{subequations}
\underline{For $\mathbf{u}^{n+1}$:}
\begin{subequations}
  \begin{equation}
    \frac{\gamma_0\mathbf{u}^{n+1}-\gamma_0\tilde{\mathbf{u}}^{n+1}}{\Delta t}
    - \nu_m\nabla^2\mathbf{u}^{n+1}
    = \nu_m \nabla\times\nabla\times \mathbf{u}^{*,n+1}
    \label{equ:vel_1}
  \end{equation}
  \begin{equation}
    \mathbf{u}^{n+1} = \mathbf{w}^{n+1},
    \quad \text{on} \ \partial\Omega_i\cup\partial\Omega_w
    \label{equ:vel_2}
  \end{equation}
  \begin{equation}
    \begin{split}
      \mathbf{n}\cdot\nabla\mathbf{u}^{n+1} =&
      \left(1 - \frac{\mu^{n+1}}{\mu_0}  \right)\mathbf{n}\cdot\mathbf{D}(\mathbf{u}^{*,n+1})
      + \frac{1}{\mu_0}\left[
        P^{n+1}\mathbf{n} + H(\vec{c}^{n+1})\mathbf{n}
        + \mathbf{E}(\mathbf{n},\mathbf{u}^{*,n+1},\rho^{n+1}) \right. \\
        &
        \left.
        + \mathbf{f}_b^{n+1} - \mu_0(\nabla\cdot\mathbf{u}^{*,n+1})\mathbf{n}
        \right]
      -\mathbf{n}\cdot(\nabla\mathbf{u}^{*,n+1})^T,
      \quad \text{on} \ \partial\Omega_o.
    \end{split}
    \label{equ:vel_3}
  \end{equation}
\end{subequations}
%

In the above equations, $\tilde{\mathbf{u}}^{n+1}$ is an auxiliary
velocity approximating $\mathbf{u}^{n+1}$,   and
$S$ is a chosen positive constant
that satisfies a condition to be specified later.
$\rho_0$ is a chosen constant that satisfies the condition
$0 < \rho_0 \leqslant \min(\tilde{\rho}_1,\dots,\tilde{\rho}_N)$.
$\nu_m$ is a chosen constant that is sufficiently large,
and we employ 
$\nu_m\geqslant \max\left(\frac{\tilde{\mu}_1}{\tilde{\rho}_1},\dots,\frac{\tilde{\mu}_N}{\tilde{\rho}_N}  \right)$
in the current paper.
$\mu_0$ is a chosen constant satisfying the condition
that $\mu_0=\tilde{\mu}_1$ if 
$\tilde{\mu}_1=\tilde{\mu}_2=\dots=\tilde{\mu}_N$, and otherwise
$\mu_0>\min(\tilde{\mu}_1,\dots,\tilde{\mu}_N)$.
In \eqref{equ:phi_6} $\left.\frac{\partial c_i}{\partial t}\right|^{n+1}_{exp}$ is
an explicit approximation of the time derivative given by
\begin{equation}
\left.\frac{\partial c_i}{\partial t}  \right|^{n+1}_{exp}=\left\{
\begin{array}{ll}
\frac{1}{\Delta t}(c_i^n-c_i^{n-1}), & J=1, \\
\frac{1}{\Delta t}(\frac{5}{2}c_i^n - 4c_i^{n-1}+\frac{3}{2}c_i^{n-2}), & J=2.
\end{array}
\right.
\end{equation}


Several comments on the above algorithm are in order at this point:
\begin{itemize}

\item
To solve the set of phase field variables, 
an extra term $\frac{S}{\eta^2}(c_i^{n+1}-c_i^{*,n+1})$ has been added to
the semi-discretized phase field equations \eqref{equ:phi_1}. 
This term is equivalent to zero to the $J$-th order
accuracy in time.
This term enables the reformulation of the $(N-1)$ semi-discretized
4th-order phase field equations into $2(N-1)$ de-coupled Helmholtz-type equations \cite{Dong2017}. This is an often-used strategy for two-phase flow
simulations (see e.g.~\cite{DongS2012,Dong2012}). It is crucial for spatial
discretizations with $C^0$ continuous spectral elements
employed in the current paper.

\item
In the discrete boundary condition \eqref{equ:phi_4} the same
extra zero term is added. This term is crucial, and without it
significant loss of mass for some fluid phases can be observed.

\item
The discrete conditions \eqref{equ:phi_6} and \eqref{equ:phi_7}
result from an explicit and an implicit treatment 
of the inertial term $\frac{\partial c_i}{\partial t}$ in
the boundary condition \eqref{equ:obc_phi_2_mod}.
These two approximations will be employed
to implement the outflow condition at different
stages of the implementation, which will become clear 
from later discussions.

\item
The equations \eqref{equ:pressure_1} and \eqref{equ:vel_1}
constitute a rotational velocity correction scheme
for the momentum equation \eqref{equ:nse_trans}.
The scheme adopts a reformulation of the 
pressure term and the viscous term in the same fashion
as in \cite{DongS2012},
\begin{equation*}
\frac{1}{\rho}\nabla P \approx \frac{1}{\rho_0}\nabla P +
\left(\frac{1}{\rho} - \frac{1}{\rho_0}  \right)\nabla P^*,
\quad
\frac{\mu}{\rho}\nabla^2\mathbf{u} \approx \nu_m\nabla^2\mathbf{u}
+ \left(\nu_m - \frac{\mu}{\rho} \right)\nabla\times\nabla\times\mathbf{u}^*
\end{equation*}
where $P^*$ and $\mathbf{u}^*$ are explicit approximations of
$P$ and $\mathbf{u}$ respectively.
These reformulations lead to time-independent coefficient matrices
for the pressure and velocity linear algebraic systems
after discretization, which is crucial for numerical efficiency.

\item
Equation \eqref{equ:pressure_4} is a discrete Dirichlet type
condition for the pressure on the outflow boundary $\partial\Omega_o$.
It results from taking the inner product between the boundary condition
\eqref{equ:obc_vel_mod} and the directional vector $\mathbf{n}$ normal
to $\partial\Omega_o$ and treating the velocity in
an explicit fashion.

\item
The discrete condition \eqref{equ:vel_3} is essentially a combination of 
the following two approximations:
\begin{equation*}
\left\{
\begin{split}
&
\mathbf{n}\cdot\nabla\mathbf{u}^{n+1} = \mathbf{n}\cdot\mathbf{D}(\mathbf{u}^{n+1})
 - \mathbf{n}\cdot(\nabla\mathbf{u}^{*,n+1})^T \\
&
\mu_0\mathbf{n}\cdot\mathbf{D}(\mathbf{u}^{n+1}) =
(\mu_0-\mu)\mathbf{n}\cdot\mathbf{D}(\mathbf{u}^{*,n+1}) \\
&
\qquad\qquad\qquad\quad
+ \left[
  P^{n+1}\mathbf{n} + H(\vec{c}^{n+1})\mathbf{n}
  + E(\mathbf{n},\mathbf{u}^{*,n+1},\rho^{n+1})
  + \mathbf{f}_b^{n+1}
\right], \quad \text{on} \ \partial\Omega_o.
\end{split}
\right.
\end{equation*}
The second approximation above stems from the 
outflow boundary condition \eqref{equ:obc_vel_mod},
but with the terms involving $\mu_0$ incorporated.
The construction with the $\mu_0$ terms was first 
introduced in \cite{Dong2014obc} for two-phase outflows.
This construction is crucial for the stability of the scheme when large
viscosity ratios among the fluids occur at the outflow/open boundary.
Note also that an extra term involving $(\nabla\cdot\mathbf{u})\mathbf{n}$
is incorporated into the discrete condition \eqref{equ:vel_3}.

\end{itemize}

\subsection{Implementation with Spectral Elements}
\label{sect: SEM}


We next implement the algorithm given by \eqref{equ:phi_1}--\eqref{equ:vel_3}
using $C^0$ continuous high-order spectral
elements~\cite{SherwinK1995,KarniadakisS2005,ZhengD2011}.
We first derive the weak forms for different flow variables
in the spatially continuous sense. Then we will specify the approximation
spaces and provide the fully discrete formulation.

Thanks to the term involving $\frac{S}{\eta^2}$,
each of the ($N-1$)
equations in \eqref{equ:phi_1} can be equivalently reformulated into
two de-coupled Helmholtz-type equations (see \cite{Dong2017} for details):
\begin{subequations}
\begin{align}
&
\nabla^2\psi_i^{n+1} - \left(\alpha + \frac{S}{\eta^2}  \right)\psi_i^{n+1} = Q_i + \nabla^2 R_i,
\quad 1\leqslant i\leqslant N-1, \label{equ:CH_psi} \\
&
\nabla^2 c_i^{n+1} + \alpha c_i^{n+1} = \psi_i^{n+1}, 
\quad 1\leqslant i\leqslant N-1, \label{equ:CH_phi}
\end{align}
\end{subequations}
where $\psi_i^{n+1}$ is an auxiliary variable defined by \eqref{equ:CH_phi},
and 
\begin{equation}
\left\{
\begin{split}
&
Q_i = \frac{1}{m_0}\left(
  g_i^{n+1} - \mathbf{u}^{*,n+1}\cdot\nabla c_i^{*,n+1} + \frac{\hat{c}_i}{\Delta t} 
\right), \quad 1\leqslant i\leqslant N-1, \\
&
R_i = -\frac{S}{\eta^2}c_i^{*,n+1} + \sum_{j=1}^{N-1}\zeta_{ij}h_j(\vec{c}^{*,n+1}), 
\quad 1\leqslant i\leqslant N-1, \\
&
\alpha = \frac{S}{2\eta^2}\left[
  -1 + \sqrt{1 - \frac{4\gamma_0}{m_0\Delta t}\left(\frac{\eta^2}{S}  \right)^2 }
\right].
\end{split}
\right.
\end{equation}
The reformulation also results in the following condition that
the chosen constant $S$ must satisfy,
$S \geqslant \eta^2\sqrt{\frac{4\gamma_0}{m_0\Delta t}}$.
It is noted that this condition implies $\alpha <0$
and $\alpha+\frac{S}{\eta^2}>0$ in \eqref{equ:CH_psi} and \eqref{equ:CH_phi}.


Let $\varphi(\mathbf{x})$ denote an arbitrary function on $\Omega$ with
sufficient regularity and satisfying the condition
\begin{equation}
\varphi(\mathbf{x}) = 0, \quad \text{on} \ \partial\Omega_i.
\label{equ:cond_0_phi}
\end{equation}
Taking the $L^2$ inner product between $\varphi$ and equation \eqref{equ:CH_psi}
leads to
\begin{equation}
\begin{split}
\int_{\Omega} \nabla\psi_i^{n+1}\cdot\nabla\varphi
&+ \left(\alpha + \frac{S}{\eta^2}  \right)\int_{\Omega} \psi_i^{n+1}\varphi
= -\int_{\Omega} Q_i\varphi + \int_{\Omega}\nabla R_i\cdot\nabla\varphi \\
&+ \int_{\partial\Omega_w}\left(\mathbf{n}\cdot\nabla\psi_i^{n+1} 
  - \mathbf{n}\cdot\nabla R_i \right)\varphi 
  + \int_{\partial\Omega_o}\left(\mathbf{n}\cdot\nabla\psi_i^{n+1}
  - \mathbf{n}\cdot\nabla R_i \right)\varphi,
  \quad \forall \varphi,
\end{split}
\label{equ:psi_weak_1}
\end{equation}
where we have used integration by part,
the divergence theorem
and the condition \eqref{equ:cond_0_phi}.
In light of \eqref{equ:CH_phi} and \eqref{equ:phi_5},
the condition \eqref{equ:phi_4} can be transformed into
\begin{equation}
\mathbf{n}\cdot\nabla\psi_i^{n+1} - \mathbf{n}\cdot\nabla R_i
 = \left(\alpha + \frac{S}{\eta^2}  \right)g_{ci}^{n+1} - g_{bi}^{n+1},
\quad \text{on} \ \partial\Omega_w.
\end{equation}
Similarly, for $\partial\Omega_o$ the condition
\eqref{equ:phi_4} can be transformed into
\begin{equation}
\mathbf{n}\cdot\nabla\psi_i^{n+1} - \mathbf{n}\cdot\nabla R_i
= \left(\alpha + \frac{S}{\eta^2}  \right)\left(
  -d_0\left.\frac{\partial c_i}{\partial t} \right|^{n+1}_{exp} + g_{ei}^{n+1}\right)- g_{bi}^{n+1},
\quad \text{on}\ \partial\Omega_o
\end{equation}
where we have used \eqref{equ:CH_phi} and \eqref{equ:phi_6}.
Substitution of the above two expression into \eqref{equ:psi_weak_1}
leads to the weak form for $\psi_i^{n+1}$,
\begin{equation}
\begin{split}
\int_{\Omega} &\nabla\psi_i^{n+1}\cdot\nabla\varphi
+ \left(\alpha + \frac{S}{\eta^2}  \right)\int_{\Omega} \psi_i^{n+1}\varphi
= -\int_{\Omega} Q_i\varphi + \int_{\Omega}\nabla R_i\cdot\nabla\varphi 
- \int_{\partial\Omega_w\cup\partial\Omega_o}g_{bi}^{n+1}\varphi \\
&+ \left(\alpha + \frac{S}{\eta^2}  \right)\int_{\partial\Omega_w}g_{ci}^{n+1}\varphi
+ \left(\alpha + \frac{S}{\eta^2}  \right)
\int_{\partial\Omega_o}\left(-d_0\left.\frac{\partial c_i}{\partial t} \right|^{n+1}_{exp} + g_{ei}^{n+1}\right)\varphi,
\quad 1\leqslant i\leqslant N-1,
\ \ \forall \varphi.
\end{split}
\label{equ:psi_weakform}
\end{equation}
%
In light of \eqref{equ:CH_phi} and \eqref{equ:phi_3}, the discrete
condition \eqref{equ:phi_2} can be
transformed into
\begin{equation}
  \psi_i^{n+1} = \alpha c_{bi}^{n+1}
  + \sum_{j=1}^{N-1}\zeta_{ij}h_j(\vec{c}_b^{n+1}) - g_{ai}^{n+1},
  \quad 1\leqslant i\leqslant N-1,
  \quad \text{on} \ \partial\Omega_i,
  \label{equ:dbc_psi}
\end{equation}
where $\vec{c}_b=(c_{b1}, c_{b2},\dots,c_{bN-1})$.


Take the $L^2$ inner product between $\varphi(\mathbf{x})$
and equation \eqref{equ:CH_phi}, and we have
\begin{equation}
  \int_{\Omega}\nabla c_i^{n+1}\cdot\nabla\varphi
  - \alpha\int_{\Omega}c_i^{n+1}\varphi
  = -\int_{\Omega}\psi_i^{n+1}\varphi
  + \int_{\partial\Omega_w}\mathbf{n}\cdot\nabla c_i^{n+1}\varphi
  + \int_{\partial\Omega_o}\mathbf{n}\cdot\nabla c_i^{n+1}\varphi,
  \quad \forall \varphi
  \label{equ:phi_weak_1}
\end{equation}
where we have used integration by part,
the divergence theorem and equation \eqref{equ:cond_0_phi}.
Substitution of the expressions
\eqref{equ:phi_5} and \eqref{equ:phi_7} into
the above equation leads to the weak form for $c_i^{n+1}$,
\begin{multline}
  \int_{\Omega}\nabla c_i^{n+1}\cdot\nabla\varphi
  - \alpha\int_{\Omega}c_i^{n+1}\varphi
  + \frac{\gamma_0 d_0}{\Delta t}\int_{\partial\Omega_o}c_i^{n+1}\varphi
  = -\int_{\Omega}\psi_i^{n+1}\varphi
  + \int_{\partial\Omega_w} g_{ci}^{n+1}\varphi \\
  + \int_{\partial\Omega_o}\left(\frac{d_0}{\Delta t}\hat{c}_i + g_{ei}^{n+1} \right)\varphi,
  \quad 1\leqslant i\leqslant N-1,
  \ \ \forall \varphi.
  \label{equ:phi_weakform}
\end{multline}


We re-write \eqref{equ:pressure_1} into
\begin{equation}
  \frac{\gamma_0}{\Delta t}\tilde{\mathbf{u}}^{n+1}
  + \frac{1}{\rho_0}\nabla P^{n+1}
  = \mathbf{G}^{n+1}
  -\frac{\mu^{n+1}}{\rho^{n+1}}\nabla\times\bm{\omega}^{*,n+1}
  \label{equ:pressure_1_trans}
\end{equation}
where the vorticity is $\bm{\omega} = \nabla\times\mathbf{u}$ and
\begin{equation}
  \begin{split}
  \mathbf{G}^{n+1} =& \frac{1}{\rho^{n+1}}\left[
    \mathbf{f}^{n+1} - \tilde{\mathbf{J}}^{n+1}\cdot\nabla\mathbf{u}^{*,n+1}
    + \nabla\mu^{n+1}\cdot\mathbf{D}(\mathbf{u}^{*,n+1})
    - \sum_{i,j=1}^{N-1}\lambda_{ij}\left(\psi_j^{n+1}-\alpha c_j^{n+1}\right)\nabla c_i^{n+1}
    \right] \\
  &
  - \mathbf{u}^{*,n+1}\cdot\nabla\mathbf{u}^{*,n+1}
  + \frac{\hat{\mathbf{u}}}{\Delta t}
  + \left(\frac{1}{\rho_0}-\frac{1}{\rho^{n+1}}  \right)\nabla P^{*,n+1}
  \end{split}
  \label{equ:G_expr}
\end{equation}
Let $q(\mathbf{x})$ denote an arbitrary function with sufficient regularity
and satisfying the condition
\begin{equation}
  q(\mathbf{x}) = 0, \quad \text{on} \ \partial\Omega_o.
  \label{equ:cond_0_p}
\end{equation}
Taking the $L^2$ inner product between equation \eqref{equ:pressure_1_trans}
and $\nabla q$ leads to
\begin{equation}
  \frac{1}{\rho_0}\int_{\Omega}\nabla P^{n+1}\cdot\nabla q
  = \int_{\Omega}\mathbf{G}^{n+1}\cdot\nabla q
  -\int_{\Omega}\frac{\mu^{n+1}}{\rho^{n+1}}\nabla\times\bm{\omega}^{*,n+1}\cdot\nabla q
  - \frac{\gamma_0}{\Delta t}\int_{\partial\Omega_i\cup\partial\Omega_w}\mathbf{n}\cdot\mathbf{w}^{n+1} q,
  \quad \forall q
\end{equation}
where we have used integration by part, the divergence theorem
and the condition \eqref{equ:cond_0_p}.
In light of the identity
$
\frac{\mu}{\rho}\nabla\times\bm{\omega}\cdot\nabla q
=\nabla\cdot\left(\frac{\mu}{\rho}\bm{\omega}\times\nabla q  \right)
-\nabla\left(\frac{\mu}{\rho} \right)\times\bm{\omega}\cdot\nabla q,
$
the above equation is transformed into
the weak form about $P^{n+1}$
\begin{equation}
  \begin{split}
  \int_{\Omega}\nabla P^{n+1}\cdot\nabla q
  =& \rho_0\int_{\Omega}\left[\mathbf{G}^{n+1}
    + \nabla\left(\frac{\mu^{n+1}}{\rho^{n+1}}  \right)\times\bm{\omega}^{*,n+1} \right]\cdot\nabla q \\
  &
  -\rho_0\int_{\partial\Omega_i\cup\partial\Omega_w\cup\partial\Omega_o}\frac{\mu^{n+1}}{\rho^{n+1}}\mathbf{n}\times\bm{\omega}^{*,n+1}\cdot\nabla q
  - \frac{\gamma_0\rho_0}{\Delta t}\int_{\partial\Omega_i\cup\partial\Omega_w}\mathbf{n}\cdot\mathbf{w}^{n+1} q,
  \quad \forall q.
  \end{split}
  \label{equ:p_weakform}
\end{equation}


Sum up equations \eqref{equ:vel_1} and \eqref{equ:pressure_1} and we get
\begin{equation}
  \frac{\gamma_0}{\nu_m\Delta t}\mathbf{u}^{n+1}
  -\nabla^2\mathbf{u}^{n+1}
  = \frac{1}{\nu_m}\left(\mathbf{G}^{n+1}-\frac{1}{\rho_0}\nabla P^{n+1}  \right)
  -\frac{1}{\nu_m}\left(\frac{\mu^{n+1}}{\rho^{n+1}}-\nu_m  \right)
  \nabla\times\bm{\omega}^{*,n+1}.
  \label{equ:vel_1_trans}
\end{equation}
Let $\varpi(\mathbf{x})$ be an arbitrary scalar function with sufficient regularity
and satisfying the condition
\begin{equation}
\varpi(\mathbf{x}) = 0, \quad \text{on} \ \partial\Omega_i\cup\partial\Omega_w.
\label{equ:cond_0_vel}
\end{equation}
Taking the $L^2$ inner product between
$\varpi(\mathbf{x})$ and equation \eqref{equ:vel_1_trans}
leads to
\begin{equation}
  \begin{split}
    \frac{\gamma_0}{\nu_m\Delta t}\int_{\Omega}\mathbf{u}^{n+1}\varpi
    &+ \int_{\Omega}\nabla\varpi\cdot\nabla\mathbf{u}^{n+1} 
    = \frac{1}{\nu_m}\int_{\Omega}\left(\mathbf{G}^{n+1}-\frac{1}{\rho_0}\nabla P^{n+1}   \right)
    \varpi \\
    &- \frac{1}{\nu_m}\int_{\Omega}\left( \frac{\mu^{n+1}}{\rho^{n+1}}-\nu_m \right)\nabla\times\bm{\omega}^{*,n+1}\varpi
    + \int_{\partial\Omega_o}\mathbf{n}\cdot\nabla\mathbf{u}^{n+1}\varpi,
    \quad \forall \varpi
  \end{split}
  \label{equ:vel_weak_1}
\end{equation}
where we have used integration by part,
the divergence theorem and the condition \eqref{equ:cond_0_vel}.
Noting the relation
\begin{equation*}
  \int_{\Omega}\left( \frac{\mu}{\rho}-\nu_m \right)\nabla\times\bm{\omega}\varpi =
  \int_{\Omega}\left( \frac{\mu}{\rho}-\nu_m \right)\bm{\omega}\times\nabla\varpi - 
  \int_{\Omega}\nabla\left(\frac{\mu}{\rho}  \right)\times\bm{\omega}\varpi
  + \int_{\partial\Omega}\left(\frac{\mu}{\rho}-\nu_m \right)\mathbf{n}\times\bm{\omega}\varpi
\end{equation*}
and in light of \eqref{equ:vel_3}, we can transform
\eqref{equ:vel_weak_1} into
\begin{equation}
  \begin{split}
   & \frac{\gamma_0}{\nu_m\Delta t}\int_{\Omega}\mathbf{u}^{n+1}\varpi
    + \int_{\Omega}\nabla\varpi\cdot\nabla\mathbf{u}^{n+1} \\  
    &= \frac{1}{\nu_m}\int_{\Omega}\left(\mathbf{G}^{n+1}
    -\frac{1}{\rho_0}\nabla P^{n+1}
    +\nabla\left(\frac{\mu^{n+1}}{\rho^{n+1}}  \right)\times\bm{\omega}^{*,n+1}   \right)\varpi \\
    & \ \ \
    - \frac{1}{\nu_m}\int_{\Omega}\left( \frac{\mu^{n+1}}{\rho^{n+1}}-\nu_m \right)\bm{\omega}^{*,n+1}\times\nabla\varpi 
    -\frac{1}{\nu_m}\int_{\partial\Omega_o}\left(\frac{\mu^{n+1}}{\rho^{n+1}}-\nu_m \right)\mathbf{n}\times\bm{\omega}^{*,n+1}\varpi \\
    & \ \ \
    + \int_{\partial\Omega_o}\left\{
    - \mathbf{n}\cdot(\nabla\mathbf{u}^{*,n+1})^T
    + \left(1 - \frac{\mu^{n+1}}{\mu_0}  \right)\mathbf{n}\cdot\mathbf{D}(\mathbf{u}^{*,n+1})
    \right. \\
    & \qquad\qquad \ \ \
    \left.
    + \frac{1}{\mu_0}\left[
      P^{n+1}\mathbf{n} + H(\vec{c}^{n+1})\mathbf{n}
      + \mathbf{E}(\mathbf{n},\mathbf{u}^{*,n+1},\rho^{n+1}) 
      + \mathbf{f}_b^{n+1} -\mu_0(\nabla\cdot\mathbf{u}^{*,n+1})\mathbf{n}
      \right]
    \right\}\varpi,
    \ \ \forall \varpi
  \end{split}
  \label{equ:vel_weakform}
\end{equation}
which is the weak form about $\mathbf{u}^{n+1}$.

Let $H^1(\Omega)$ denote the set of globally continuous
square-integrable functions on $\Omega$ with square-integrable
derivatives.
Define
\begin{equation}
\left\{
\begin{split}
&
H_{c0}^1(\Omega) = \left\{\
  v\in H^1(\Omega) \ : \ v|_{\partial\Omega_i} = 0
\ \right\}, \\
&
H_{p0}^1(\Omega) = \left\{\
  v\in H^1(\Omega) \ : \ v|_{\partial\Omega_o} = 0
\ \right\}, \\
&
H_{u0}^1(\Omega) = \left\{\
  v\in H^1(\Omega) \ : \ v|_{\partial\Omega_i\cup\partial\Omega_w} = 0
\ \right\}.
\end{split}
\right.
\label{equ:def_space_0}
\end{equation} 
We require that the equations \eqref{equ:psi_weakform} and
\eqref{equ:phi_weakform} hold for all $\varphi \in H_{c0}(\Omega)$,
and that equation \eqref{equ:p_weakform} holds for all
$q \in H_{p0}(\Omega)$, and that
equation \eqref{equ:vel_weakform} holds for
all $\varpi \in H_{u0}(\Omega)$.

To discretize these equations using $C^0$ spectral
elements, we first partition the domain $\Omega$
using a spectral element mesh.
Let $\Omega_h$ denote the discretized $\Omega$,
$
\Omega_h = \cup_{e=1}^{N_{el}}\Omega_h^e,
$
where $\Omega_h^e$ ($1\leqslant e\leqslant N_{el}$) denotes the element $e$ and
$N_{el}$ is the number of elements in the mesh.
Let $\partial\Omega_h$, $\partial\Omega_{ih}$, $\partial\Omega_{wh}$
and $\partial\Omega_{oh}$ denote the 
discretized boundaries of different types,
$
\partial\Omega_h = \partial\Omega_{ih}\cup\partial\Omega_{wh}\cup\partial\Omega_{oh}.
$
Let $d$ ($d=2$ or $3$) denote
the dimension in space, and $\Pi_{K}(\Omega_h^e)$ denote
the linear space of polynomials defined on $\Omega_h^e$
whose degrees are characterized by $K$ ($K$ is referred
to the element order hereafter). 
Define
\begin{equation}
  \left\{
  \begin{split}
    &
    X_{h} = \{\ v\in H^1(\Omega_h) \ :\ v|_{\Omega_h^e}\in \Pi_{K}(\Omega_h^e),
    \ 1\leqslant e\leqslant N_{el}  \ \}, \\
    &
    X_{h0}^{u} = \{\ v\in X_{h} \ :\ v|_{\partial\Omega_{ih}\cup\partial\Omega_{wh}}=0    \ \}, \\
    &
    X_{h0}^p = \{\ v\in X_h \ :\ v|_{\partial\Omega_{oh}}=0    \ \}, \\
    &
    X_{h0}^c = \{\ v\in X_h \ :\ v|_{\partial\Omega_{ih}}=0    \ \},
  \end{split}
  \right.
\end{equation}
In the following we use subscript in $(\cdot)_h$ to represent
the discretized version of $(\cdot)$.
The fully discretized equations consists of
the following: \\
\underline{For $\psi_{ih}^{n+1}$:}
find $\psi_{ih}^{n+1} \in X_h$ such that
\begin{equation}
\begin{split}
\int_{\Omega_h} &\nabla\psi_{ih}^{n+1}\cdot\nabla\varphi_h
+ \left(\alpha + \frac{S}{\eta^2}  \right)\int_{\Omega_h} \psi_{ih}^{n+1}\varphi_h
= -\int_{\Omega_h} Q_{ih}\varphi_h + \int_{\Omega_h}\nabla R_{ih}\cdot\nabla\varphi_h \\
&
- \int_{\partial\Omega_{wh}\cup\partial\Omega_{oh}}g_{bih}^{n+1}\varphi_h 
+ \left(\alpha + \frac{S}{\eta^2}  \right)\int_{\partial\Omega_{wh}}g_{cih}^{n+1}\varphi_h \\
&
+ \left(\alpha + \frac{S}{\eta^2}  \right)
\int_{\partial\Omega_{oh}}\left(-d_0\left.\frac{\partial c_{ih}}{\partial t} \right|^{n+1}_{exp} + g_{eih}^{n+1}\right)\varphi_h,
\quad 1\leqslant i\leqslant N-1,
\quad \forall \varphi_h \in X_{h0}^{c},
\end{split}
\label{equ:psi_weakform_disc}
\end{equation}
and
\begin{equation}
  \psi_{ih}^{n+1} = \alpha c_{bih}^{n+1}
  + \sum_{j=1}^{N-1}\zeta_{ij}h_j(\vec{c}_{bh}^{n+1}) - g_{aih}^{n+1},
  \quad 1\leqslant i\leqslant N-1,
  \quad \text{on} \ \partial\Omega_{ih}.
  \label{equ:dbc_psi_disc}
\end{equation}
\underline{For $c_{ih}^{n+1}$:} find $c_{ih}^{n+1} \in X_h$ such that
\begin{multline}
  \int_{\Omega_h}\nabla c_{ih}^{n+1}\cdot\nabla\varphi_h
  - \alpha\int_{\Omega_h}c_{ih}^{n+1}\varphi_h
  + \frac{\gamma_0 d_0}{\Delta t}\int_{\partial\Omega_{oh}}c_{ih}^{n+1}\varphi_h
  = -\int_{\Omega_h}\psi_{ih}^{n+1}\varphi_h
  + \int_{\partial\Omega_{wh}} g_{cih}^{n+1}\varphi_h \\
  + \int_{\partial\Omega_{oh}}\left(\frac{d_0}{\Delta t}\hat{c}_{ih} + g_{eih}^{n+1} \right)\varphi_h,
  \quad 1\leqslant i\leqslant N-1,
  \quad \forall \varphi_h \in X_{h0}^c,
  \label{equ:phi_weakform_disc}
\end{multline}
and
\begin{equation}
  c_{ih}^{n+1} = c_{bih}^{n+1},
  \quad 1\leqslant i\leqslant N-1,
  \quad \text{on} \ \partial\Omega_{ih}.
  \label{equ:dbc_phi_disc}
\end{equation}
\underline{For $P^{n+1}_h$:} find $P_h^{n+1}\in X_h$ such that
\begin{equation}
  \begin{split}
  &\int_{\Omega_h}\nabla P_h^{n+1}\cdot\nabla q_h
  = \rho_0\int_{\Omega_h}\left[\mathbf{G}_h^{n+1}
    + \nabla\left(\frac{\mu_h^{n+1}}{\rho_h^{n+1}}  \right)\times\bm{\omega}_h^{*,n+1} \right]\cdot\nabla q_h \\
  &
  -\rho_0\int_{\partial\Omega_{ih}\cup\partial\Omega_{wh}\cup\partial\Omega_{oh}}\frac{\mu_h^{n+1}}{\rho_h^{n+1}}\mathbf{n}_h\times\bm{\omega}_h^{*,n+1}\cdot\nabla q_h
  - \frac{\gamma_0\rho_0}{\Delta t}\int_{\partial\Omega_{ih}\cup\partial\Omega_{wh}}\mathbf{n}_h\cdot\mathbf{w}_h^{n+1} q_h,
  \quad \forall q_h\in X_{h0}^p,
  \end{split}
  \label{equ:p_weakform_disc}
\end{equation}
and
\begin{equation}
P_h^{n+1} = \mu_h^{n+1}\mathbf{n}_h\cdot\mathbf{D}(\mathbf{u}_h^{*,n+1})\cdot\mathbf{n}_h
-H(\vec{c}_h^{n+1}) - \mathbf{n}_h\cdot\mathbf{E}(\mathbf{n}_h,\mathbf{u}_h^{*,n+1},\rho_h^{n+1})
-\mathbf{f}_{bh}^{n+1}\cdot\mathbf{n}_h,
\quad \text{on} \ \partial\Omega_{oh}.
\label{equ:dbc_p_disc}
\end{equation}
\underline{For $\mathbf{u}_h^{n+1}$:} find
$\mathbf{u}_h^{n+1} \in [X_h]^d$ such that
\begin{equation}
  \begin{split}
   & \frac{\gamma_0}{\nu_m\Delta t}\int_{\Omega_h}\mathbf{u}_h^{n+1}\varpi_h
    + \int_{\Omega_h}\nabla\varpi_h\cdot\nabla\mathbf{u}_h^{n+1} \\  
    &= \frac{1}{\nu_m}\int_{\Omega_h}\left(\mathbf{G}_h^{n+1}
    -\frac{1}{\rho_0}\nabla P_h^{n+1}
    +\nabla\left(\frac{\mu_h^{n+1}}{\rho_h^{n+1}}  \right)\times\bm{\omega}_h^{*,n+1}   \right)\varpi_h \\
    & \ \ \
    - \frac{1}{\nu_m}\int_{\Omega_h}\left( \frac{\mu_h^{n+1}}{\rho_h^{n+1}}-\nu_m \right)\bm{\omega}_h^{*,n+1}\times\nabla\varpi_h 
    -\frac{1}{\nu_m}\int_{\partial\Omega_{oh}}\left(\frac{\mu_h^{n+1}}{\rho_h^{n+1}}-\nu_m \right)\mathbf{n}_h\times\bm{\omega}_h^{*,n+1}\varpi_h \\
    & \ \ \
    + \int_{\partial\Omega_{oh}}\left\{
    - \mathbf{n}_h\cdot(\nabla\mathbf{u}_h^{*,n+1})^T
    + \left(1 - \frac{\mu_h^{n+1}}{\mu_0}  \right)\mathbf{n}_h\cdot\mathbf{D}(\mathbf{u}_h^{*,n+1})
    \right. \\
    & \qquad\qquad 
    \left.
    + \frac{1}{\mu_0}\left[
      P_h^{n+1}\mathbf{n}_h + H(\vec{c}_h^{n+1})\mathbf{n}_h
      + \mathbf{E}(\mathbf{n}_h,\mathbf{u}_h^{*,n+1},\rho_h^{n+1}) 
      + \mathbf{f}_{bh}^{n+1} -\mu_0(\nabla\cdot\mathbf{u}_h^{*,n+1})\mathbf{n}_h
      \right]
    \right\}\varpi_h, \\
    & \quad \forall \varpi_h \in X_{h0}^u,
  \end{split}
  \label{equ:vel_weakform_disc}
\end{equation}
and
\begin{equation}
  \mathbf{u}_h^{n+1} = \mathbf{w}_h^{n+1},
  \quad \text{on} \ \partial\Omega_{ih}\cup\partial\Omega_{wh}.
  \label{equ:dbc_vel_disc}
\end{equation}


So the final solution procedure is as follows.
Given $(\mathbf{u}_h^n, P_h^n, \psi_{ih}^{n},c_{ih}^n)$,
we compute $\psi_{ih}^{n+1}$, $c_{ih}^{n+1}$,
$P_h^{n+1}$ and $\mathbf{u}_{h}^{n+1}$
successively through these steps:
\begin{itemize}

\item
  Solve \eqref{equ:psi_weakform_disc}, together with
  the Dirichlet condition \eqref{equ:dbc_psi_disc}, for $\psi_{ih}^{n+1}$;

\item
  Solve \eqref{equ:phi_weakform_disc}, together with
  the Dirichlet condition \eqref{equ:dbc_phi_disc},
  for $c_{ih}^{n+1}$;

\item
  Solve \eqref{equ:p_weakform_disc}, together with
  the Dirichlet condition \eqref{equ:dbc_p_disc},
  for $P_h^{n+1}$;

\item
  Solve \eqref{equ:vel_weakform_disc}, together with
  the Dirichlet condition \eqref{equ:dbc_vel_disc},
  for $\mathbf{u}_h^{n+1}$.

\end{itemize}
%
When implementing the Dirichlet condition \eqref{equ:dbc_p_disc},
it should be noted that a projection of
the computed pressure data onto $H^1(\partial\Omega_{oh})$
is needed with $C^0$ elements because of the
spatial derivatives involved in
the $\mathbf{D}(\mathbf{u})$ term.



It can be noted that the final algorithm only requires
the solution of a number of de-coupled individual
Helmholtz-type equations (including Poisson) within
a time step. The linear algebraic systems
after discretization involve only constant and time-independent
coefficient matrices for all flow variables,
even though large density contrasts and large viscosity
contrasts may be present with the different fluids.
Therefore these coefficient matrices can be pre-computed,
which makes the computation very efficient in cases with
large density ratios and large viscosity ratios.


\section{Representative Numerical Examples}
\label{sec:tests}

In this section we provide extensive numerical results for several flow problems
involving multiple fluid components and inflow/outflow boundaries
in two dimensions
to test the set of open/outflow boundary conditions and the numerical
algorithm developed in the previous section. The results demonstrate that
the proposed method can serves as an accurate and reliable tool for the
investigation of multi-phase flow problems in unbounded domains.
Note that all numerical simulations presented here are performed
by using the volume fractions $c_i\,(1\leq i \leq N-1)$ as the order parameters, as defined in \eqref{equ:varphi_expr}.

To begin with, we briefly comment on the normalization of physical variables  and parameters, which has been addressed in detail in the previous works \cite{Dong2014,Dong2015,Dong2017}. Let $L$ denote a length scale, $U_0$ denote a velocity scale and $\varrho_d$ denote a density scale. By consistently normalizing the physical variables and parameters
based on the normalization constants given in Table \ref{table:normalization}, the resultant non-dimensionalized problem (governing equations, boundary/initial conditions) will retain the same form as its dimensional problem. Hereafter, all the flow variables and parameters
have  been appropriately normalized based on Table \ref{table:normalization}, unless otherwise specified.

\begin{table}[t]
\centering 
\begin{tabular}{l c| l c} 
\hline 
Variables/parameters & normalization constant & Variables/parameters & normalization constant \\  
\hline
$\bs x,\, \eta$& $L$ &$t,\,\Delta t$& $L/U_0$\\
$\bs u,\,\bs w$ &$U_0$ & $\rho,\,\rho_i,\,\tilde \rho_i,\rho_0,\,\varphi_i$ & $\varrho_d$\\
$S,\,c_i,c_{bi}$& 1&$g_i$&$U_0/L$\\
$\alpha,\,g_{ai}$&$1/L^2$&$g_{bi}$&$1/L^3$\\
$g_{ci},\,g_{ei}$&1/L&$d_0$&$1/U_0$\\
$\bs g_r$&$U_0^2/L$ &$\bs f$&$\varrho_d U_0^2/L$\\
$P,\,p,\, {\bs f}_b$&$\varrho_dU_0^2$&$\mu,\tilde \mu_i,\,\mu_0$&$\varrho_d U_0 L$\\
$\Gamma_{\mu},\,\nu_m$&$U_0L$&$\tilde \gamma_i,\, \Gamma$&$1/\varrho_d$\\
$\lambda_{ij}$&$\varrho_dU_0^2L^2$&$\sigma_{ij}$&$\varrho_d U_0^2L$\\
$m_0$&$U_0L^3$&$\zeta_{ij}$&$1/\varrho_dU_0^2L^2$\\
\hline 
\end{tabular}
\caption{Normalization of flow variables and simulation parameters. $L$ is a length scale, $U_0$ is a velocity scale, and $\varrho_d$ is a density scale. }
\label{table:normalization} 
\end{table}

\subsection{Convergence Rates}

\begin{figure}[tbp]
\centering
 \subfigure[Spatial and temporal convergence test]{ \includegraphics[scale=.50]{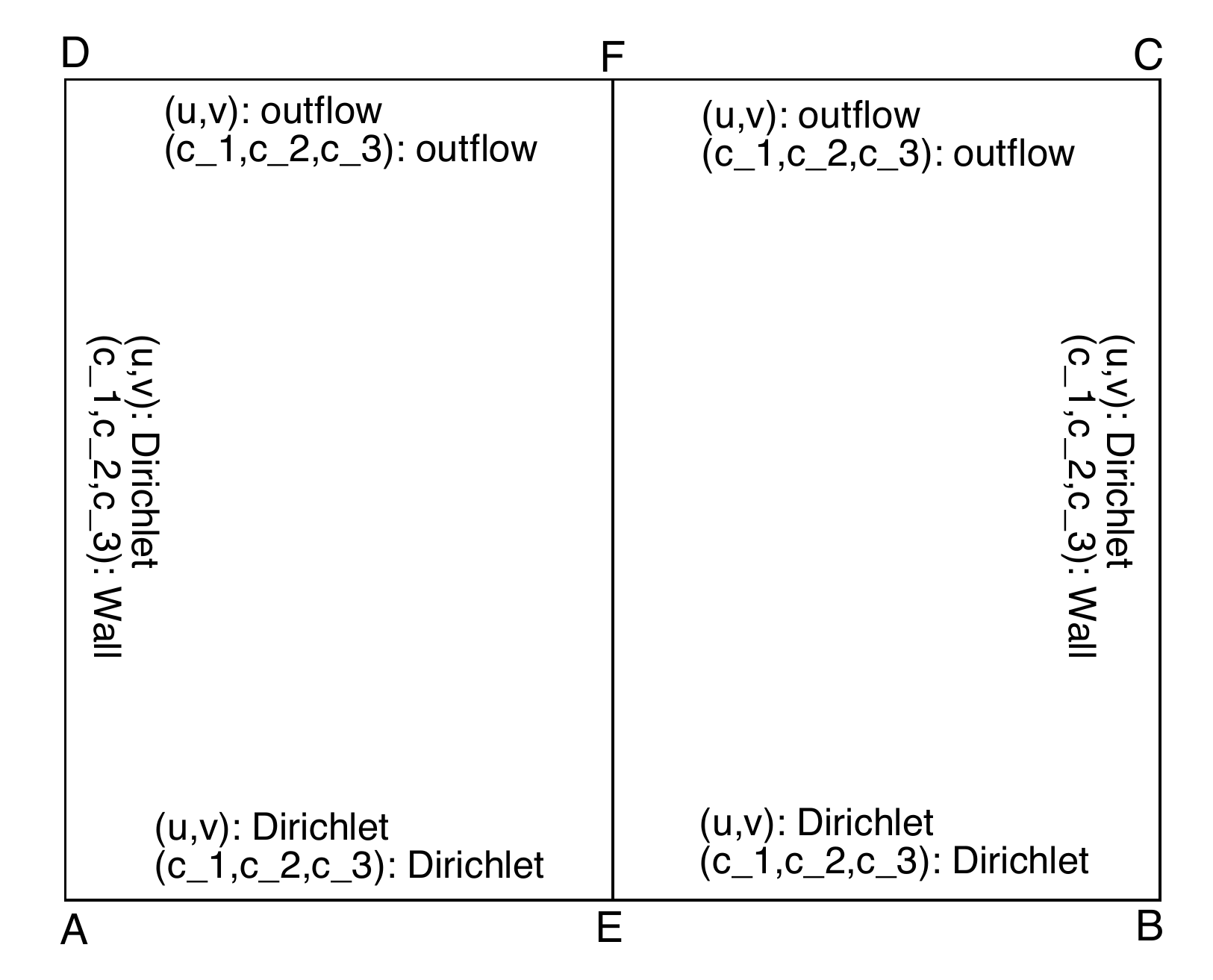}}\\
\subfigure[$L^2$ errors vs element order]{ \includegraphics[scale=.42]{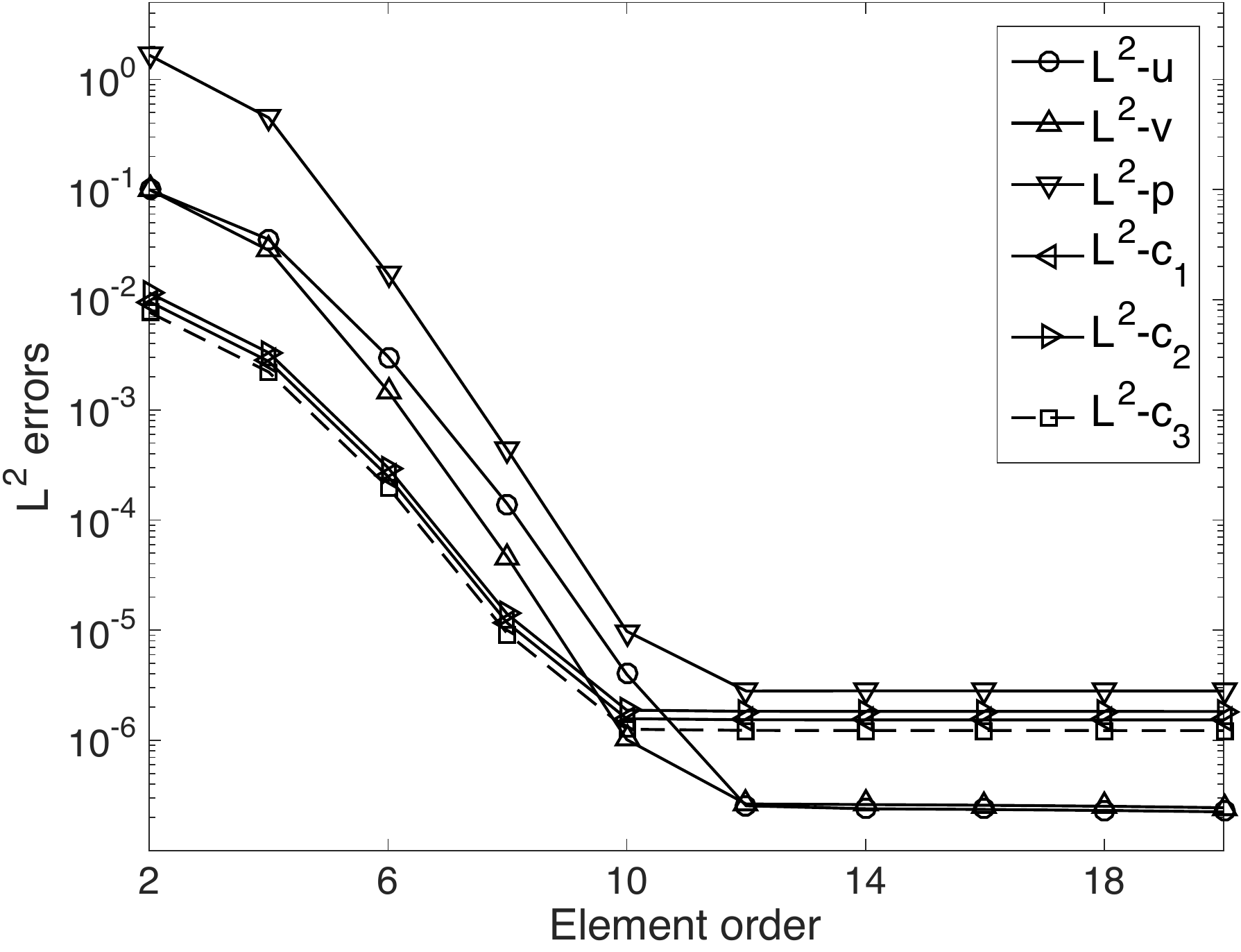}} \qquad
\subfigure[$L^2$ errors vs $\Delta t$]{ \includegraphics[scale=.42]{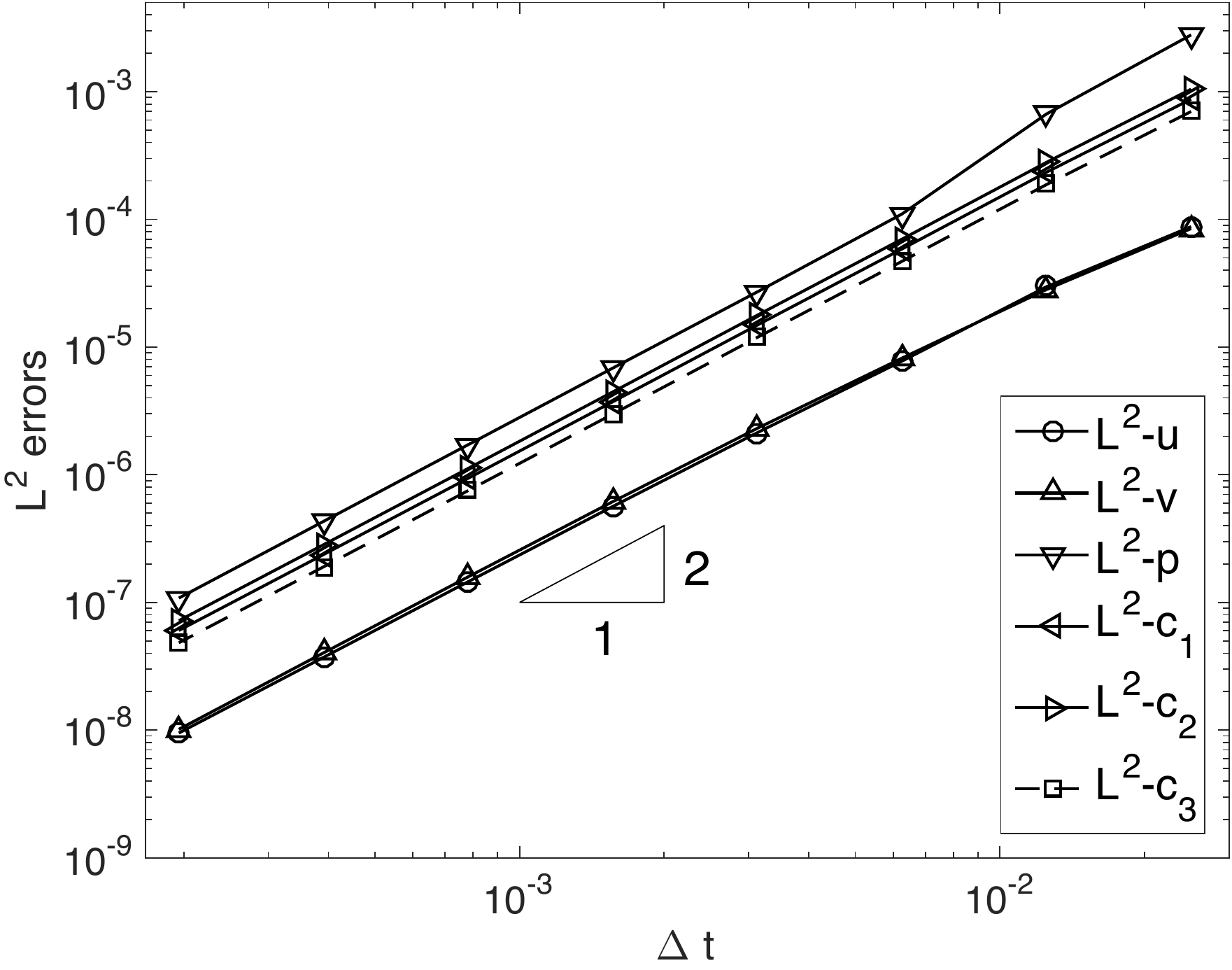}}
\caption{Spatial/temporal convergence tests: (a) Problem configuration; (b) $L^{2}$ errors of flow variables versus element order (fixed $\Delta t=0.001$ and $t_f=0.1$); (c) $L^{2}$ errors of flow variables versus $\Delta t$ (fixed element order $16$ and $t_f=0.1$). }
\label{standtest}
\end{figure}

The goal of this subsection is to demonstrate numerically the spatial and temporal convergence rates of the  method developed herein  using a contrived analytic solution with the proposed N-phase energy-stable open boundary conditions.

Consider the computational domain $\Omega=\overline{ABCD}:=\{(x,y)|  0\leq x \leq 2,  -1 \leq y \leq 1   \}$ shown in Fig. \ref{standtest}(a) and a four-fluid (i.e., $N=4$) mixture contained in this domain. We assume the following analytic expressions for the flow variables of this four-phase system,
\begin{equation}\label{equ:contrivedsolu}
\begin{cases}
&u=A_0 \sin(ax) \cos(\pi y) \sin(\omega_0 t)\, ,\\
&v=-(A_0a/\pi)\cos(ax)\sin(\pi y) \sin (\omega_0 t)\,,\\
&P=A_0 \sin(ax)\sin(\pi y) \cos(\omega_0 t)\,,\\
& c_1=\cfrac{1}{6}\big[1+A_1\cos(a_1x)\cos(b_1 y)\sin(\omega_1 t)\big]\, ,\\
& c_2=\cfrac{1}{6}\big[1+A_2\cos(a_2x)\cos(b_2 y)\sin(\omega_2 t)\big] \, ,\\
& c_3=\cfrac{1}{6}\big[1+A_3\cos(a_3x)\cos(b_3 y)\sin(\omega_3 t)\big]\, ,\\
&c_4=1-c_1-c_2-c_3,
\end{cases}
\end{equation}
where $(u,v)$ are the two components of the velocity $\bs u.$  The above expressions satisfy the system of equations with appropriate choice of the source terms. The source term $\bs f$ in \eqref{equ:nse_trans} is chosen such that the analytic expressions given in \eqref{equ:contrivedsolu} satisfy equation \eqref{equ:nse_trans}. We choose $ g_i\, (i=1,2,3)$ in equations \eqref{equ:CH} such that \eqref{equ:contrivedsolu} satisfies each of the equations \eqref{equ:CH}. The initial conditions \eqref{equ:ic_vel}-\eqref{equ:ic_phi} are imposed for the velocity and phase field functions, respectively, where $\bs u^{\rm in}$ and $c_i^{\rm in}\, (i=1,2,3)$ are chosen by letting $t=0$ at the contrived solution \eqref{equ:contrivedsolu}.

The flow domain $\Omega$ is discretized using two quadrilateral spectral elements of equal size ($\overline{AEFD}$ and $\overline{EBCF}$). On the sides $\overline{AD},\,\overline{AB},\,\overline{BC},$ we impose Dirichlet boundary condition \eqref{equ:dbc_vel} for the velocity field, where the boundary velocity $\bs w$ is chosen according to the analytic expressions in \eqref{equ:contrivedsolu}. For the phase field functions, we impose the wall contact-angle conditions \eqref{equ:bc_chempot} and \eqref{equ:wbc_phi_1_mod} on $\overline{AD}$ and $\overline{BC},$ and impose the Dirichlet conditions \eqref{equ:ibc_phi_1} and \eqref{equ:ibc_phi_2_mod} on $\overline{AB}.$ On the side $\overline{DC}$ we impose the open boundary conditions \eqref{equ:obc_vel_mod} with $(\theta,\alpha_1,\alpha_2)=(1,1,0)$  for the momentum equations,
and \eqref{equ:bc_chempot} and \eqref{equ:obc_phi_2_mod} for the phase field functions. The source terms $g_{ai},\, g_{bi},\, g_{ci},\,g_{ei},\, c_{bi},\, (i=1,2,3)$ and $\bs f_b$ therein are chosen such that the contrived solution in \eqref{equ:contrivedsolu} satisfies the boundary conditions.

\begin{table}[tb]
\centering 
\begin{tabular}{l c| l c} 
\hline 
Parameter & Value & Parameter & Value \\  
\hline
$a,\,a_1,\,a_2,\,a_3$ & $\pi$ & $b_1,\,b_2,\,b_3$ & $\pi$ \\ 
$A_0$ & 2.0 & $A_1,\,A_2,\,A_3$ & 1.0 \\
$\omega_0,\,\omega_1$ & 1.0 & $\omega_2$ & 1.2 \\
$\omega_3$ & 0.8 & $\eta$, $t_f$ & 0.1 \\
${\tilde \rho}_1$    & 1.0 &${\tilde \rho}_2$  & 3.0  \\ 
 ${\tilde \rho}_3$   &2.0  &${\tilde \rho}_4$  & 4.0   \\ 
${\tilde \mu}_1$    &0.01  &${\tilde \mu}_2$  &0.02   \\ 
${\tilde \mu}_3$     &0.03  &${\tilde \mu}_4$   &0.04   \\ 
 $\sigma_{12}$   &$6.236\times 10^{-3}$  & $\sigma_{13} $&$7.265\times 10^{-3}$   \\ 
 $\sigma_{14} $  &$3.727\times 10^{-3}$  &$\sigma_{23} $ &  $8.165\times 10^{-3}$ \\ 
$\sigma_{24} $   &$5.270\times 10^{-3}$  &$\sigma_{34} $ & $ 6.455\times 10^{-3}$  \\ 
$\alpha_1,\, \theta$&1.0 & $\alpha_2$& 0 \\
$ \delta$&0.05 &
$d_0$& 0.2\\
$m_0$    &$1.0\times 10^{-5}$  &$\mu_0$  & ${\rm max}(\tilde \mu_1,\cdots, \tilde \mu_4)$   \\ 
$\rho_0$ & ${\rm min}(\tilde \rho_1,\cdots,\tilde \rho_4) $   &$\nu_m$ & $\frac{1}{2} \big[{\rm max}\{ \frac{\mu_i}{\rho_i} \}_{i=1}^4+{\rm min}\{\frac{\mu_i}{\rho_i} \}_{i=1}^4\big]$ \\
 $J$ (temporal order)   &  2& Number of elements  & 2  \\ 
$\Delta t$    & (varied) & Element order  & (varied)   \\ 
\hline 
\end{tabular}
\caption{Simulation parameter values for the convergence-rate tests.}
\label{table:standtest} 
\end{table}

The numerical algorithm from Section \ref{sec:algorithm}-\ref{sect: SEM} is employed to integrate in time the governing equations for this four-phase system from $t=0$ to $t=t_f.$ Then the numerical solution and the exact solution as given by \eqref{equ:contrivedsolu} at $t=t_f$ are compared, and the errors in the  $L^2$  norm for various flow variables are computed.
All the physical and numerical parameters involved in the simulation of this problem, including the values of constants  $A_i$ and $\omega_i$ ($i=0,\cdots,3$), $a,$ $a_i$ and $b_i$ ($i=1,2,3$) in the contrived solution \eqref{equ:contrivedsolu},  are tabulated in Table \ref{table:standtest}. 

Both spatial and temporal convergence tests have been performed to demonstrate the reliability of the proposed algorithm. In the first test, we fix the integration time at $t_f=0.1$ and the time step size at $\Delta t=0.001$ ($100$ time steps), and vary the element order systematically between $2$ and $20.$ The same element order has been used for these two spectral elements. Fig. \ref{standtest}(b) plots the numerical errors at $t=t_f$ in $L^2$ norm for different flow variables as a function of the element order. It is evident that within a specific range of the element order (below around $12$), the errors decrease exponentially when increasing element order, displaying an exponential convergence rate in space. Beyond the element order of about $12,$ the error curves level off as the element order further increases, showing a saturation caused by the temporal truncation error. 

In the second test, we fix the integration time at $t_f=0.1$ and the element order at a large value $16,$ and vary the time step size systematically between $\Delta t=1.953125\times 10^{-4}$ and $\Delta t=0.025.$ Fig. \ref{standtest}(c) shows the numerical errors at $t=t_f$ in $L^2$ norm for different variables as a function of $\Delta t$ in logarithmic scales. It can be observed that the numerical errors exhibit a second order convergence rate in time.

The above numerical results indicate that the numerical algorithm developed herein has a spatial exponential convergence rate and a temporal second-order convergence rate for multi-phase problems with energy-stable open boundary conditions.

\subsection{A Three-Phase Capillary Wave Problem}

\begin{figure}[tbp]
\centering
 \subfigure[Configuration for three-phase capillary wave problem ]{ \includegraphics[scale=.4]{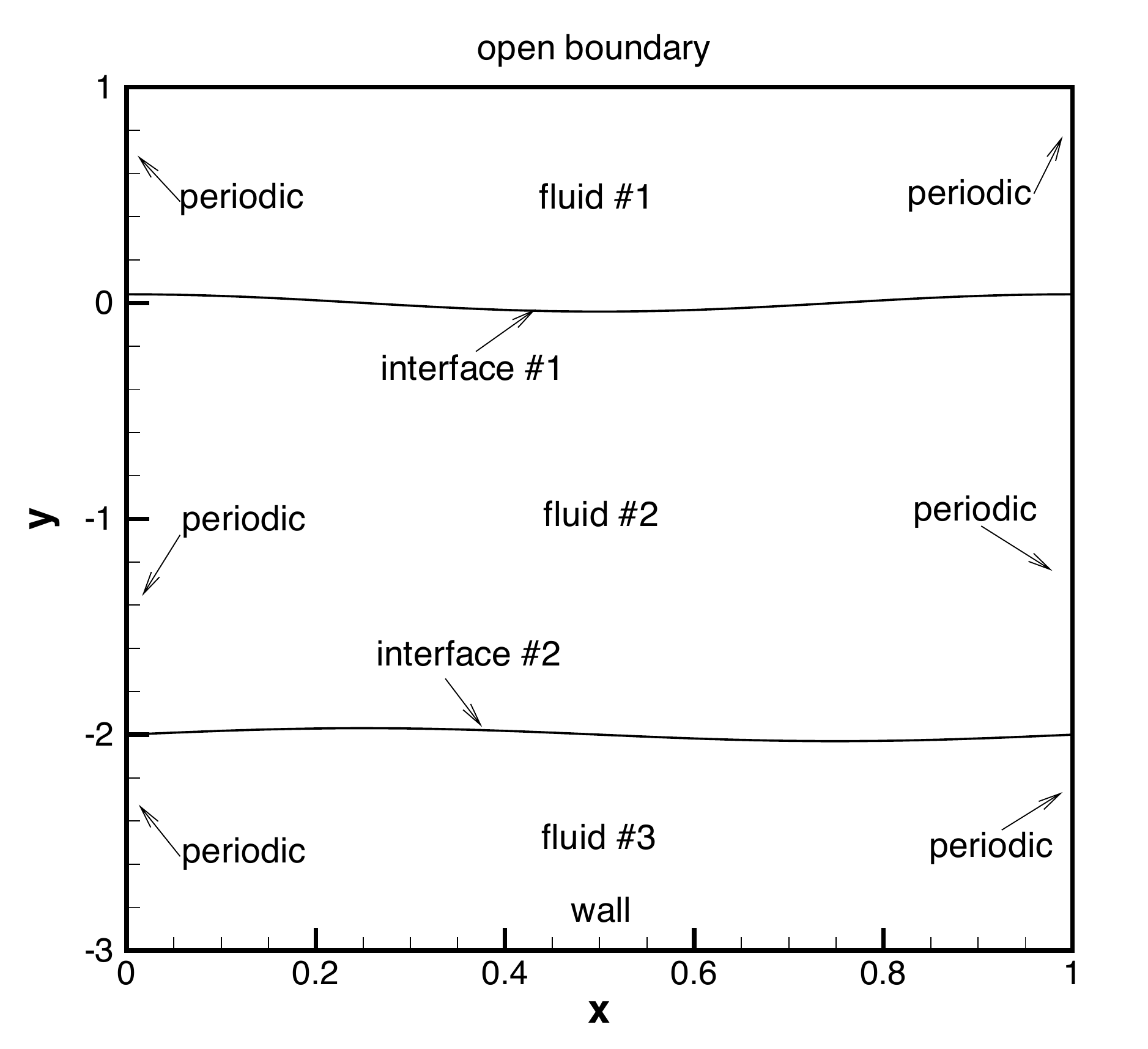}} \qquad
 \subfigure[Spectral element mesh ]{ \includegraphics[scale=.4]{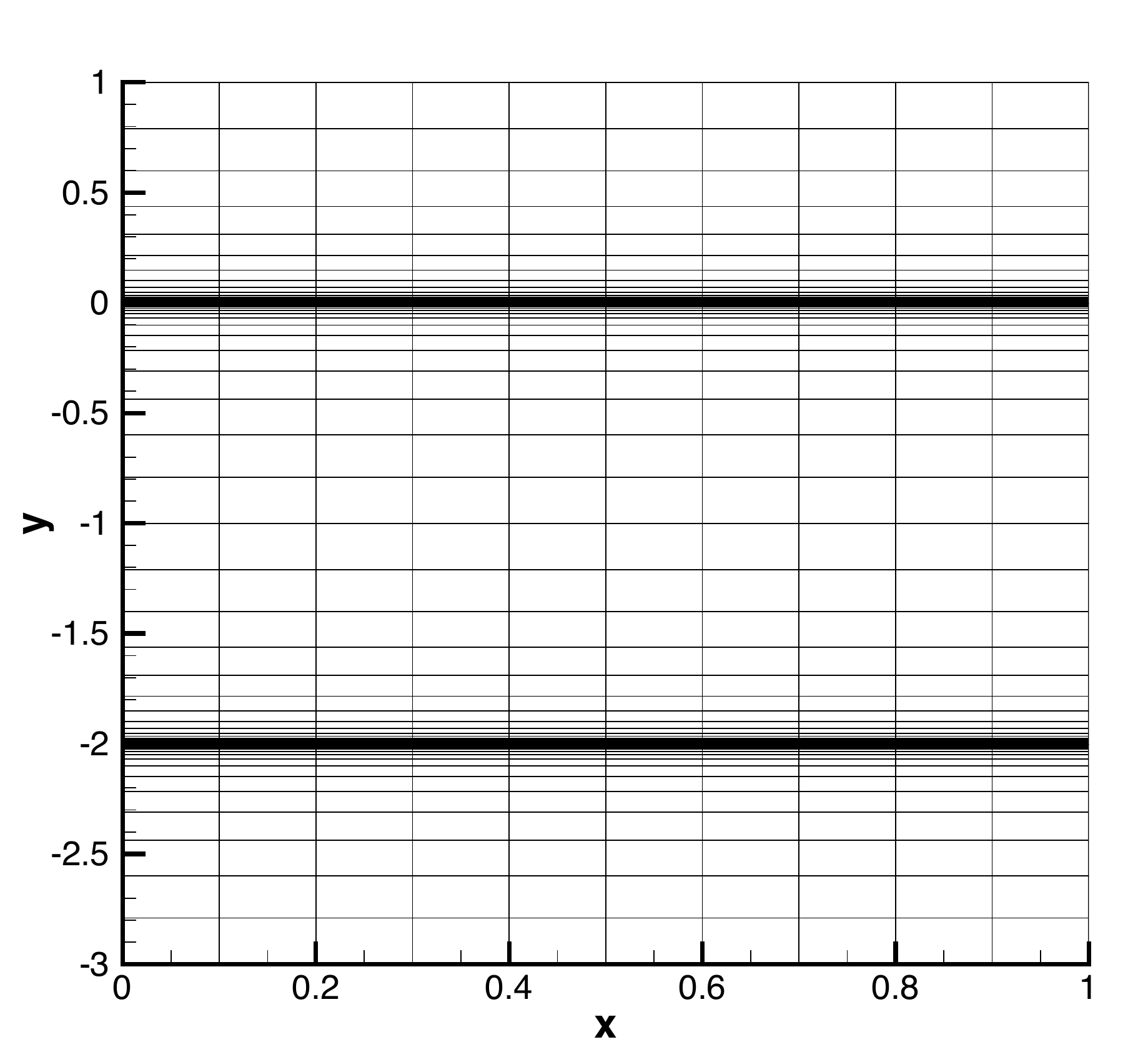}} 
\caption{Three-phase capillary wave problem: (a) Computational domain and configuration. (b) Spectral element mesh of $800$ quadrilateral elements.}
\label{CapillaryConfig}
\end{figure}

In this subsection, we use a three-phase capillary wave problem as a benchmark to
test the physical accuracy of the current method  with energy-stable open boundary conditions.

The problem setting is as follows.
We consider three immiscible incompressible fluids contained in an infinite domain
(see Fig.~\ref{CapillaryConfig}(a) for an illustration).
The upper portion of the domain is occupied by the lightest fluid (fluid $\#1$), and the lower portion of the domain is occupied by the heaviest fluid (fluid $\#3$), and the middle is occupied by fluid $\#2.$ The gravity is assumed to be in the downward direction.
The interfaces formed between fluid $\#1$ and fluid $\#2$ (interface $\#1$) and between fluid $\#2$ and fluid $\#3$ (interface $\#2$) are perturbed from their horizontal equilibrium positions by a small amplitude sinusoidal wave form, and start to oscillate at $t=0.$
The objective here is to study the motion of the interfaces over time. 

Although this is a three-phase problem,
if the two interfaces are far part and the capillary-wave amplitudes are sufficiently
small compared with the distance between the interfaces and
the dimension of the domain in the vertical direction,
the interaction between the interfaces will be weak.
The motion of each interface will therefore be essentially the same as
that of the interface alone in a two-phase setting, i.e.~with
the third fluid absent.
This allows us to compare qualitatively and quantitatively
the numerical results for the three-phase capillary-wave simulations
with e.g.~Prosperetti's exact physical solution
(see \cite{Prosperetti1981}) for two-phase capillary-wave problems.
In \cite{Prosperetti1981} an exact time-dependent standing-wave solution to the
two-phase capillary-wave problem was derived, given that the two fluids must have  matched kinematic viscosities (but their densities and dynamic viscosities can be different).

In what follows, we will simulate the three-phase capillary-wave problem
under the following settings:
(i) the two interfaces are far apart;
(ii) the capillary amplitudes
are small compared with both the distance between the interfaces and the
vertical dimension of the domain; and (iii)
the kinematic viscosity $\nu$ satisfies $ \nu=\frac{\tilde \mu_1}{\tilde \rho_1}=\frac{\tilde \mu_2}{\tilde \rho_2}=\frac{\tilde \mu_3}{\tilde \rho_3}$.

Specifically, the simulation setting is illustrated in Fig.~\ref{CapillaryConfig}(a).
We consider the computational domain $\Omega=\{(x,y)| 0\leq x \leq 1, -3 \leq y \leq 1   \}.$ The bottom side of the domain is a solid  wall of neutral wettability,
and the top side is open where the fluid can freely leave (or enter) the domain.
On the left and right sides, all the variables are assumed to be periodic at $x=0$ and $x=1.$ The equilibrium positions of the fluid interface $\#1$ and interface $\#2$ are assumed to coincide with $y=0$ and $y=-2$, respectively. The initial perturbed profile of the fluid interface $\#1$ and interface $\#2$ are given by $y=H_0 \cos(k_w x)$ and $y=y_1+H_0\cos(k_w x),$ respectively, where $y_1=-2,$ $H_0=0.01$ is the initial amplitude, $\lambda_w=1$ is the wavelength of the perturbation profiles, and $k_w=2\pi/\lambda_w$ is the wave number. Note that the initial capillary amplitude $H_0$ is small compared with the dimension of the domain in the vertical direction and the distance between the two fluid interfaces.
Therefore, the effect of the wall at the domain bottom and
the influence of the third fluid on the motion of
the fluid interface will be small.

\begin{table}[tbp]
\centering 
\begin{tabular}{l c| l c} 
\hline 
Parameter & Value & Parameter & Value \\  
\hline 
 $H_0$         &0.01            &   $k_w$ (wave number)          &  $2\pi$             \\
  $\sigma_{ij}\, (1\leq i\neq j\leq 3)$        &   1.0         &    $|\bs g_r|$ (gravity)          &     1.0          \\
 $\tilde \rho_1$         & 1.0           &   $\tilde \rho_2,\,\tilde \rho_3$           &  (varied)             \\
 $\tilde \mu_1$          & 0.01           & $\tilde \mu_2$             & $\tilde \mu_1 \frac{\tilde \rho_2}{\tilde \rho_1}$              \\
 $\tilde \mu_3$           &  $\tilde \mu_1 \frac{\tilde \rho_3}{\tilde \rho_1}$               & $\nu=\frac{\tilde \mu_1}{\tilde \rho_1}=\frac{\tilde \mu_2}{\tilde \rho_2}=\frac{\tilde \mu_3}{\tilde \rho_3}$ (kinematic viscosity)           & $0.01$              \\
 $\delta$&0.05&$\mu_0$ & ${\rm max}(\tilde \mu_1,\tilde \mu_2, \tilde \mu_3)$\\
$\theta$ &1.0&$\alpha_1,$ $\alpha_2$ & $0$ \\
$d_0$& 0 & Element order  & 8 \\
$\rho_0$ & ${\rm min}(\tilde \rho_1,\cdots,\tilde \rho_3) $   &$\nu_m$ & 0.01  \\
 $J$ (temporal order)   &  2& Number of elements  & 800  \\ 
$m_0$   &(varied)            &    $\eta$         & (varied)              \\
$\Delta t$    & (varied)   \\ 
\hline 
\end{tabular}
\caption{Simulation parameter values for the three-phase capillary wave problem.}
\label{table:capillary} 
\end{table}

The computational domain is partitioned with $800$  quadrilateral elements,
with $10$ and $80$ elements respectively in $x$ and $y$ directions (Fig.~\ref{CapillaryConfig}(b)). The elements are uniform in the $x$ direction, and are non-uniform and clustered around the regions $-0.012\leq y \leq 0.012$ and $-2.012 \leq y \leq -1.988.$
In the simulations, the external body force $\bs f$ in equation  \eqref{equ:nse_trans} is set to $\bs f=\rho \bs g_r,$ where $\bs g_r$ is the gravitational acceleration, and
the source terms in \eqref{equ:CH} are set to $g_i=0 \,(i=1,2).$ On the bottom wall, the boundary condition \eqref{equ:dbc_vel} with $\bs w=\bs 0$ is imposed for the velocity, and the boundary conditions \eqref{equ:bc_chempot} and \eqref{equ:wbc_phi_1_mod} with $g_{bi}=g_{ci}=0\,(i=1,2)$ are imposed for the phase field functions.
On the top domain boundary, the energy-stable open boundary condition \eqref{equ:obc_vel_mod}
with $\bs f_b=\bs 0$  and $(\theta,\alpha_1,\alpha_2)=(1,0,0)$ is imposed for the momentum equation, and the conditions
\eqref{equ:bc_chempot} and \eqref{equ:obc_phi_2_mod}
with $g_{bi}=g_{ei}=0\,(i=1,2)$ and $d_0=0$ are imposed for the phase field functions.
The initial velocity is set to zero, and the initial volume fractions are prescribed as follows:
\begin{equation}\label{equ:capiinitial}
\begin{cases}
&c_1=\cfrac{1}{2}\Big[ 1+\tanh \cfrac{y-H_0 \cos ( k_w x)}{\sqrt{2}  \eta}   \Big],\\
&c_2=\cfrac{1}{2}\Big[ \tanh \cfrac{y-y_1-H_0 \cos( k_w x)}{\sqrt{2} \eta}  - \tanh \cfrac{y-H_0 \cos( k_w x)}{\sqrt{2} \eta}                      \Big],\\
&c_3=1-c_1-c_2.
\end{cases}
\end{equation}
We list in Table \ref{table:capillary} the values for the physical and numerical parameters involved in this problem.

\begin{figure}[tbp]
\centering
\subfigure[Interface $\#1$ with various element orders]{ \includegraphics[scale=.4]{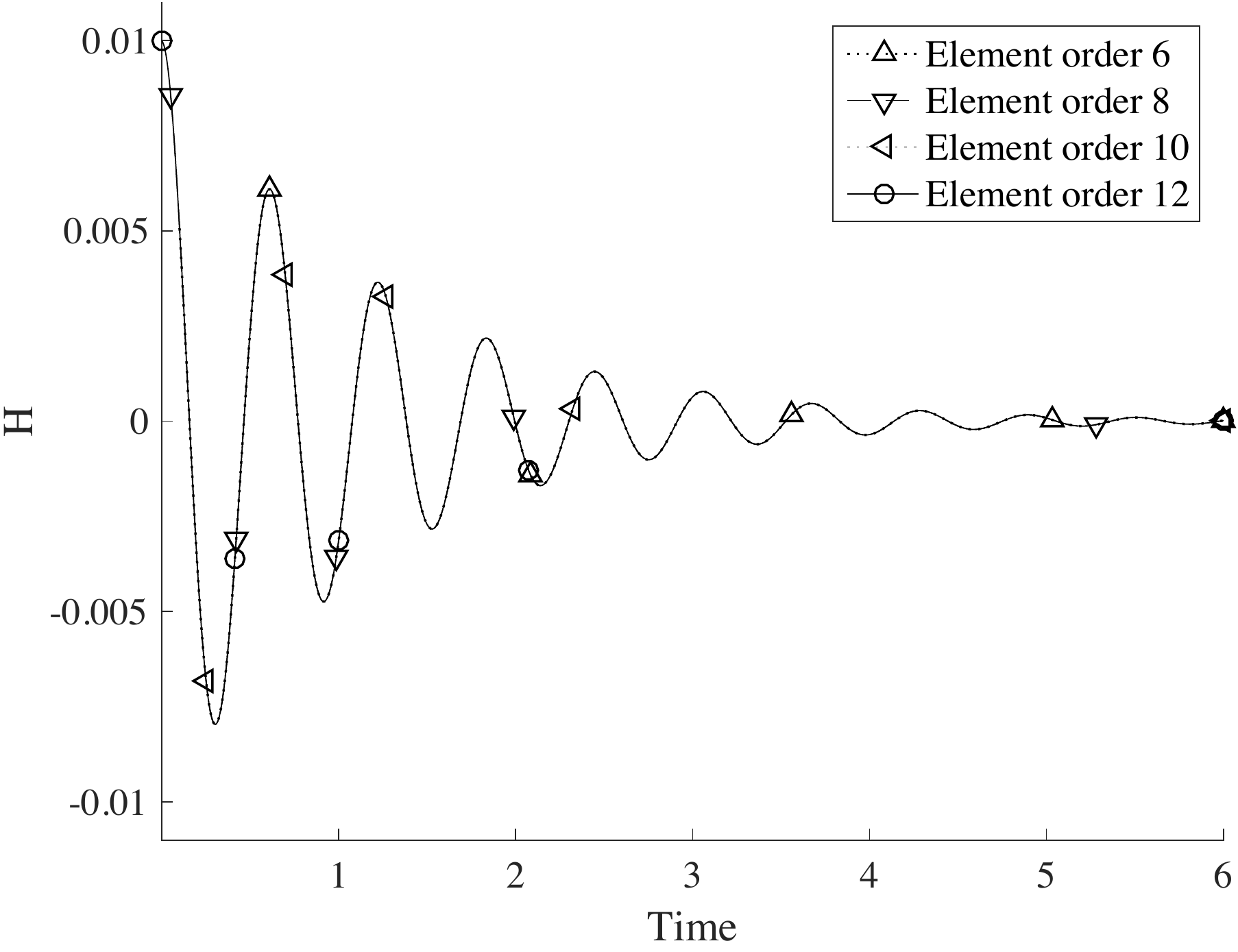}} 
\qquad
\subfigure[Interface $\#2$ with various element orders]{ \includegraphics[scale=.4]{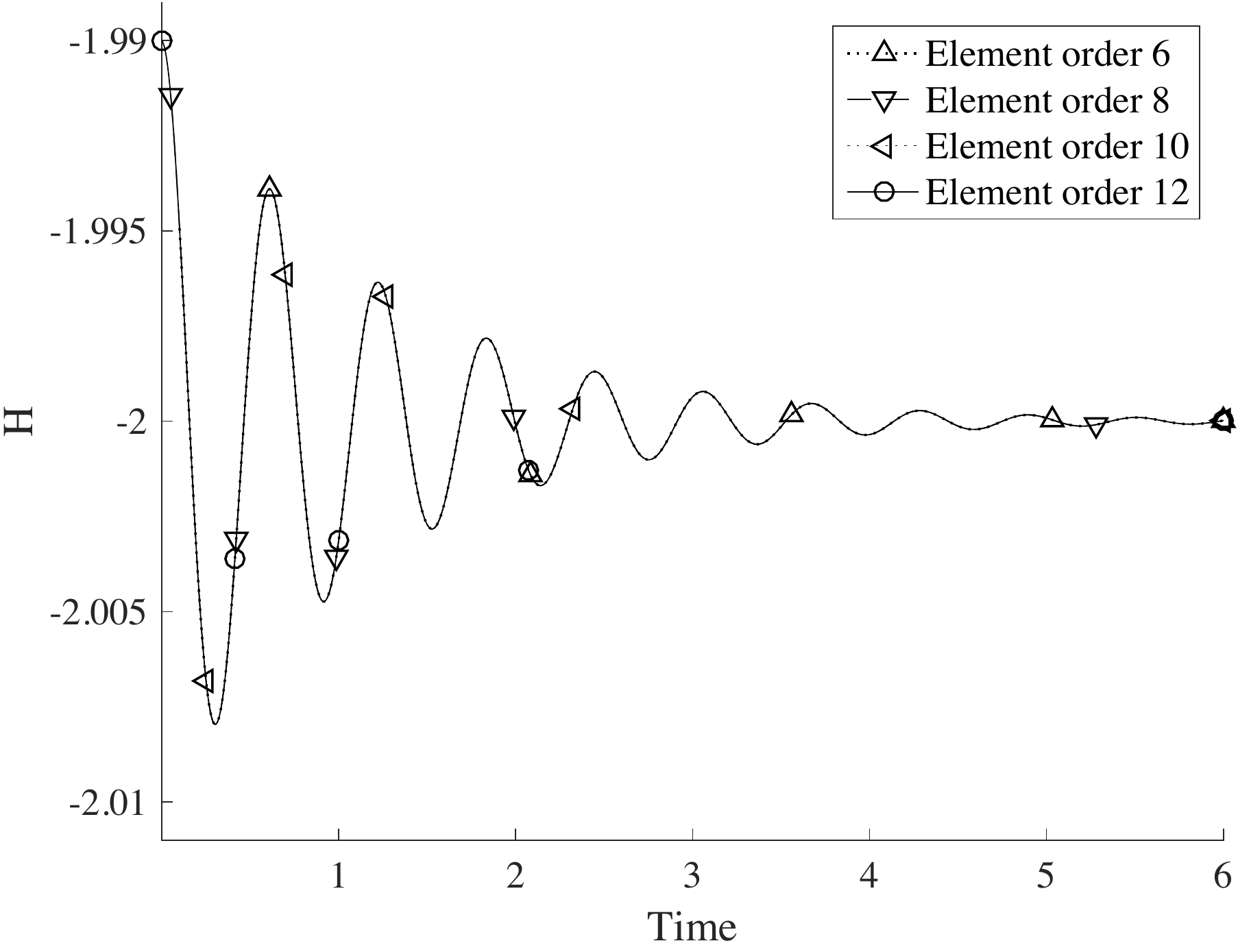}} 
 \subfigure[Interface $\#1$ with various $\Delta t$]{ \includegraphics[scale=.4]{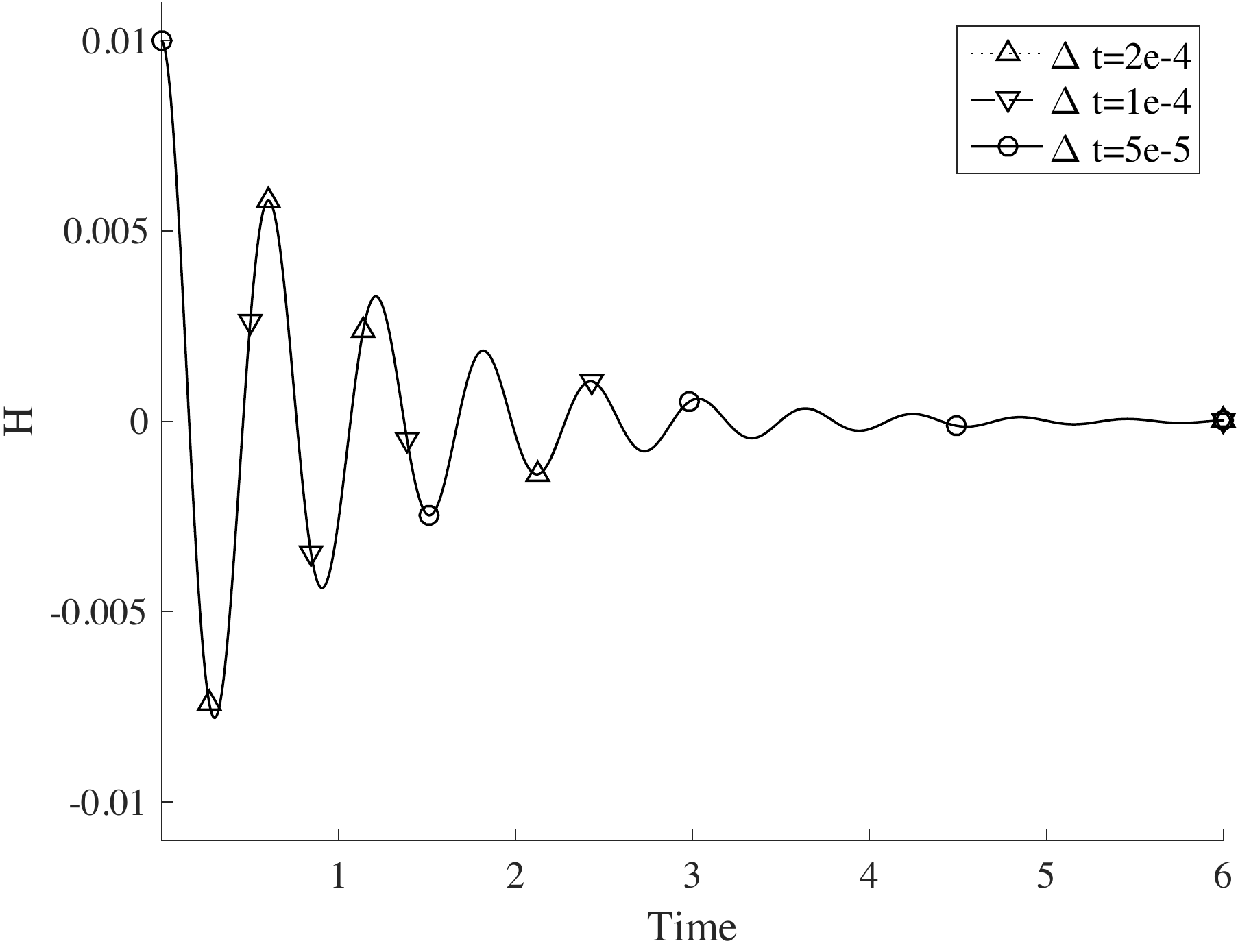}} 
\qquad
\subfigure[Interface $\#2$ with various $\Delta t$]{ \includegraphics[scale=.4]{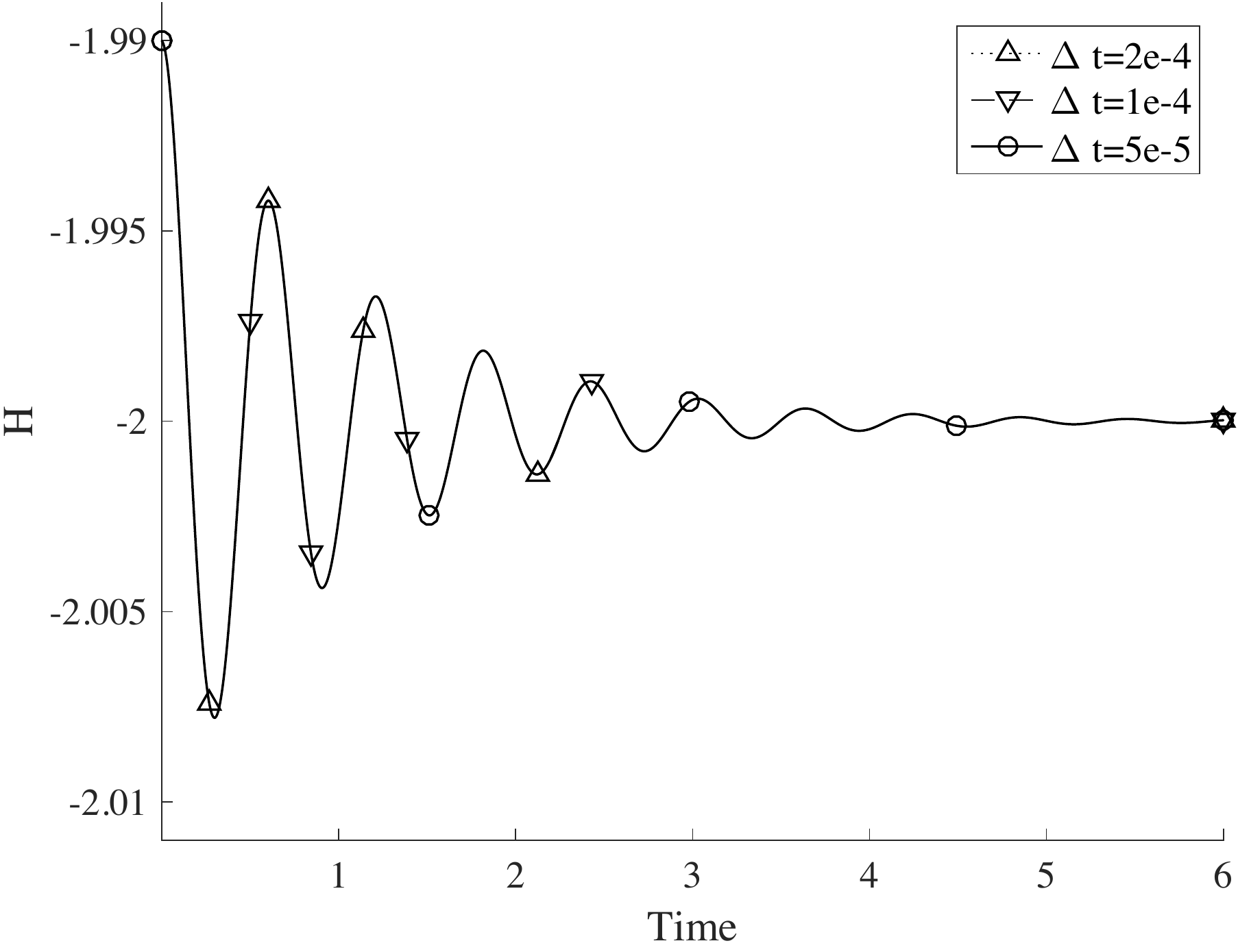}} 
\caption{Three-phase capillary wave problem  (matched density $\tilde \rho_1=\tilde \rho_2=\tilde \rho_3=1$). (a)-(b): Effect of spatial resolution (element order) on the capillary amplitude history. Simulation results are obtained with a fixed time step size $\Delta t=10^{-4},$ interfacial thickness $\eta=0.01,$ mobility $m_0=10^{-5}$ and various element orders.
  (c)-(d): Effect of time step size on the capillary amplitude history. Simulation results are obtained with a fixed element order 8, interfacial thickness $\eta=0.005$, mobility $m_0=10^{-5}$ and various time step sizes $\Delta t$. }
\label{varyordertime}
\end{figure}

Let us first focus on a matched density for the three fluids, i.e., $\tilde \rho_1=\tilde \rho_2=\tilde \rho_3=1$, and study the effects of several parameters on
the simulation results.
We have performed extensive tests to ensure that our simulation results
have converged with respect to the spatial and temporal resolutions.
Fig.~\ref{varyordertime}(a)-(b) show a spatial resolution test.
Here we compare the time histories of the capillary wave amplitudes of the interfaces $\#1$ and $\#2$ obtained with several element orders ranging from 6 to 12 in the simulations.
The history curves corresponding to different element orders overlap with one another,
suggesting independence of the results with respect to the grid resolution.
Fig.~\ref{varyordertime}(c)-(d) show a temporal resolution test.
We compare the capillary wave amplitude histories obtained using several time step sizes.
The results obviously indicate the convergence with respect to $\Delta t.$ 

These resolution tests indicate that an element order $8$
and a time step size $\Delta t=10^{-4}$ will be sufficient for the spatial
and temporal resolutions with current spectral element mesh.
Therefore, the majority of subsequent simulations will be conducted
using these parameter values.


\begin{figure}[tbp]
\centering
\subfigure[Interface $\#1$ with various $m_0$]{ \includegraphics[scale=.4]{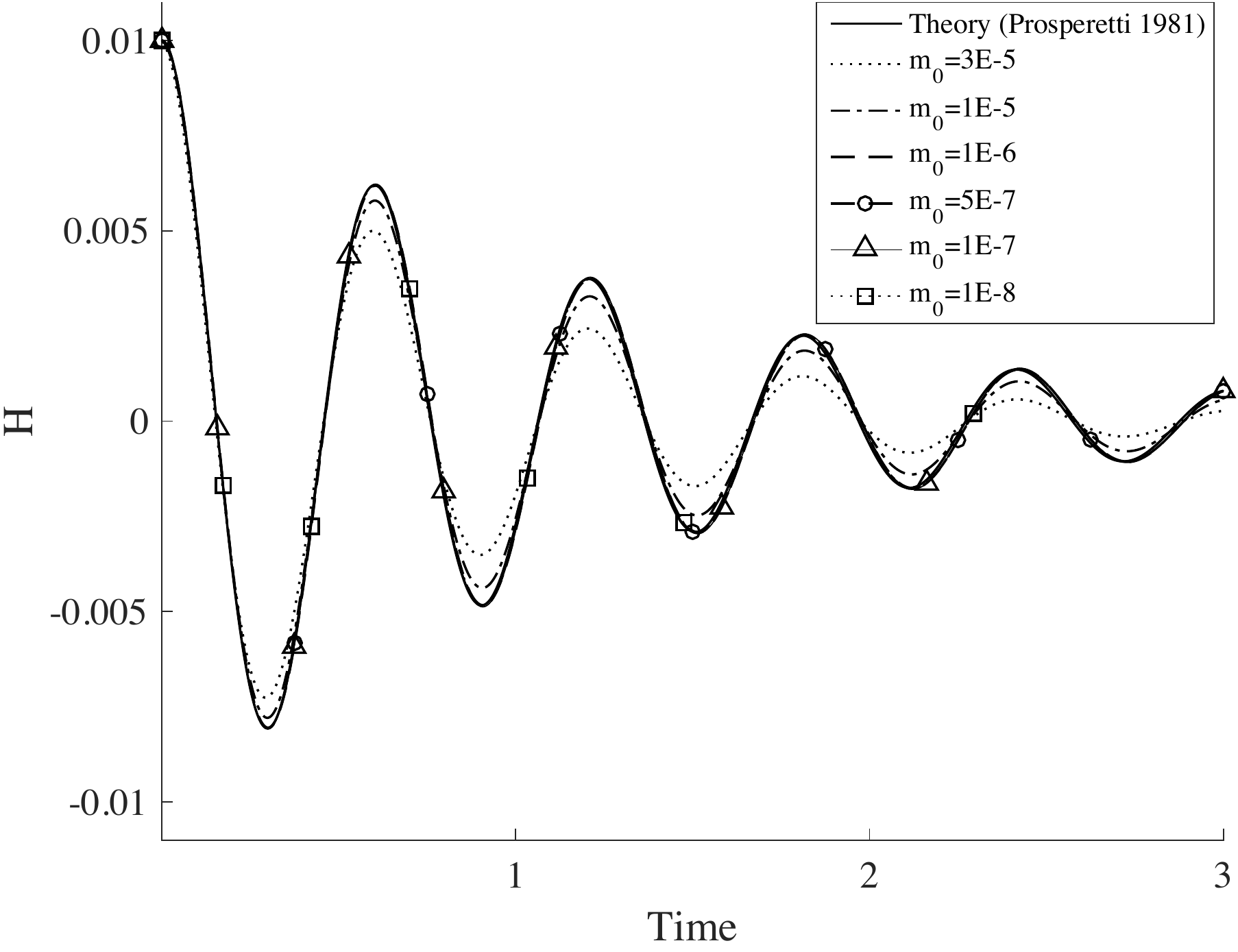}} 
\qquad
\subfigure[Interface $\#2$ with various $m_0$]{ \includegraphics[scale=.4]{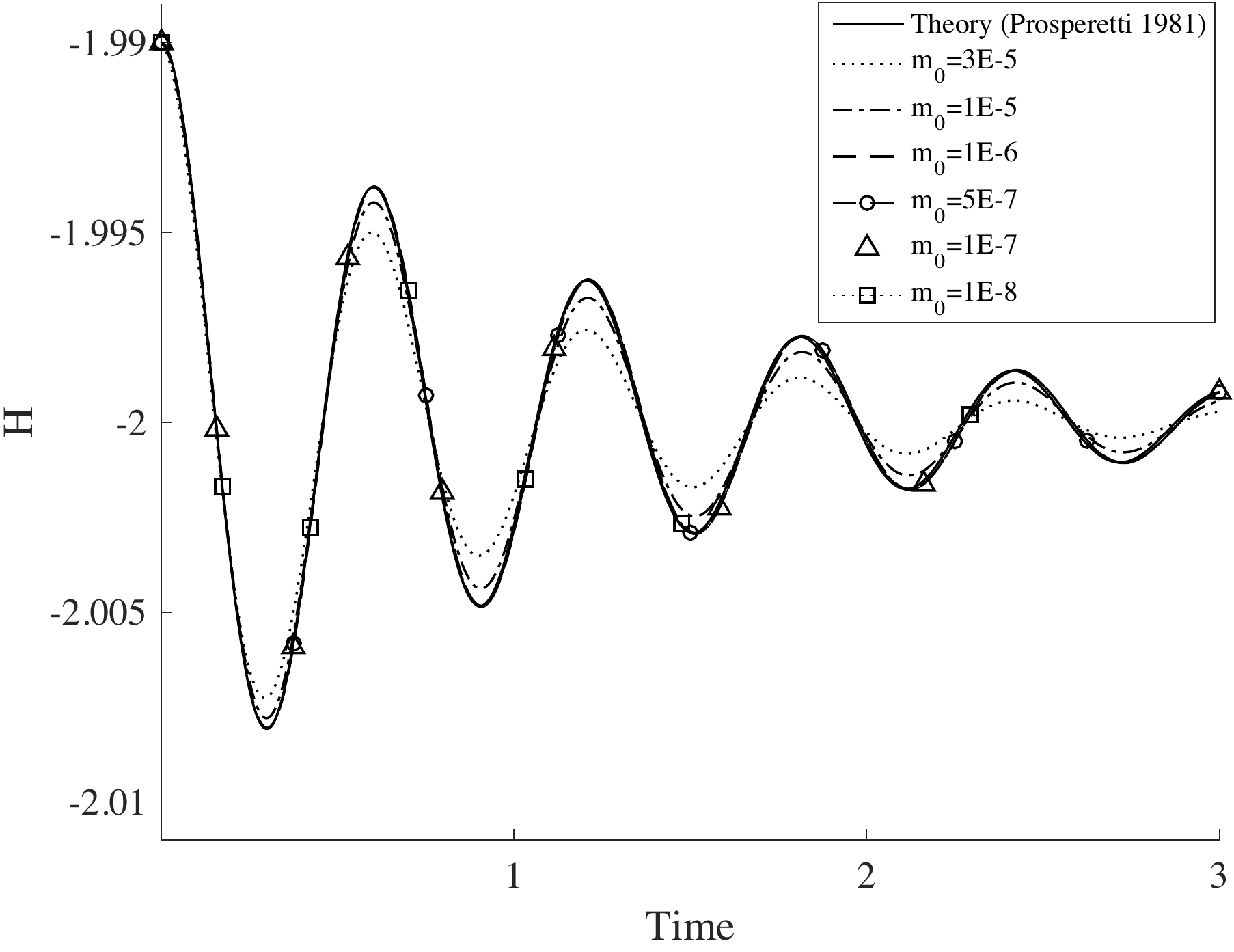}} 
 \subfigure[Interface $\#1$ with various $\eta$]{ \includegraphics[scale=.4]{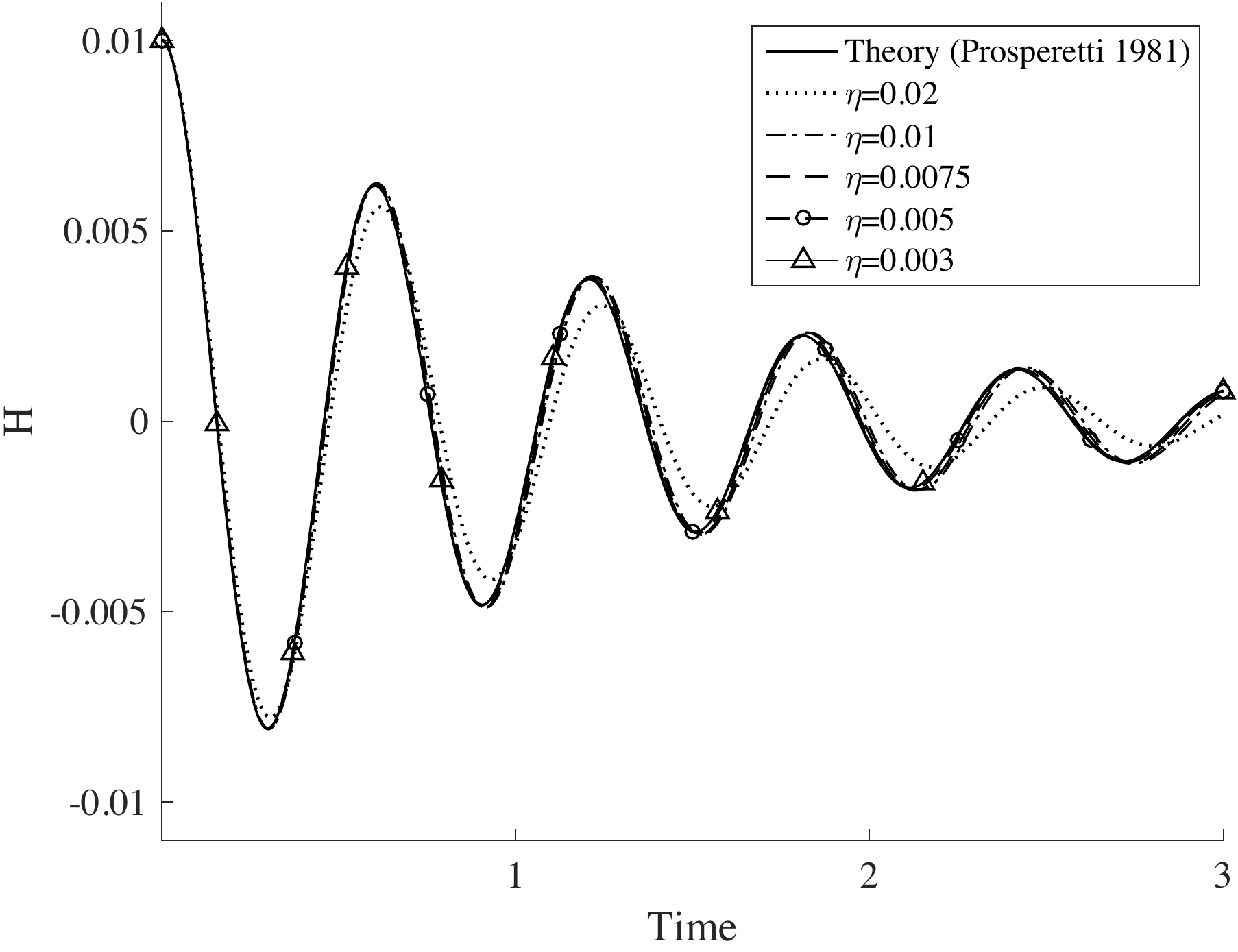}} 
\qquad
\subfigure[Interface $\#2$ with various $\eta$]{ \includegraphics[scale=.4]{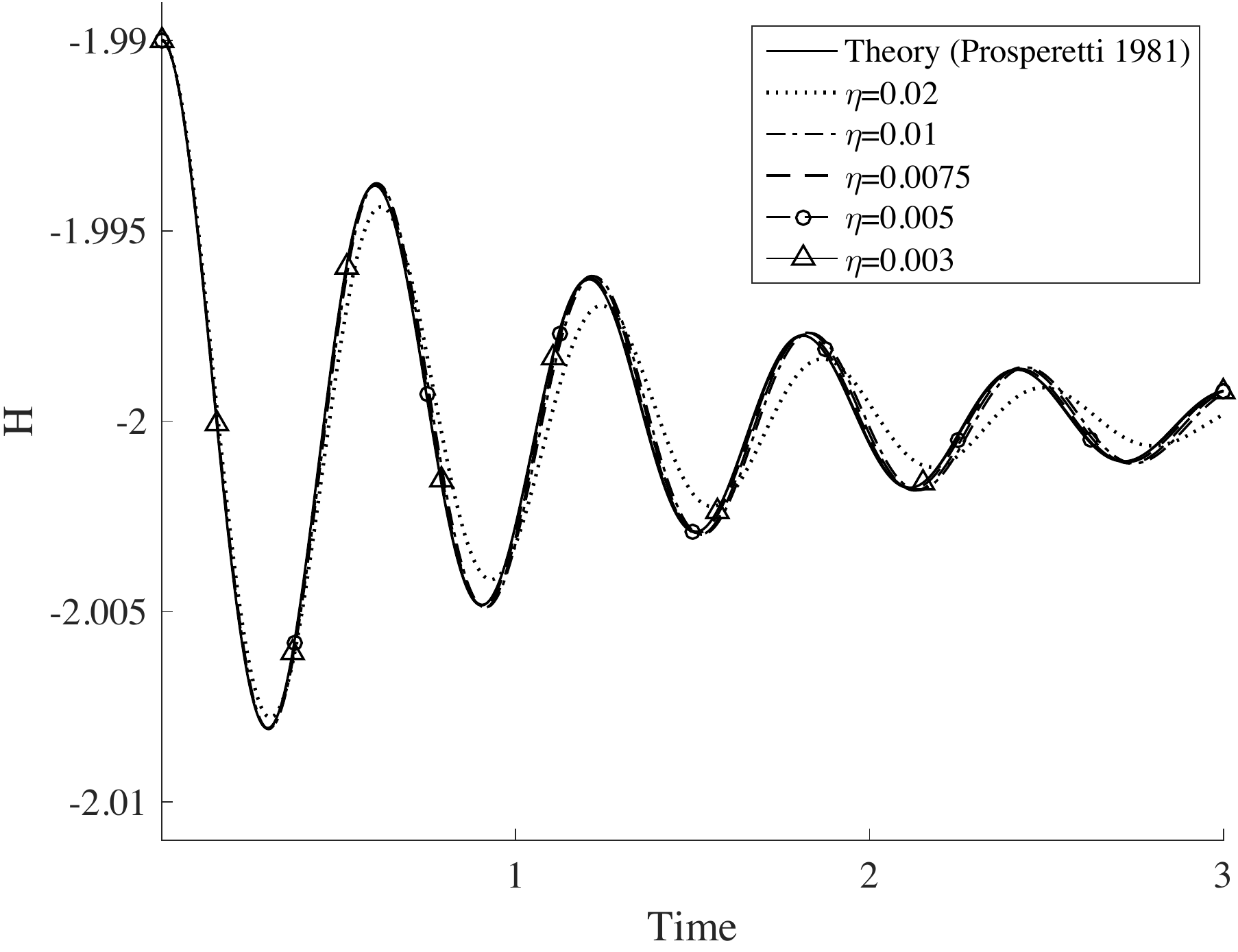}} 
\caption{Capillary wave problem  (matched density $\tilde \rho_1=\tilde \rho_2=\tilde \rho_3=1$). (a)-(b): Comparison of capillary amplitude histories corresponding to
different  mobility $m_0$ values and Prosperetti's exact solution~\cite{Prosperetti1981}.
Simulation results correspond to a time step size $\Delta t=10^{-4},$ element order $8$,
and interfacial thickness $\eta=0.005.$
(c)-(d): Comparison of capillary amplitude  histories corresponding to
different interfacial thickness $\eta$ values and Prosperetti's exact solution.
Simulation results correspond to a time step size $\Delta t=10^{-4},$ element order $8$,
and mobility $m_0=5 \times 10^{-7}.$}
\label{varymobeta}
\end{figure}

The effect of the mobility coefficient $m_0$ on the simulation results
is shown by Fig.~\ref{varymobeta}(a)-(b), in which we
compare the time histories of the capillary wave amplitudes of the two interfaces
obtained with a fixed interfacial thickness scale $\eta=0.005$ and
various mobility values ranging between $m_0=3\times 10^{-5}$ and $m_0=10^{-8}$.
The exact physical solution given by \cite{Prosperetti1981} for
this case is also
included in the figure for comparison.
It is observed that the computation becomes unstable
if $m_0$ is too large (larger than around $m_0=3\times 10^{-5}$).
As $m_0$ decreases from $3\times 10^{-5}$ to $10^{-8}$, we initially observe
an effect on the amplitude and phase of the history signals obtained from
the simulations.
But as $m_0$ becomes sufficiently small,
the difference in the simulated capillary amplitude histories becomes very small,
and the history curves converge to the exact solution by \cite{Prosperetti1981}.
In fact, when $m_0$ decreases below $10^{-6}$, the difference between the numerical results and the theoretic solution is negligible.

Fig.~\ref{varymobeta}(c)-(d) show the effect of the interfacial thickness
scale $\eta$ on the simulation results.
In this figure we compare time histories of the capillary amplitude obtained with the interfacial thickness scale parameter $\eta$ ranging from $0.02$ to $0.003$ with a fixed mobility $m_0=5\times 10^{-7}.$ The exact physical solution is also included in the plots.
Some influence on the amplitude and the phase of the history curves can be observed
as $\eta$ decreases from $0.02$ to $0.01.$ As $\eta$ decreases further to $\eta=0.0075$ and below, on the other hand, the history curves essentially overlap with one another and little difference can be discerned among them, suggesting a convergence of the results with respect to $\eta.$

\begin{figure}[tbp]
  \caption{Three-phase capillary wave (different density ratios): Comparison of time histories of the capillary wave amplitude between simulation and Prosperetti's exact solution  for densities $(\tilde \rho_1,\tilde \rho_2,\tilde \rho_3)=(1,10,10)$ ((a)-(b)), $(1,10,100)$ ((c)-(d)),
    $(1,100,100)$ ((e)-(f)), $(1,10,1000)$ ((g)-(h)), and $(1,1,1000)$ ((i)-(j)).
    The simulation results are obtained with a time step size $\Delta t=10^{-4}$ for (a)-(f), $\Delta t=2\times 10^{-5}$ for (g)-(j), an  element order 8, interfacial thickness $\eta=0.003$, and mobility $m_0=5 \times 10^{-7}.$}
\centering
 \subfigure[Interface $\#1$ with $(\tilde \rho_1,\tilde \rho_2,\tilde \rho_3)=(1,10,10)$]{ \includegraphics[scale=.4]{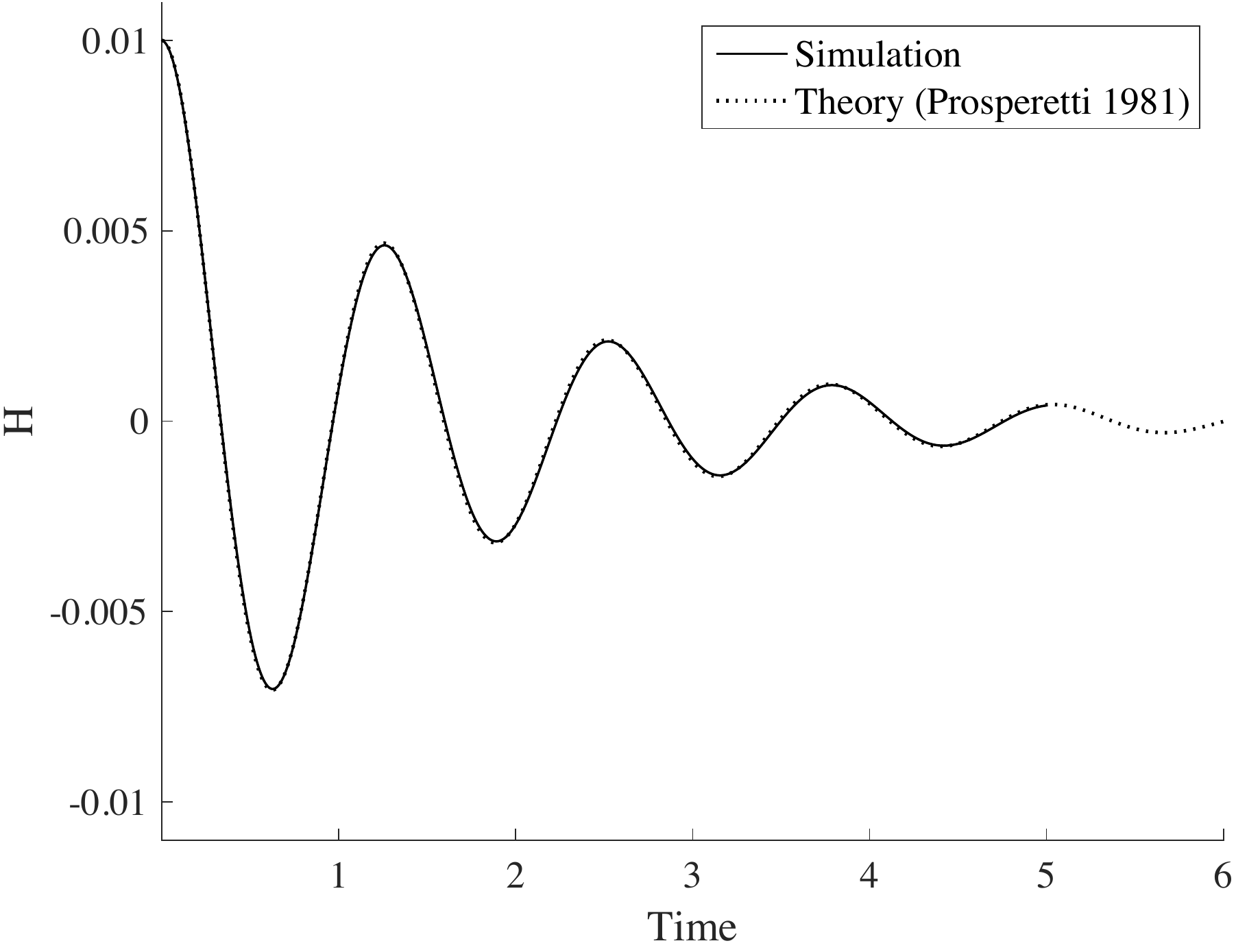}} 
\qquad
 \subfigure[Interface $\#2$ with $(\tilde \rho_1,\tilde \rho_2,\tilde \rho_3)=(1,10,10)$]{ \includegraphics[scale=.4]{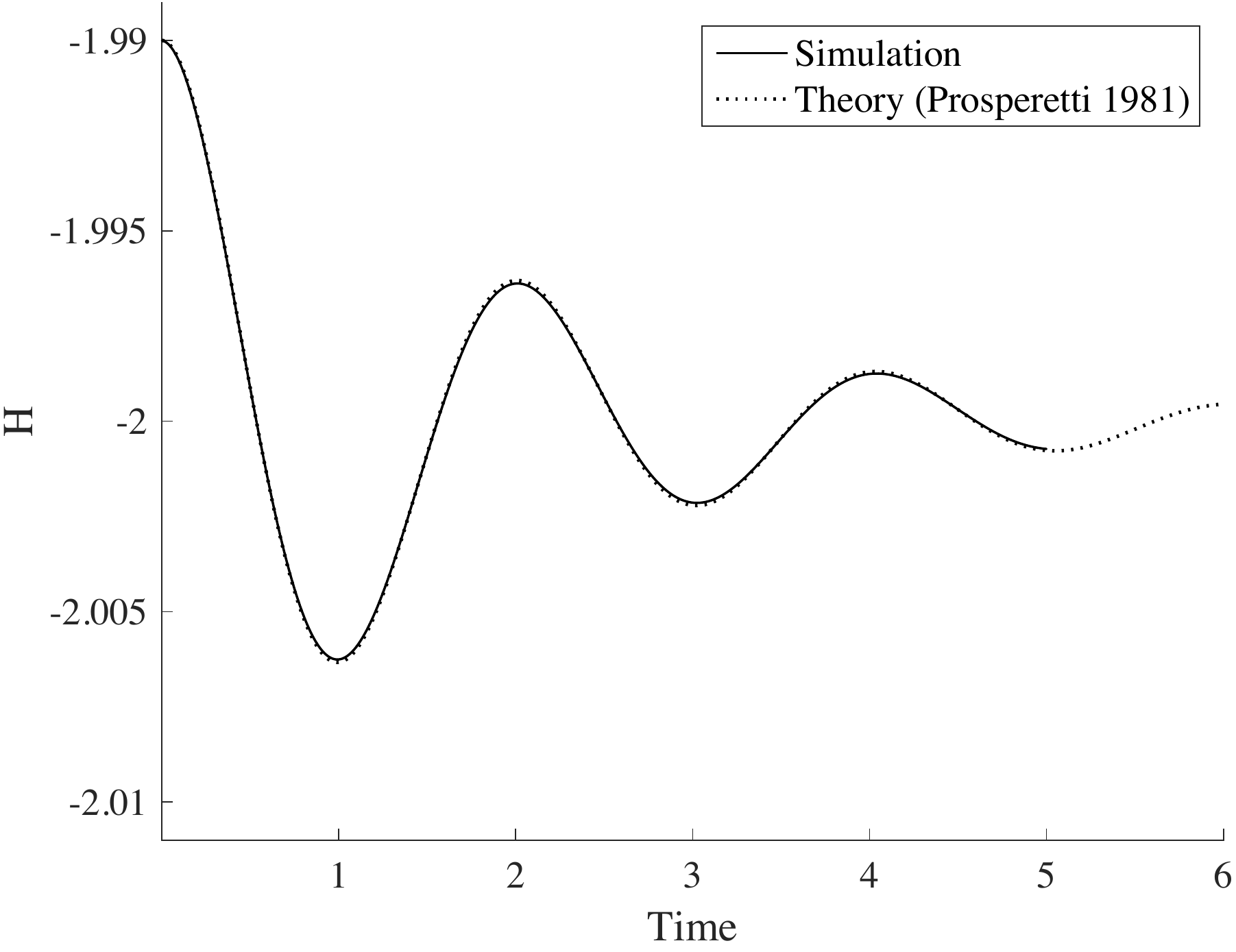}} 
 \subfigure[Interface $\#1$ with $(\tilde \rho_1,\tilde \rho_2,\tilde \rho_3)=(1,10,100)$]{ \includegraphics[scale=.4]{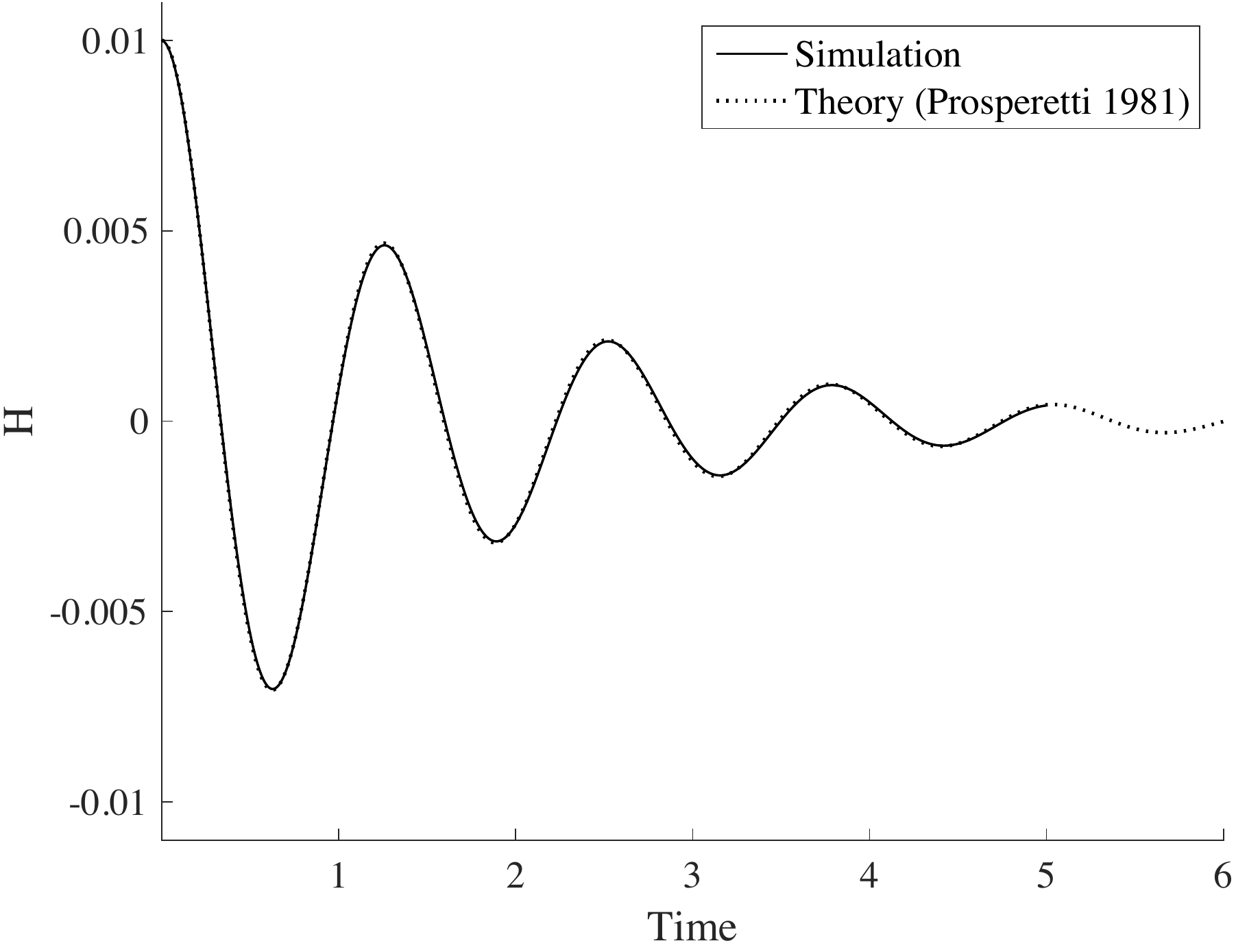}} 
\qquad
 \subfigure[Interface $\#2$ with $(\tilde \rho_1,\tilde \rho_2,\tilde \rho_3)=(1,10,100)$]{ \includegraphics[scale=.4]{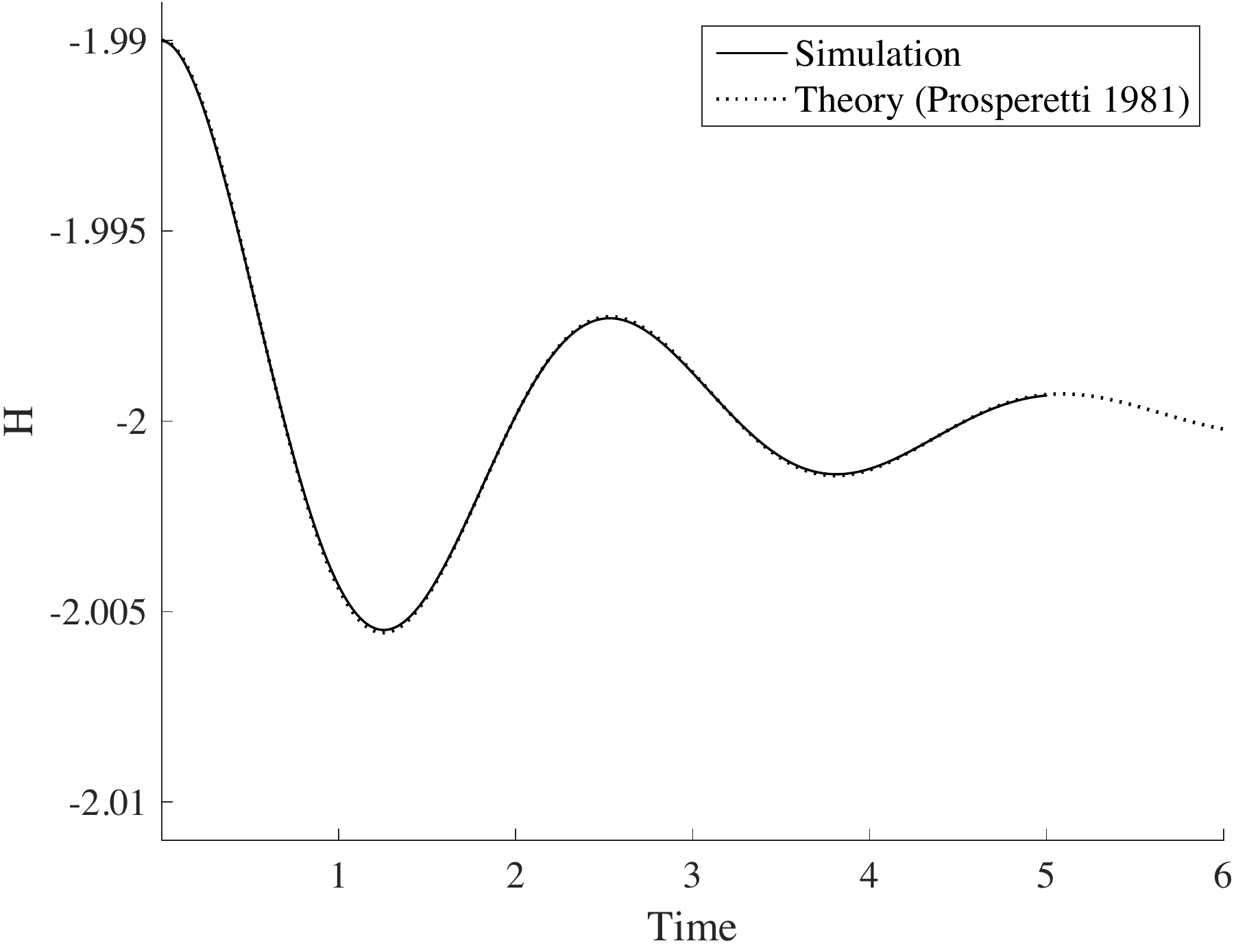}} 
  \subfigure[Interface $\#1$ with $(\tilde \rho_1,\tilde \rho_2,\tilde \rho_3)=(1,100,100)$]{ \includegraphics[scale=.4]{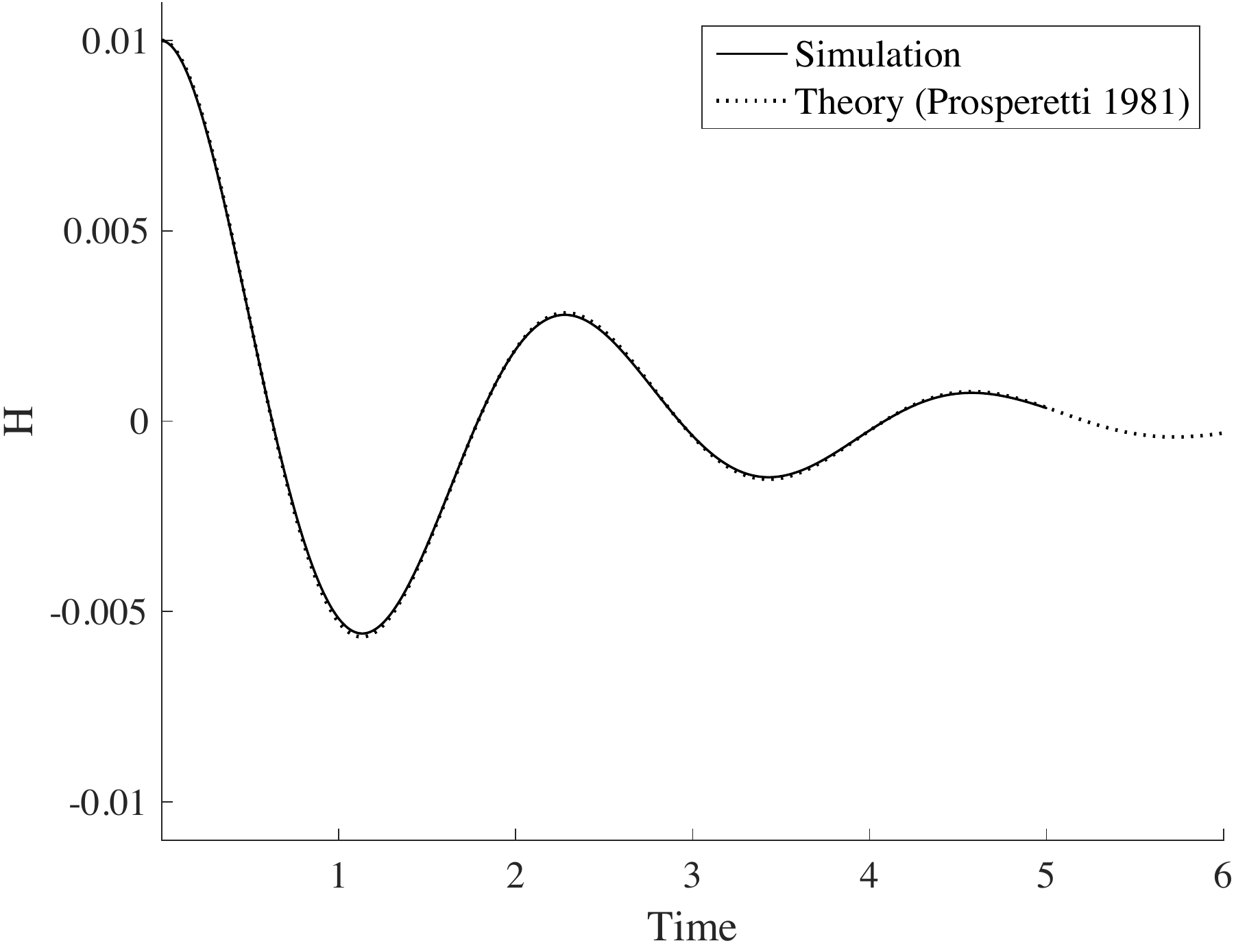}} 
\qquad
\subfigure[Interface $\#2$ with $(\tilde \rho_1,\tilde \rho_2,\tilde \rho_3)=(1,100,100)$]{ \includegraphics[scale=.4]{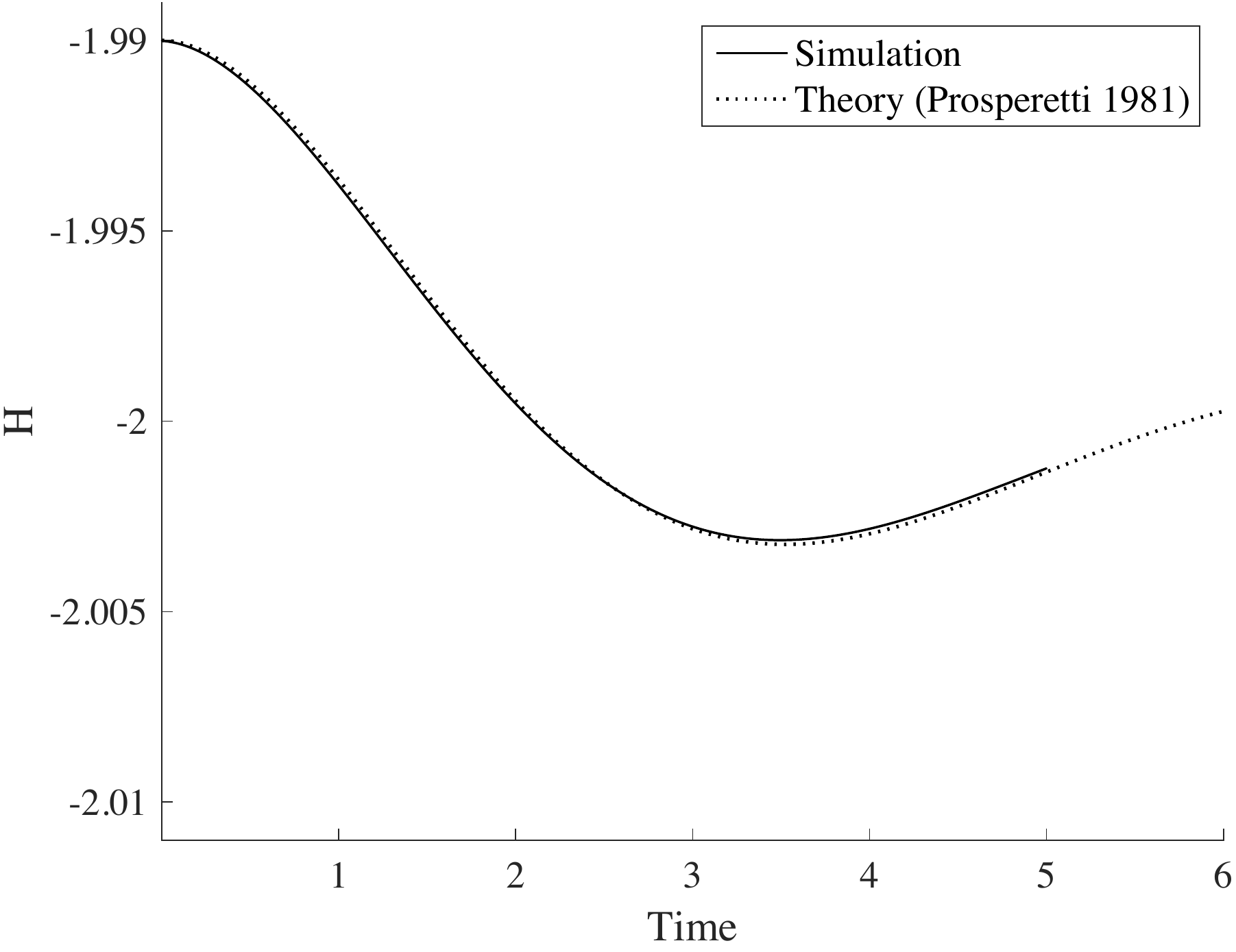}} 
\label{densityvary1}
\end{figure}
\begin{figure}[htbp]\ContinuedFloat
\centering
\subfigure[interface $\#1$ with $(\tilde \rho_1,\tilde \rho_2,\tilde \rho_3)=(1,10,1000)$]{ \includegraphics[scale=.4]{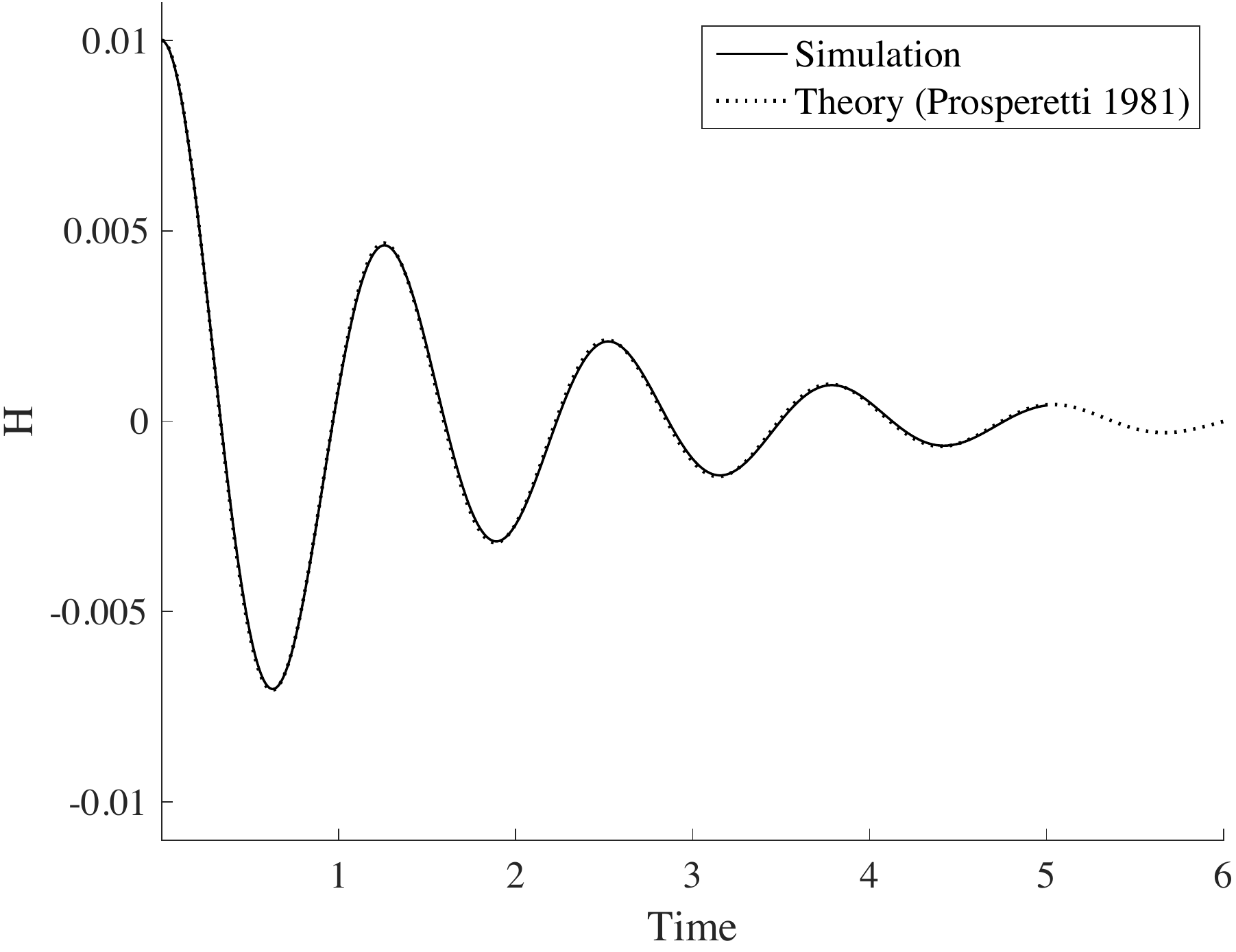}} 
\qquad
\subfigure[interface $\#2$ with $(\tilde \rho_1,\tilde \rho_2,\tilde \rho_3)=(1,10,1000)$]{ \includegraphics[scale=.4]{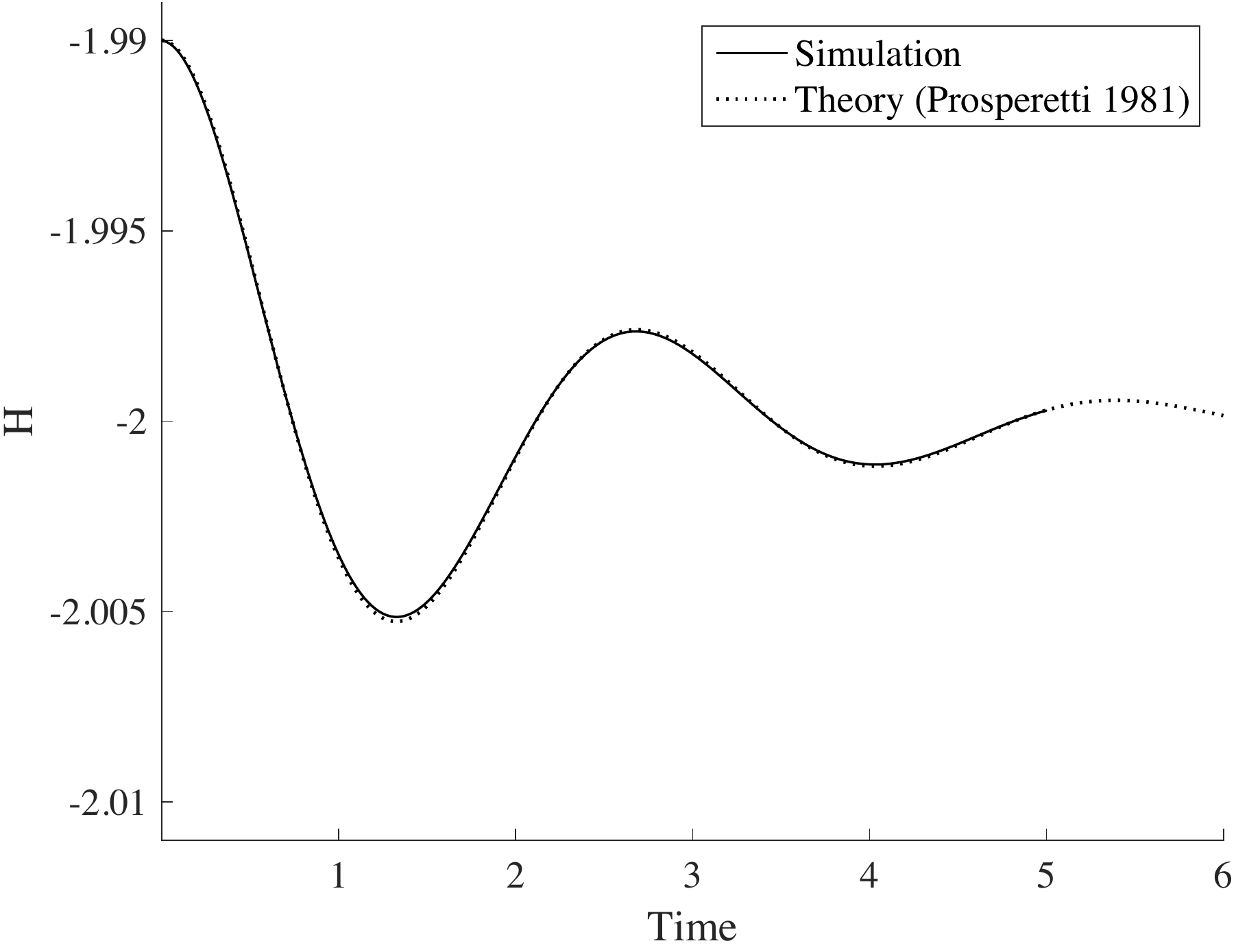}} 
 \subfigure[interface $\#1$ with $(\tilde \rho_1,\tilde \rho_2,\tilde \rho_3)=(1,1,1000)$]{ \includegraphics[scale=.4]{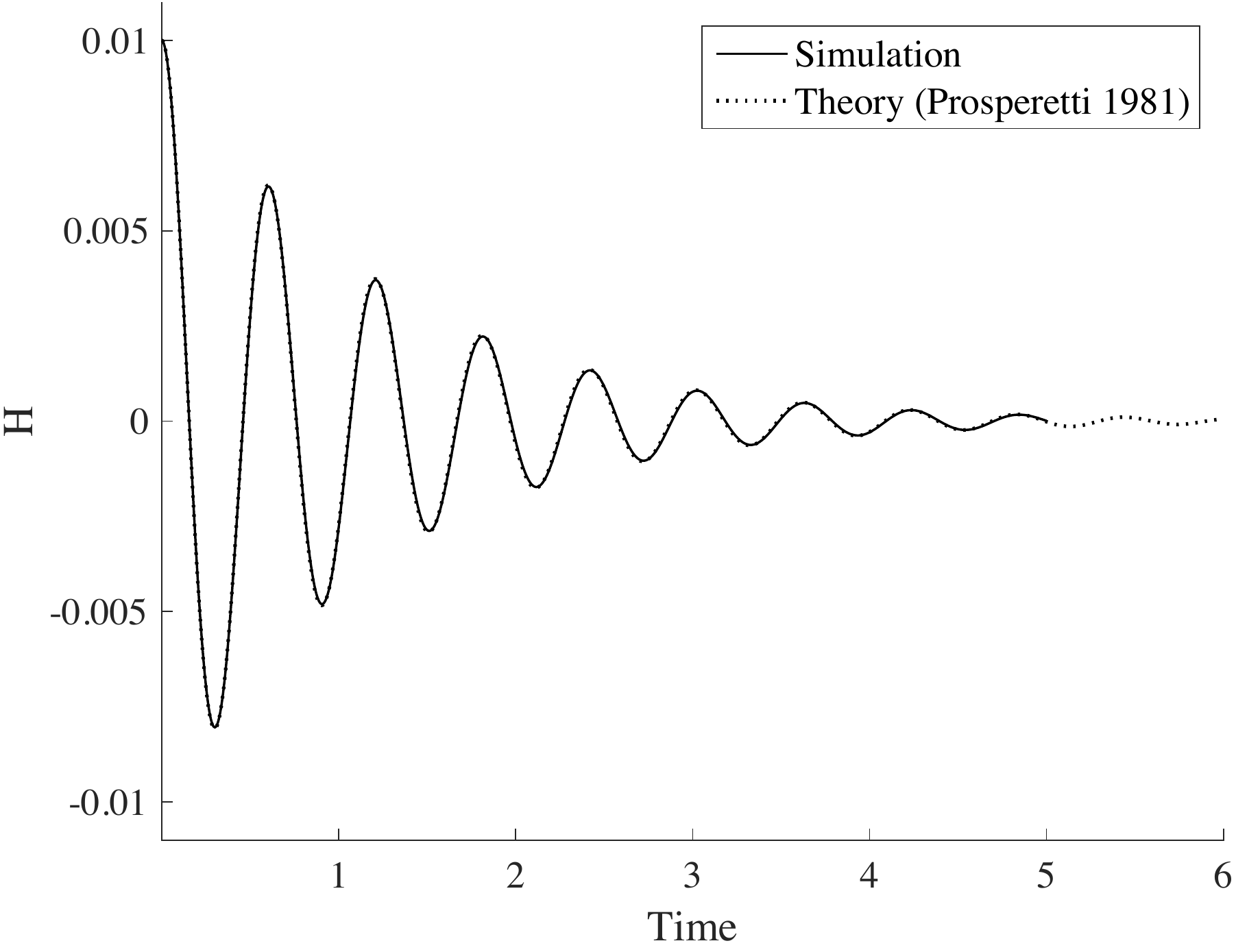}} 
\qquad
\subfigure[interface $\#2$ with $(\tilde \rho_1,\tilde \rho_2,\tilde \rho_3)=(1,1,1000)$]{ \includegraphics[scale=.4]{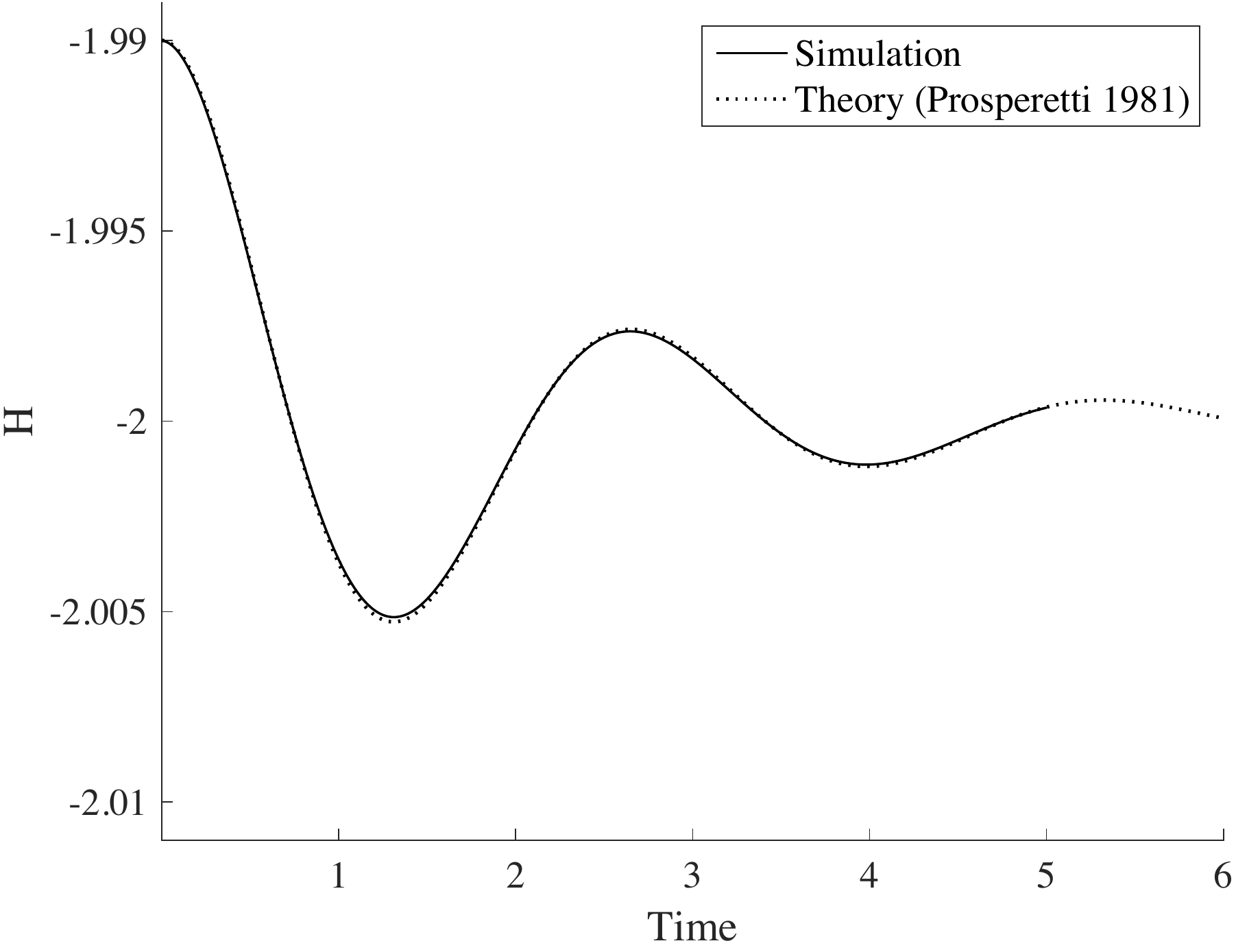}} 
\end{figure}

Let us next investigate the effect of density ratios
on the motion of the fluid interfaces.
In these tests we vary the densities and dynamic viscosities of the fluid $\#2$ and fluid $\#3$
($\tilde \rho_2$, $\tilde \rho_3$ and $\tilde \mu_2$, $\tilde \mu_3$) systematically
while the relation $\nu=\frac{\tilde \mu_1}{\tilde \rho_1}=\frac{\tilde \mu_2}{\tilde \rho_2}=\frac{\tilde \mu_3}{\tilde \rho_3}$ is maintained as required by the theoretic solution in \cite{Prosperetti1981}.
In Fig. \ref{densityvary1}, we show the time histories of the  capillary amplitudes
corresponding to five  density contrasts,
$(\tilde \rho_1,\tilde \rho_2,\tilde \rho_3)$ equal (a)-(b): $(1,10,10)$, (c)-(d): $(1,10,100)$,
(e)-(f): $(1,100,100)$, (g)-(h): $(1,10,1000)$, and (i)-(j): $(1,1,1000)$,
and compare them with the theoretic solutions from \cite{Prosperetti1981}.
The  simulation results are obtained with an
element order 8, interfacial thickness $\eta=0.003$, and mobility $m_0=5 \times 10^{-7}$.
The time step size in the simulations is $\Delta t=10^{-4}$ for the plots (a)-(f),
and a smaller $\Delta t=2\times 10^{-5}$ for the cases involving
$\tilde{\rho}_3=1000$ (plots (g)-(j)) in order to ensure the stability of simulations.
We observe that the density contrasts have a dramatic effect on the motions of
the interfaces, and the dynamics of the two interfaces have become very different.
Under the same density ratio, increase in the density values appears to cause
the period of oscillation to increase
and the attenuation of the oscillation amplitude to be more pronounced;
see e.g.~Figs.~\ref{densityvary1}(c) and (d).
Increase in the density ratio seems to have a similar effect with respect to
the oscillation amplitude and period; compare e.g.~Figs.~\ref{densityvary1}(a) and (e).
It can also be observed that
the history curves from the simulations essentially overlap with those of the
exact solutions for all the density contrasts and little difference can be perceived,
indicating that our method has captured the dynamics of the
fluid interfaces correctly.
 
%

The three-phase capillary wave problem and in particular the comparisons
with Prosperetti's exact solution for this problem demonstrate that the N-phase formulation with the proposed open boundary conditions and the numerical method
developed herein (with $N=3$) have produced physically accurate results for a wide range of density ratios (up to density ratio $1000$ tested here).

\subsection{Interaction of Two Liquid Jets in Ambient Water}

\begin{figure}[tb]
  \centering
    \includegraphics[width=0.35\textwidth]{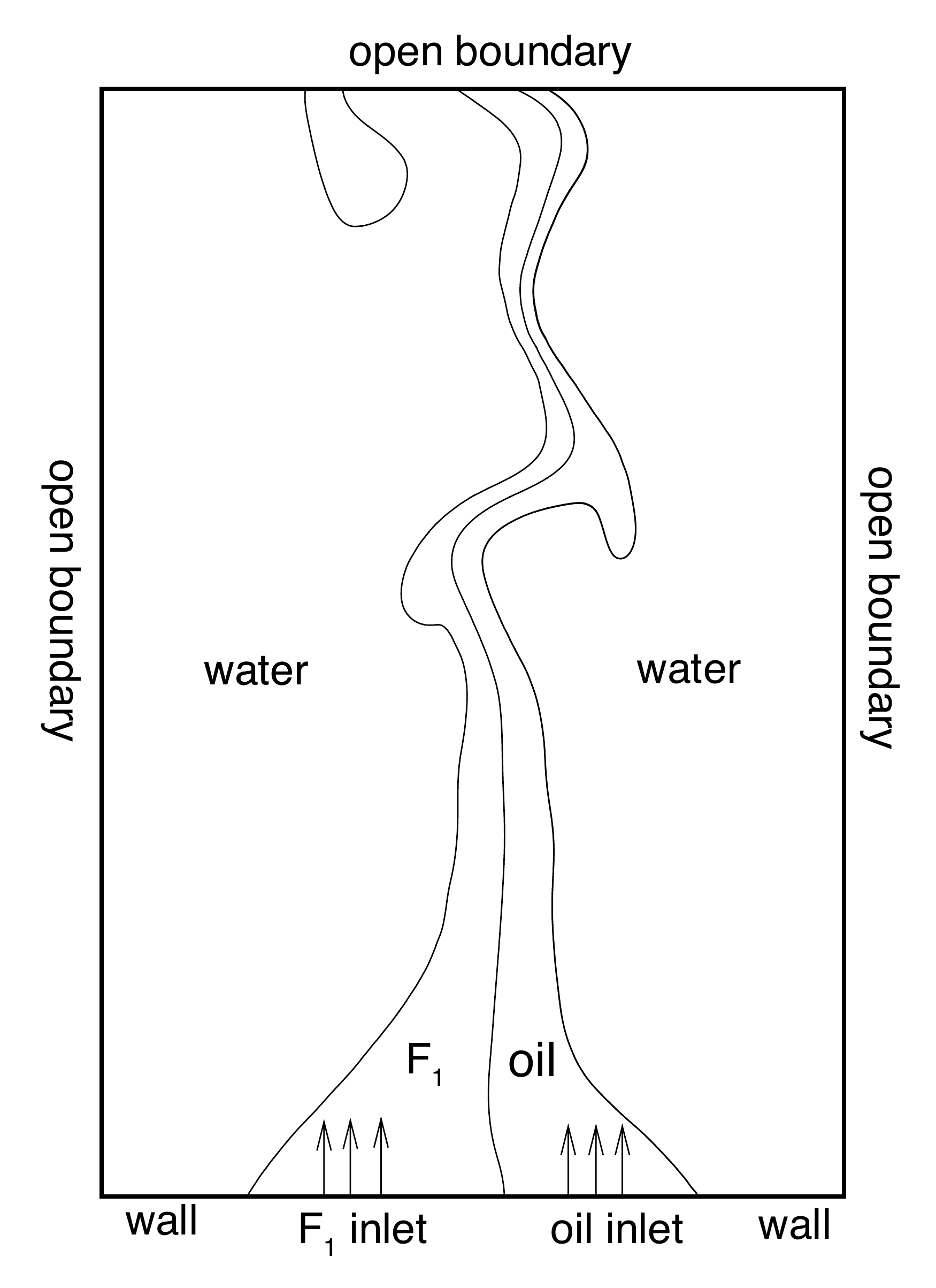}
   \caption{\small Configuration of the interaction of $F_1$-oil jets  in water. }
    \label{oiljetconfig}
\end{figure}

In this subsection, we test the proposed open boundary conditions and
the numerical method by considering the interactions of two fluid jets
in an infinite expanse of ambient water. 
The two jets consists of two different liquids. One of the jets is oil,
and the other is a liquid referred to as ``F$_1$''. The F$_1$ liquid
is assumed to be lighter than water and immiscible with both oil and water.
This test problem involves multiphase inflow/outflow boundaries. How to
deal with such boundaries is critical to the successful simulation of this
problem.

Specifically,
we consider a rectangular flow domain $\Omega=\{(x,y)|-0.5L \leq x\leq 0.5L, 0 \leq y \leq 1.5L  \}$ where $L=6cm$, as shown in Fig.~\ref{oiljetconfig}. The bottom side of the domain ($y=0$) is a solid wall of neutral wettability. The other three sides of the domain are all open
where the fluids can enter or leave the domain freely.
The domain initially contains water inside. The bottom wall has two orifices, each having a diameter $0.2L$. The centers of two orifices are located at $(x_1,y_1)=(-0.2L,0)$ and $(x_2,y_2)=(0.2L,0)$, respectively.  A jet of a certain fluid labeled by F$_1$ enters the domain through the left orifice, and
a jet of oil is introduced into the domain through the right orifice.
The gravity $\bs g_r$ is assumed to point downward ($-y$ direction).
The configuration of this problem models the motion of the $F_1$ jet and the oil jet
in an infinite expanse of water.
The two jets rise through the water due to buoyancy,
interact with each other, and move out of the domain through the open boundaries.
The goal here is to investigate the long-time behavior of this three-phase system.

\begin{table}[tbp]
\centering 
\begin{tabular}{l l l l} 
\hline 
Density [$kg/m^3$]:& $F_1$ - 600 & water - 998.2071& oil - 400 or 100\\
Dynamic viscosity [$kg/(m\cdot s)$]: &$F_1$ - $2\times 10^{-2}$   & water - $1.002\times 10^{-3}$   & oil - $9.15\times 10^{-2}$\\
Surface tension [$kg/s^2$]:&$F_1/$water - $4.5\times 10^{-2}$ &$F_1/$oil - $4.8\times 10^{-2}$ & oil/water - $4.4\times 10^{-2}$  \\
Gravity [$m/s^2$]:&9.8&&\\
\hline 
\end{tabular}
\caption{Physical property values of fluids $F_1$, water and oil.}
\label{table:oiljet} 
\end{table}

The physical properties (including the densities, viscosities, pair-wise surface tensions) of F$_1$, water, and oil employed in this problem, as well as the gravitational acceleration, are listed in Table \ref{table:oiljet}. We choose $L=6cm$ as the length scale,  the density of F$_1$ as the density scale $\varrho_d,$ and the centerline velocity at
the orifices as the velocity scale $U_0$.
Then the problem is non-dimensionalized based on Table \ref{table:normalization}. In what follows, all physical and numerical parameters have been properly normalized.

\begin{table}[tbp]
\centering 
\begin{tabular}{l c| l c} 
\hline 
Parameter & Value & Parameter & Value \\  
\hline 
$x_1$&-0.2&$x_2$&0.2\\
$R$&0.1&$d_0$&0.5\\
$\alpha_1,$ $\theta$ &1 & $\alpha_2$& 0\\
 $\delta$&0.01&$\mu_0$ & ${\rm max}(\tilde \mu_1, \tilde \mu_2, \tilde \mu_3)$\\
$\rho_0$ & ${\rm min}(\tilde \rho_1,\cdots,\tilde \rho_3) $   &$\nu_m$ & $1.56\times 10^{-2}$  \\
$m_0$   &$1\times 10^{-8}$            &    $\eta$         & 0.01              \\
 $J$ (temporal order)   &  2& Number of elements  & 600  \\ 
$\Delta t$    & $2\times 10^{-5}$ & Element order  & 6   \\ 
\hline 
\end{tabular}
\caption{Simulation parameter values for the interaction of two liquid jets in ambient water.}
\label{table:oiljet2} 
\end{table}

In the numerical experiments, we specify $F_1$, water and oil as the first, second, and the third fluids, with the normalized densities $\tilde \rho_1$, $\tilde \rho_2$, and $ \tilde \rho_3,$ respectively. We discretize the computational domain  with a mesh of $600$ quadrilateral elements of uniform size, with $20$ elements in the $x$ direction and $30$ elements in the $y$ direction. The element order is $6$ for all the elements.  The time step size is chosen as $\Delta t=2\times 10^{-5}$ and all the simulation results afterwards are obtained with  interfacial thickness $\eta=0.01$, and mobility $m_0=10^{-8}.$
To balance the gravity of the water, in the simulations we also apply an external pressure gradient pointing upward ($y$ direction) in the whole domain with a magnitude $\rho_w |\bs g_r|,$ where $\rho_w$ is the density of water. As a result, the region occupied by water has no net external body force exerted on it. 
The external body force $\bs f$ in equation  \eqref{equ:nse_trans} is set to $\bs f=\rho \bs g_r-\tilde \rho_2 \bs g_r,$ where $\tilde \rho_2$ and $\bs g_r$ are the normalized density of water and gravitational acceleration, respectively. The source term in \eqref{equ:CH} are set to $g_i=0 \,(i=1,2).$ On the bottom wall (excluding the fluid inlets), we impose the Dirichlet boundary  condition \eqref{equ:dbc_vel} for the velocity with $\bs w=\bs 0$ and the boundary conditions \eqref{equ:bc_chempot} and \eqref{equ:wbc_phi_1_mod} with $g_{bi}=g_{ci}=0\,(i=1,2)$  for the phase field variables. At the F$_1$ and oil inlets, we assume a parabolic profile for
the velocity, i.e.~$\bs w=(0,w_y)$ in \eqref{equ:dbc_vel} with 
\begin{equation}\label{equ:wxy}
  w_y=U_0\left[1-\Big(\frac{x-x_1}{R}\Big)^2\right], \;\;x\in (x_1-R,x_1+R);\;\;\;
  w_y=U_0\left[1-\Big(\frac{x-x_2}{R}\Big)^2\right], \;\;x\in (x_2-R,x_2+R),\quad
\end{equation}
where $R=0.1L$ is the radius of the orifice and $U_0=24.49cm/s$ is the centerline velocity at the orifices.
For the phase field functions we impose the following distributions at
the two fluid inlets,
\begin{equation}\label{equ:oiljetphase}
c_1=1,\;\;c_2=0, \ x\in (x_1-R,x_1+R);\;\;\; c_1=0,\;\;c_2=0, \;\;x\in (x_2-R,x_2+R).
\end{equation}
This distribution means that only the F$_1$ fluid is present at the left inlet,
and only the oil is present at the right inlet.
On the other three sides, we impose the open boundary
conditions \eqref{equ:obc_vel_mod}, \eqref{equ:bc_chempot}
and \eqref{equ:obc_phi_2_mod}, respectively for the velocity and the phase field functions, where $\bs f_b=\bs 0$, $g_{bi}=g_{ei}=0,\ (i=1,2)$, and $(\theta,\alpha_1,\alpha_2)=(1,1,0)$.
In \eqref{equ:obc_vel_mod}, the $d_0$ value is determined by the following procedure.
We first perform a preliminary simulation using $d_0=0$, and then
estimate a convection velocity scale at the outlet boundary. The $d_0$
is then set as the inverse of this convection velocity scale.
For the current problem, $d_0$ is set to $0.5$ based on this procedure.

The initial velocity is set to zero, and the initial volume fractions are set as follows:
\begin{equation}\label{equ:oilphaseinit}
\begin{cases}
&c_1=\big[ H(x- x_1+R)-H(x-x_1-R)\big] \big[H(y)-H(y-2R)\big],\\
&c_2=1-c_1-c_3 ,\\
&c_3=\big[ H(x-x_2+R)-H(x-x_2-R)\big] \big[H(y)-H(y-2R)\big],
\end{cases}
\end{equation}
where $H(x)$ is the Heaviside step function, taking the unit value if $x\geqslant 0$
and vanishing otherwise.
It should be noted that these initial distributions for
the phase field functions and the velocity have no effect on
the long-term behavior of the system. Any transient influence
will be convected out of the domain eventually.
The values for the simulation parameters in this problem
are collected in Table \ref{table:oiljet2}. 

We have considered two cases, corresponding to two different density
values for the oil: $400kg/m^3$ in the first case, and
$100 kg/m^3$ in the second case.

\begin{figure}[tbp]
  \centering
    \includegraphics[width=0.4\textwidth,height=0.4\textwidth]{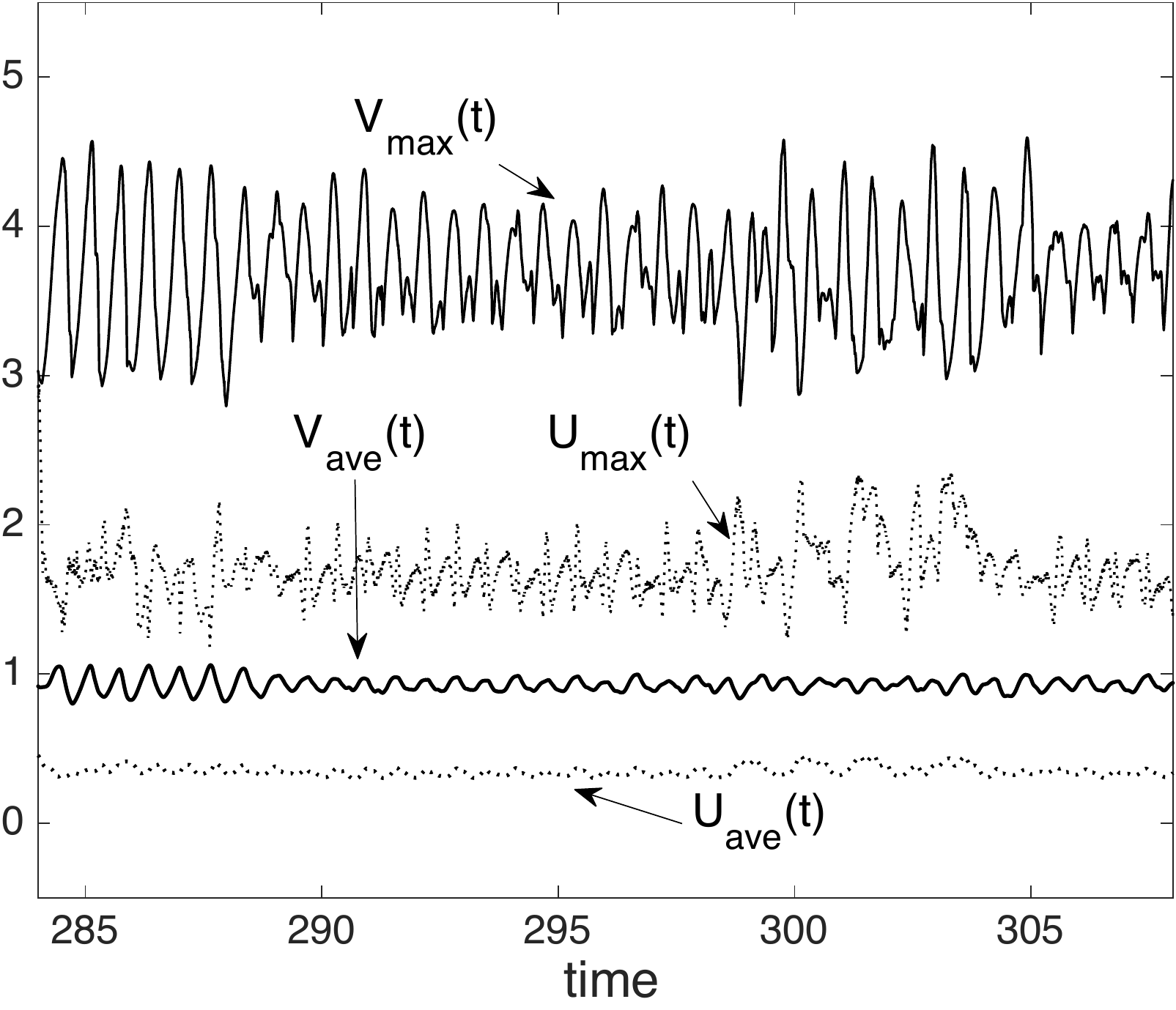}
    \caption{\small \small Time histories of the maximum and average velocity magnitudes for
      the case $(\tilde \rho_1, \tilde \rho_2, \tilde \rho_3)=(1,1.664, 0.667)$, showing that the flow has reached a statistically stationary state.}
    \label{velositymagrho1}
\end{figure}

Let us first consider the case with an oil density $400kg/m^3$. 
The normalized densities for $F_1,$ water and oil are $(\tilde \rho_1,\tilde \rho_2,\tilde \rho_3)=(1,1.664, 0.667)$ for this case.
We have performed  a long-time simulation of the problem so that the flow has reached a statistically stationary  state.
We have monitored the following maximum magnitudes $U_{\rm max}, V_{\rm max}$ and average magnitudes $(U_{\rm ave}, V_{\rm ave})$ of the $x$ and $y$ components of velocity at each time step:
\begin{equation}\label{equ:velmagnitude}
\begin{aligned}
&U_{\rm max}(t)= {\rm max}_{\bs x\in \Omega}|u(\bs x, t)|,\quad V_{\rm max}(t)= {\rm max}_{\bs x\in \Omega}|v(\bs x, t)|; \\
&U_{\rm ave}(t)=\Big(\frac{1}{V_{\Omega}}\int_{\Omega}|u|^2 d{\Omega}   \Big)^{\frac{1}{2}},\quad V_{\rm ave}(t)=\Big(\frac{1}{V_{\Omega}}\int_{\Omega}|v|^2 d{\Omega}   \Big)^{\frac{1}{2}},
\end{aligned}
\end{equation}
where $V_{\Omega}=\int_{\Omega}d\Omega$ is the volume of the domain. Fig.~\ref{velositymagrho1}
shows a temporal window of the time histories of these velocity magnitudes.
It can be observed that while these physical quantities fluctuate over time,  their fluctuations
are all around some constant mean values,
indicating that the flow has reached a statistically stationary state.

\begin{figure}[tbp]
\centering
 \subfigure[$t=303.62$]{ \includegraphics[scale=.22]{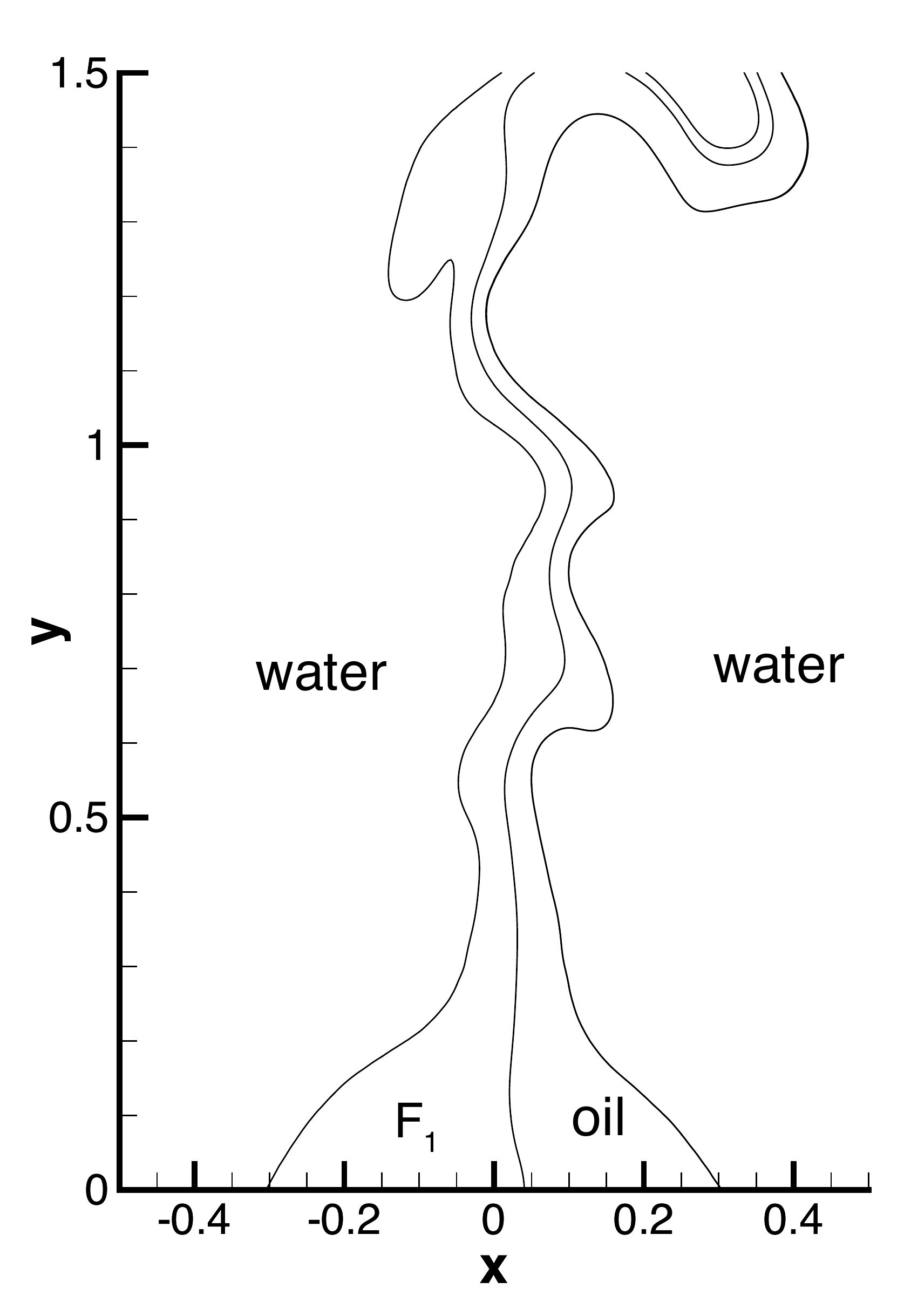}}  \hspace*{-10pt}
\subfigure[$t=303.72$]{ \includegraphics[scale=.22]{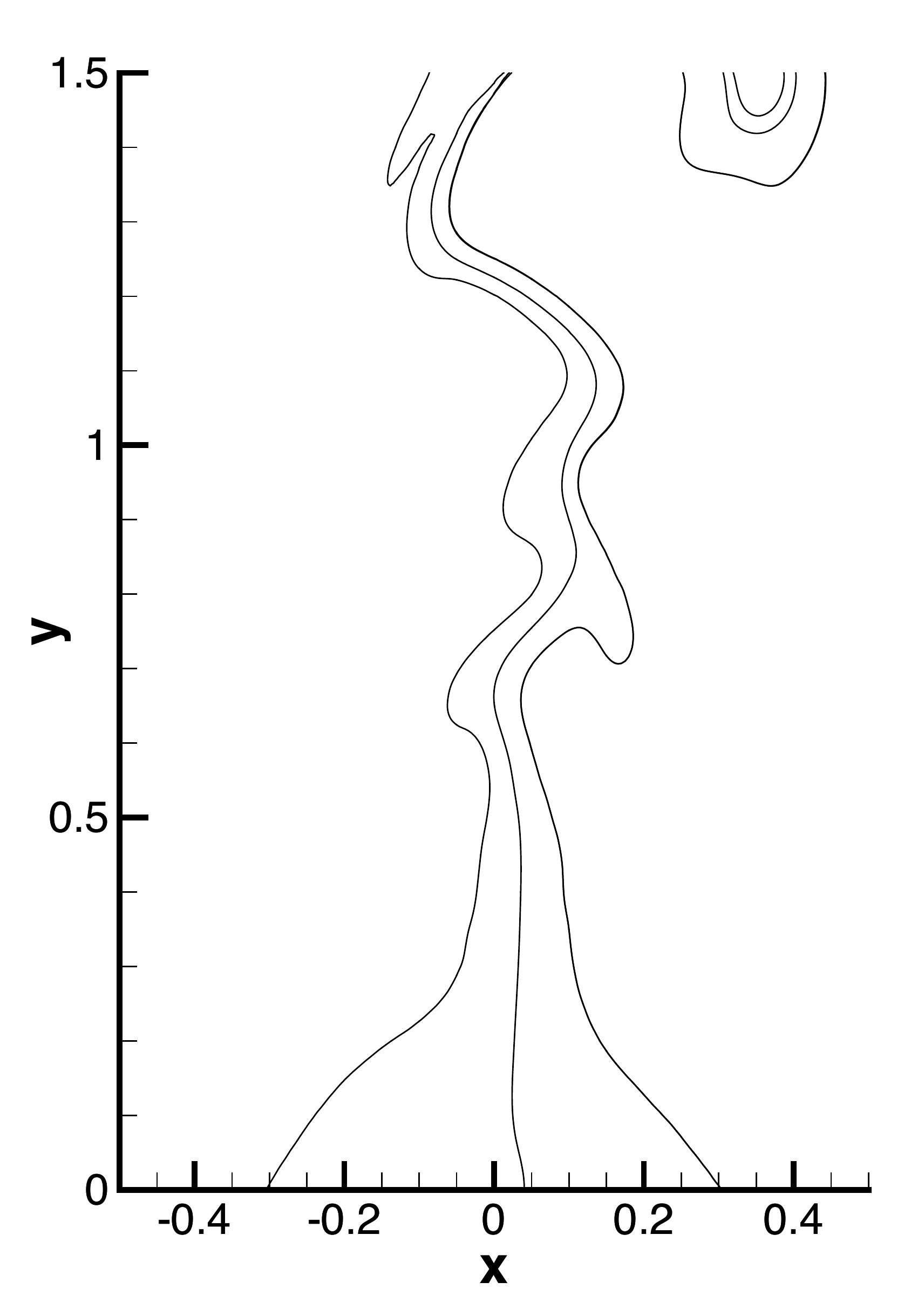}}  \hspace*{-10pt}
 \subfigure[$t=303.82$]{ \includegraphics[scale=.22]{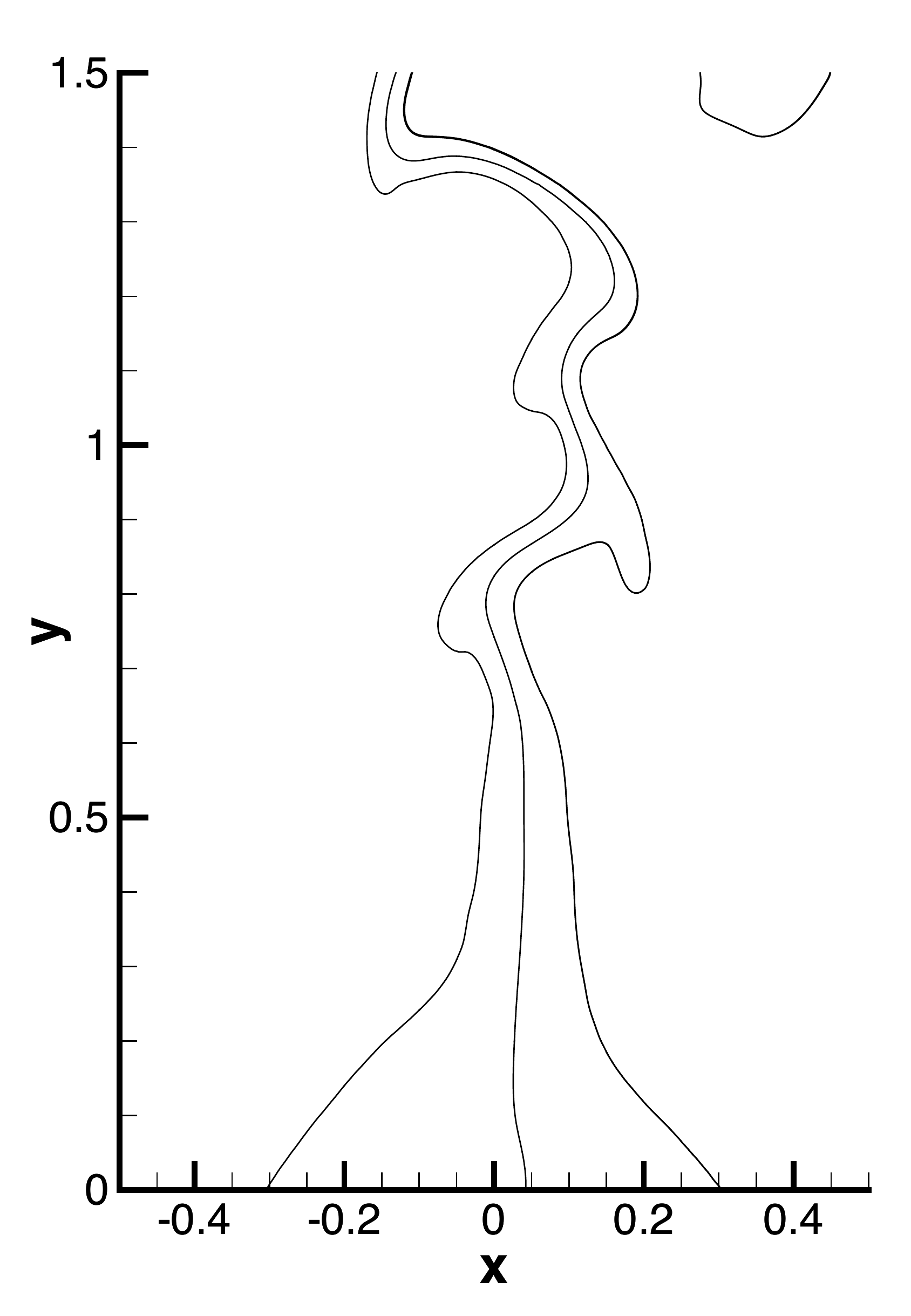}}  \hspace*{-10pt}
\subfigure[$t=303.92$]{ \includegraphics[scale=.22]{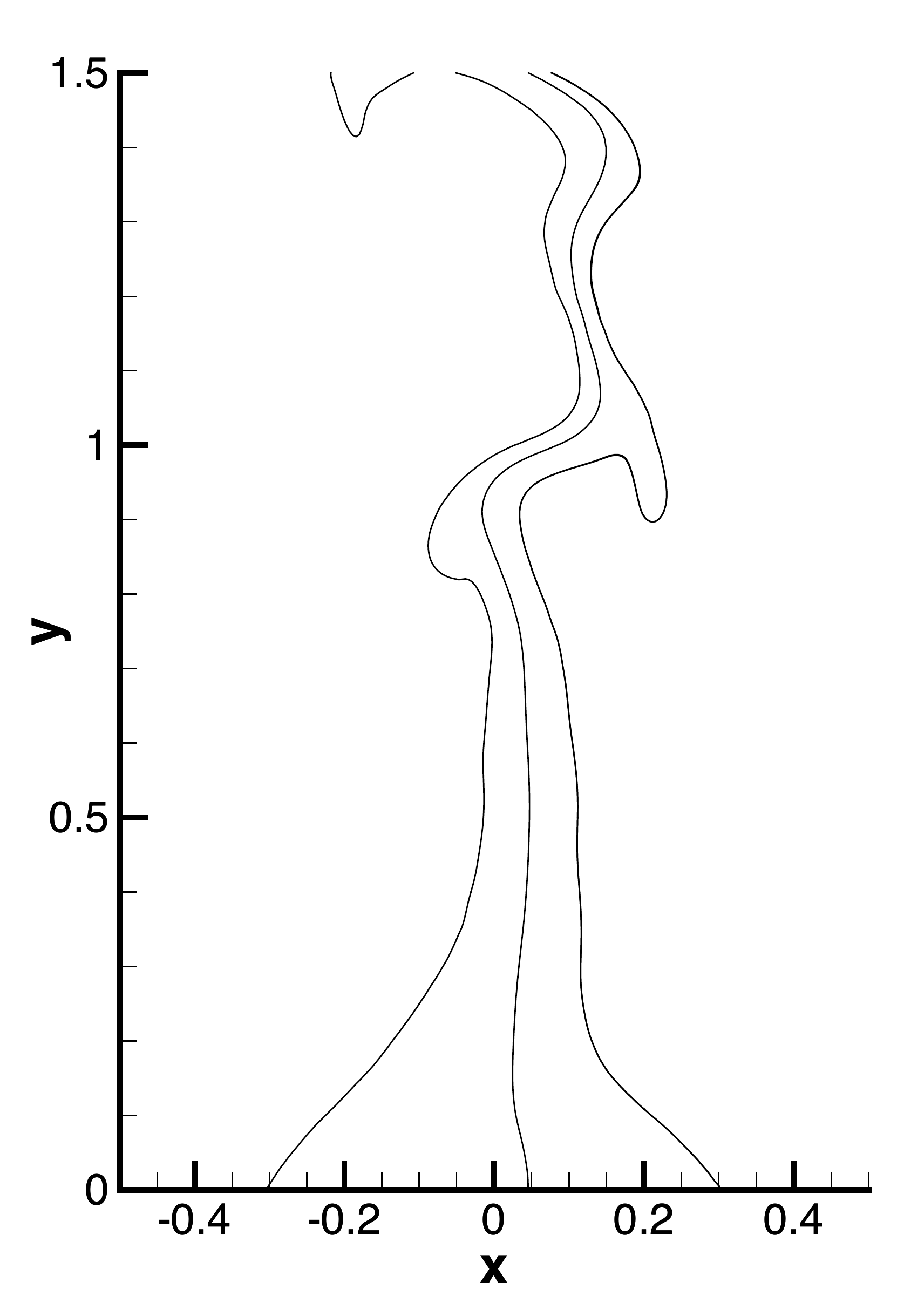}} \\
 \subfigure[$t=304.02$]{ \includegraphics[scale=.22]{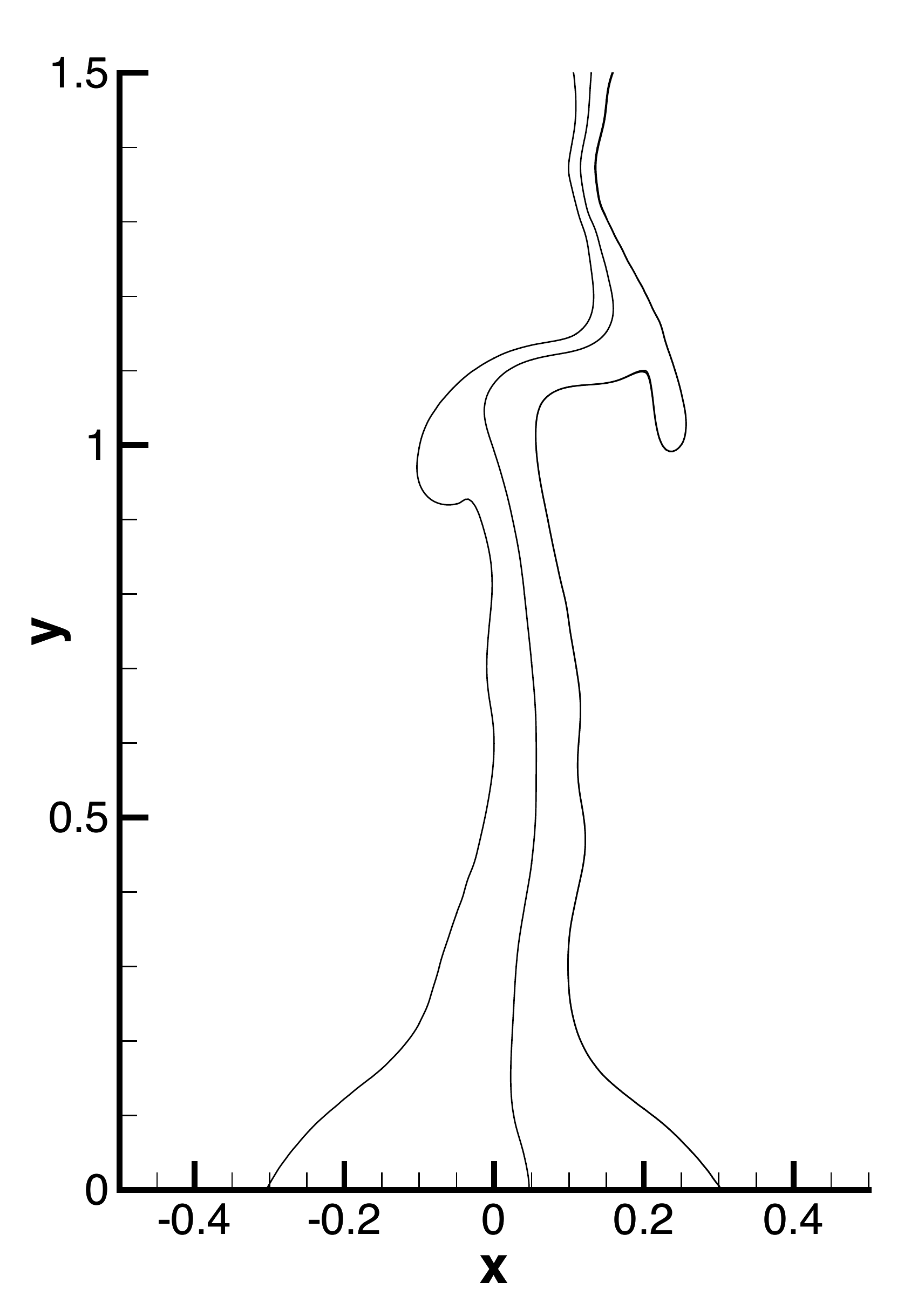}} \hspace*{-10pt}
\subfigure[$t=304.12$]{ \includegraphics[scale=.22]{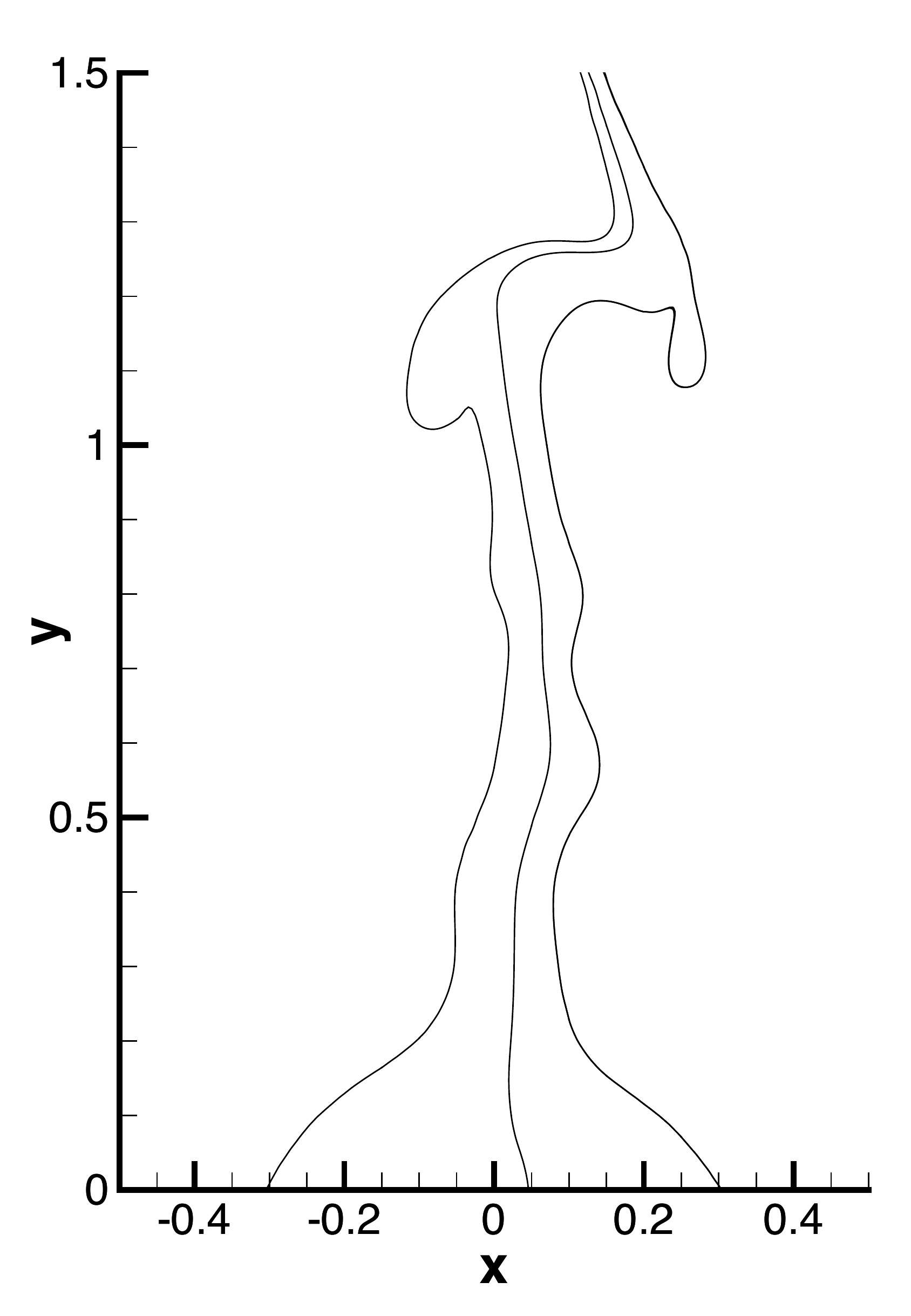}} \hspace*{-10pt}
 \subfigure[$t=304.22$]{ \includegraphics[scale=.22]{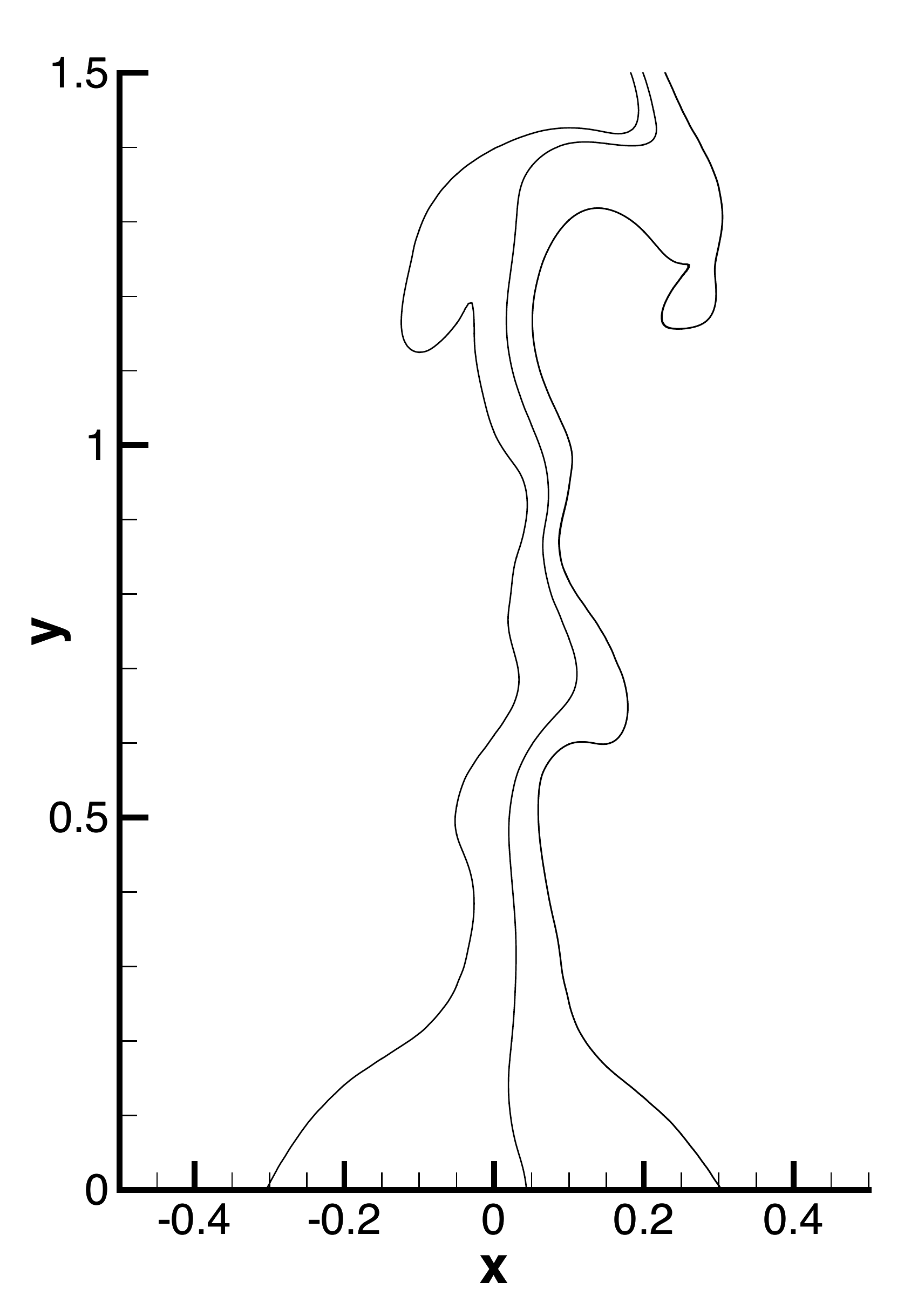}} \hspace*{-10pt}
\subfigure[$t=304.32$]{ \includegraphics[scale=.22]{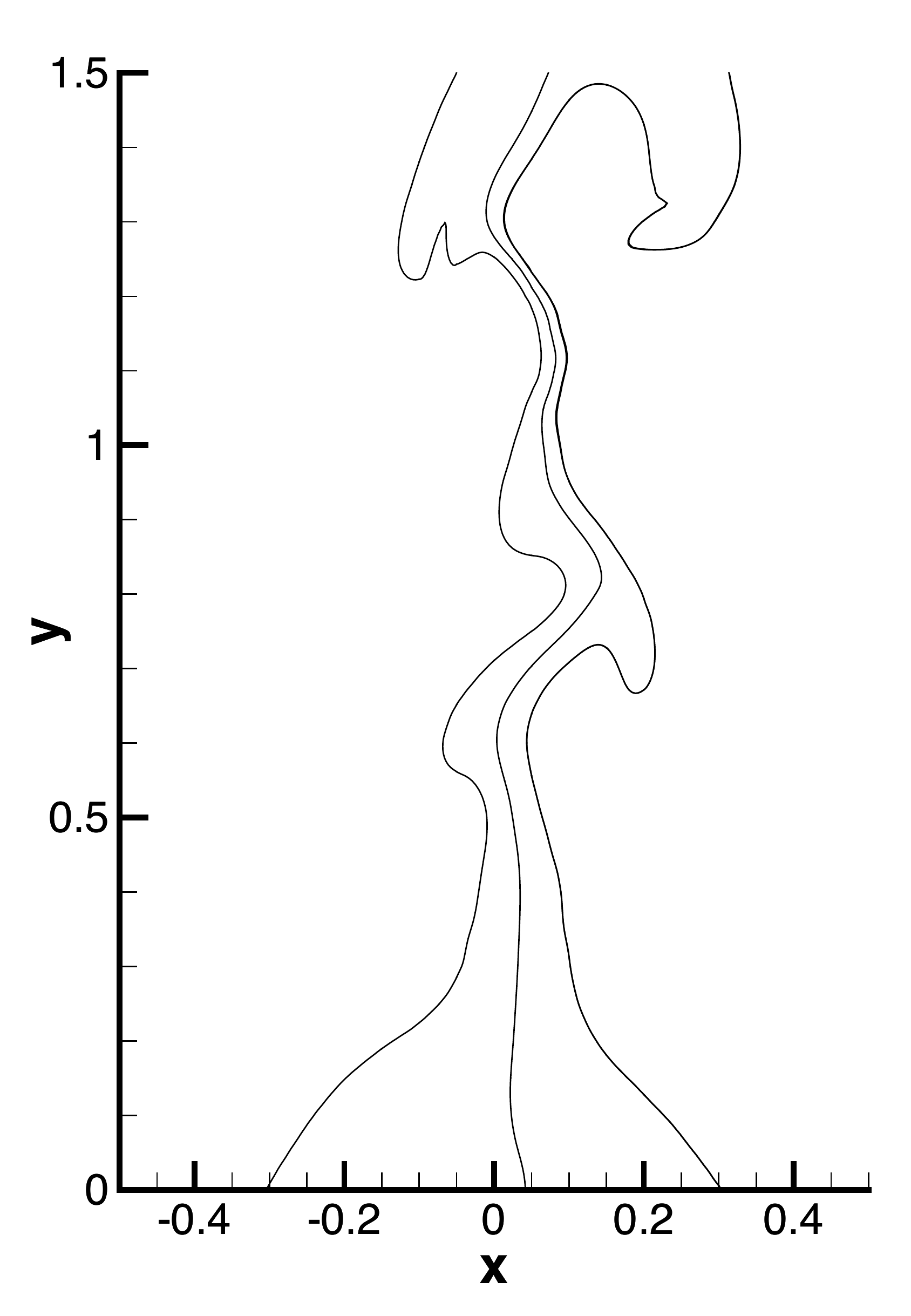}} \\
 \subfigure[$t=304.42$]{ \includegraphics[scale=.22]{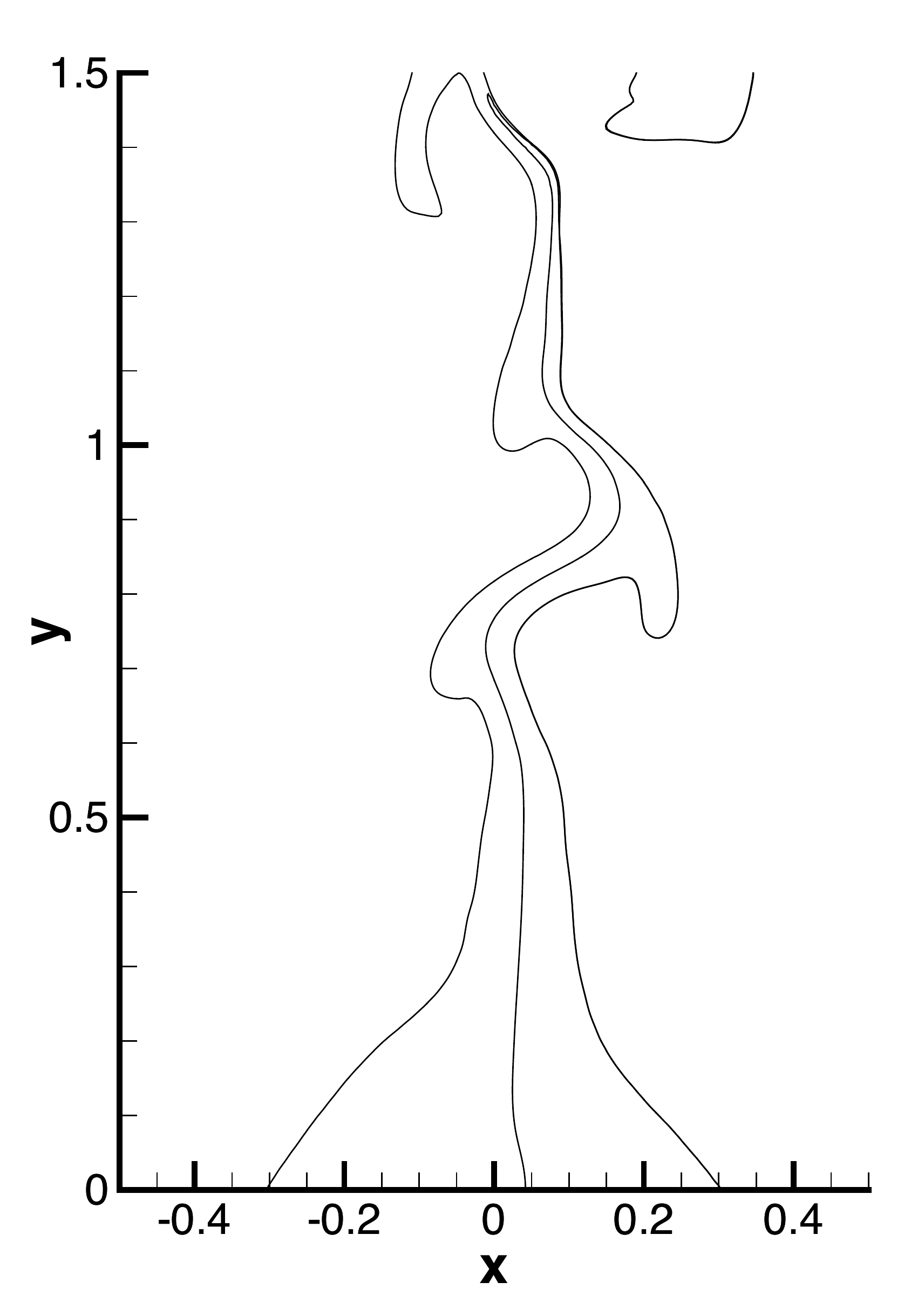}}  \hspace*{-10pt}
\subfigure[$t=304.52$]{ \includegraphics[scale=.22]{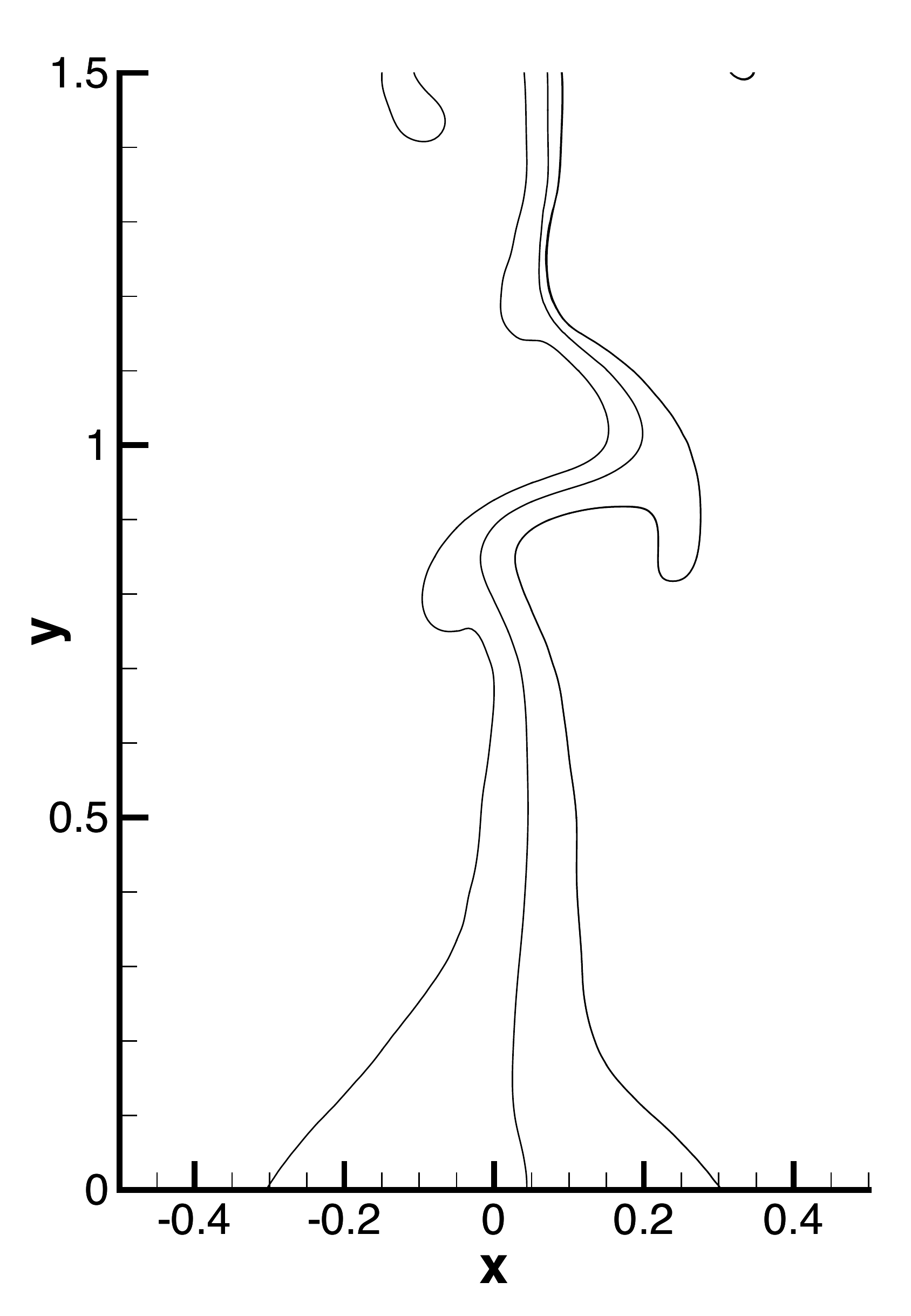}}  \hspace*{-10pt}
 \subfigure[$t=304.62$]{ \includegraphics[scale=.22]{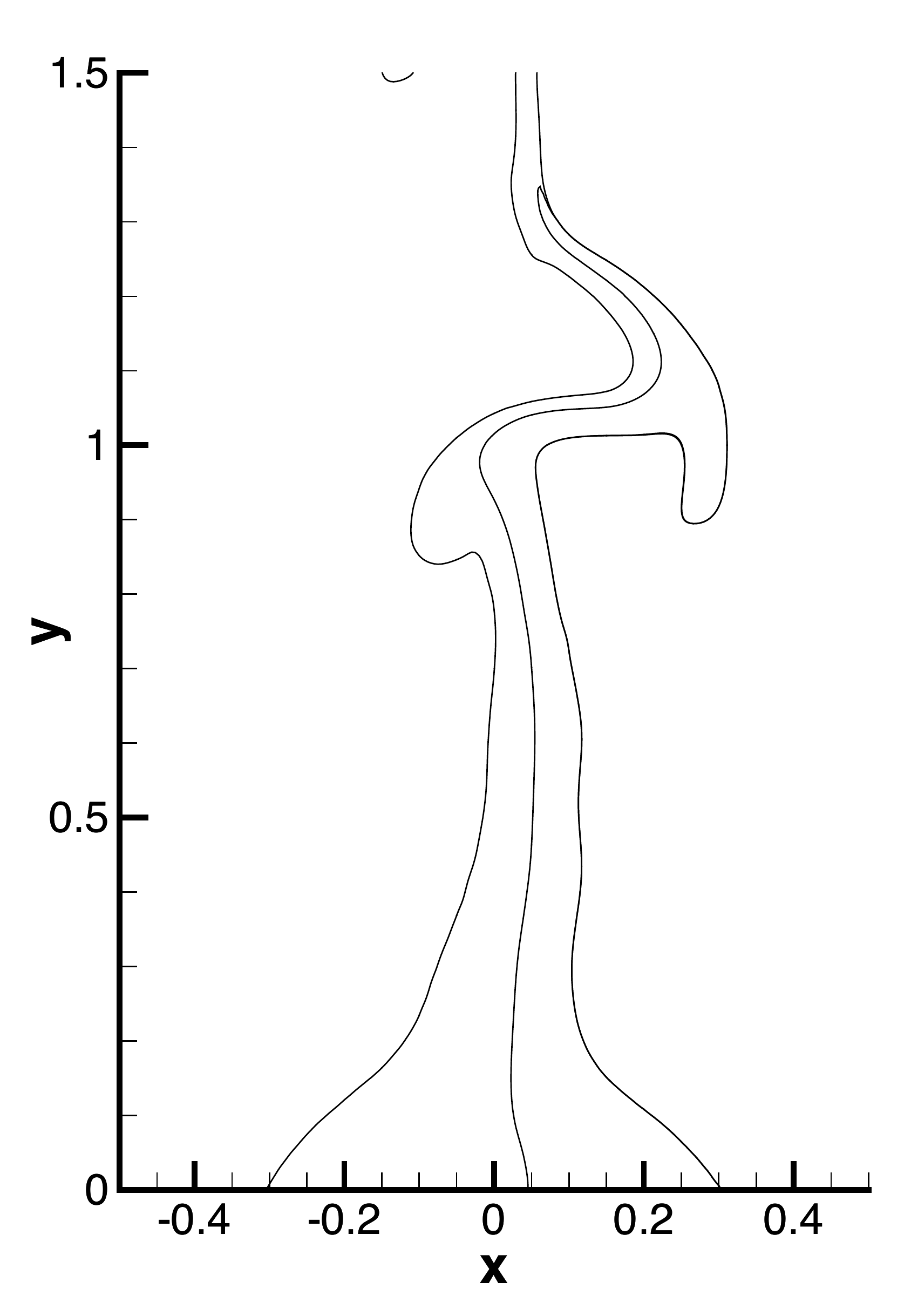}}  \hspace*{-10pt}
\subfigure[$t=304.72$]{ \includegraphics[scale=.22]{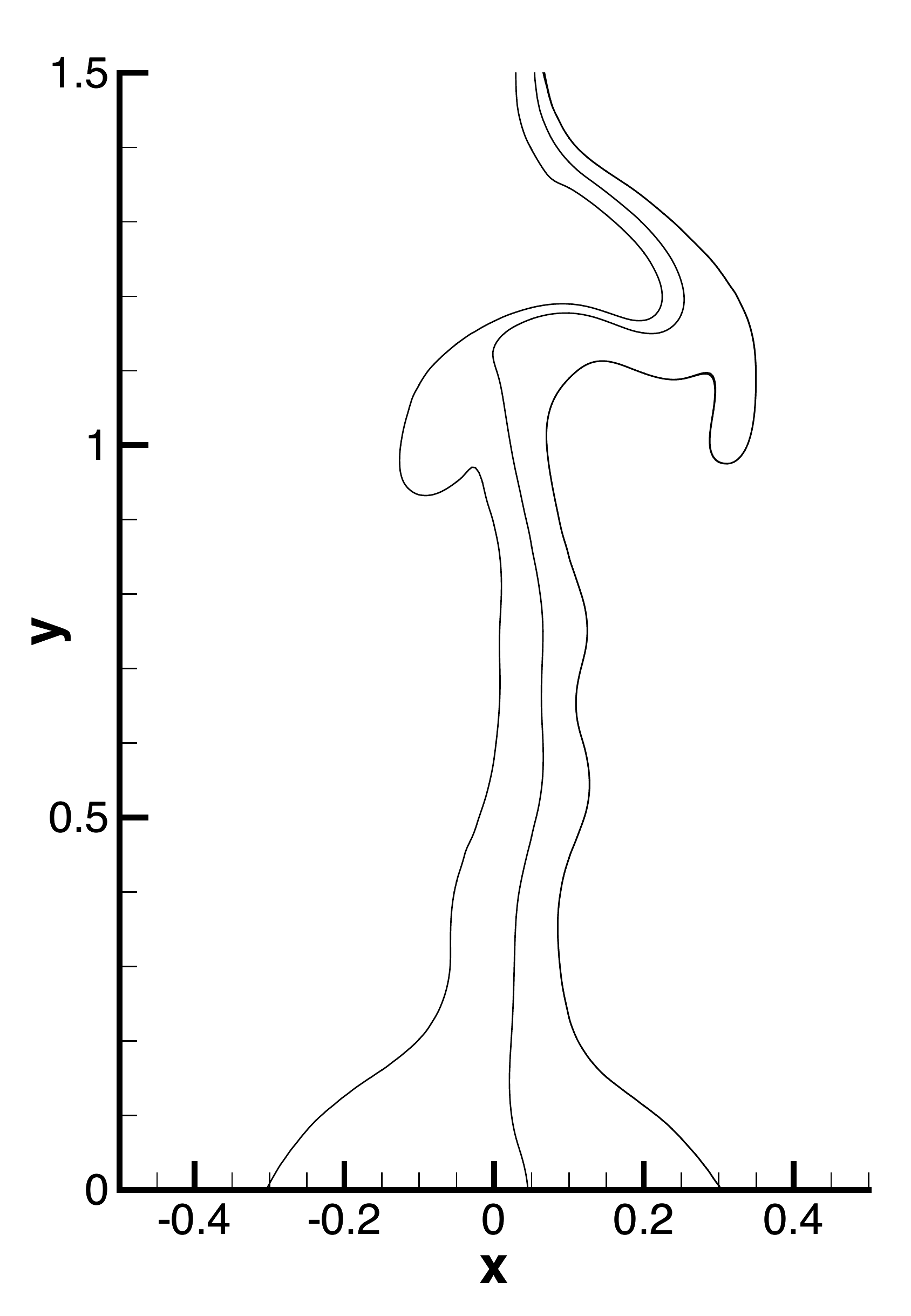}} 
\caption{Temporal sequence of snapshots of fluid interfaces, visualized by the volume-fraction contours $c_i=1/2\, (i=1,2,3),$ showing the interaction of two liquid jets of $F_1$  and oil in water, with the normalized densities $(\tilde \rho_1,\tilde \rho_2,\tilde \rho_3 )=(1,1.664,0.667)$.
  The jet inlets are centered at $x=-0.2,\,0.2,$ respectively with radius $0.1$.
}
\label{oiljetrho1}
\end{figure}

We look into the dynamical characteristics of the fluid $F_1$ and oil jets in water.
Fig.~\ref{oiljetrho1} shows a temporal sequence of snapshots of the fluid interfaces,
visualized by contours of the volume fractions $c_i=1/2 \,(i=1,2,3)$ for
the three fluids.
First we observe that at the bottom wall the F$_1$ fluid and the oil coming out of
the orifices spread on the wall and fill up the space in between. As a result, the
two fluids touch each other on the bottom wall and form a compound oil-$F_1$ jet.
Note that the base of the compound oil-F$_1$ jet is broader than
the combined size of the two orifices.
The compound jet exhibits distinct characteristics in different regions.
In the region near the orifices ($y/L \lesssim 0.5$ in this case), the compound
jet maintains a relatively stable configuration. The jet tapers off
along the vertical direction in this region,
due to the velocity increase caused by the buoyancy.
This is reminiscent of the behavior of a single oil jet in ambient water
studied in \cite{Dong2014obc}.
Beyond this stable region, the compound jet exhibits  a wavy pattern in its profile. The jet diameter modulates along the vertical direction, and bulges form around
the jet continually and periodically
(Figs.~\ref{oiljetrho1}(a)-(b), (f)-(h)) due to a Plateau-Rayleigh
instability~\cite{Plateau1873,Rayleigh1892}.
Further downsteam, the dynamics of the jet becomes very complicated.
The compound jet and the bulges along its profile appear to fold back
in certain regions at times, causing very large deformations of the jet;
see e.g.~Figs.~\ref{oiljetrho1}(e)-(g) and (j)-(l).
We observe that the regions occupied by the F$_1$ fluid and by the oil
in the compound jet are not symmetric.
It can also be observed that our method allows the compound oil-F$_1$ jet
and the fluid interfaces to
exit the domain through the open boundary in a fairly natural fashion;
see e.g.~Figs.~\ref{oiljetrho1}(a)-(d) and (h)-(k).

\begin{figure}[tbp]
\centering
 \subfigure[$t=303.62$]{ \includegraphics[scale=.22]{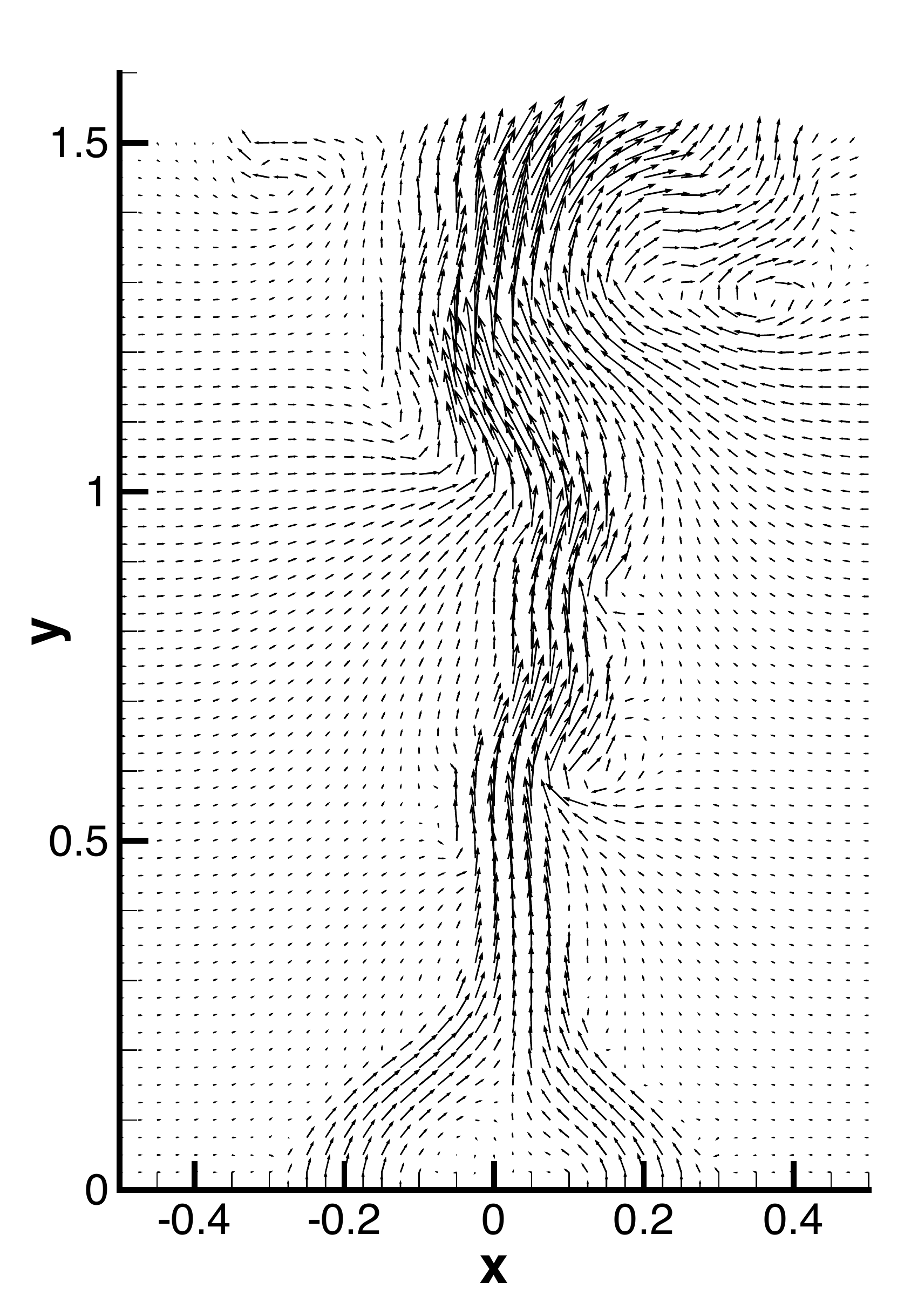}}  \hspace*{-10pt}
\subfigure[$t=303.72$]{ \includegraphics[scale=.22]{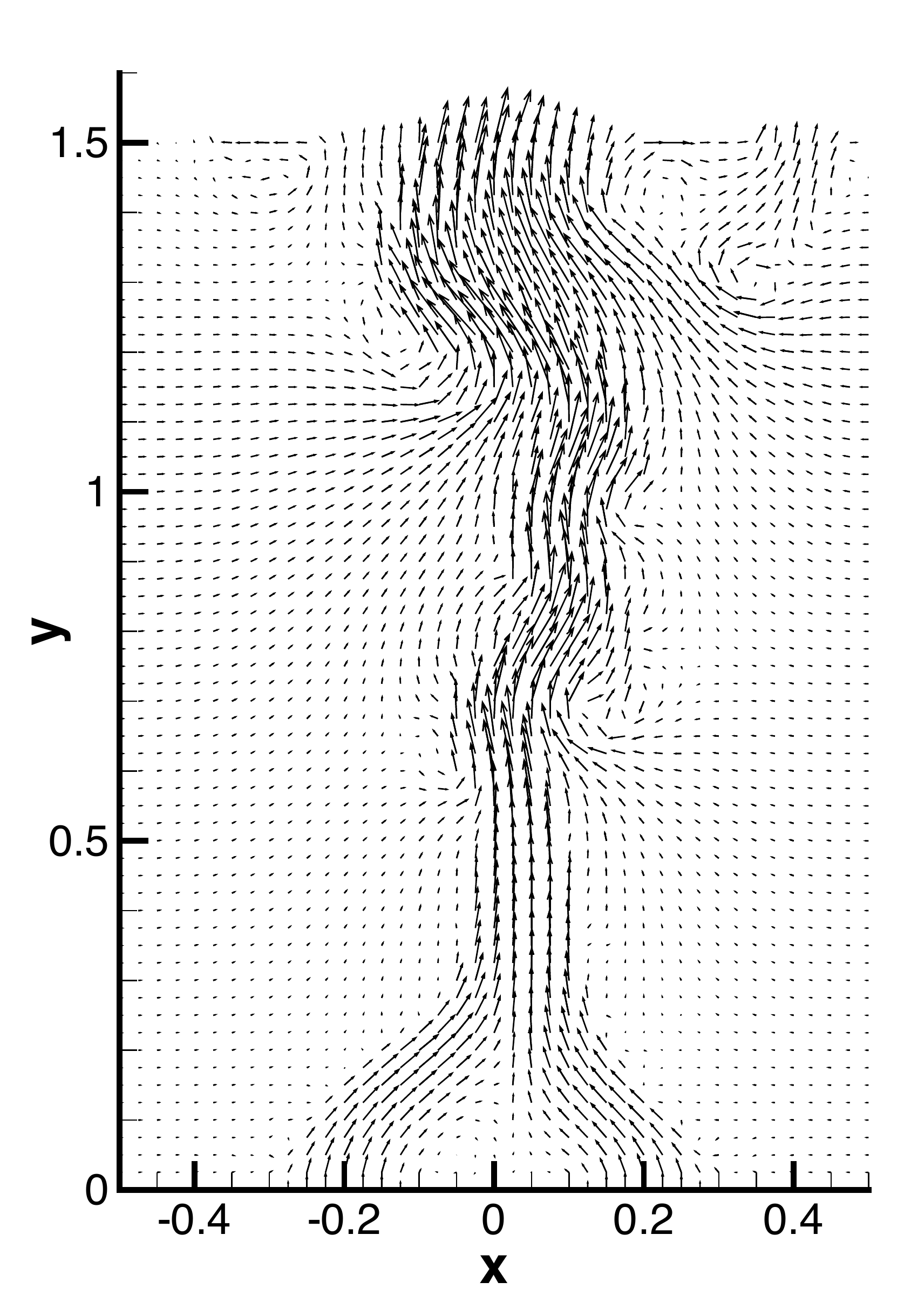}}  \hspace*{-10pt}
 \subfigure[$t=303.82$]{ \includegraphics[scale=.22]{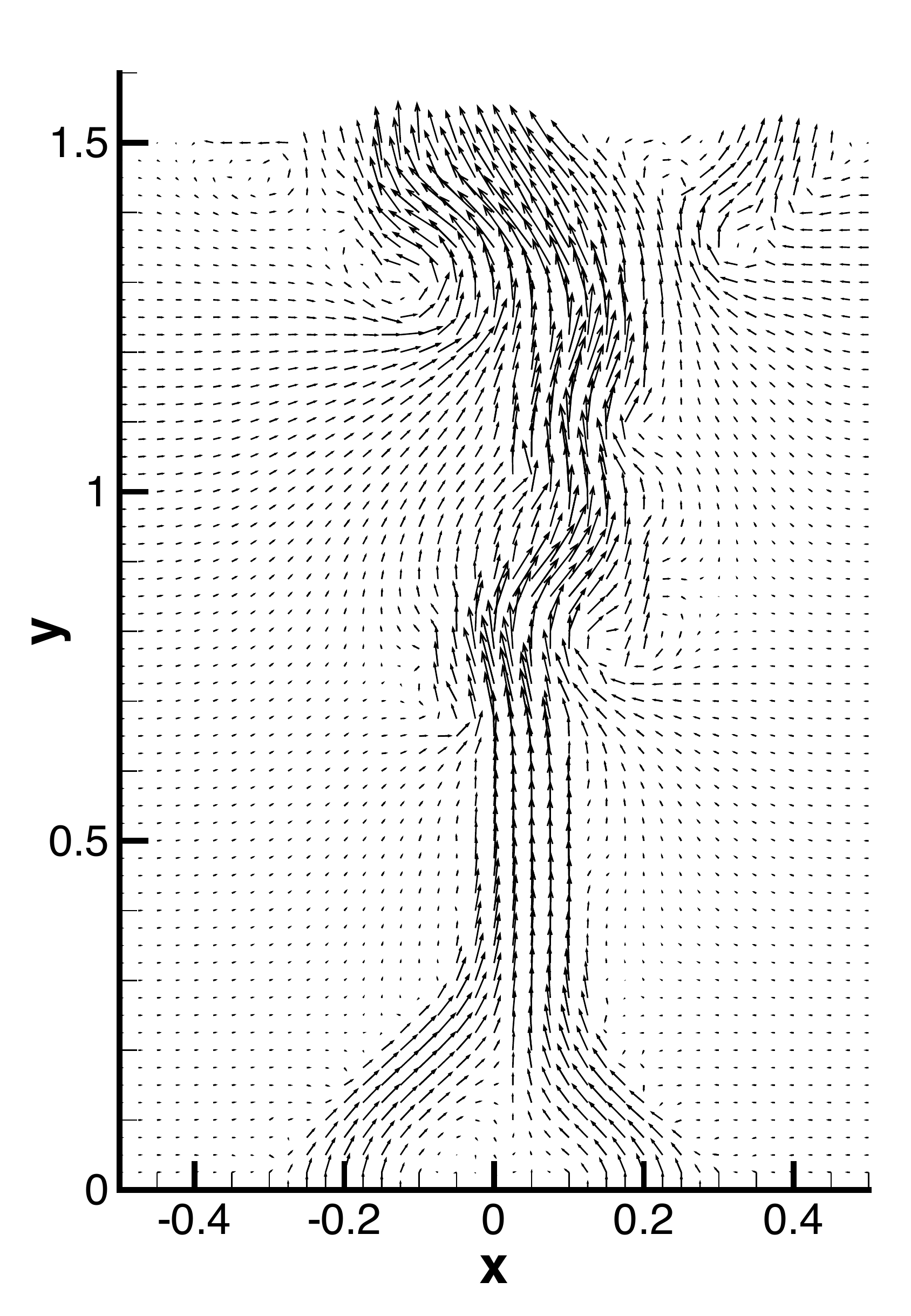}}  \hspace*{-10pt}
\subfigure[$t=303.92$]{ \includegraphics[scale=.22]{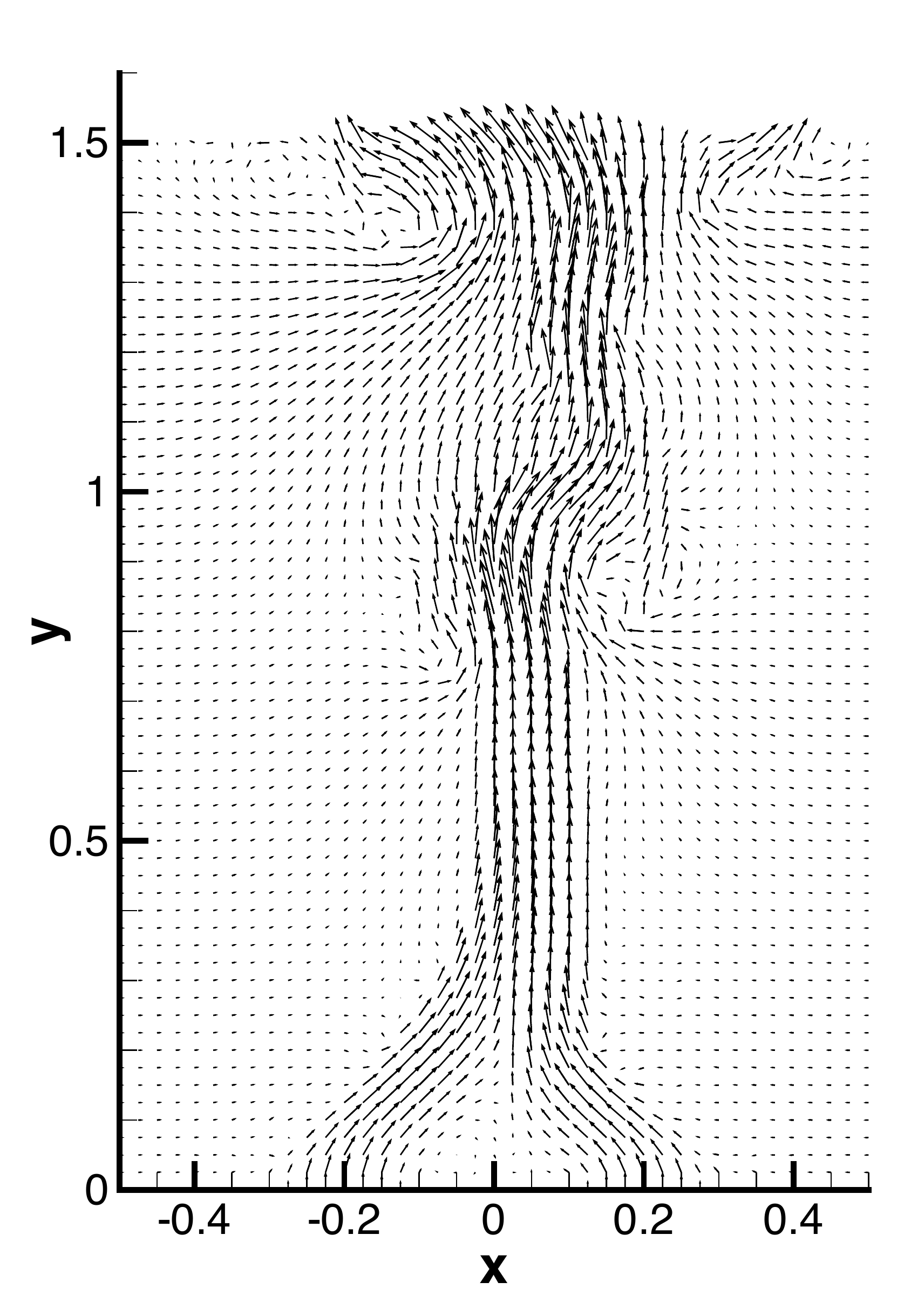}} \\
 \subfigure[$t=304.02$]{ \includegraphics[scale=.22]{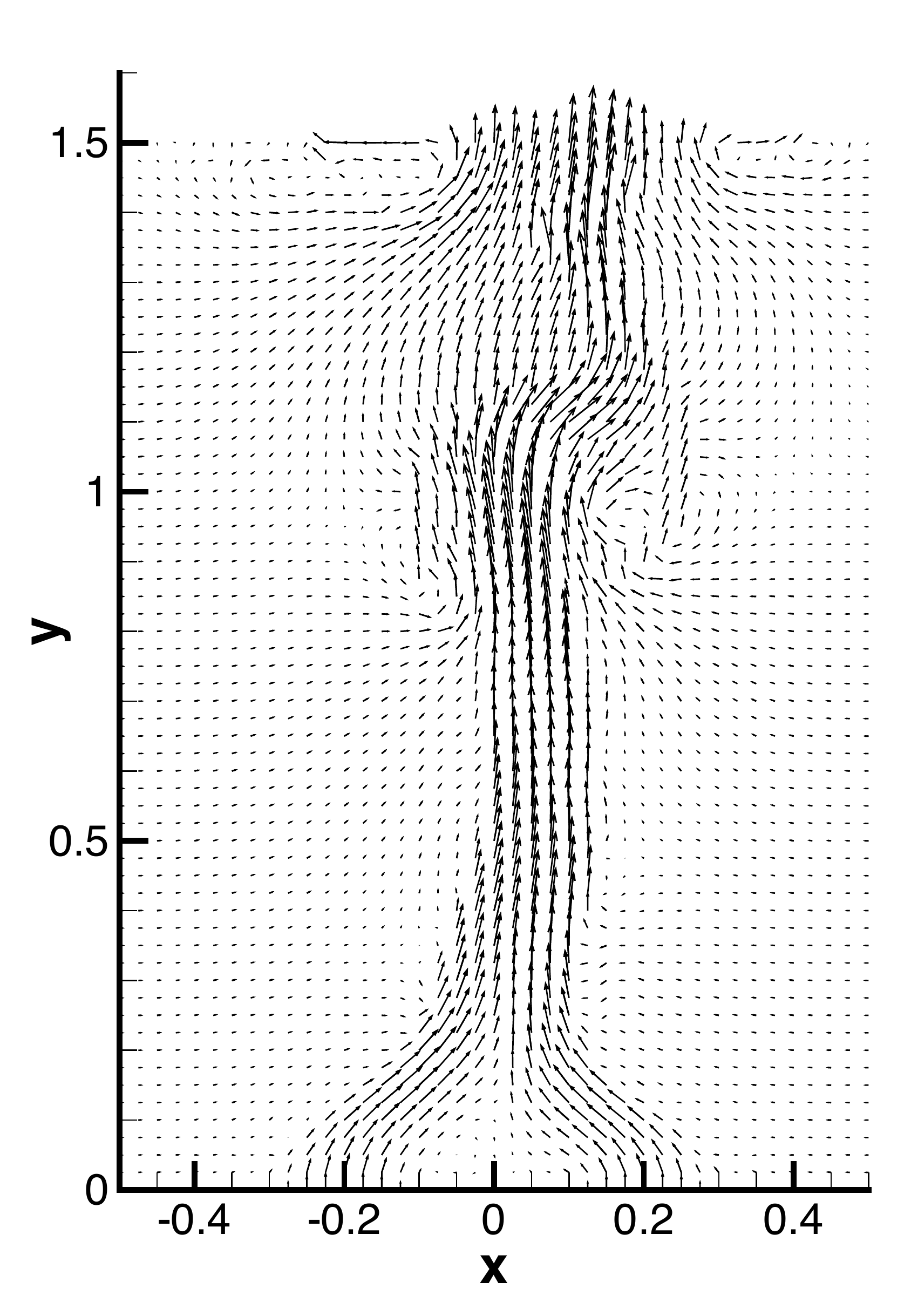}}  \hspace*{-10pt}
\subfigure[$t=304.12$]{ \includegraphics[scale=.22]{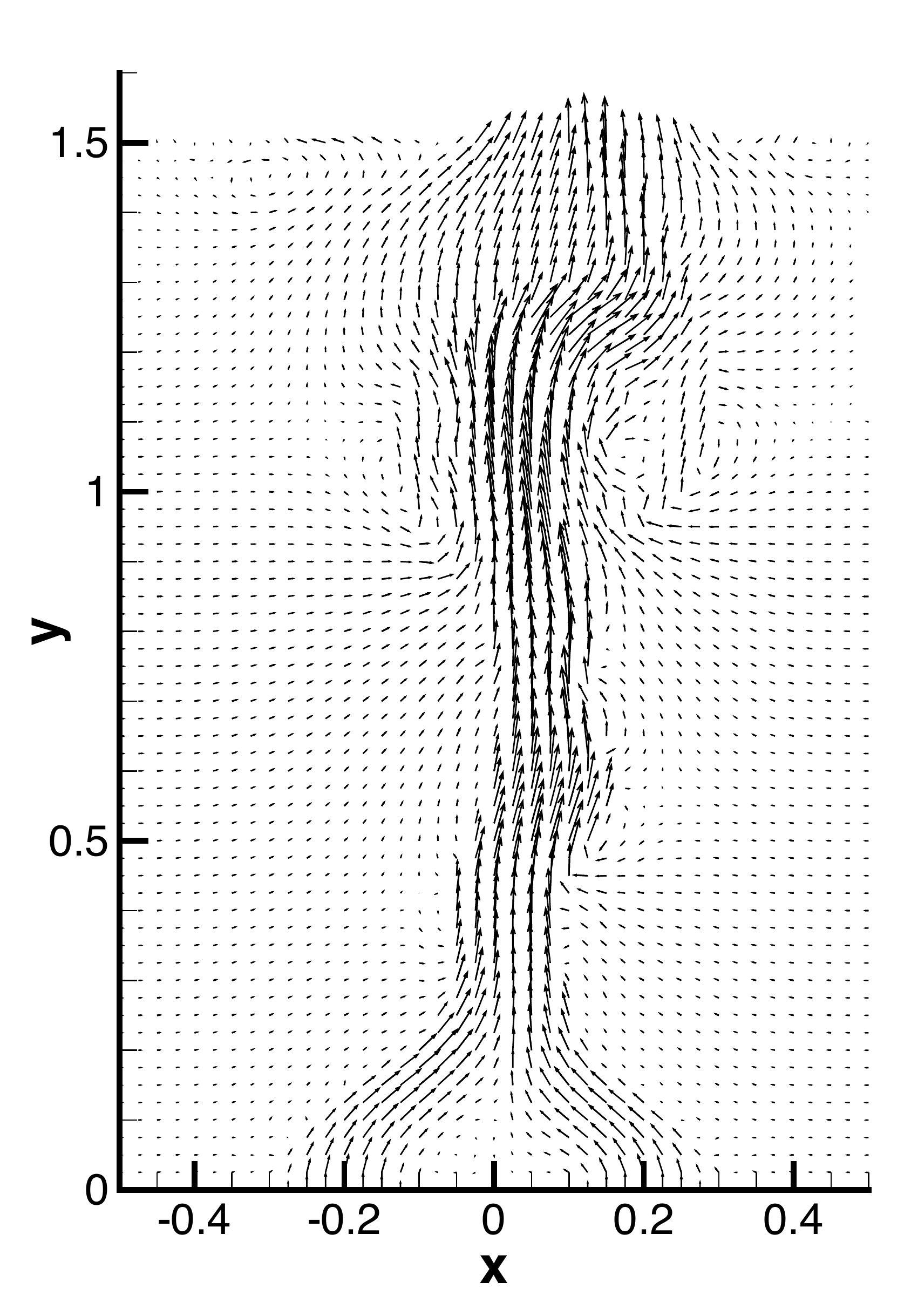}}  \hspace*{-10pt}
 \subfigure[$t=304.22$]{ \includegraphics[scale=.22]{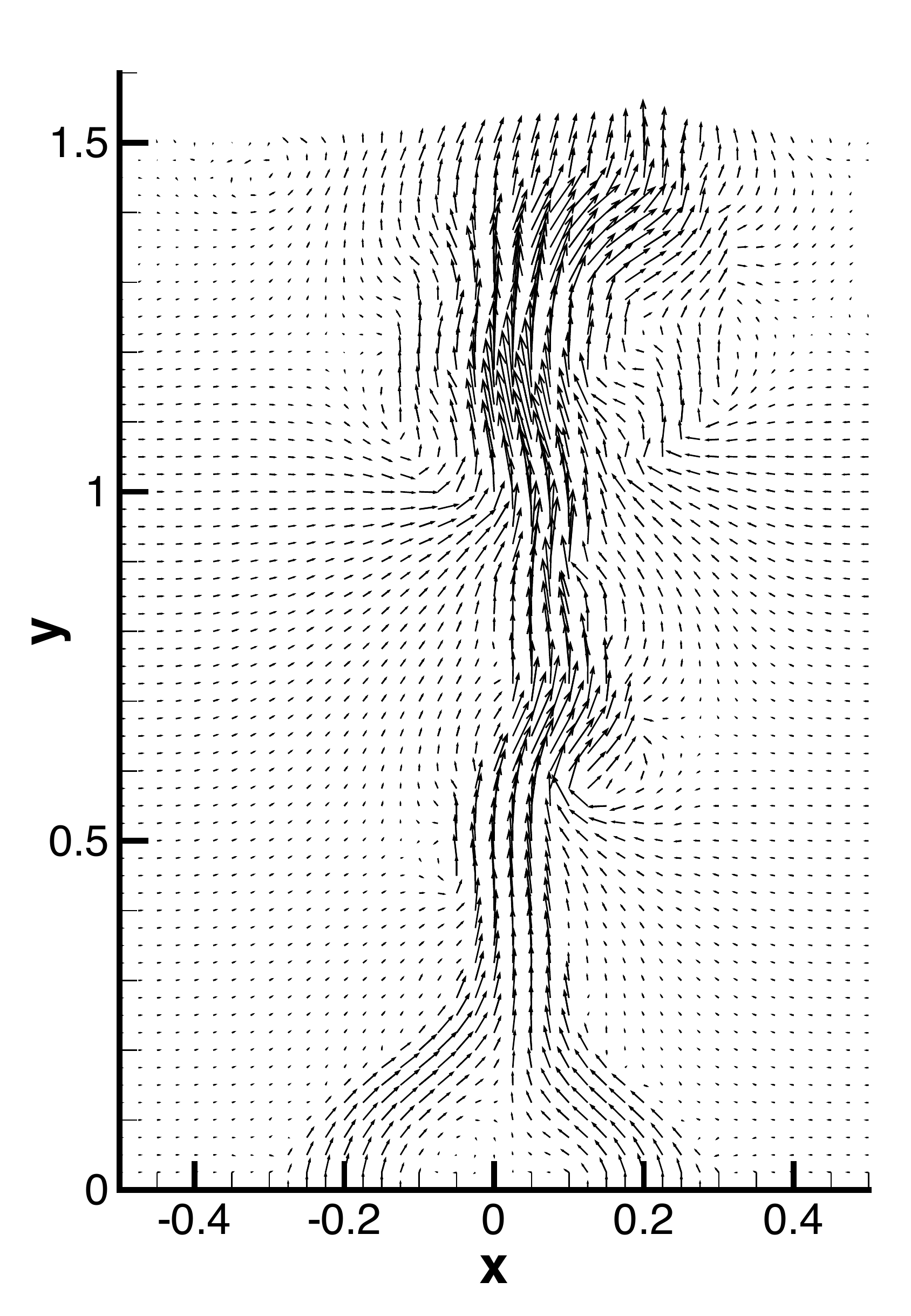}}  \hspace*{-10pt}
\subfigure[$t=304.32$]{ \includegraphics[scale=.22]{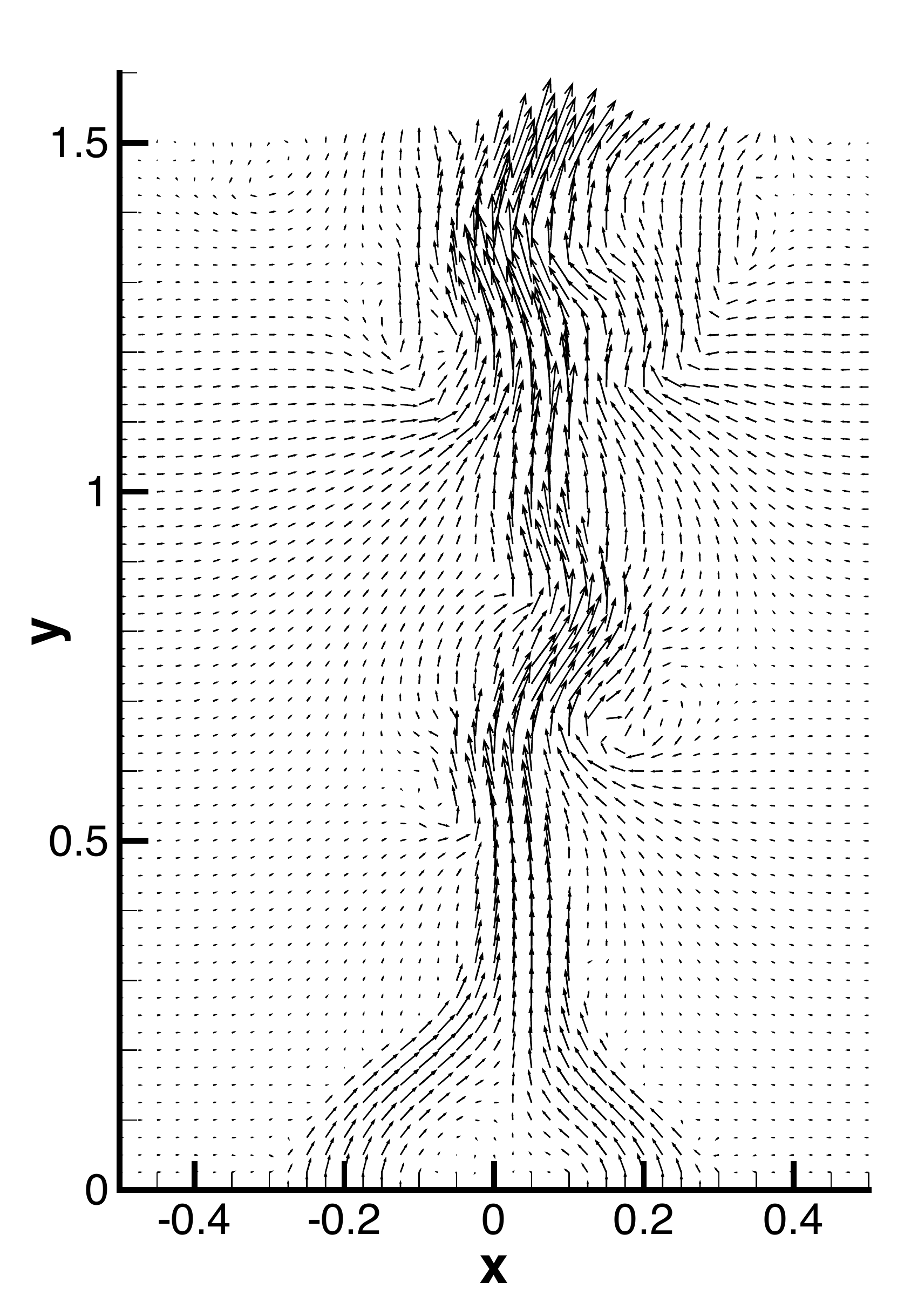}} \\
 \subfigure[$t=304.42$]{ \includegraphics[scale=.22]{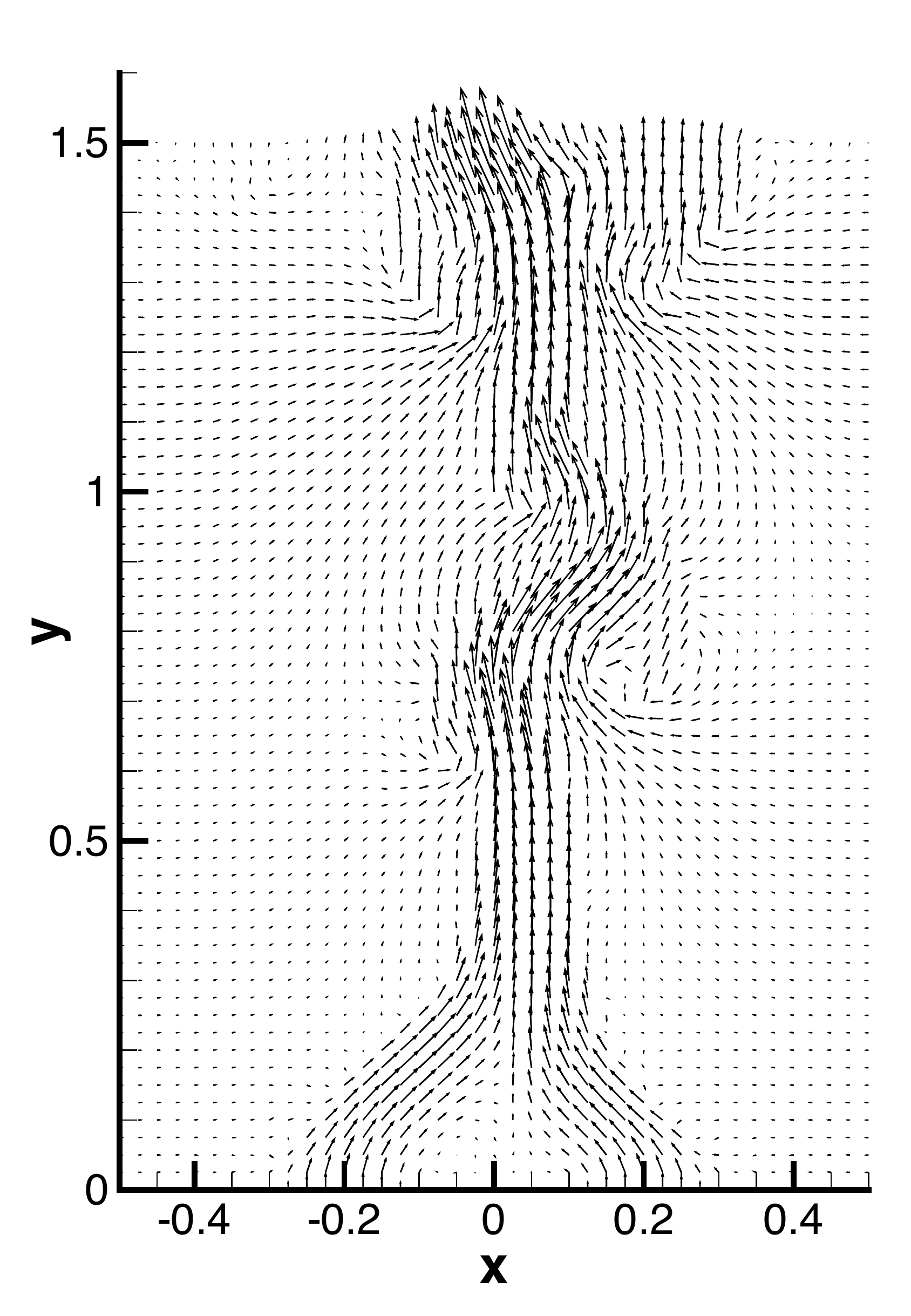}}  \hspace*{-10pt}
\subfigure[$t=304.52$]{ \includegraphics[scale=.22]{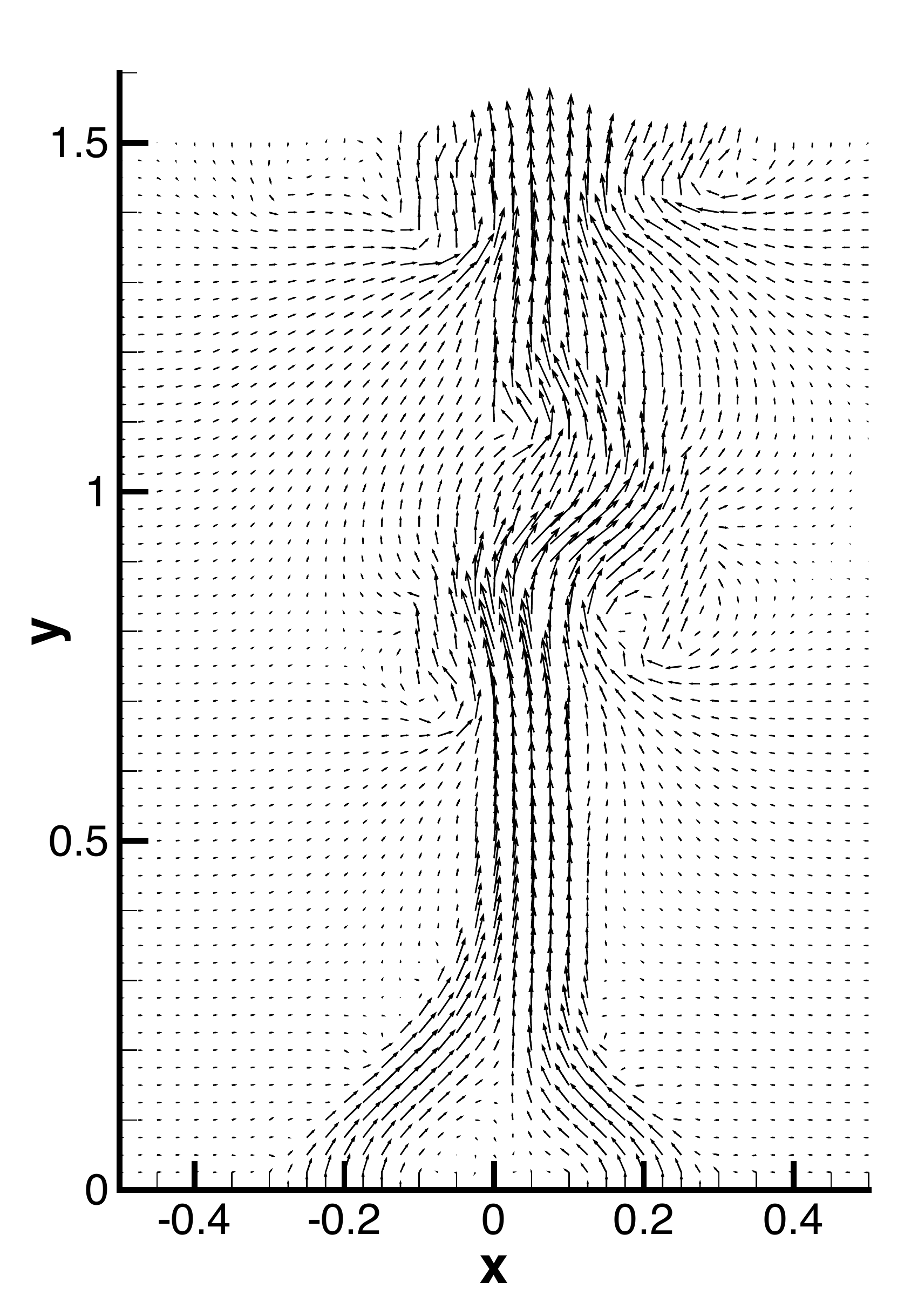}}  \hspace*{-10pt}
 \subfigure[$t=304.62$]{ \includegraphics[scale=.22]{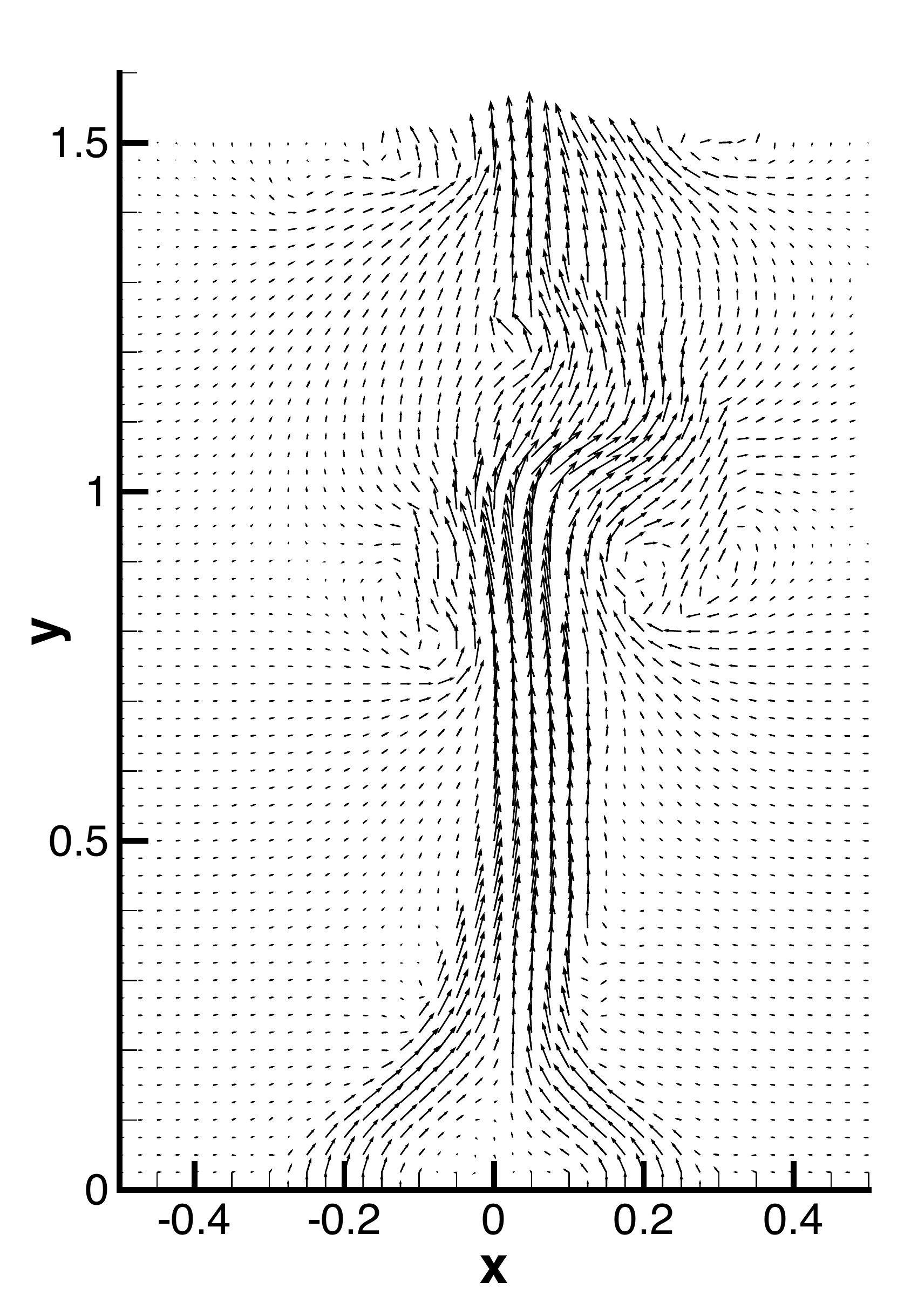}}  \hspace*{-10pt}
\subfigure[$t=304.72$]{ \includegraphics[scale=.22]{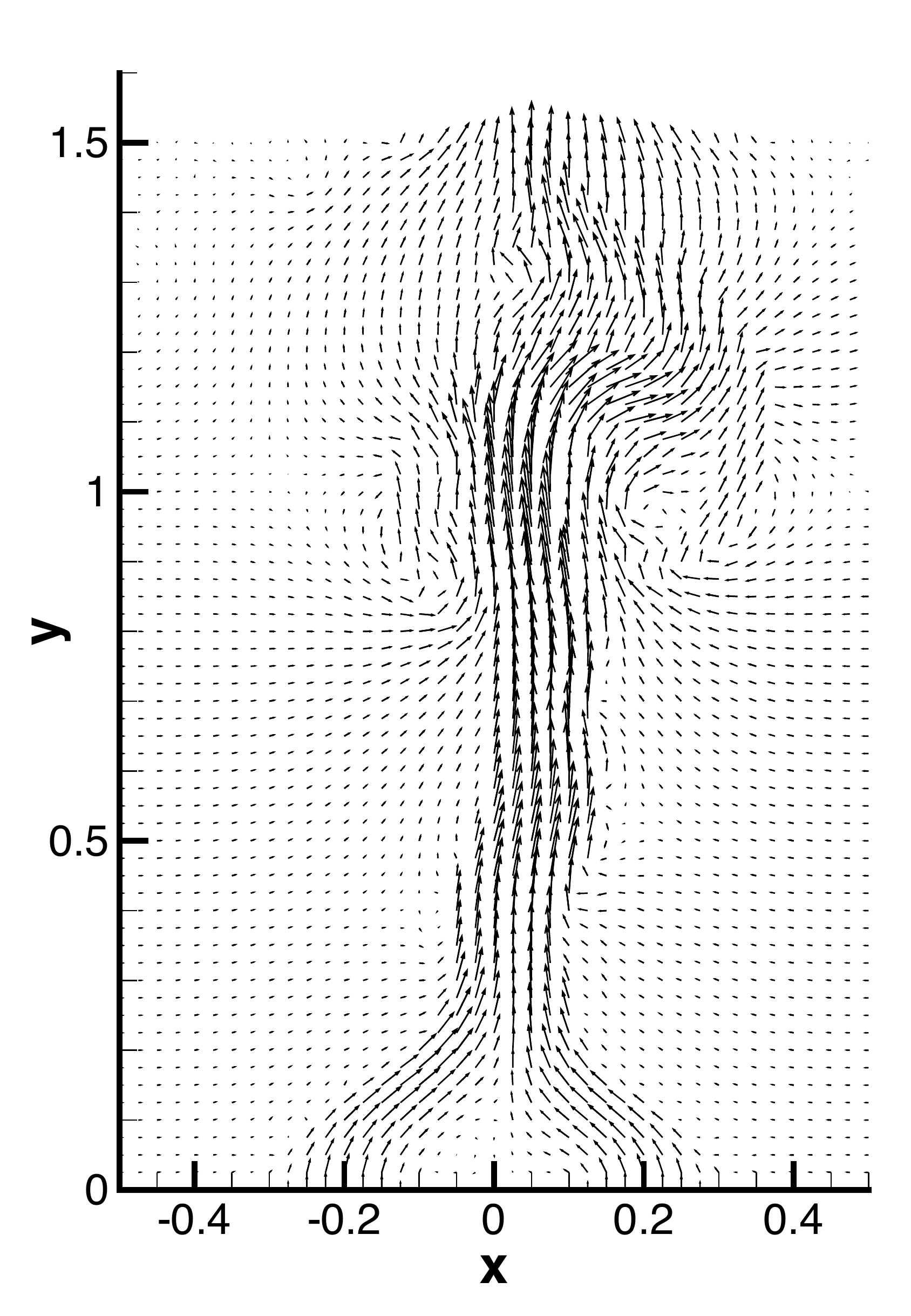}} 
\caption{Temporal sequence of snapshots of velocity distributions of
  two liquid jets in water with normalized densities $(\tilde \rho_1,\tilde \rho_2,\tilde \rho_3 )=(1,1.664,0.667)$.
  Velocity vectors are plotted on every eighth quadrature points in each direction within each element.
}
\label{oiljetrhovector1}
\end{figure}

Fig.~\ref{oiljetrhovector1} shows a temporal sequence of snapshots of the velocity fields of this flow, taken at identical time instants as those of the volume-fraction plots of Fig.~\ref{oiljetrho1}.
Several characteristics are evident from these plots.
First, the velocity patterns clearly indicate that
the streams of the F$_1$ fluid and the oil bend toward each other
after exiting the orifices, and merge to form a flow stream of
the compound jet. The velocity in the region between the two
orifices near the wall is very weak. Note that this region is occupied
by the F$_1$ and the oil.
Second,
the region occupied by the compound jet stream,
as shown by the velocity patterns, is wider than the actual region the material oil/F$_1$ occupy (see Fig.~\ref{oiljetrho1}),
especially in the regions more downstream and near
the upper open boundary.
This suggests that the water in the vicinity of compound $F_1$-oil jet has been accelerated to form a wider high-speed region.
Third,
the jet stream exhibits a lateral spread along the streamwise direction, as can be observed from the velocity patterns, and
pairs of vortices can be observed to form along the jet profile.
These vortices reside behind the $F_1$-oil bulges,
form periodically as new bulges emerge,
and travel downstream along with the bulges.
Finally, we note that on the side boundaries
the velocity generally points into the domain,
indicating that the water has in general been sucked into the domain from both sides.
The velocity patterns of Fig.~\ref{oiljetrhovector1}
indicate that the method developed herein allows the flow
to pass through the open/outflow boundaries in a smooth
and natural way.

\begin{figure}[tbp]
  \centering
    \includegraphics[height=0.4\textwidth]{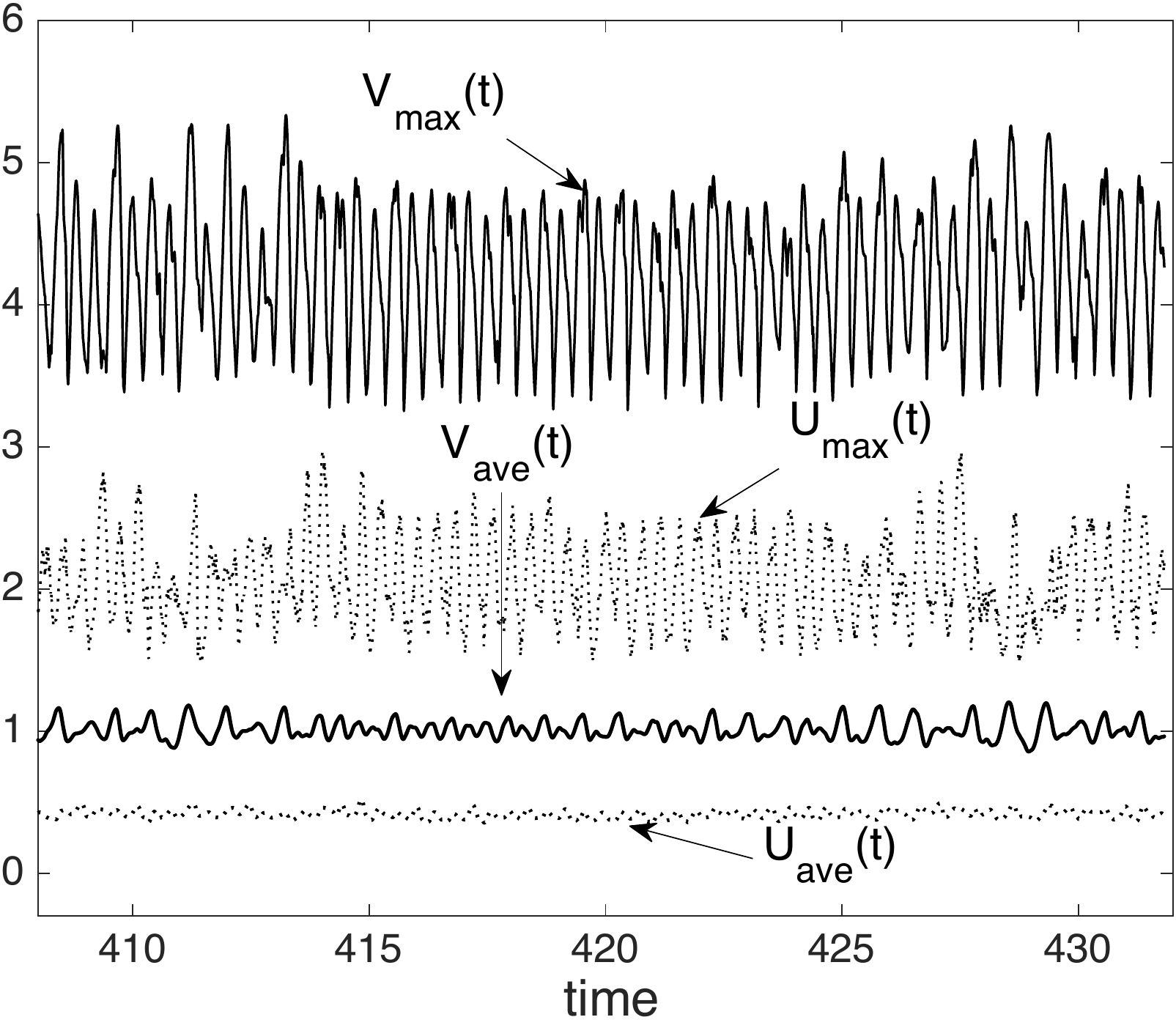}
    \caption{\small Two liquid jets in water: time histories of the
      maximum and average velocity magnitudes  with normalized densities
      $(\tilde \rho_1, \tilde \rho_2, \tilde \rho_3)=(1,1.664, 0.1664)$, showing that the flow has reached a statistically stationary state.}
    \label{velositymagrho2}
\end{figure}

\begin{figure}[tbp]
\centering
 \subfigure[$t=426.92$]{ \includegraphics[scale=.22]{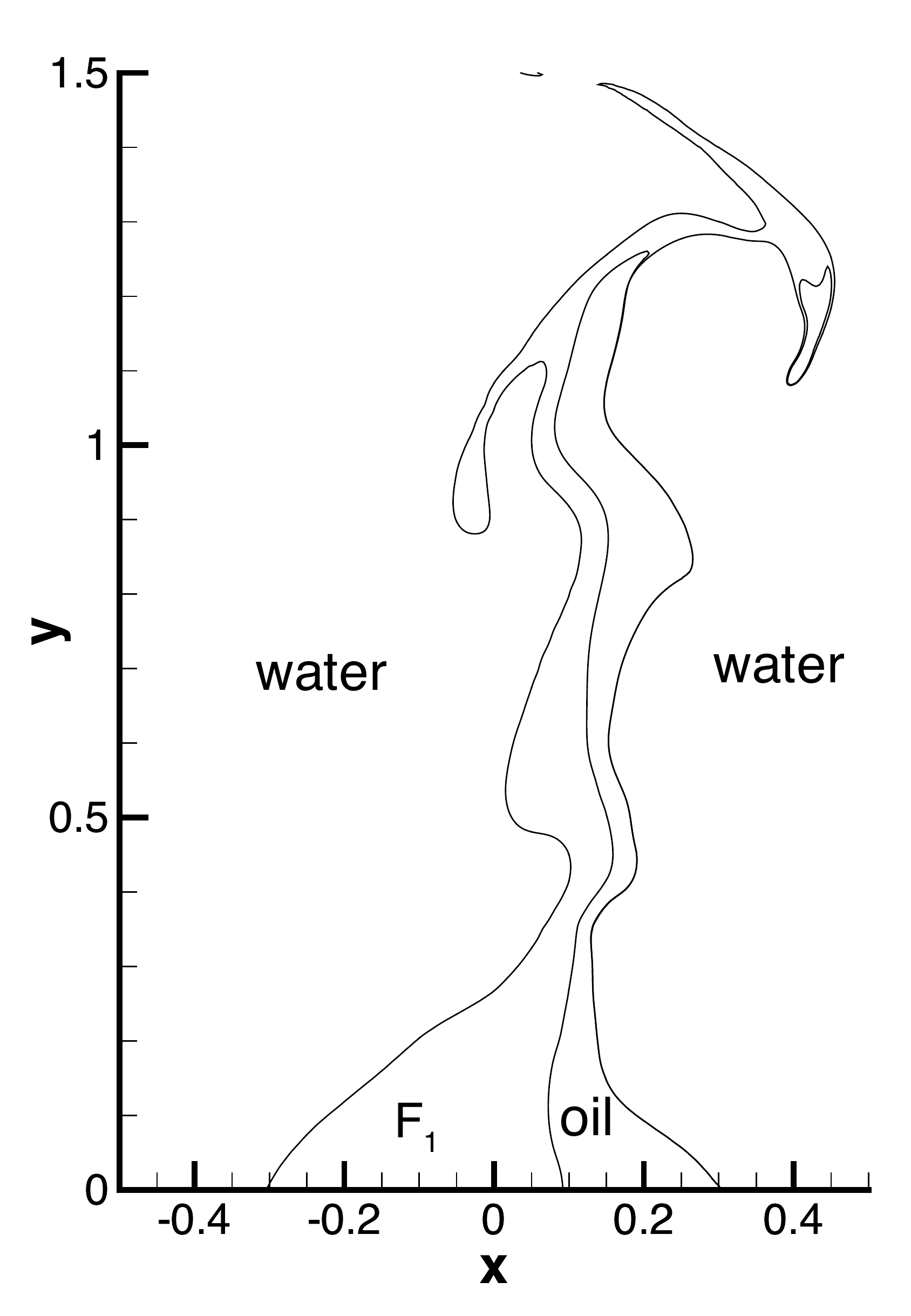}} \hspace*{-10pt}
\subfigure[$t=427.02$]{ \includegraphics[scale=.22]{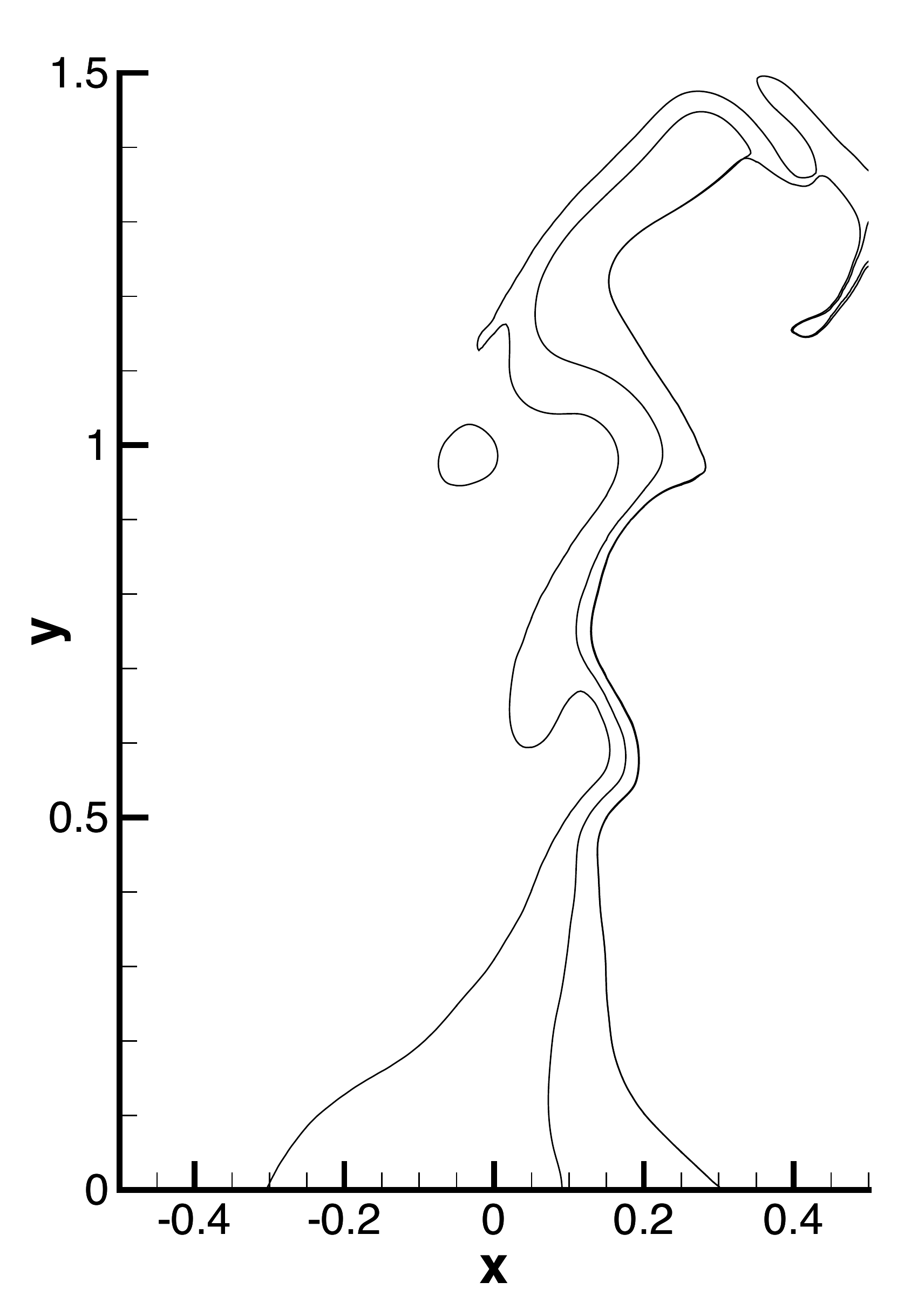}} \hspace*{-10pt}
 \subfigure[$t=427.12$]{ \includegraphics[scale=.22]{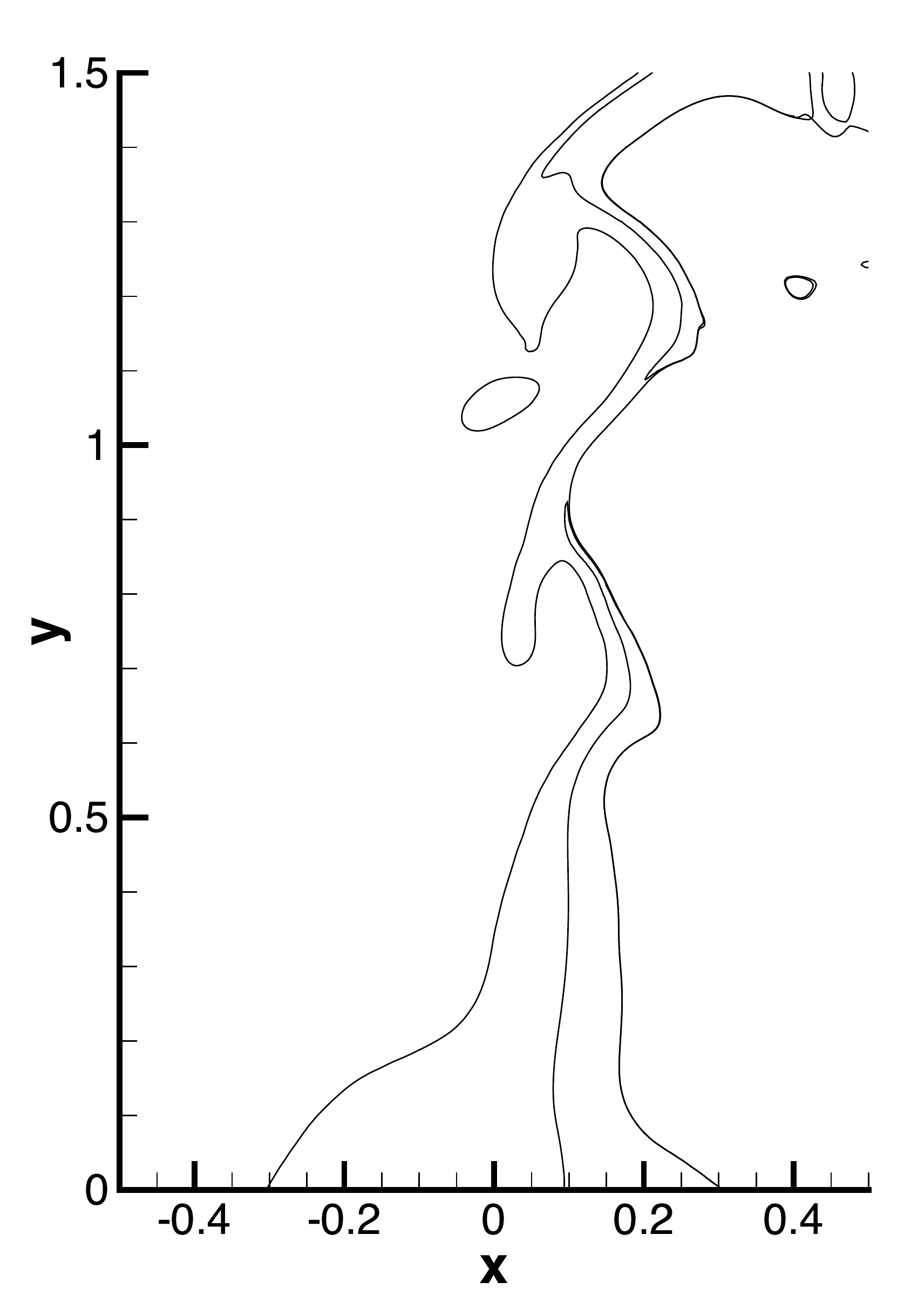}} \hspace*{-10pt}
\subfigure[$t=427.22$]{ \includegraphics[scale=.22]{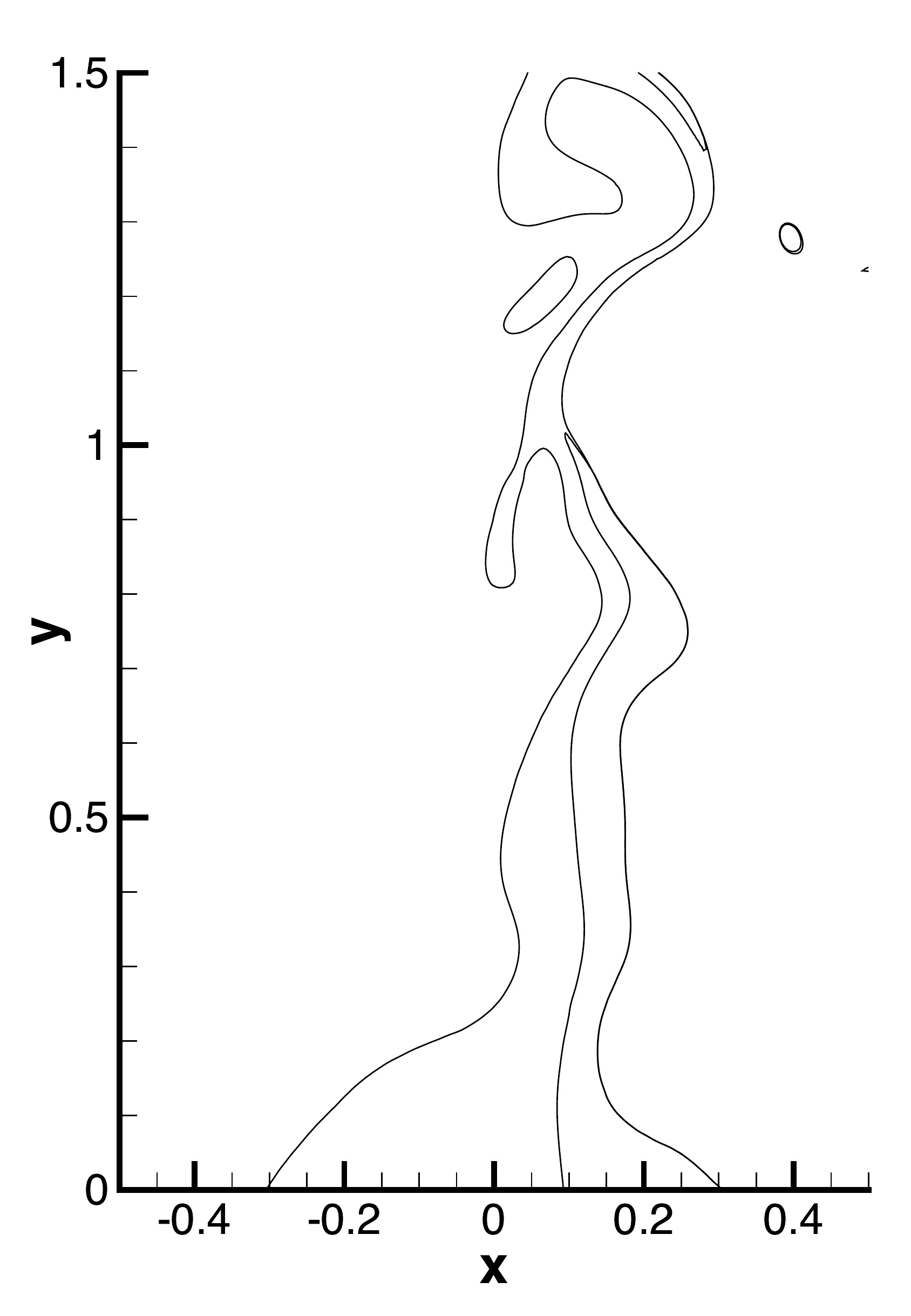}} \\
 \subfigure[$t=427.32$]{ \includegraphics[scale=.22]{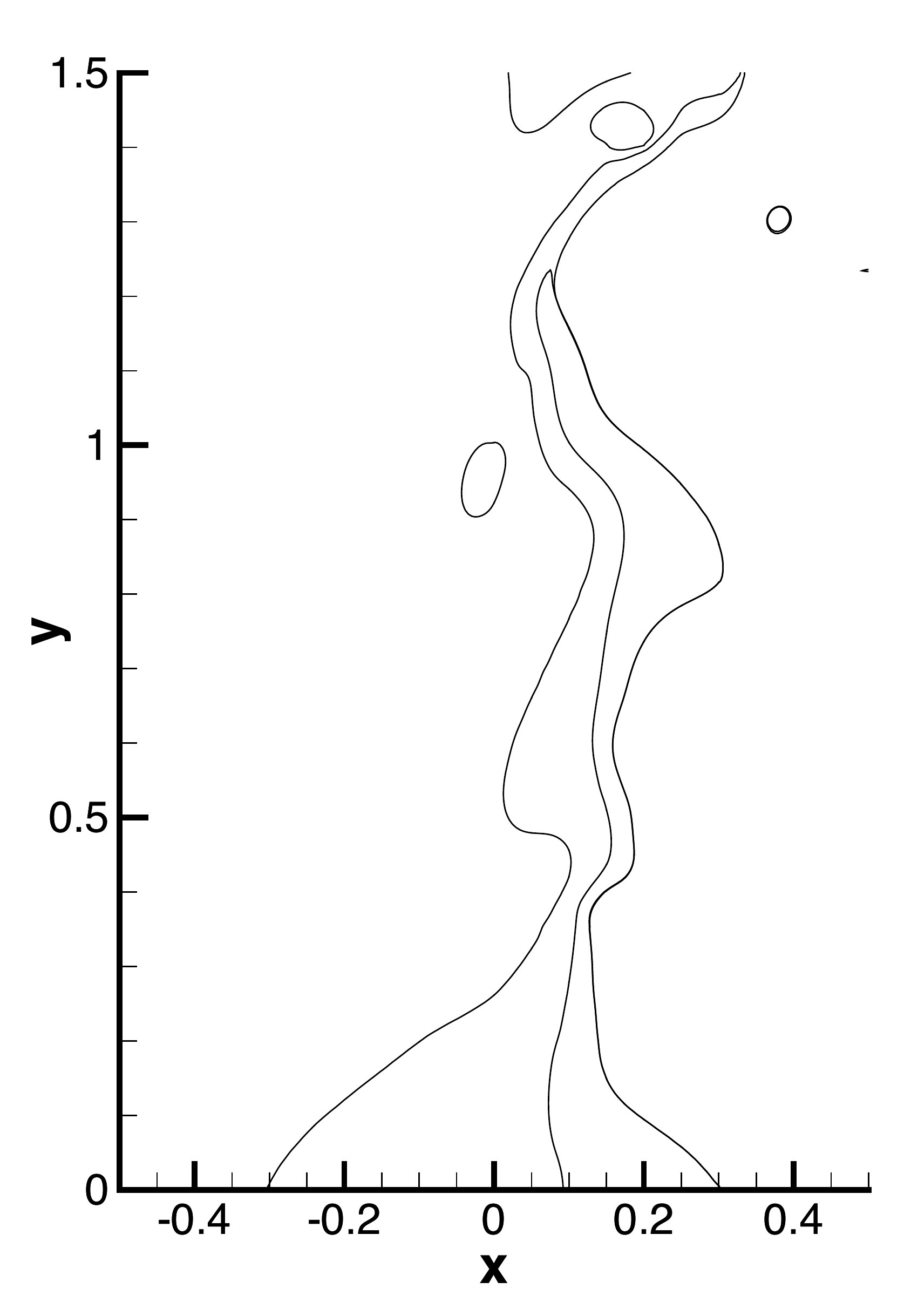}} \hspace*{-10pt}
\subfigure[$t=427.42$]{ \includegraphics[scale=.22]{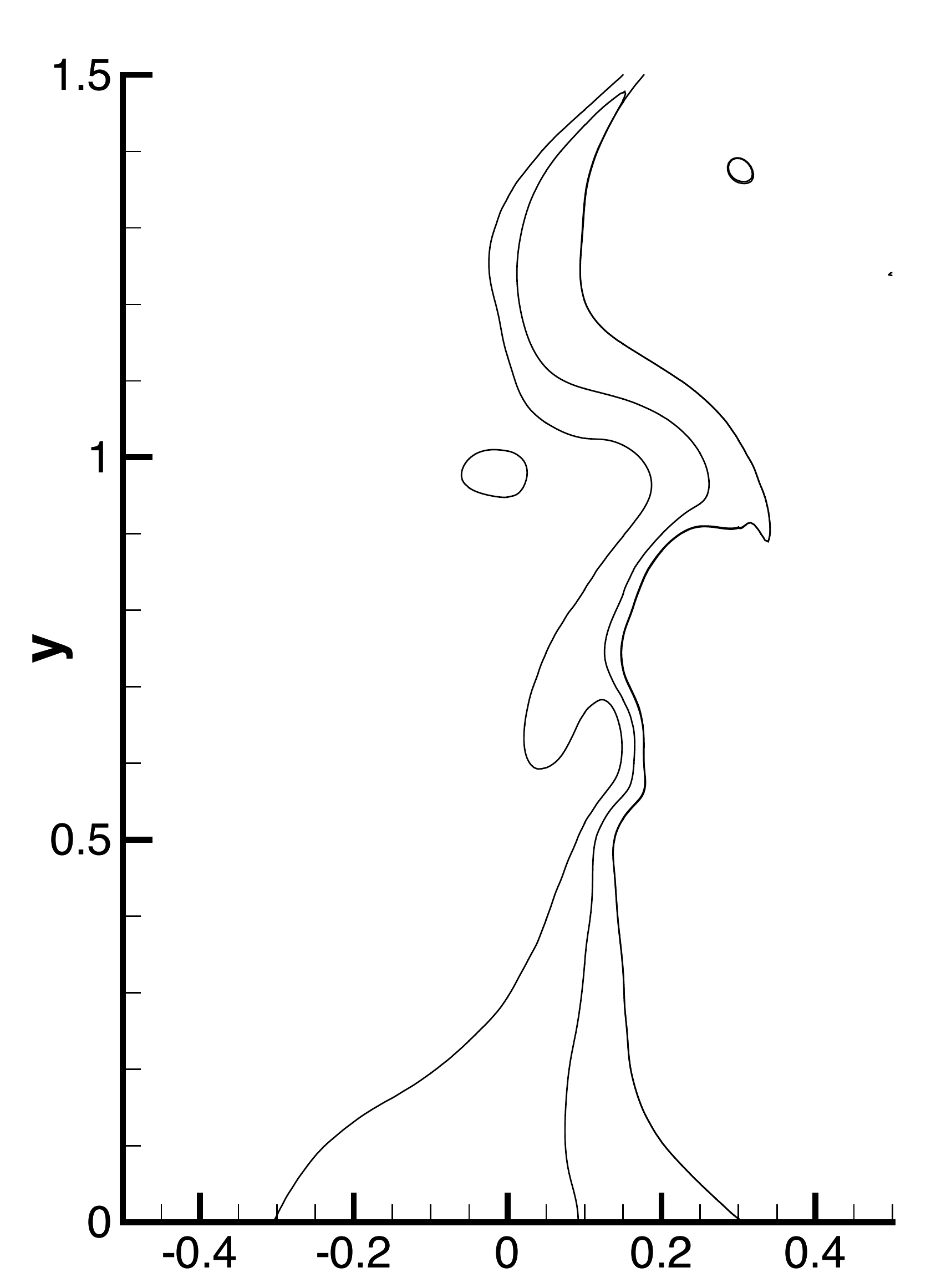}} \hspace*{-10pt}
 \subfigure[$t=427.52$]{ \includegraphics[scale=.22]{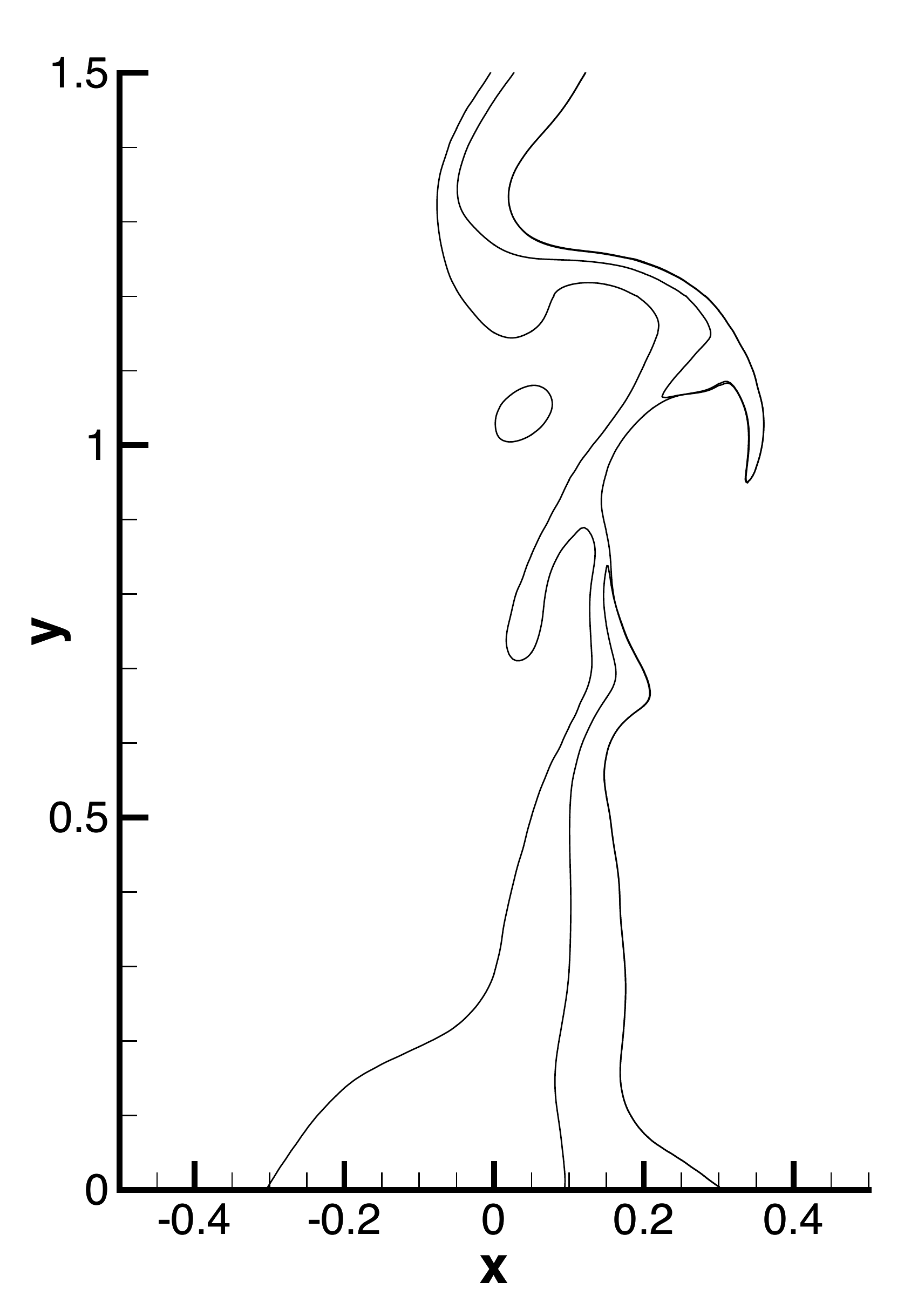}} \hspace*{-10pt}
\subfigure[$t=427.62$]{ \includegraphics[scale=.22]{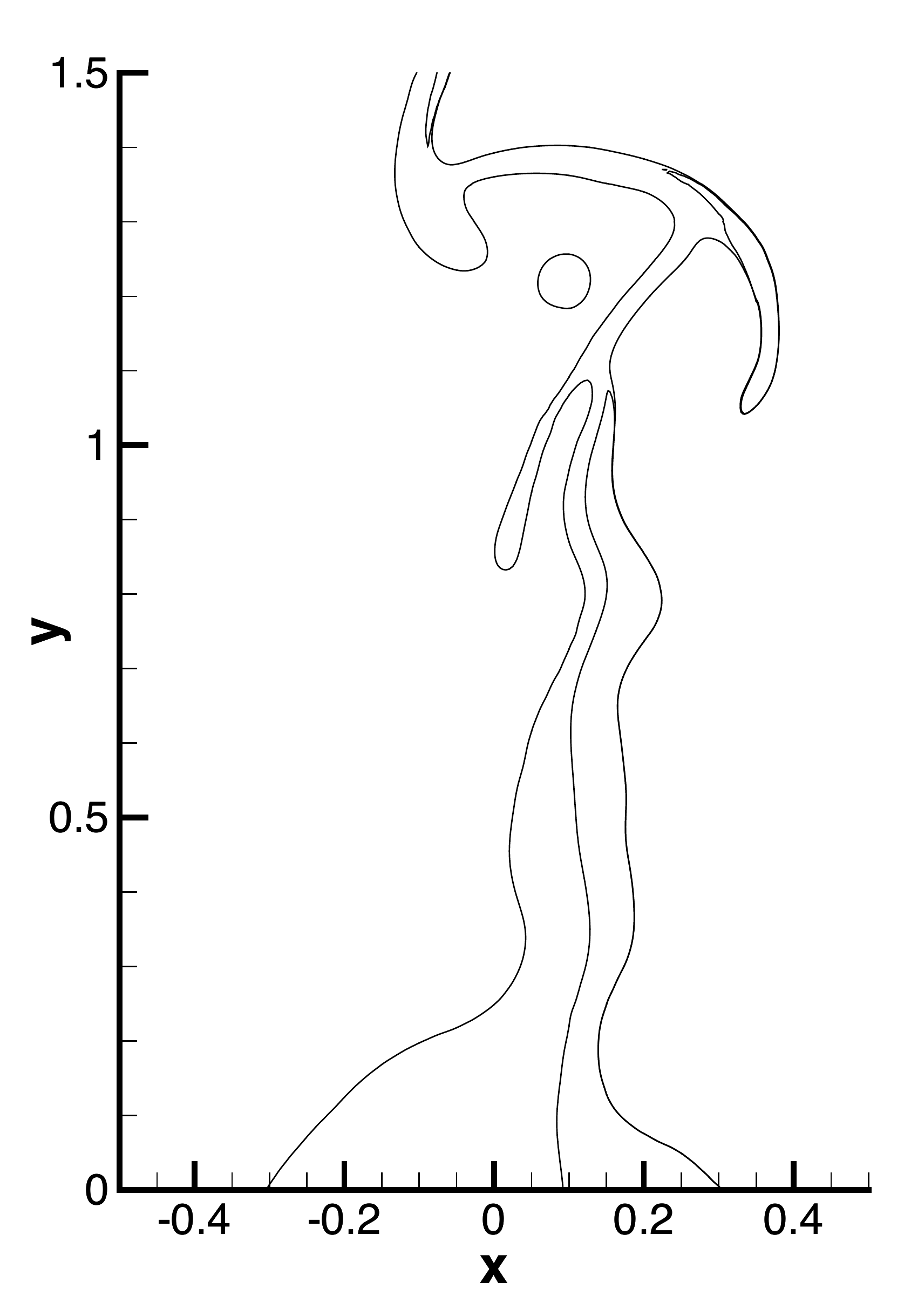}} \\
 \subfigure[$t=427.72$]{ \includegraphics[scale=.22]{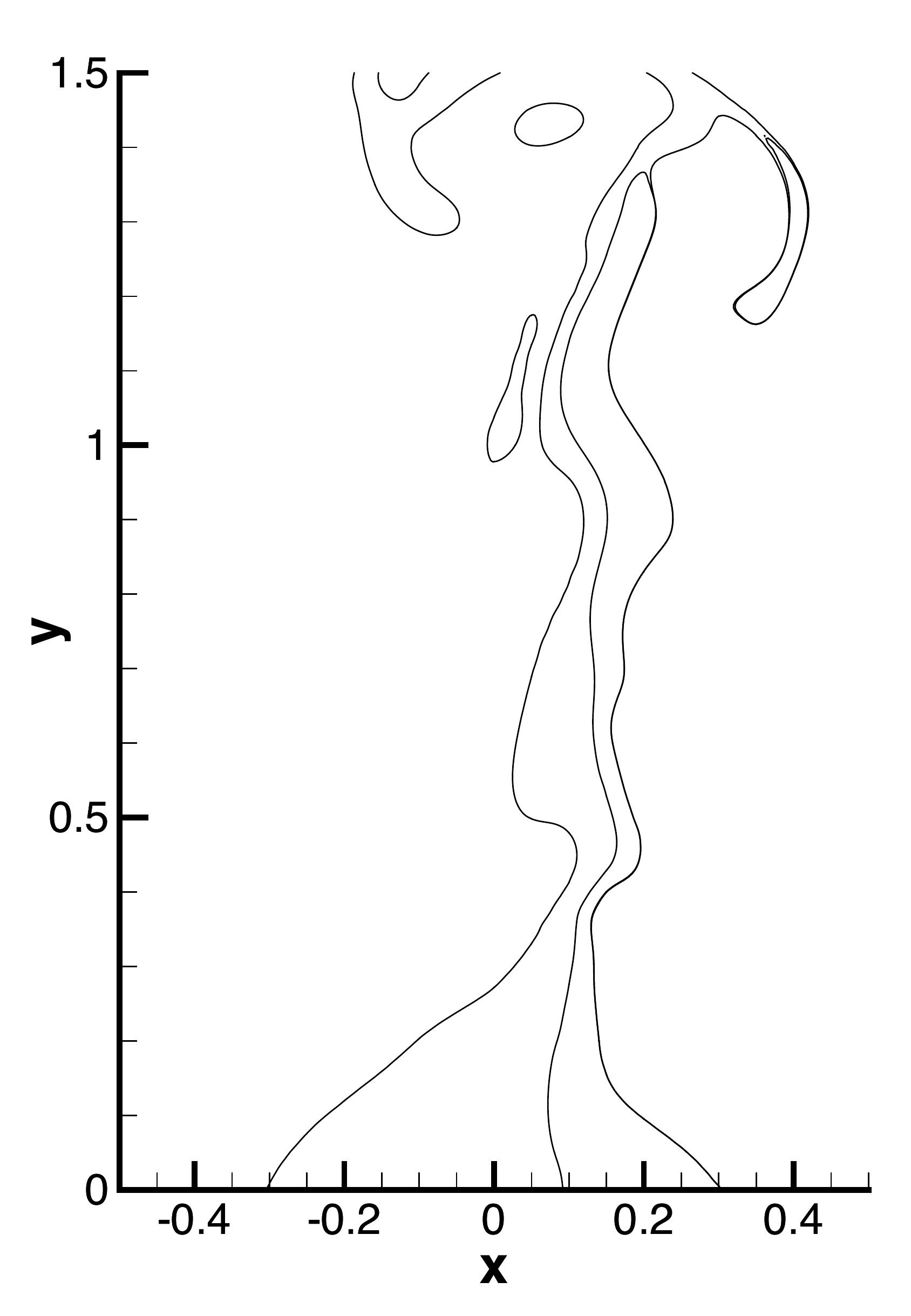}}  \hspace*{-10pt}
\subfigure[$t=427.82$]{ \includegraphics[scale=.22]{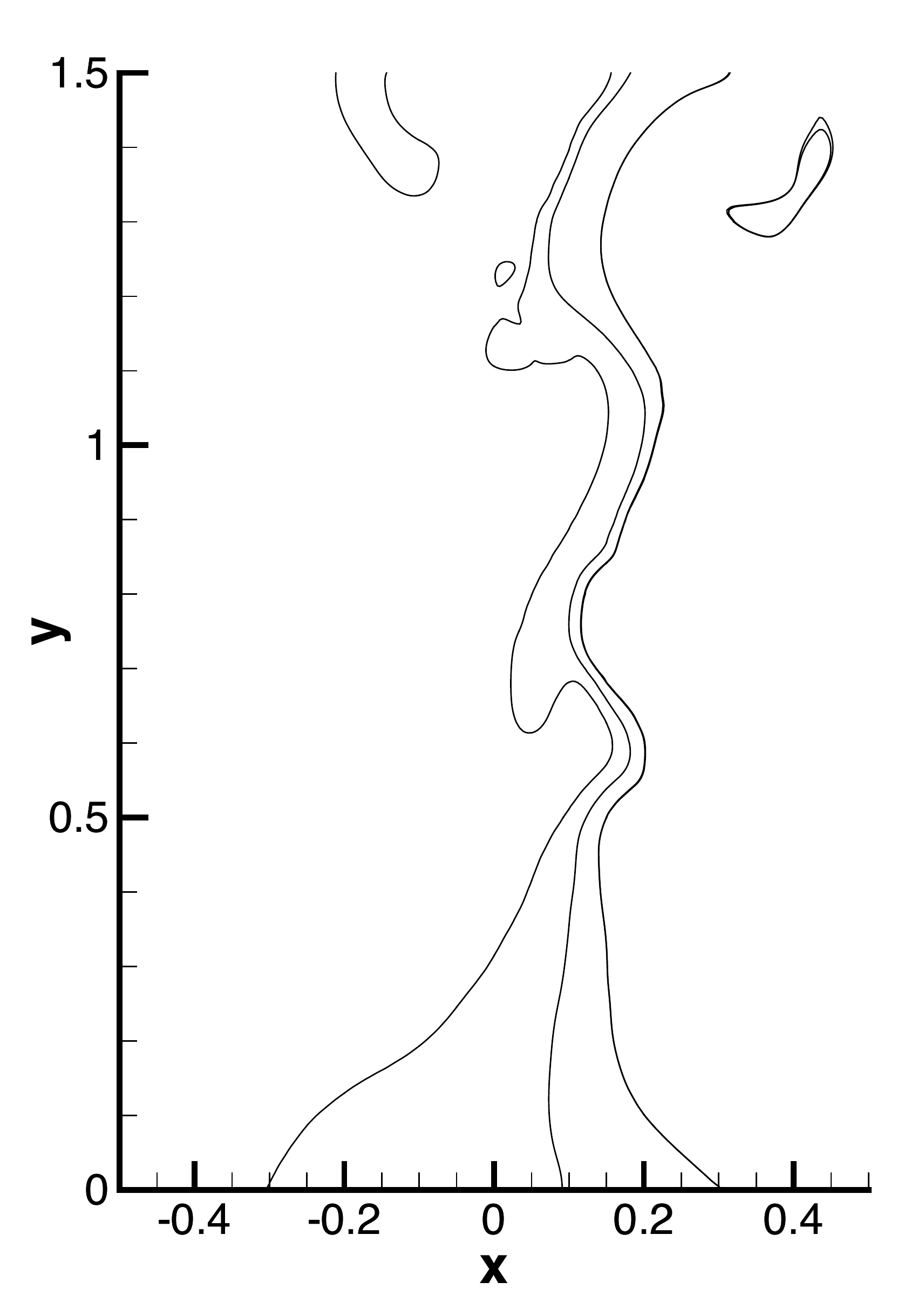}}  \hspace*{-10pt}
 \subfigure[$t=427.92$]{ \includegraphics[scale=.22]{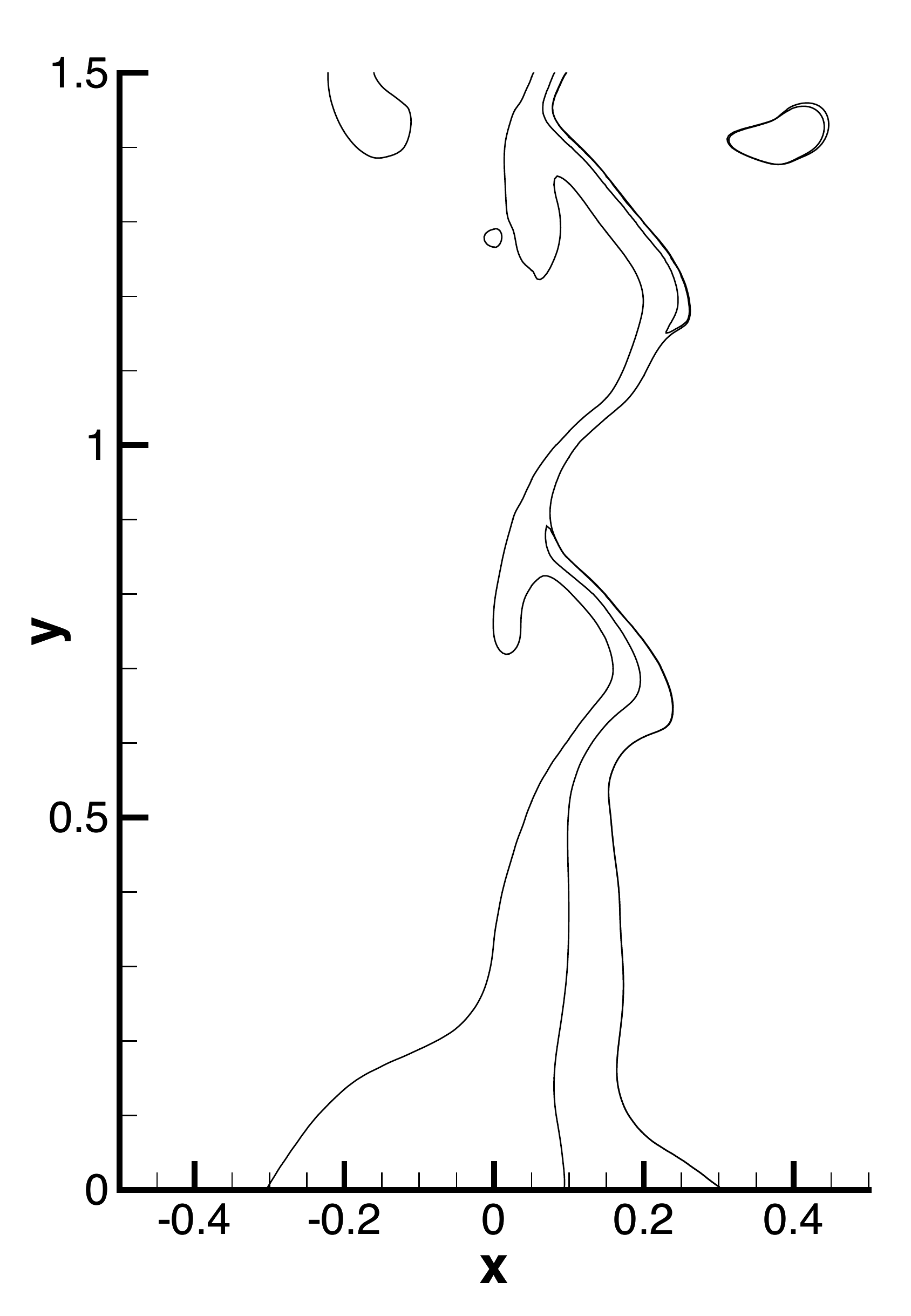}}  \hspace*{-10pt}
\subfigure[$t=428.02$]{ \includegraphics[scale=.22]{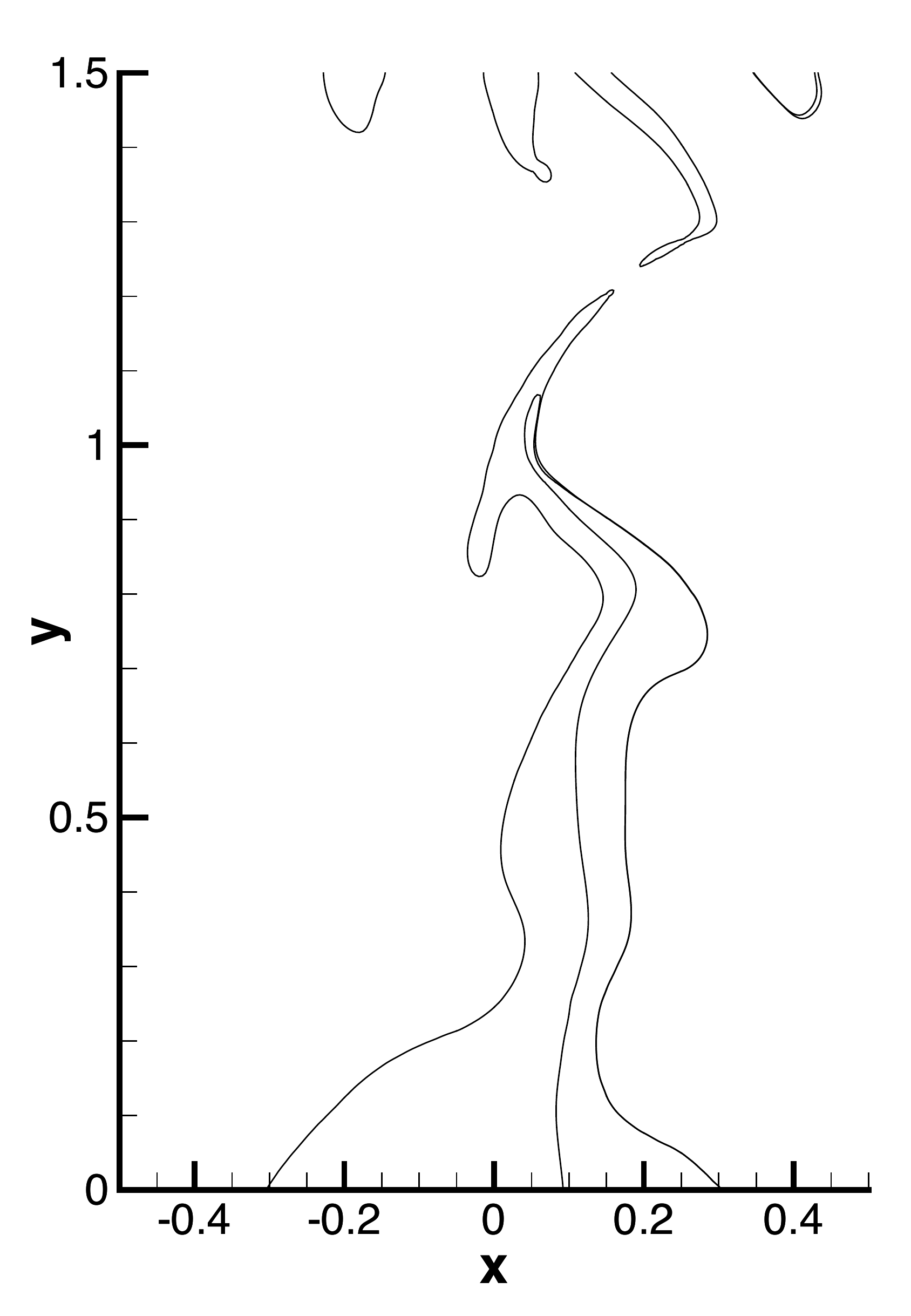}} 
\caption{ Temporal sequence of snapshots of fluid interfaces, visualized by the volume-fraction contours $c_i=1/2\, (i=1,2,3)$,
  showing the interaction of two liquid jets in water,
  with the normalized densities $(\tilde \rho_1,\tilde \rho_2,\tilde \rho_3 )=(1,1.664,0.167)$.
}
\label{oiljetrho2contour}
\end{figure}

\begin{figure}[tbp]
\centering
 \subfigure[$t=426.92$]{ \includegraphics[scale=.22]{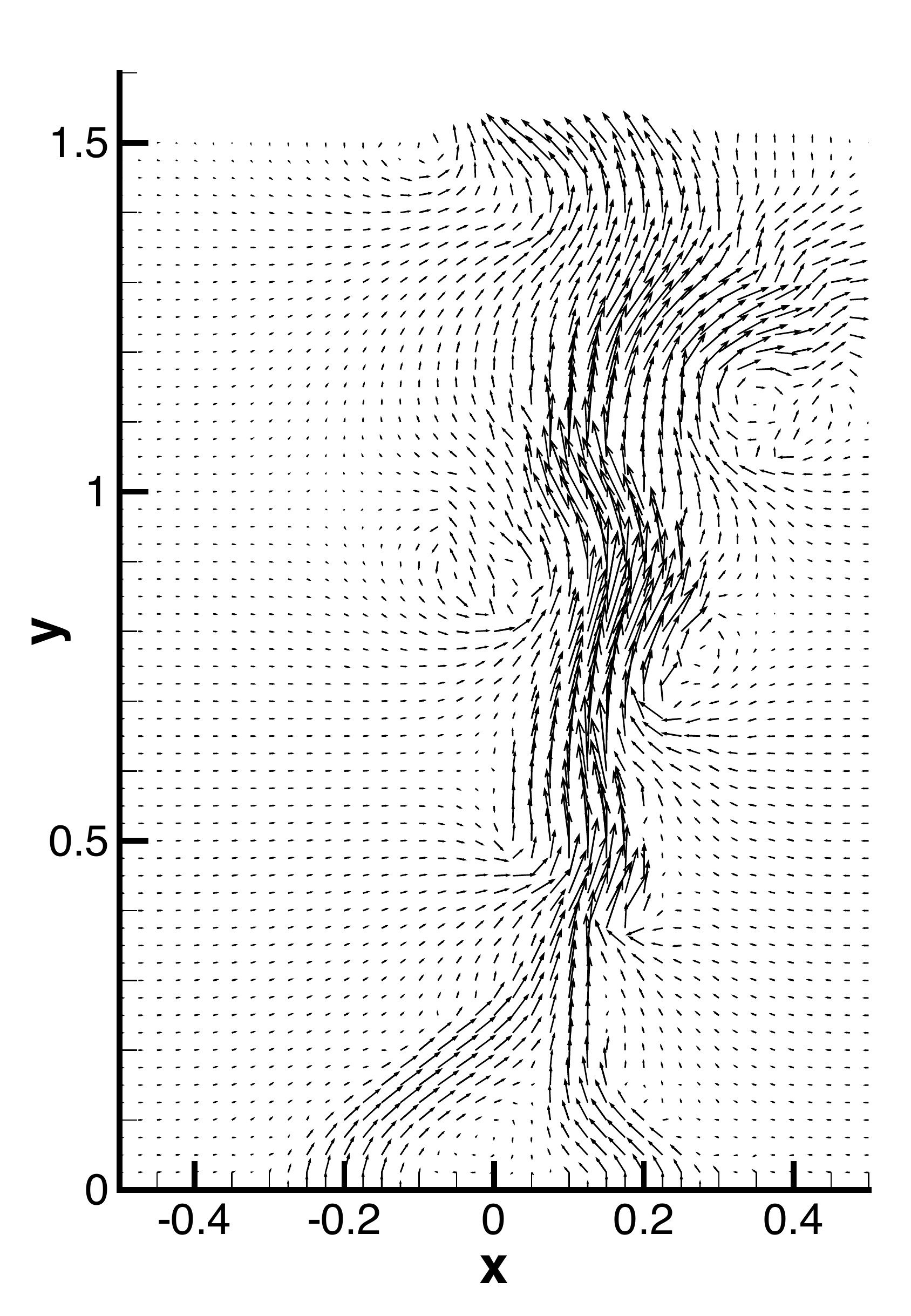}} \hspace*{-10pt}
\subfigure[$t=427.02$]{ \includegraphics[scale=.22]{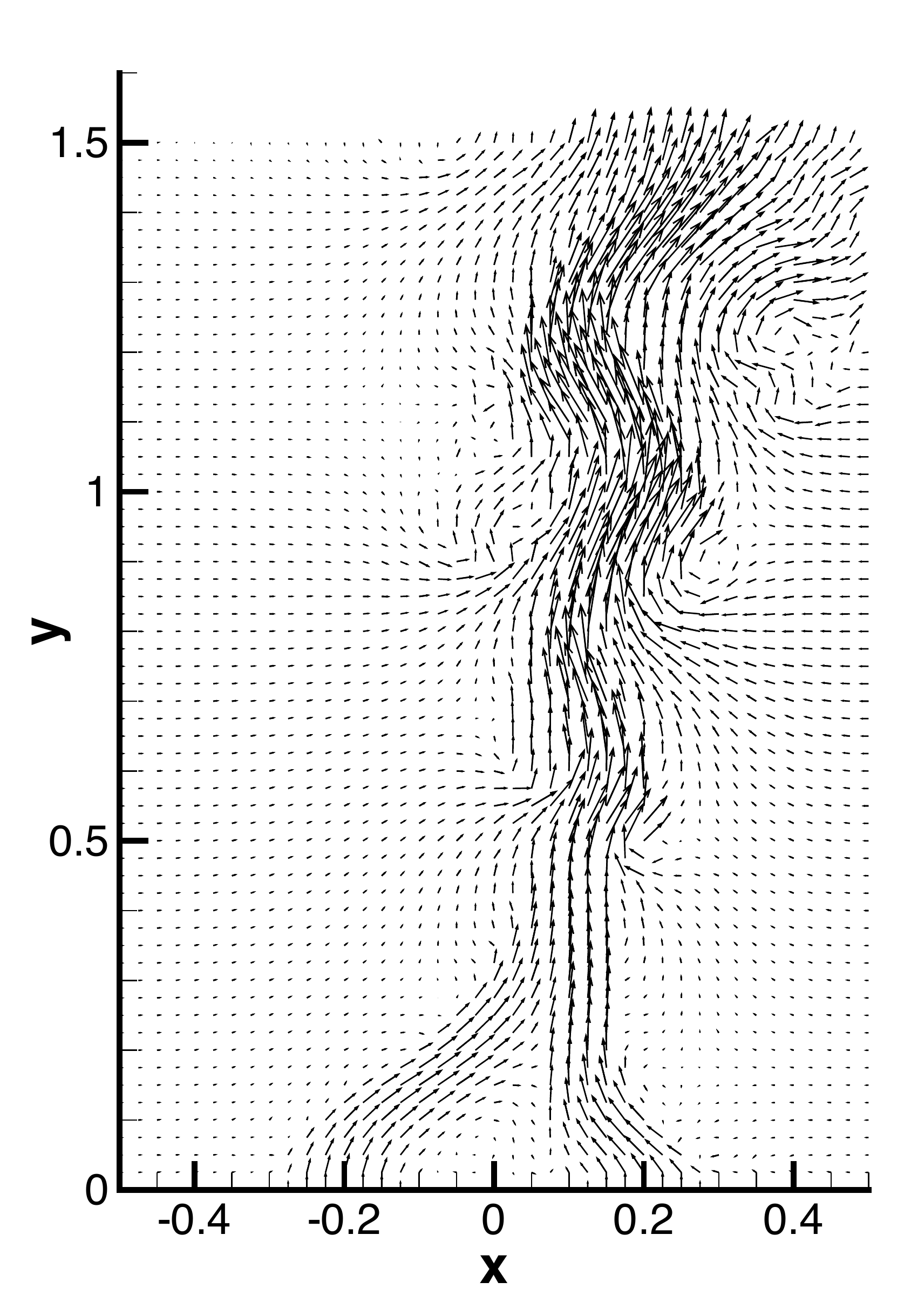}} \hspace*{-10pt}
 \subfigure[$t=427.12$]{ \includegraphics[scale=.22]{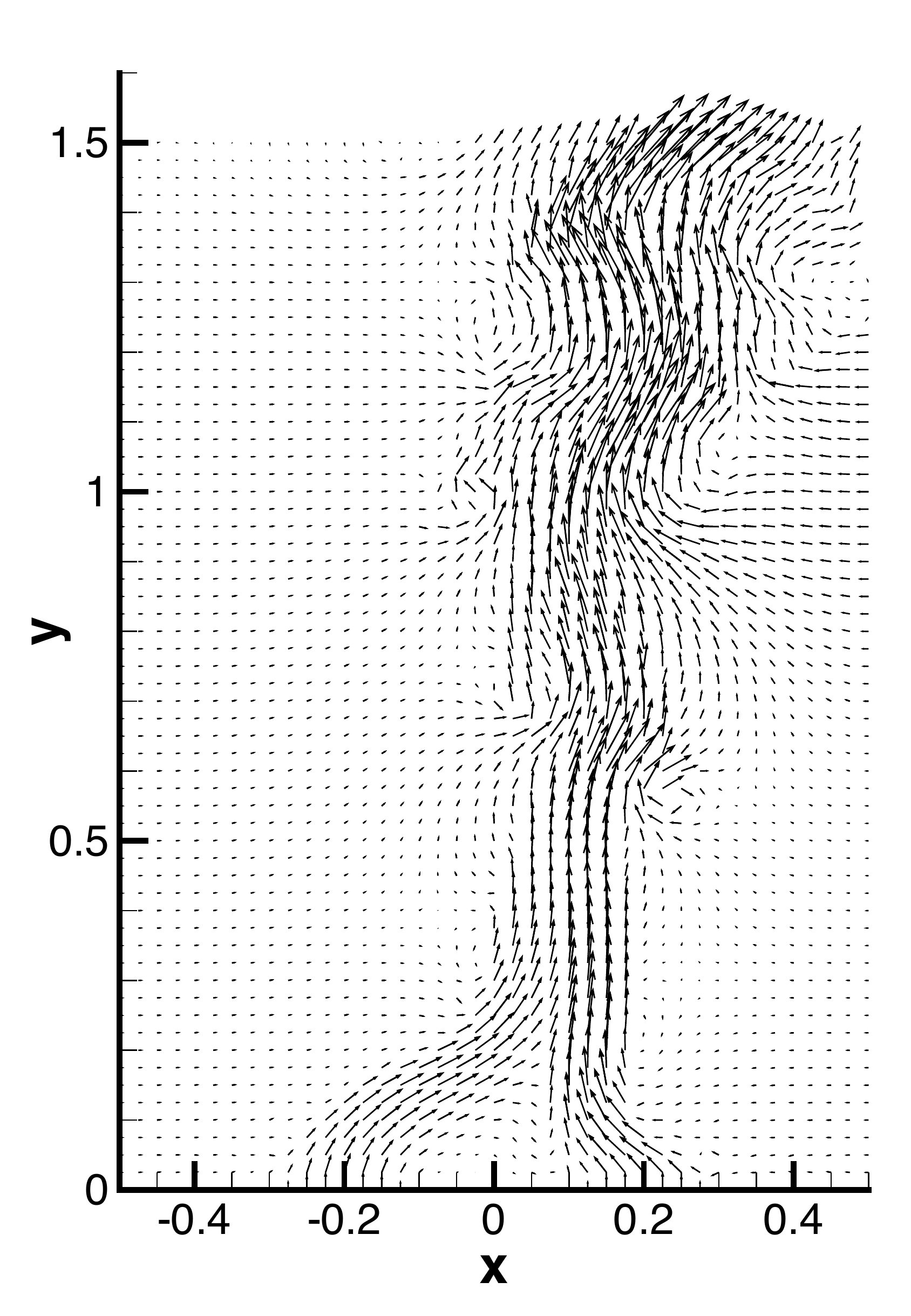}} \hspace*{-10pt}
\subfigure[$t=427.22$]{ \includegraphics[scale=.22]{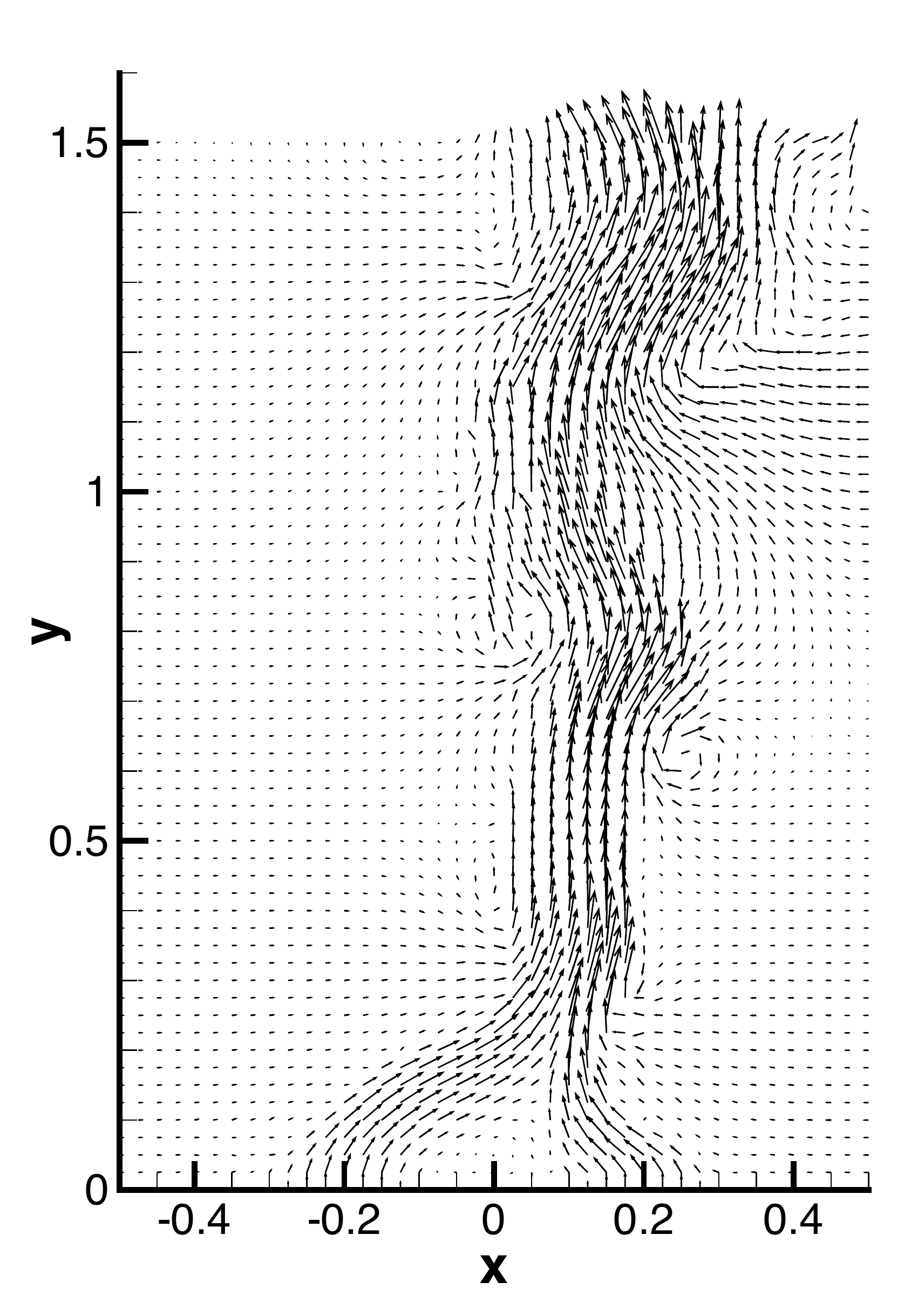}} \\
 \subfigure[$t=427.32$]{ \includegraphics[scale=.22]{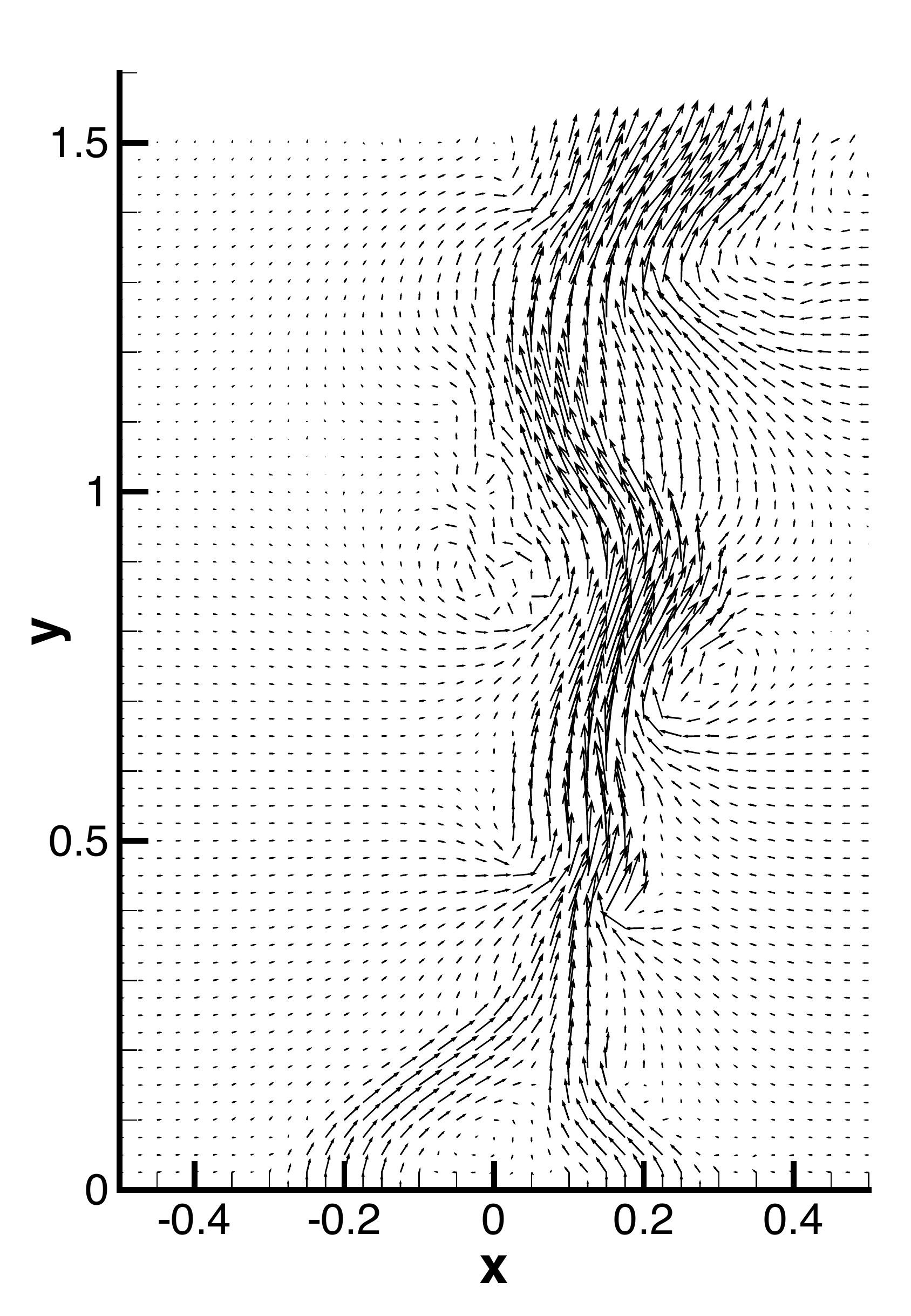}} \hspace*{-10pt}
\subfigure[$t=427.42$]{ \includegraphics[scale=.22]{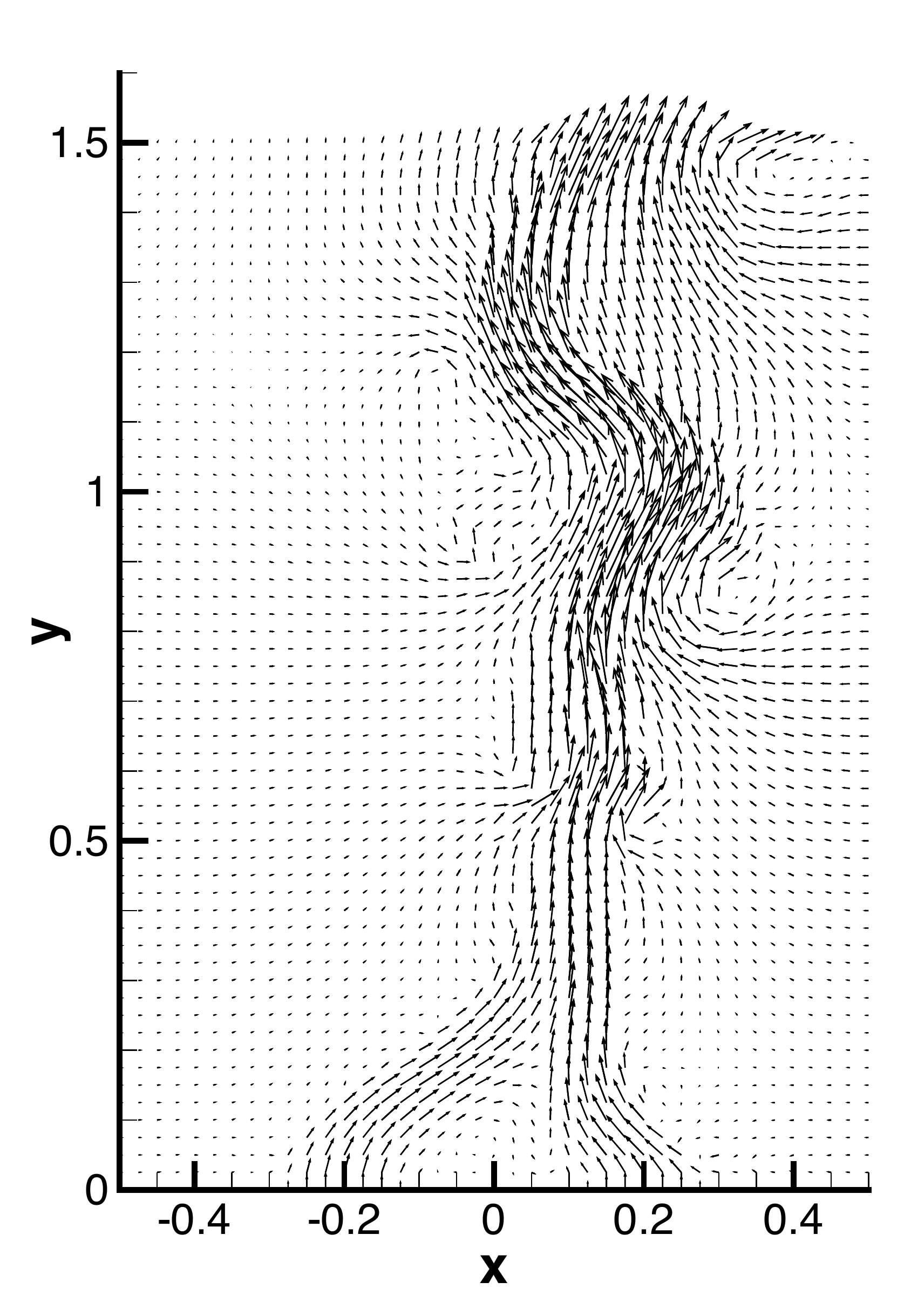}} \hspace*{-10pt}
 \subfigure[$t=427.52$]{ \includegraphics[scale=.22]{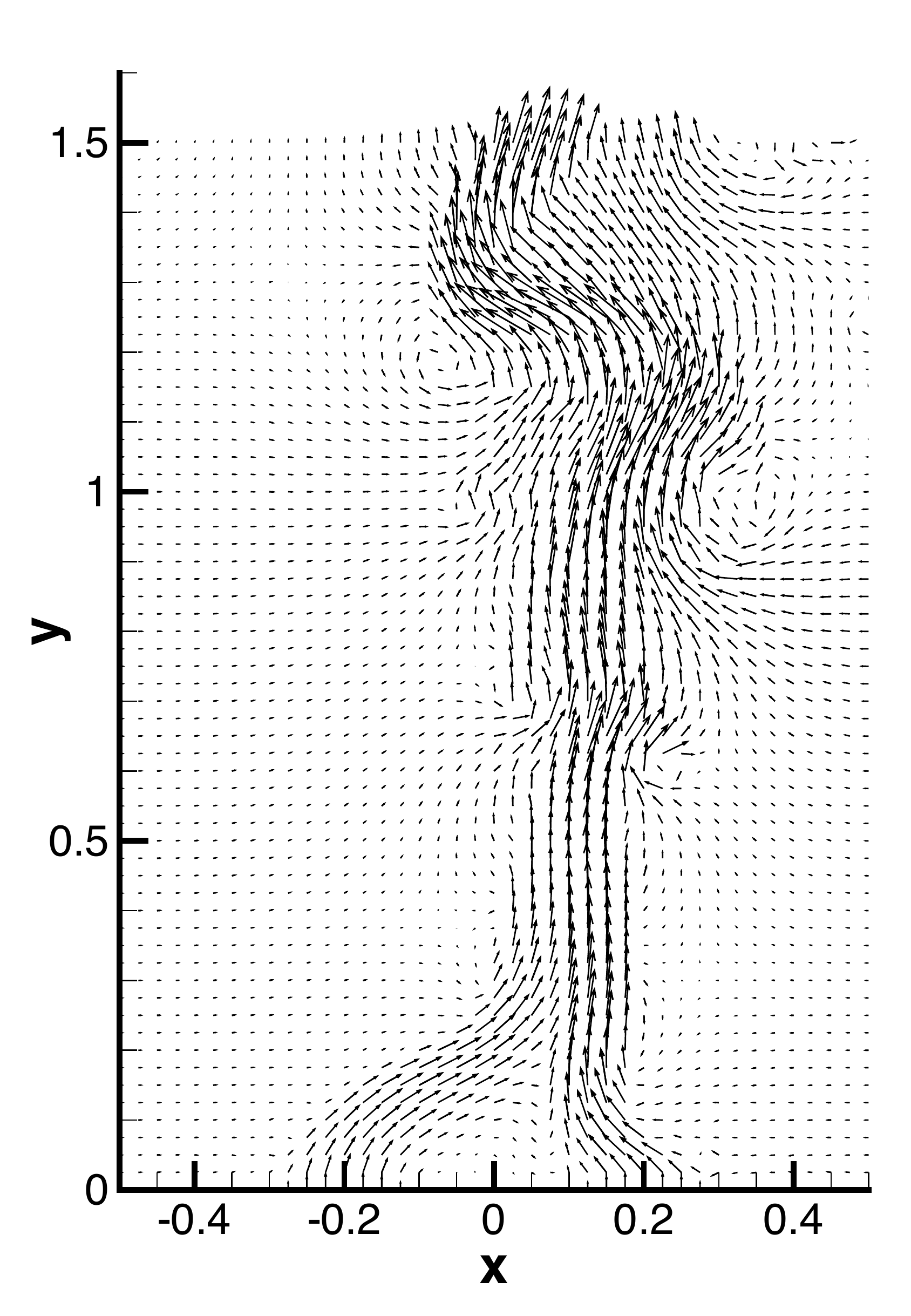}} \hspace*{-10pt}
\subfigure[$t=427.62$]{ \includegraphics[scale=.22]{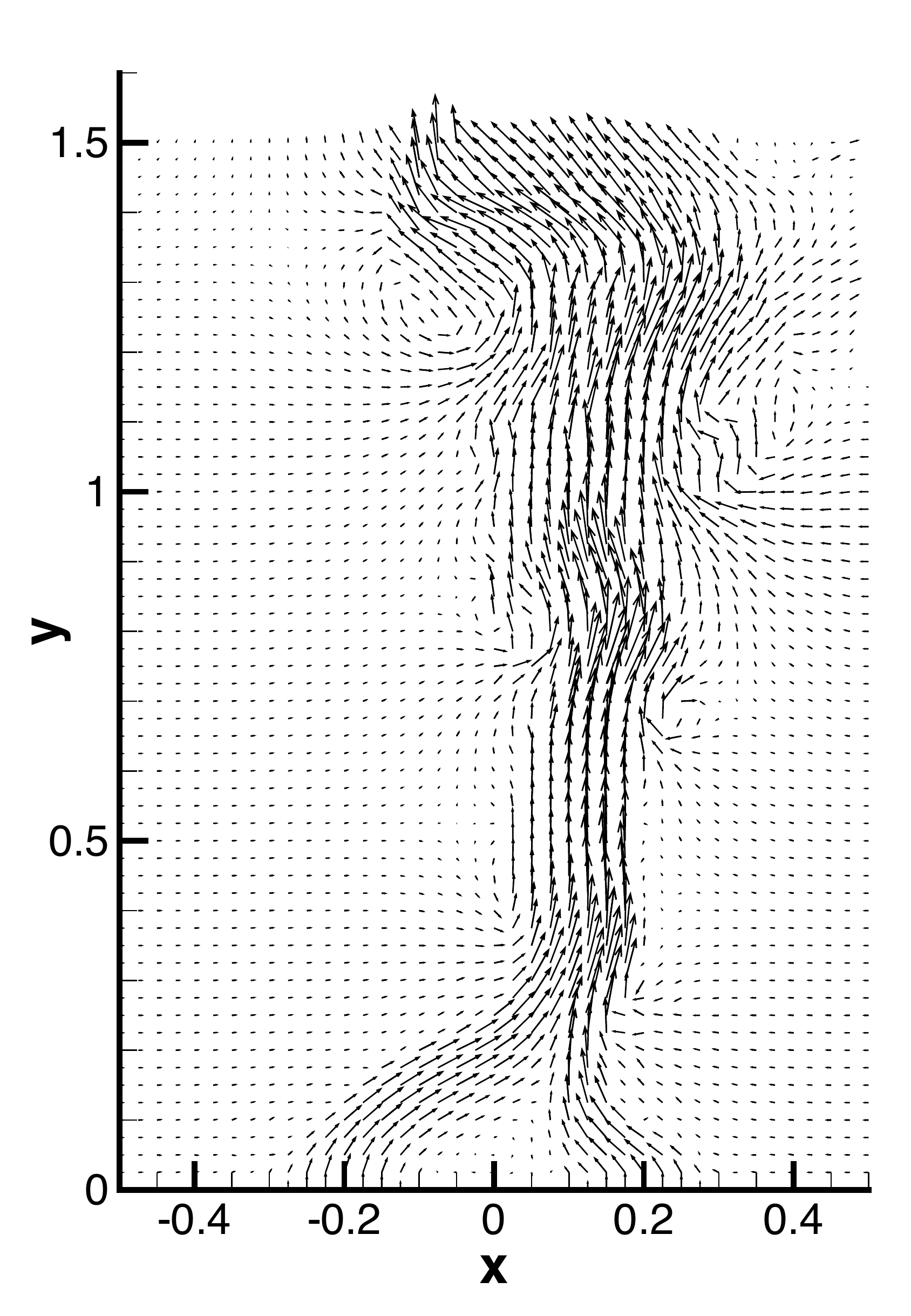}} \\
 \subfigure[$t=427.72$]{ \includegraphics[scale=.22]{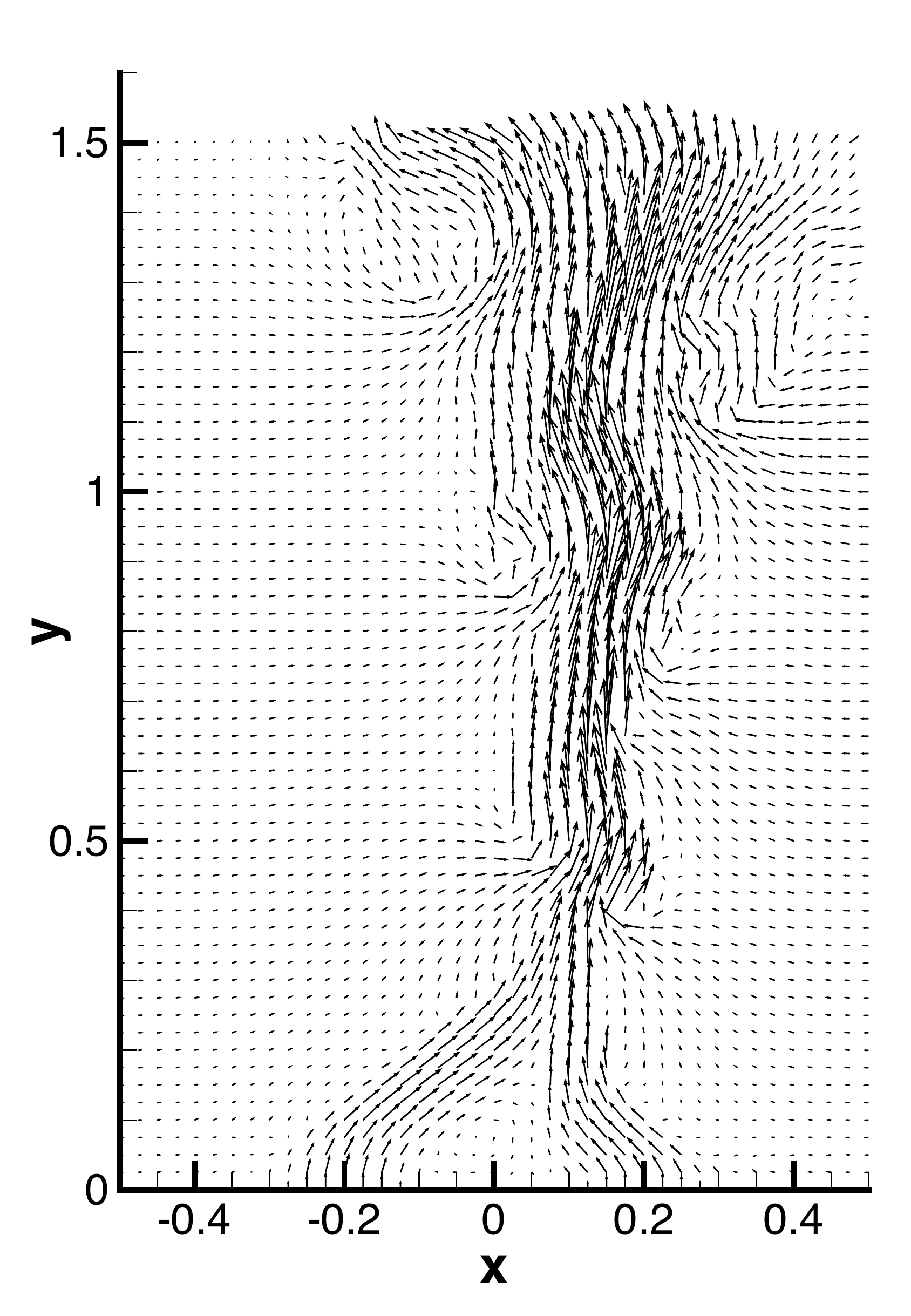}}  \hspace*{-10pt}
\subfigure[$t=427.82$]{ \includegraphics[scale=.22]{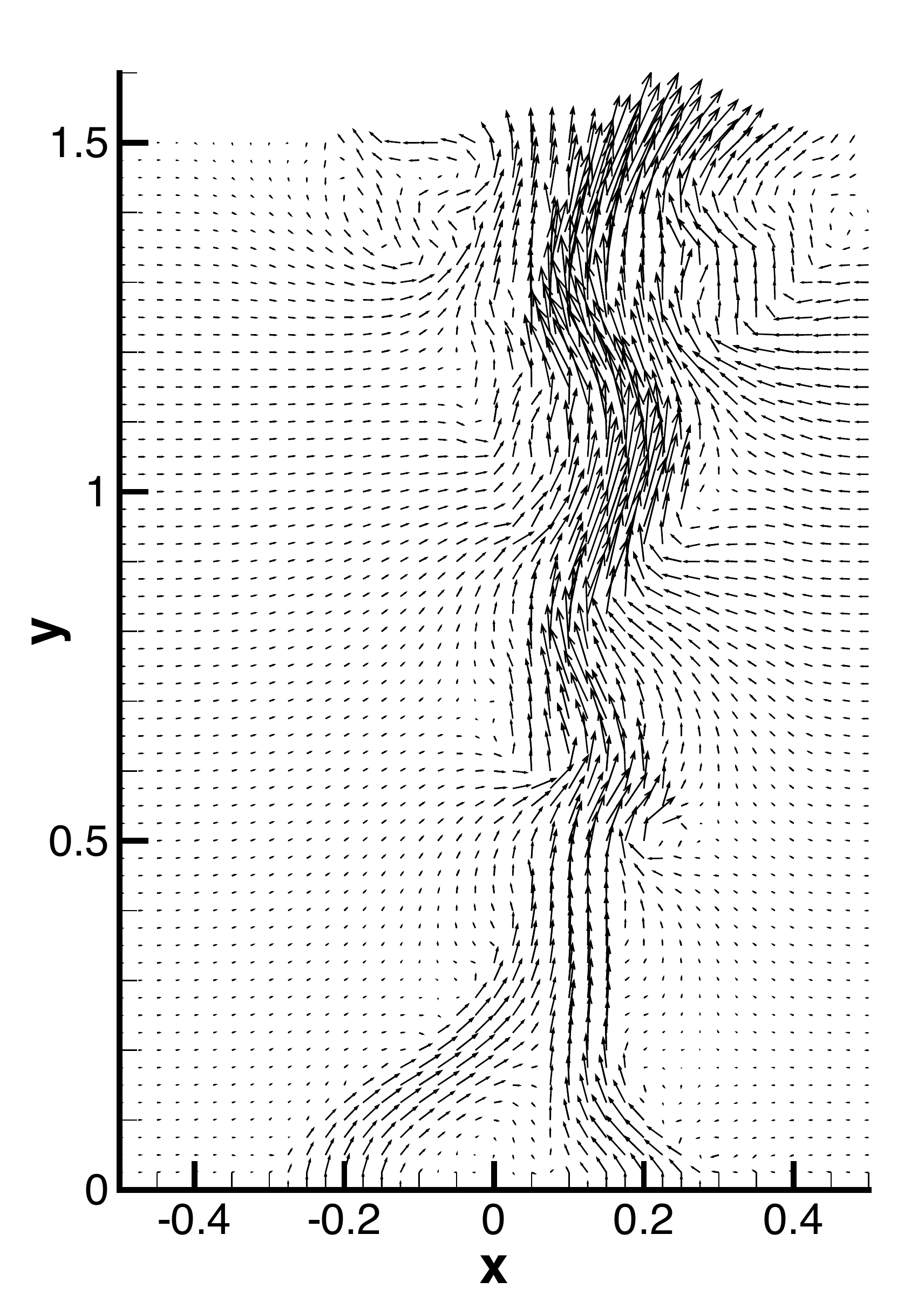}}  \hspace*{-10pt}
 \subfigure[$t=427.92$]{ \includegraphics[scale=.22]{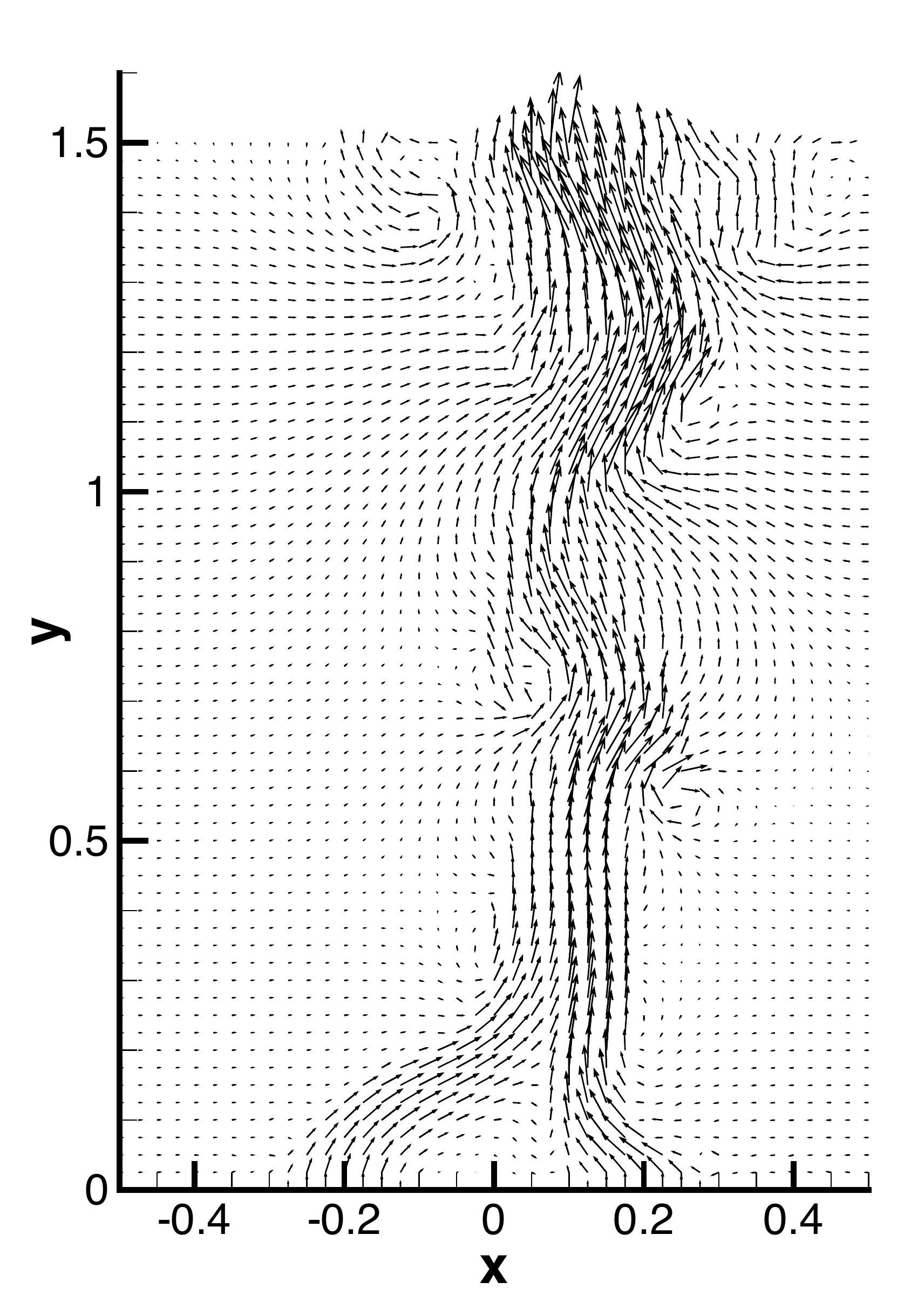}}  \hspace*{-10pt}
\subfigure[$t=428.02$]{ \includegraphics[scale=.22]{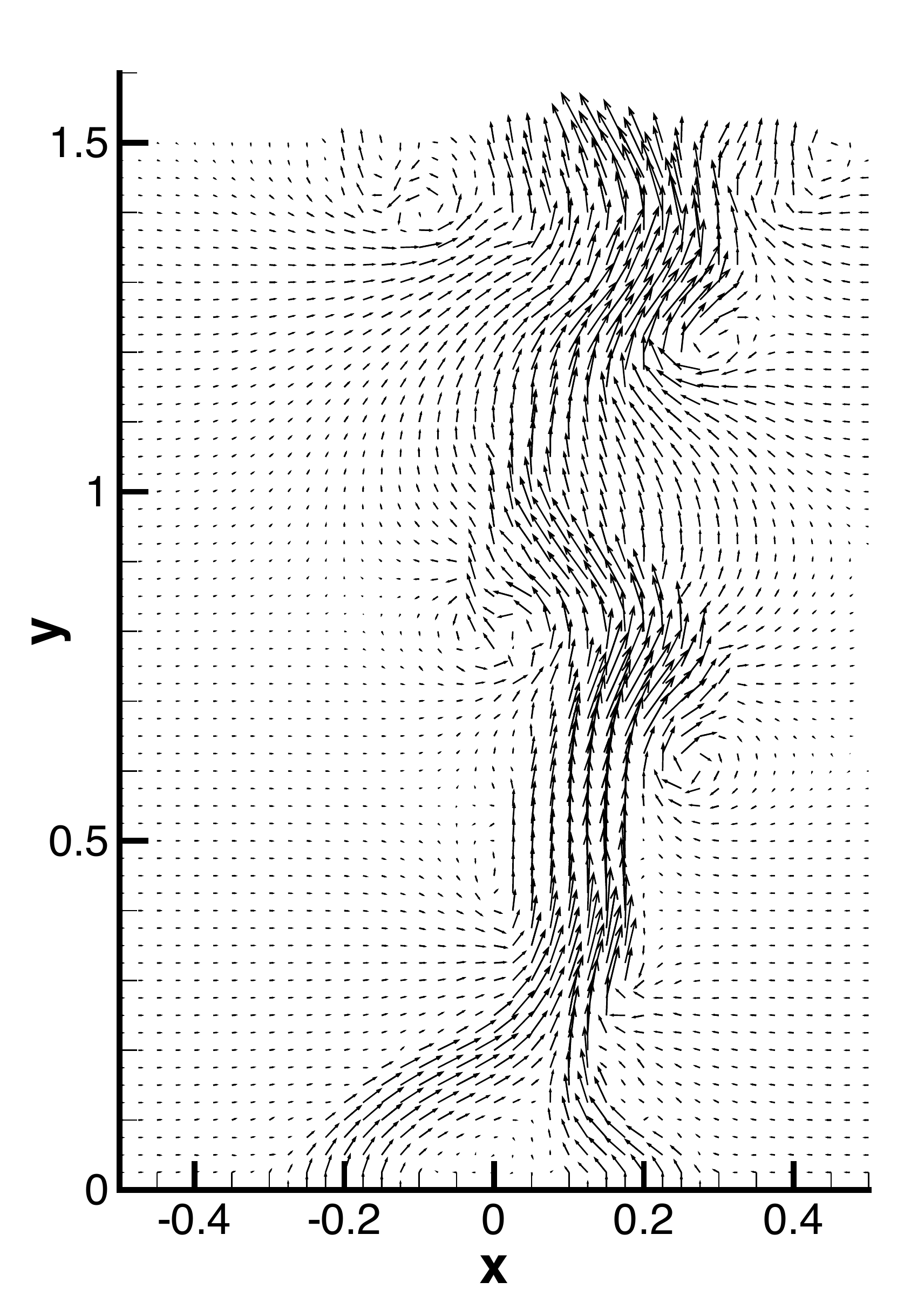}} 
\caption{ Temporal sequence of snapshots of velocity distributions of two liquid jets in
  water problem,
  with normalized densities $(\tilde \rho_1,\tilde \rho_2, \tilde \rho_3 )=(1,1.664,0.167).$
  Velocity vectors are plotted on every eighth quadrature points in each direction within each element.
}
\label{oiljetrho2vector}
\end{figure}

Let us next consider the second case, with an oil density
$100kg/m^3$. 
The normalized densities for $F_1$, water and oil are
$(\tilde \rho_1, \tilde \rho_2, \tilde \rho_3)=(1,1.664, 0.1664)$.
All the other physical parameters are the same as in the first case.
Long-time simulations have been performed for this case, and
Fig.~\ref{velositymagrho2} shows time histories of the maximum and average
velocity magnitudes defined in \eqref{equ:velmagnitude}, indicating that
the flow has reached a statistically stationary state.
Figs.~\ref{oiljetrho2contour} and \ref{oiljetrho2vector}
are the temporal sequence of snapshots of the fluid interfaces and
the velocity fields corresponding to this case.
The general characteristics of the dynamics of jets and
the velocity distributions are similar to those of
the first case. But some marked differences can be noticed.
The compound oil-F$_1$ jet becomes notably more unstable because
of the stronger buoyancy force in the oil region.
We observe
a smaller region ($y/L\gtrsim 0.3$) with a relatively stable jet profile
near the base of the jet.
Downstream of this region, the deformation of the jet profiles
is much more pronounced than in the first case,
and droplets of the oil and F$_1$ fluid are observed
to break off from the compound jet.
The velocity field in the region occupied by the compound oil-F$_1$ jet
appear stronger and more violent compared with that of the first case.
Vortices and backflows can also be observed at the upper or side
boundaries at times; see Fig.~\ref{oiljetrho2vector}(a)-(d).
The results indicate that with the proposed method the fluid interfaces
and the flow structures appear to be able to pass through
the open/outflow boundaries smoothly and seamlessly.

\section{Concluding Remarks}
\label{sec:summary}


We have developed a set of effective outflow/open boundary
conditions (and also inflow boundary conditions) for
simulating multiphase flows consisting of
$N$ ($N\geqslant 2$) immiscible incompressible fluids
in domains involving outflow and inflow boundaries.
These boundary conditions are designed to satisfy
two properties: energy stability and reduction
consistency. The proposed boundary conditions ensure that,
at the continuum level,
their contributions to the N-phase energy balance
will not cause the total system energy to increase over time,
regardless of the flow state at the outflow/open boundary. 
In other words, this property holds even in situations
where strong vortices or backflows occur at the outflow/open
boundary.
This is the reason why the proposed boundary conditions are
effective in overcoming the backflow instability in
N-phase flow problems.
The reduction consistency of the boundary conditions is
a physical consistency requirement for N-phase formulations~\cite{Dong2017}.
This property means that the boundary conditions
honor the inherent equivalence relations
between N-phase systems and the resultant smaller multiphase systems
when some fluid components were absent from the N-phase
system.

We have also presented an efficient numerical algorithm for
 the proposed outflow/inflow boundary
conditions together with the N-phase governing equations.
The main issue lies in the numerical treatments of
the inertia term in the open boundary conditions
for the phase field equations and the variable viscosity
in the open boundary condition for the momentum equation.
With appropriate reformulations and treatments of such terms
in our algorithm, the computations for different flow variables
and the computations for different phase field variables
have been completely de-coupled. The proposed
algorithm involves only the solution of a number of
Helmholtz-type equations within each time step. 
The linear algebraic systems resulting from discretizations involve
only constant and time-independent coefficient matrices,
which can be pre-computed,
even though large density contrasts and large viscosity contrasts
may be present in the N-phase system.
These characteristics make the algorithm computationally very
efficient and attractive.

We have tested the proposed method with extensive
numerical experiments for several problems involving multiple
fluid components and in domains with outflow and inflow boundaries.
In particular, we have compared in detail our simulation results
for the three-phase capillary wave problem
with Prosperetti's exact physical solution~\cite{Prosperetti1981}
under various physical and simulation parameters.
These comparisons demonstrate that the proposed method
produces physically accurate results.


Multiphase flows involving inflow/outflow boundaries are
an important class of problems, which have widespread applications
in oil/gas industries, carbon sequestration, microfluidics
and optofluidics~\cite{HuppertN2014,PsaltisQY2006,RodriguezSMG2015}. 
These problems are also critical to the study of long-time behaviors and
statistical features of multiphase flows.
The key technique for simulating multiphase inflow/outflow
problems lies in how to deal with the multiphase
outflow/open boundaries.
The method developed in the current work provides an effective
and powerful tool for simulating this class of problems.
We anticipate that it will be useful and instrumental in
the investigation of long-time statistics of multiphase problems
and in the development a number of related areas.


While the outflow/open boundary conditions proposed here
ensure the energy stability of the N-phase system at
the continuum level, this property is not guaranteed
by the numerical algorithm presented here at the discrete level.
The current algorithm is only conditionally stable,
and requires sufficient spatial resolution and small enough
time step size to achieve stable and accurate simulations.
An interesting question is how to devise an algorithm
for these outflow/open boundary conditions together with
the N-phase governing equations to guarantee the energy
stability at the discrete level.
This problem seems to be highly non-trivial.
It would be an interesting problem to contemplate
for future research.


\section*{Acknowledgement}
This work was partially supported by
NSF (DMS-1318820, DMS-1522537).

\bibliographystyle{plain}
\bibliography{nphase,obc,mypub,nse,sem,contact_line,interface,multiphase}

\end{document}